\documentclass{article}

\PassOptionsToPackage{authoryear,round}{natbib}
\usepackage[eandd, final]{neurips_2026}

\usepackage{amsmath,amsfonts,bm}

\def\eqref#1{equation~\ref{#1}}

\def\1{\bm{1}}

\DeclareMathAlphabet{\mathsfit}{\encodingdefault}{\sfdefault}{m}{sl}
\SetMathAlphabet{\mathsfit}{bold}{\encodingdefault}{\sfdefault}{bx}{n}
\newcommand{\tens}[1]{\bm{\mathsfit{#1}}}

\def\tB{{\tens{B}}}

\def\tF{{\tens{F}}}

\def\tU{{\tens{U}}}

\usepackage[utf8]{inputenc} 
\usepackage[T1]{fontenc}    
\usepackage{hyperref}       
\usepackage{url}            
\usepackage{booktabs}       
\usepackage{amsfonts}       
\usepackage{nicefrac}       
\usepackage{microtype}      
\usepackage[table,dvipsnames]{xcolor}         
\usepackage{amsmath}    
\usepackage{graphicx}
\usepackage{pdflscape}
\usepackage{pifont}
\usepackage{siunitx}
\usepackage{verbatim}
\usepackage{tikz}
\usepackage{cancel}
\usepackage{float}
\usepackage{tabularx}
\usepackage{array}
\usepackage{makecell}
\usepackage{multirow}
\usepackage{setspace}
\usepackage[T1]{fontenc}
\usepackage[scaled=0.95]{inconsolata}
\usepackage{algorithm}
\usepackage{algpseudocode}
\usepackage{amsmath}
\usepackage{amssymb}

\usepackage{minitoc}

\usetikzlibrary {arrows.meta}
\definecolor{bluelow}{HTML}{64B5F6}
\definecolor{bluehigh}{RGB}{239,243,255}
\newcommand{\cw}[2]{\cellcolor{bluelow!#1!bluehigh}#2}
\definecolor{sdorangemin}{HTML}{FCE6CF}
\definecolor{sdorangemax}{HTML}{F28E2B}
\newcommand{\sdcw}[2]{\cellcolor{sdorangemin!\number\numexpr100-#1\relax!sdorangemax}#2}
\definecolor{sabwgreenmin}{HTML}{EAF6E3}
\definecolor{sabwgreenmax}{HTML}{59A14F}
\newcommand{\sabw}[2]{\cellcolor{sabwgreenmin!\number\numexpr100-#1\relax!sabwgreenmax}#2}
\definecolor{ssoopurplemin}{HTML}{C5B0D5}
\definecolor{ssoopurplemax}{HTML}{9467BD}
\newcommand{\ssoo}[2]{\cellcolor{ssoopurplemin!\number\numexpr100-#1\relax!ssoopurplemax}#2}
\title{Neptuna: A Comprehensive Machine Learning Framework for Benchmarking Complex Multiphase Flows}

\author{
  Harish Ramachandran \\
  Technical University of Munich\\
  \texttt{harish.ramachandran@tum.de} \\
  \And
  Björn Kimpel \\
  Technical University of Munich \\
  \texttt{bjoern.kimpel@tum.de} \\
  \And
  Thomas Paula \\
  Technical University of Munich \\
  \texttt{thomas.paula@tum.de} \\
  \And
  Josef Winter \\
  Technical University of Munich \\
  \texttt{josef.winter@tum.de} \\
  \And
  Steffen Schmidt \\
  Technical University of Munich \\
  \texttt{steffen.schmidt@tum.de} \\
  \And
  Nikolaus Adams \\
  Technical University of Munich \\
  \texttt{nikolaus.adams@tum.de} \\
}

\begin{document}

\doparttoc 
\faketableofcontents 
\part{}

\maketitle

\begin{abstract}
Compressible multiphase flows involving shocks and material interfaces arise in applications such as bubble collapse and droplet breakup, where strong nonlinear interactions produce complex interface deformation, mixing, and multiscale dynamics. Developing reliable machine learning surrogates for these flows remains challenging due to the simultaneous presence of compressibility, sharp discontinuities, and multiphase effects. In this work, we introduce the first large-scale benchmark specifically designed for shock-driven compressible multiphase flows, comprising 3 TB of high-fidelity 2D and 3D datasets featuring shock-induced bubble collapse and droplet breakup. We evaluate diverse surrogate model families on our benchmarking framework: Neptuna, including convolutional, spectral, transformer-based, and pre-trained PDE foundation models. Beyond standard MSE training, we investigate composite losses that combine MSE with Sobolev, interface-aware, and structure-aware terms, along with adaptive loss balancing using SoftAdapt and GradNorm. Evaluation includes pointwise, spectral, feature-focused, structural, and physics-informed metrics. Results show that no single model performs best across all datasets and metrics, while composite losses significantly improve interface preservation and spectral fidelity. Among adaptive weighting strategies, SoftAdapt provides the most consistent improvements with almost no overhead compared to MSE-only training. The benchmarking repository and the dataset are made available at \url{https://tumaer.github.io/Neptuna/}
\end{abstract}

\section{Introduction}


Compressible multiphase flows involving shock waves and material interfaces arise in a wide range of natural phenomena and engineering applications. Strong nonlinear interactions between shocks and phase boundaries lead to complex dynamics such as interface deformation, vorticity generation, mixing, and topological changes. 


Shock-induced air bubble collapse in water represents an important regime of compressible multiphase flows. In gas–liquid configurations, such as air bubbles in mechanical equilibrium with water, the large acoustic impedance mismatch leads to strong compression and energy focusing during collapse, often producing high-pressure regions and jet formation. In contrast, gas–gas configurations, such as heavy gas (e.g. R22) bubbles in air, isolate compressibility-driven effects and provide a setup to analyze shock refraction, vorticity evolution, and mixing without the influence of surface tension or cavitation dynamics. Such phenomena have numerous real-world applications, including bubble curtains for protecting submerged structures \citep{wursig2000development} and shockwave lithotripsy \citep{leighton2012acoustic}.

Another important class of problems is shock-induced droplet breakup in air. In this case, the interaction between the shock and a liquid droplet leads to a range of deformation and breakup mechanisms governed by the competition between inertial and surface tension forces. Depending on the Weber number, the droplet may undergo different breakup modes, such as Rayleigh Taylor Piercing (RTP) and Shear Induced Entrainment (SIE), each associated with a distinct instability mechanism \citep{theofanous2008physics}. These processes are highly relevant in applications such as fuel atomization, spray dynamics, and high-speed combustion \citep{lefebvre2017atomization}.

Despite significant progress in Computational Fluid Dynamics (CFD) solvers, simulating compressible multiphase flows with shocks remains computationally expensive due to the need for high spatial and temporal resolution to accurately capture the flow physics. These challenges have led to increasing interest in machine learning-based surrogate models for faster approximations. While several existing dataset repositories capture important aspects of compressible flow physics \citep{takamoto2022pdebench, herde2024poseidon} or multiphase dynamics \citep{hassan2023bubbleml,hassan2025bubbleformer}, they do not jointly address compressibility, shock interactions, and multiphase effects in a comprehensive benchmark setting. To the best of our knowledge, this work provides the first benchmark datasets specifically designed for compressible multiphase flows with shocks, thereby extending the complexity of flow regimes in existing publically available datasets.


We provide a \textbf{3.0 TB collection of high-fidelity 2D and 3D datasets} and study reliable surrogate modeling for highly nonlinear, multiscale, and regime-dependent flows. We evaluate SOTA baselines across convolutional, spectral, and attention-based architectures, including ConvNeXt (\textbf{CNeXt}) \citep{liu2022convnet}, \textbf{CNO} \citep{raonic2023convolutional}, \textbf{FFNO} \citep{tran2021factorized}, and \textbf{ScOT} \citep{herde2024poseidon}, and fine-tune two pre-trained models: \textbf{Poseidon} \citep{herde2024poseidon} and \textbf{DPOT} \citep{hao2024dpot}.

Beyond evaluating a broad spectrum of baselines, we train each model under two configurations: one using only Mean Squared Error (MSE), and another using a composite loss that augments MSE with additional terms designed to capture important flow characteristics. Specifically, we include gradient-based penalties via the H$^1$-seminorm \citep{SobolevTrainingWithHigherOrder}, interface-focused metrics such as the Interface-focused Root Mean Square Error (IRMSE) \citep{hassan2023bubbleml}, and structure-aware measures like the Structural Similarity Index (SSIM) \citep{nilsson2020understanding}. These additional loss terms were selected empirically after extensive experimentation on the shock-induced air bubble collapse dataset, where their inclusion consistently improved either that said metric or a broad range of metrics during inference.  To balance the contributions of these loss components, we employ two adaptive weighting strategies: SoftAdapt (SA) \citep{heydari2019softadapt} and Gradient Normalization (GN) \citep{chen2017gradnorm}, which periodically sample each loss term and its gradient during training and adjust their respective weights accordingly. The framework also supports custom curriculum training blocks, enabling users to specify different training and evaluation strategies at different stages of training depending on the epoch.

During inference, we report a comprehensive set of evaluation metrics beyond those used for training. These include additional pointwise measures such as Mean Absolute Error (MAE), Variance Scaled Root Mean Squared Error (VRMSE) \citep{ohana2024well}, and the H$^2$-seminorm \citep{SobolevTrainingWithHigherOrder}; spectral metrics like Multilevel Wavelet Loss (MLW) \citep{WaveletBasedLoss} and wavelet frequency-binned RMSE (wfRMSE) \citep{takamoto2022pdebench}; and feature-focused metrics that emphasize sharp gradients, such as shock-focused RMSE. We also include additional structure-aware measures like the Pearson Correlation Coefficient (PCC) \citep{hu2026realpdebench}, as well as physics-informed metrics that track integral quantities of interest, including mass, momentum, kinetic-energy for bubble datasets, center of mass displacement in case of droplets and enstrophy \citep{takamoto2022pdebench} over each rollout step.


    
    
    
A summarized contribution of this work is:
\begin{itemize}
    \setlength{\itemsep}{0pt}
    \setlength{\parskip}{0pt}
    \setlength{\topsep}{0pt}
    \setlength{\partopsep}{0pt}
    \item A \textbf{3.0 TB high-fidelity benchmark dataset} for shock-driven compressible multiphase flows, covering bubble collapse and droplet breakup in 2D and 3D.
    
    \item A \textbf{systematic benchmark of diverse surrogate models}, including convolutional, operator-based, transformer-based, and pre-trained PDE foundation models.
    
    \item An investigation of \textbf{physics- and structure-aware composite losses}, combining MSE with derivative-, interface-, and structure-focused terms and adaptive loss balancing.
    
    \item A \textbf{reliability-focused evaluation framework} extending beyond pointwise errors to spectral, structural, flow feature-focused, structure-aware and physics-informed metrics.
\end{itemize}


\section{Related Work}


Developing reliable machine learning surrogates depends not only on the quality and diversity of the training data, but also on the model architecture, training strategy and loss functions used for optimization. 


Datasets capturing PDE dynamics continue to expand. Early publicly available collections, such as PDEBench \citep{takamoto2022pdebench} and PDEArena \citep{gupta2022towards}, focused on canonical PDEs that are relatively efficient to generate computationally. More recently, the field has shifted toward increasingly complex and sophisticated datasets that often require domain expertise for their curation, as showcased by the Well \citep{ohana2024well}, BlastNet \citep{chung2023turbulence}, REALM \citep{mao2025benchmarking}, BubbleFormer \citep{hassan2025bubbleformer}, and ClimSim \citep{yu2025climsim}. Complementing these efforts, RealPDEBench \citep{hu2026realpdebench} introduces experimental datasets for fine-tuning pre-trained models to better capture real-world physical effects. While related datasets such as BubbleML \citep{hassan2023bubbleml} and BubbleFormer focus on incompressible, low-Mach-number boiling flows governed by heat transfer and phase change, the present work addresses a distinct physical regime involving compressible, shock-driven multiphase flows characterized by propagating shocks, bubble collapse, and fragmentation. Additionally, our dataset encompasses droplet fragmentation dynamics, further extending the range of multiphase phenomena represented. 

Many dataset papers now release their own benchmarking code alongside the data. Among recent efforts, APEBench \citep{koehler2024apebench} provides more rigorous training strategies aimed at achieving stable long-horizon rollouts for neural operators. Related ideas also include perturbing inputs with Gaussian noise to improve robustness, as explored in \cite{sanchez2020learning}. Several works have further introduced curriculum-style training strategies, such as the pushforward trick in \cite{brandstetter2022message}, or training schedulers that transition efficiently between single-step and multistep autoregressive training over the course of epochs, as in \cite{takamoto2023learning}.

Furthermore, there is a growing interest in training surrogate models with objectives that go beyond standard pointwise losses. Instead of relying only on MSE or MAE, several studies optimize weighted combinations of auxiliary, domain-specific metrics. For example, WaveLiT \citep{sankaranwavelit} uses a combination of MSE and a wavelet-domain L$^1$ loss. PINO \citep{li2021physics} combines an L$^2$ loss term with a PDE-residual loss. Other works, such as \cite{cho2024sobolev}, explore Sobolev losses during training, while CRONet \citep{olabiyi2025cronet}, from the structural topology optimization domain, uses an SSIM loss to guide the optimization.


\section{Datasets}
\label{section:main_datasets}

\begin{figure}[h!]
    \centering
    \includegraphics[width=0.95\textwidth]{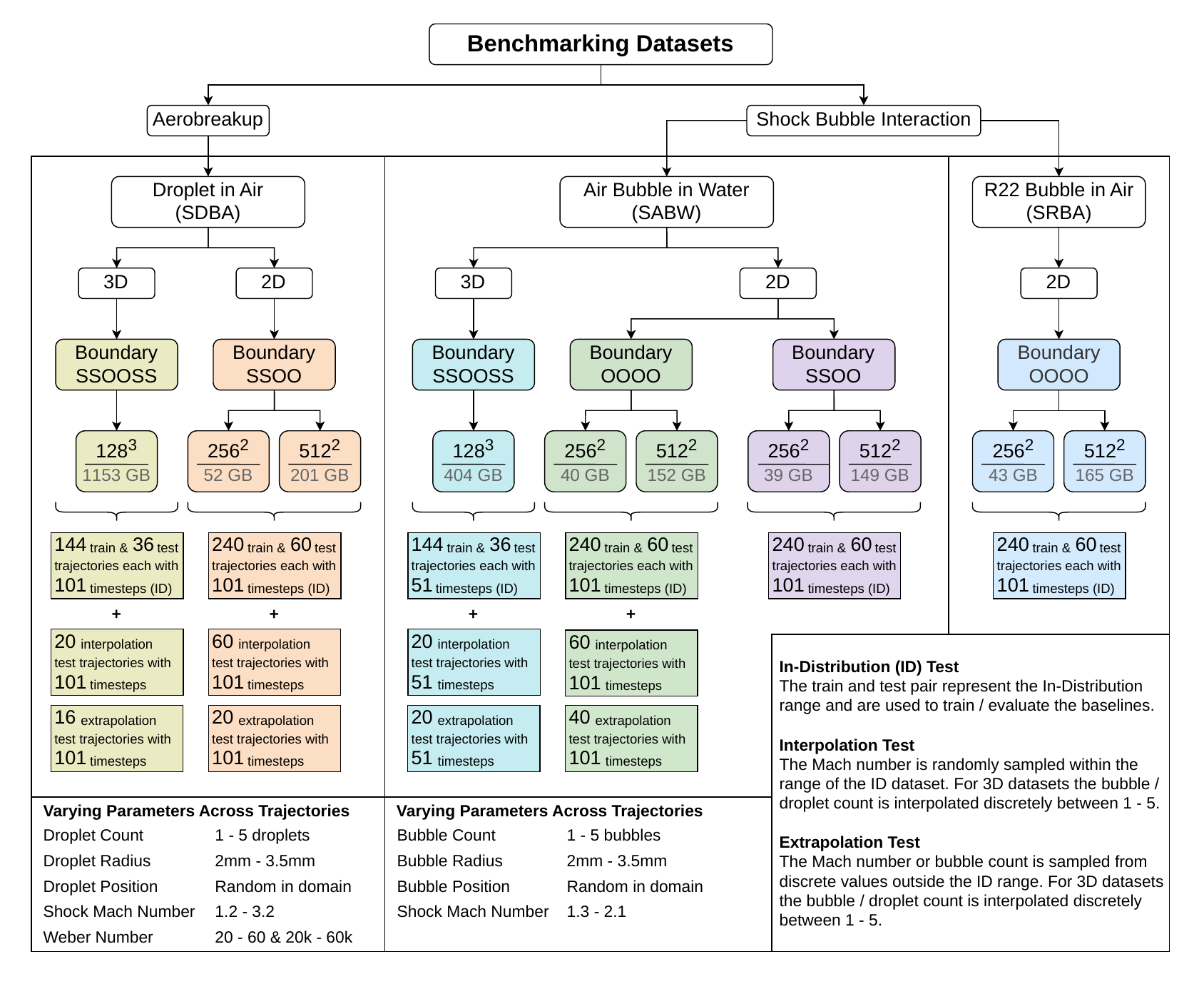}
    \caption{Dataset overview summarizing the provided datasets by dimensionality, boundary conditions, resolution, storage size, number of trajectories, and simulation-defined initial conditions. Boundary conditions are denoted using acronyms, where \texttt{S} indicates symmetry and \texttt{O} indicates open boundaries. For example, in the 3D-SDBA dataset, \texttt{SSOOSS} represents symmetry boundaries on the North, South, Top, and Bottom faces, with open boundaries on the East and West. Further details on the selection of governing parameters, including dimensionless numbers and bubble or droplet counts defining each trajectory, are provided in section   \ref{section:governing_parameter_coverage}. }
    \label{fig:dataset_overview}
\end{figure}

In this work, we introduce six datasets related to multiphase flows, generated by solving the compressible Euler equations. The generated datasets, as shown in Figure \ref{fig:dataset_overview}, span a wide variety of scenarios, including bubbles and droplets in both two and three dimensions, involving different materials and boundary conditions. To produce these datasets, we employ the high-fidelity finite-volume solver ALPACA \citep{hoppe2022alpaca}, incorporating the Robust Discrete Equations Method (RDEMIC) \citep{paula2023robust} for multiphase and interfacial treatment. Further details are described in section \ref{app_section: dataset_specs}.

\begin{figure}[h!]
    \centering
    \includegraphics[width=0.6\textwidth]{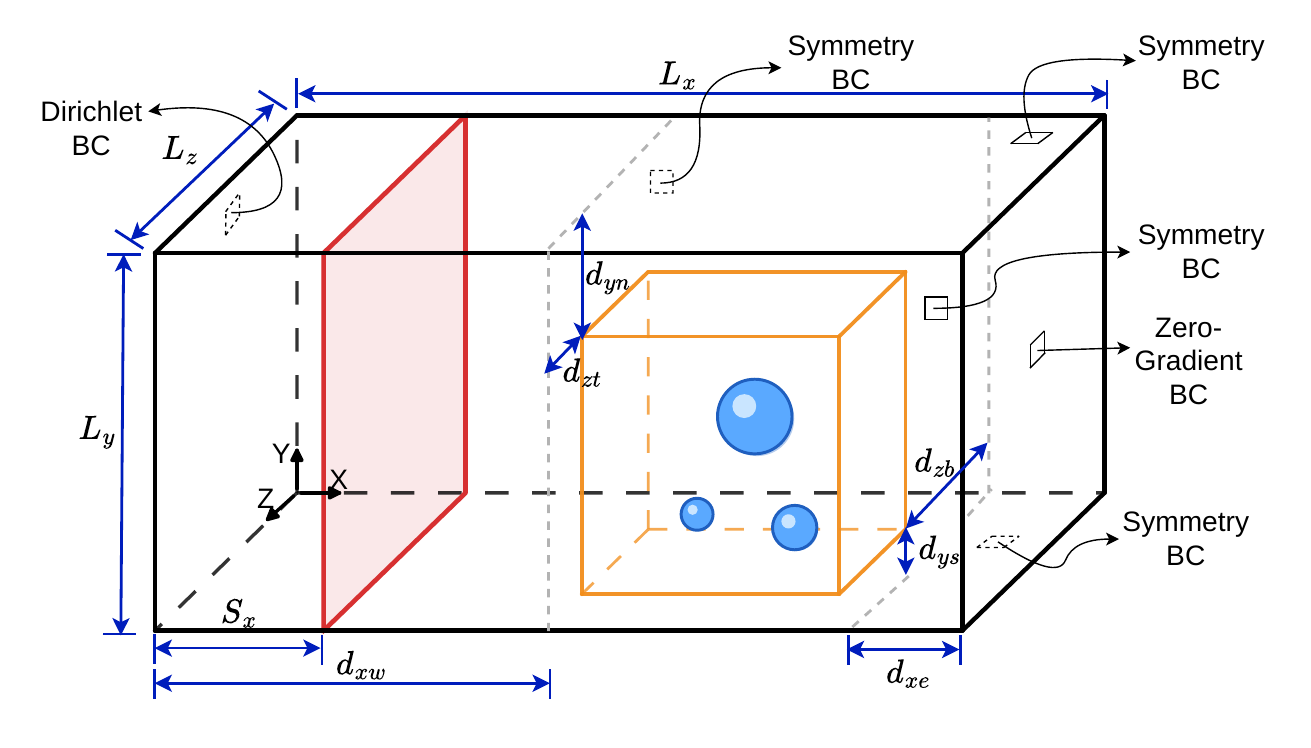}
    \caption{Schematic of the dataset generation setup. The full simulation domain is outlined in black, while the orange cube highlights the extracted subdomain used as the provided dataset. The red plane indicates the planar shock wave which interacts with the downstream bubbles or droplets.}
    \label{fig:IC}
\end{figure}

Figure \ref{fig:IC} shows the initial and boundary conditions used to generate the datasets with the dimensions mentioned in Table \ref{tab:ic_setup}. The complete simulation domain is depicted by the black cuboid, while the orange cube denotes the cropped subdomain used for the dataset. We enrich the flow dynamics within the subdomain by considering multiple boundary-condition configurations. Symmetry boundaries behave as inviscid walls, reflecting the incident shock back into the subdomain. As an alternative, we obtain a subdomain free from boundary effects by running the simulation on a larger computational domain and subsequently cropping it to the desired region. We refer to these cropped, boundary-effect-free subdomains as having open boundaries.



\begin{table}[h!]
    \centering
    \scriptsize
    \setlength{\tabcolsep}{2pt}
    \renewcommand{\arraystretch}{1.0}
    \caption{Initial condition setup for different datasets with reference to Figure~\ref{fig:IC}.}
    \label{tab:ic_setup}
    \begin{tabularx}{\linewidth}{>{\centering\arraybackslash}p{1.55cm}*{10}{>{\centering\arraybackslash}X}}
      \toprule
      \multirow{2}{*}{\makecell{Dataset}}
      & $L_x$ & $L_y$ & $L_z$ & $S_x$ & $d_{xw}$ & $d_{xe}$ & $d_{ys}$ & $d_{yn}$ & $d_{zt}$ & $d_{zb}$ \\
      & {[mm]} & {[mm]} & {[mm]} & {[mm]} & {[mm]} & {[mm]} & {[mm]} & {[mm]} & {[mm]} & {[mm]} \\
      \midrule
      \makecell{2D-SABW\\OOOO}   & 76.8  & 96.0 & -    & 25.8 & 28.2 & 10.2 & 28.8 & 28.8 & -   & -   \\
      \makecell{2D-SABW\\SSOO}   & 76.8  & 38.4 & -    & 25.8 & 28.2 & 10.2 & 0.00 & 0.00 & -   & -   \\
      \makecell{3D-SABW\\SSOOSS} & 76.8  & 38.4 & 38.4 & 25.8 & 28.2 & 10.2 & 0.00 & 0.00 & 0.00 & 0.00 \\
      \makecell{2D-SRBA\\OOOO}   & 76.8  & 96.0 & -    & 25.8 & 28.2 & 10.2 & 28.8 & 28.8 & -   & -   \\
      \makecell{2D-SDBA\\SSOO}   & 105.8 & 52.9 & -    & 25.8 & 28.2 & 24.7 & 0.00 & 0.00 & -   & -   \\
      \makecell{3D-SDBA\\SSOOSS} & 105.8 & 52.9 & 52.9 & 25.8 & 28.2 & 24.2 & 0.00 & 0.00 & 0.00 & 0.00 \\
      \bottomrule
    \end{tabularx}
\end{table}

\subsection{Shock-induced Air Bubble Collapse in Water (SABW)}
In the shock-induced air bubble collapse in water dataset, a moving planar shockwave impacts several bubbles resulting in complete breakdown of the bubbles into small fragments which get advected downstream and results in a series of interacting shockwaves emanating from the collapsed bubbles. A more detailed description of the physics of this dataset is provided in section \ref{section:app_sabw_description}.

The dataset is provided in two variants with different boundary conditions. In the first variant (OOOO), all boundaries are set to open, enabling the study of shock–bubble interactions and inter-bubble dynamics in isolation. In the second variant (SSOO), symmetry boundary conditions are applied at the north and south boundaries, causing shockwave reflections, while the east and west boundaries remain open. This setup captures not only the direct interaction between the shock and the bubbles, as well as inter-bubble effects, but also the influence of reflected shockwaves from the symmetric boundaries, acting as reflective walls. For the 3D dataset, only the symmetry-based variant (SSOOSS) is provided, as ensuring a sufficiently large simulation domain such that shockwaves from all boundaries do not reach the dataset region is computationally expensive. 

\begin{figure}[h!]
    \centering
    \includegraphics[width=0.9\textwidth]{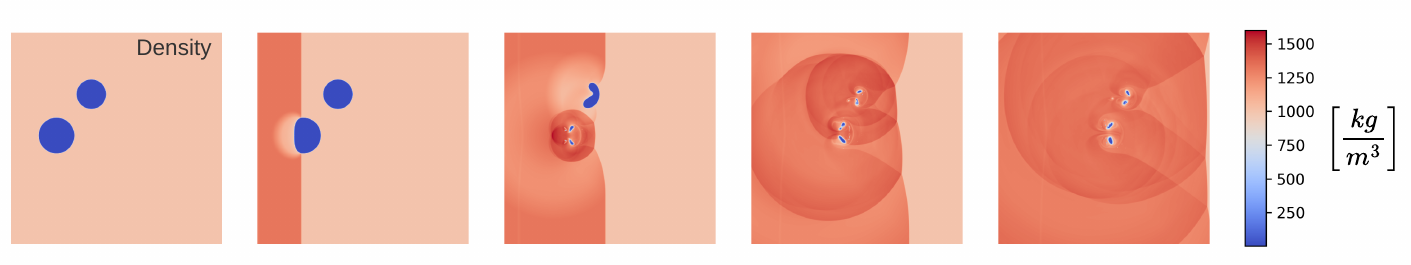}
    \caption{Uniformly spaced time snapshots of 2D-SABW at shock Mach number 1.70, with open boundary conditions on all sides (OOOO), showing the evolution of density from $t_0=0.0 \ \mu s$ to $t_{\textit{end}}=15 \ \mu s$.}
    \label{fig:2d_sabw_oooo_dp}
\end{figure}

\begin{figure}[h!]
    \centering
    \includegraphics[width=0.9\textwidth]{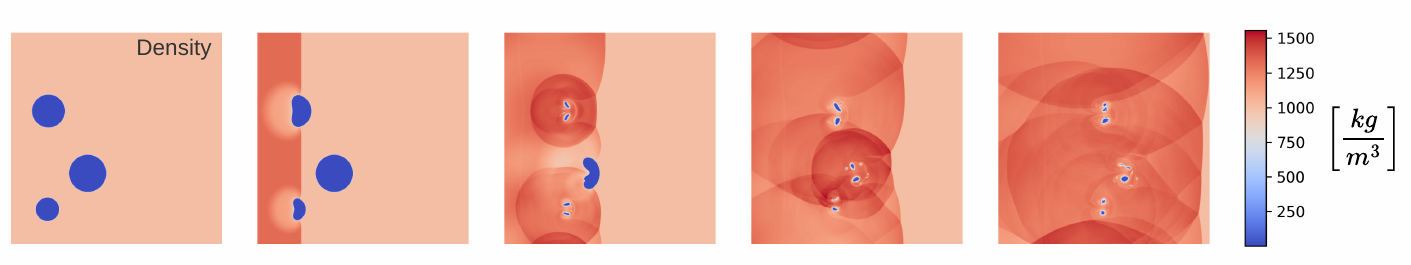}
    \caption{Uniformly spaced time snapshots of 2D-SABW at shock Mach number 1.70, with North–South symmetric and East–West open boundary conditions (SSOO), showing the evolution of density from $t_0=0.0 \ \mu s$ to $t_{\textit{end}}=15 \ \mu s$.}
    \label{fig:2d_sabw_ssoo_dp}
\end{figure}

The temporal evolution of the density field for each variant is illustrated in Figures \ref{fig:2d_sabw_oooo_dp} and \ref{fig:2d_sabw_ssoo_dp}. Additional fields of interest, including pressure, velocity, volume fraction, and schlieren are provided in the Appendix (Figures \ref{fig:2d_sabw_oooo_6fields} and \ref{fig:2d_sabw_ssoo_6fields}).

\subsection{Shock-induced Droplet Breakup in Air (SDBA)}

\begin{figure}[h!]
    \centering
    \includegraphics[width=0.9\textwidth]{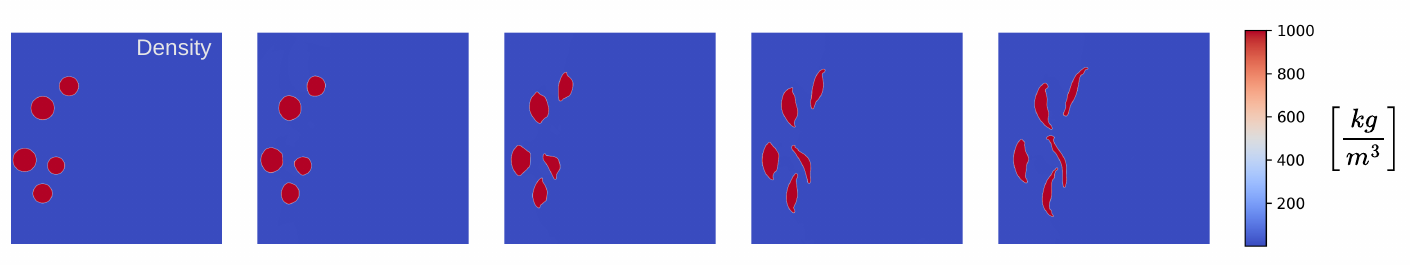}
    \caption{Uniformly spaced time snapshots of 2D-SDBA at shock Mach number 2.0, with all droplets in the RTP regime (Weber < 100) having a surface tension coefficient $\sigma=60.514$, under North–South symmetric and East–West open boundary conditions (SSOO), showing the evolution of density from $t_0=0.0 \ \mu s$ to $t_{\textit{end}}= 215.5 \ \mu s$.}
    \label{fig:2d_sdba_rtp_dp}
\end{figure}

\begin{figure}[h!]
    \centering
    \includegraphics[width=0.9\textwidth]{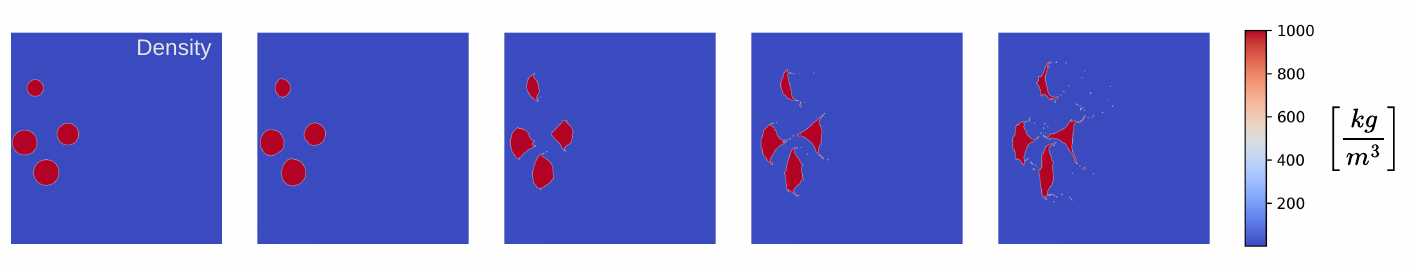}
    \caption{Uniformly spaced time snapshots of 2D-SDBA at shock Mach number 2.0, with all droplets in the SIE regime (Weber >1000) having a surface tension coefficient $\sigma=0.0719$, under North–South symmetric and East–West open boundary conditions (SSOO), showing the evolution of density from $t_0=0.0 \ \mu s$ to $t_{\textit{end}}= 215.5 \ \mu s$.}
    \label{fig:2d_sdba_sie_dp}
\end{figure}

This dataset captures the dynamics of an external shock interacting with liquid droplets suspended in air. During the interaction, the shockwave is reflected, transmitted, and diffracted at the droplets, giving rise to surface instabilities. Depending on the Weber number—a dimensionless quantity representing the ratio of aerodynamic forces to surface tension, mainly determined by droplet diameter and surface tension coefficient—the interaction results in two distinct breakup regimes: Rayleigh–Taylor Piercing (RTP) (Figures \ref{fig:2d_sdba_rtp_dp} and \ref{fig:2d_sdba_RTP_6fields}), for Weber number below 100 and Shear-Induced Entrainment (SIE) (Figures \ref{fig:2d_sdba_sie_dp} and \ref{fig:2d_sdba_SIE_6fields}) for Weber number above 1000. The detailed flow evolution and characterization of these regimes are provided in section \ref{section:app_sdba_description}. Because droplet deformation requires long simulation times, implementing absorbing boundary conditions would require an excessively large computational domain to ensure that reflected shock waves do not re-enter the subdomain. Keeping computational constraints in mind, the dataset is generated using symmetry boundary conditions, which are less computationally demanding, for both the two-dimensional SSOO and three-dimensional SSOOSS configurations. To further increase the complexity, both breakup modes are combined into a single dataset in equal proportion. This poses a significant challenge for the surrogate model, which must learn to infer the correct regime from the conditioning parameters during training and accordingly evolve its predictions at inference time.

\subsection{Shock-induced R22 Bubble Collaspe in Air (SRBA)}

\begin{figure}[h!]
    \centering
    \includegraphics[width=0.9\textwidth]{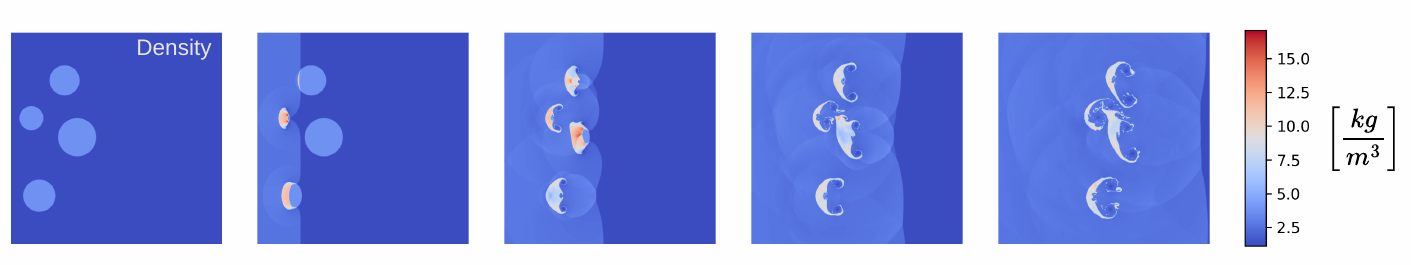}
    \caption{Uniformly spaced time snapshots of 2D-SRBA at shock Mach number 1.70, with open boundary conditions on all sides (OOOO), showing the evolution of density from  $t_0=0.0 \ \mu s$ to $t_{\textit{end}}=7.0 \ \mu s$.}
    \label{fig:2d_srba_oooo_dp}
\end{figure}

In contrast to the SABW and SDBA datasets, the shock-interaction with R22 bubbles in air represents a compressible gas-gas configuration. This setup isolates density-driven interface dynamics and enables the study of shock refraction, baroclinic vorticity generation, and Richtmyer–Meshkov-type instabilities without the confounding effects of surface tension as observed in the droplet breakup datasets. Unlike the air bubble collapse, a mushroom shaped deformation is observed as seen in Figure \ref{fig:2d_srba_oooo_dp} owing to the higher density of the R22 bubble. For this dataset, all boundaries are set to Open BC, a complete set of field variables is depicted in Figure \ref{fig:2d_srba_6fields} and the flow physics is described in section \ref{section:app_srba_description}.

\section{Benchmarking}
\label{section:main_benchmarking}
In this work, we provide a collection of trained two- and three-dimensional baseline models, including both randomly initialized networks and models fine-tuned from pre-trained weights. For the 2D setting, the randomly initialized baselines consist of 50M-parameter variants of ConvNeXt, CNO, FFNO, and ScOT, while the pre-trained baselines include Poseidon-B (122M parameters) and DPOT-M (150M parameters). For the 3D setting, we benchmark 10M-parameter variants of FFNO and ConvNeXt. Both the 2D and 3D baselines are trained under a fixed 24-hour computational budget. Detailed model-specific hyperparameters are provided in Section \ref{section:app_baseline_models}, while the overall training hyperparameters are summarized in Table \ref{tab:training-pipeline}.

For models initialized from random weights, the learning problem is defined to extract four temporal snapshots from a dataset trajectory as input and predict the subsequent state. We further condition the model on the simulation parameters associated with the dataset, summarized in Figure \ref{fig:dataset_overview}, by using these parameters to modulate the normalization layers of the baseline architectures \citep{perez2017visual}. The training and evaluation protocols are detailed in sections \ref{section:app_training_protocol} and \ref{section:app_evaluation_protocol}. For finetuning the pre-trained Poseidon-B model, we retain the original 1-to-1 input–output snapshot setting \citep{herde2024poseidon}, while for finetuning DPOT-M, we follow the original 10-to-1 input–output configuration \citep{hao2024dpot}. DPOT-M is compared against an FFNO baseline trained from random initialization, using a similar input–output configuration to ensure a fair comparison.



\section{Results and Conclusion}
\label{section:results_and_conclusion}
In this section, we present and compare benchmark results obtained using three training-loss strategies across the six datasets introduced in section \ref{section:main_datasets}. All baseline models described in section \ref{section:main_benchmarking} are trained at a resolution of $256^2$ for the 2D datasets and $128^3$ for the 3D datasets. In our ablations, we consider three loss configurations: (i) training with mean squared error (MSE) alone; (ii) training with a composite objective whose component weights are adapted using SA; and (iii) training with the same composite objective, but with adaptive loss-weighting performed using GN. The composite objective combines MSE with additional terms designed to capture complementary aspects of the flow: the H$^1$-seminorm (H1) penalizes errors in spatial gradients, the Structural Similarity Index (SSIM) encourages preservation of structure inside local windows and the interface-focused RMSE (IRMSE) emphasizes errors near bubble and droplet phase boundaries. Both SA and GN dynamically adjust the relative weights of these loss components during training. The additional computational overhead of these training strategies are analyzed from two perspectives: peak GPU memory requirement and time required for one epoch. The findings are reported in section \ref{section:app_metric_performance_comparison}.


We perform inference using a broader set of 12 evaluation metrics, as described in Table \ref{tab:loss_metrics_list}. Figures \ref{fig:radar_chart_mse_training}, \ref{fig:radar_chart_sa_training}, and \ref{fig:radar_chart_gn_training} summarize representative results averaged across the test trajectories and the rollout steps for the three training-loss strategies. For each dataset, we report five of the twelve metrics, viz. VRMSE, MLW, IRMSE, SSIM, and RMSE on the domain-integrated quantity of interest: enstrophy (iqRMSE-enstrophy). These metrics are selected to cover the five evaluation categories considered in this work: pointwise, spectral, feature-focused, structure-aware, and physics-informed metrics. To enable comparison across metrics with different scales, the raw metric values are first min-max normalized across all baseline models and training-loss strategies for a given dataset. The normalized values are then transformed by subtracting them from one, thereby reversing the scale so that lower original errors correspond to higher scores in the radar charts. Further details on the metrics are provided in section \ref{section:app_metrics}, while the loss-weighting strategies are described in section \ref{section:app_loss_balancing}. The evolution of the errors as the baselines are autoregressively rolled out are presented in section \ref{section:app_rollout_metrics}. The complete set of overall metrics aggregated across trajectories and rollouts, on all datasets and baselines are presented in section \ref{section:app_overall_inference_metrics}. Selective baseline predictions during autoregressive rollout on the presented datasets are shown in section \ref{section:app_rollout_viz}.


\subsection{Comparison of training loss strategies}

In this section we compare the inference results from the above-mentioned three different training loss strategies for the in-distribution test datasets, more generalization studies are presented in section \ref{section:generalization ablations}. Among all the 2D baselines, fine-tuned DPOT-M consistently outperforms the other baselines on the presented metrics. This could be attributed to either the rich pre-trained weights or the historic input sequence of 10 snapshots. We omit the fine-tuned Poseidon-B baseline from our discussion in this section because it performs substantially worse than the other baselines across nearly all metrics, despite having the largest parameter count. This could be attributed to the model not having a mechanism to incorporate conditioning, either through conditioning parameters or through inclusion of historic snapshots which help in distinguishing the different simulation trajectories.

\begin{figure}[h!]
    \centering
    \includegraphics[width=0.95\textwidth]{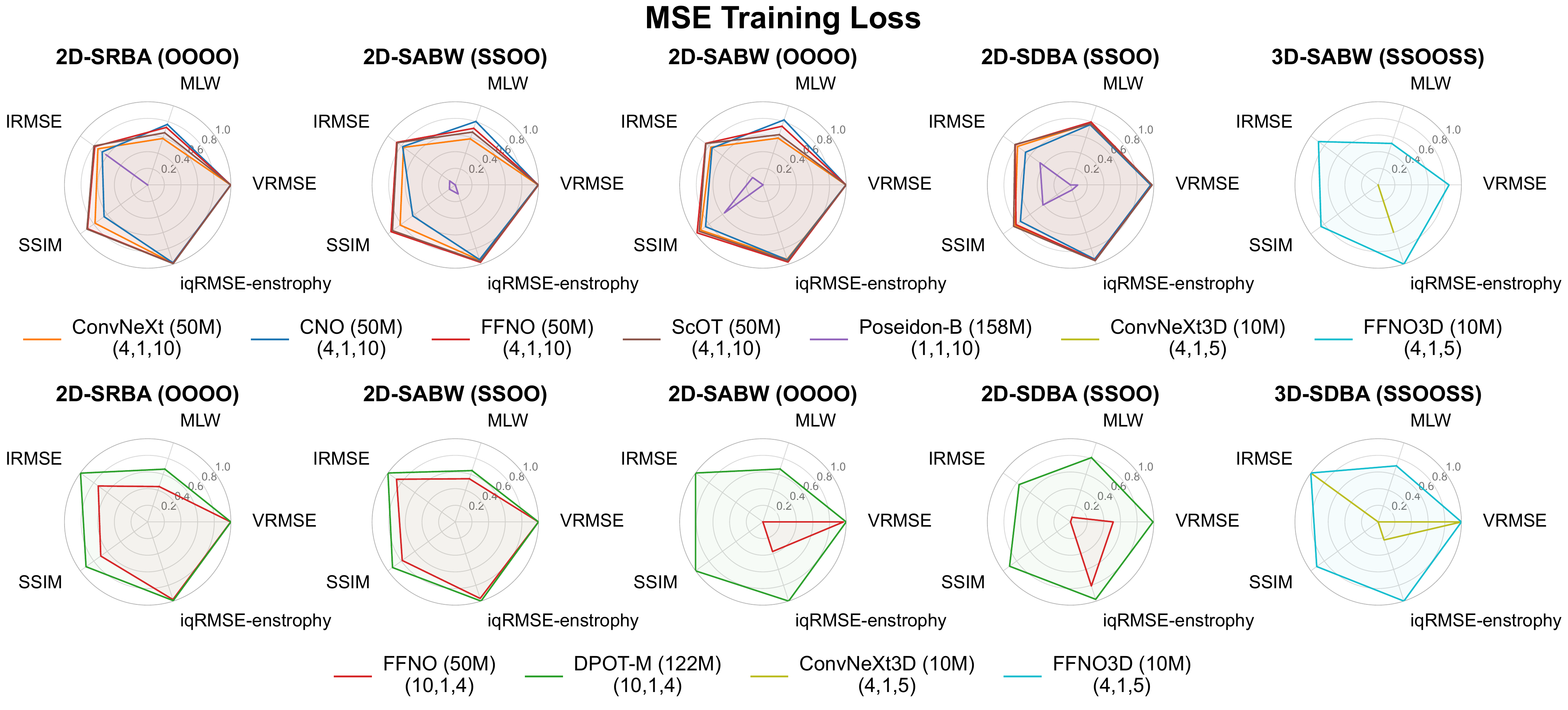}
    \caption{Radar plot showcasing performance of MSE-trained baselines across six in-distribution test datasets on five inference metrics: VRMSE, MLW, IRMSE, SSIM, and iqRMSE-enstrophy.}
    \label{fig:radar_chart_mse_training}
\end{figure}

\begin{figure}[h!]
    \centering
        \includegraphics[width=0.95\textwidth]{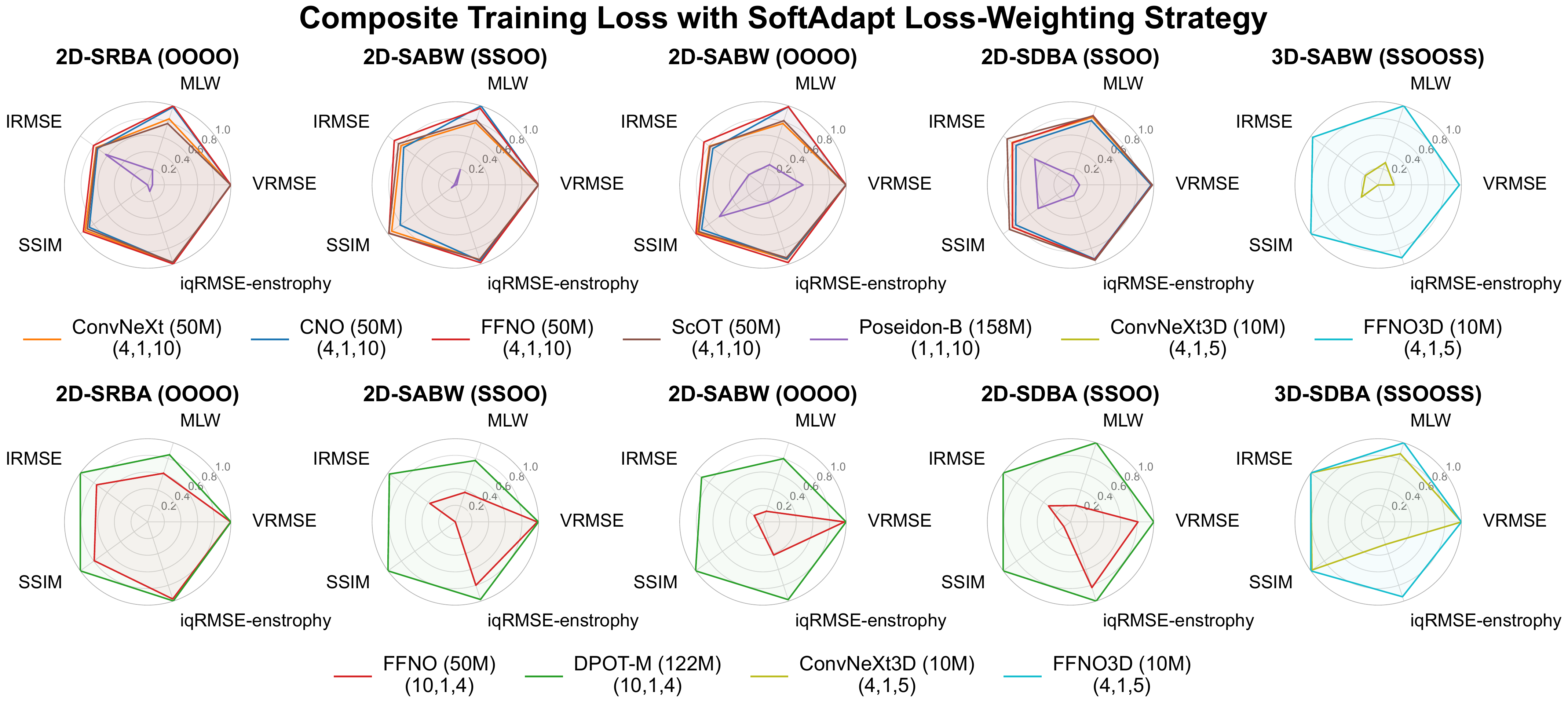}
    \caption{Radar plot showcasing performance of baselines trained with composite loss and SA loss-weighting across six in-distribution test datasets, evaluated on five inference metrics: VRMSE, MLW, IRMSE, SSIM, and iqRMSE-enstrophy}
    \label{fig:radar_chart_sa_training}
\end{figure}

\begin{figure}[h!]
    \centering
    \includegraphics[width=0.95\textwidth]{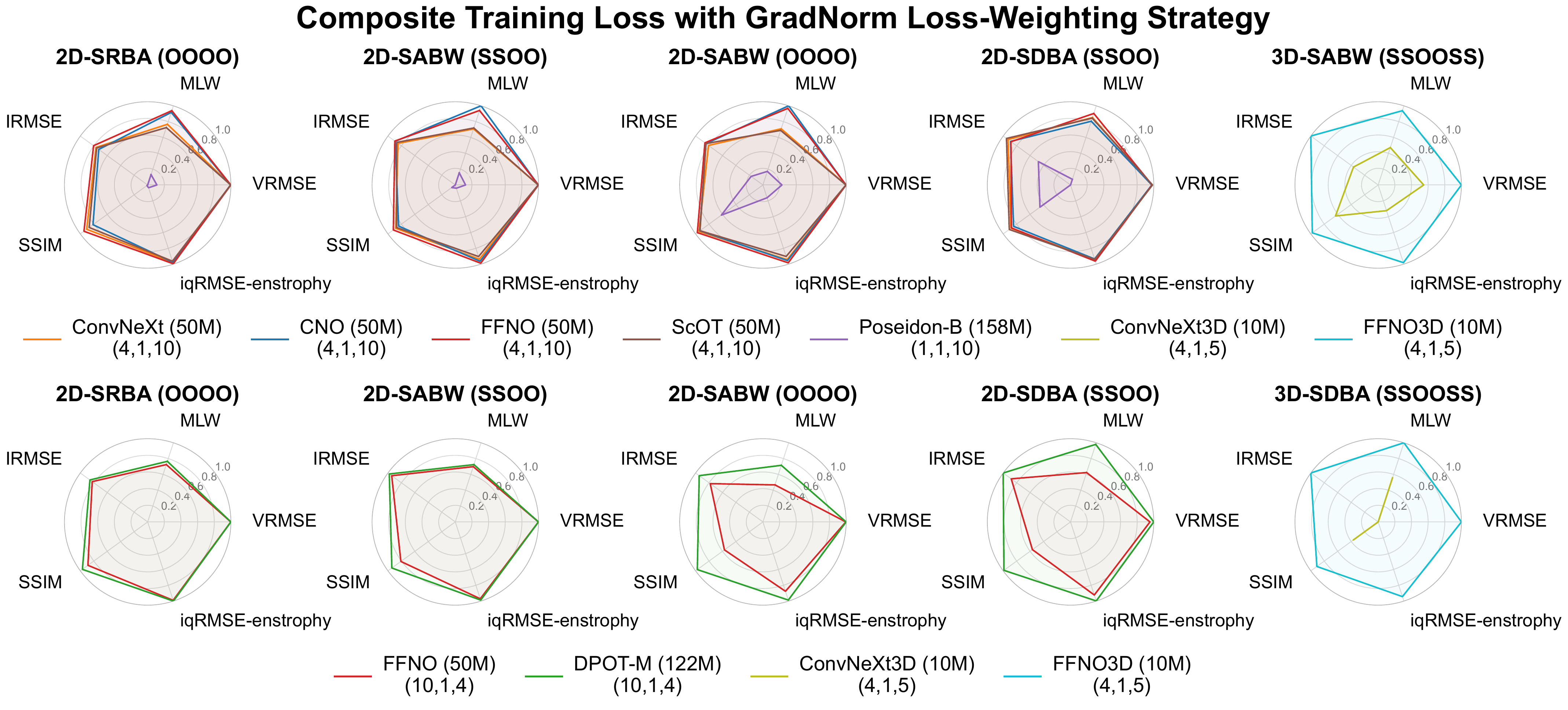}
    \caption{Radar plot showcasing performance of baselines trained with composite loss and GN loss-weighting across six in-distribution test datasets, evaluated on five inference metrics: VRMSE, MLW, IRMSE, SSIM, and iqRMSE-enstrophy.}
    \label{fig:radar_chart_gn_training}
\end{figure}

A further observation from Figures \ref{fig:radar_chart_mse_training}, \ref{fig:radar_chart_sa_training}, and \ref{fig:radar_chart_gn_training} is that FFNO consistently outperforms CNeXt on the 2D bubble-collapse datasets. This suggests that spectral operator models may be better suited to these cases, since the bubbles collapse within the first few timesteps and the dynamics are no longer dominated by a well-defined interface. As a result, the advantages of CNeXt as a convolutional edge detector become less pronounced. In contrast, for the shock-droplet breakup datasets, the droplets retain coherent interfaces as they deform and fragment, making the dynamics more localized and structure-preserving. This may explain why CNeXt performs slightly better than FFNO in these cases. Between the two composite-loss strategies, SA provides a more consistent improvement across the baselines and datasets compared to GN. 

We now examine these trends in more detail by analyzing the representative metrics selected from each evaluation category for the 2D baselines. The models trained with MSE alone perform comparably to those trained with the two composite-loss strategies on the pointwise metric VRMSE (Equation~\ref{eq:vrmse}). As shown in Tables \ref{tab:2d_sabw_oooo_pointwise}, \ref{tab:2d_sabw_ssoo_pointwise}, \ref{tab:2d_r22_sb_pointwise}, and \ref{tab:2d_sd_pointwise}, most baselines achieve $\mathrm{VRMSE}$ below 1. This indicates that the models are able to capture the mean background flow reasonably well, and that the composite objectives do not provide a substantial advantage for this pointwise metric.

The effect of composite training is more pronounced for the interface-focused RMSE. For the bubble-collapse datasets, GN-based composite training improves IRMSE by approximately 5-7\% on average across the baselines (Refer Tables \ref{tab:2d_sabw_oooo_feature_focused}, \ref{tab:2d_sabw_ssoo_feature_focused}, \ref{tab:2d_r22_sb_feature_focused}). In contrast, for the shock-droplet interaction dataset, the improvement is substantially larger, reaching approximately 40-50\% when using composite losses with SA and GN (Refer Table \ref{tab:2d_sd_feature_focused}). This difference can be explained by the nature of the interface evolution in the two problem classes. After the bubble collapse, the interface, obtained by thresholding the density, becomes difficult to identify because the bubbles fragment into very small structures. In the shock-droplet case, however, the droplet interface remains more coherent during deformation and breakup, making it easier to detect and therefore more responsive to interface-aware training.

Significant improvements were observed for all baselines for the spectral metric-MLW across all baselines when trained with composite loss. There is also a small but consistent improvement in the high-frequency component of the wavelet frequency-binned RMSE when trained with composite loss as seen in Tables \ref{tab:2d_sabw_oooo_spectral}, \ref{tab:2d_sabw_ssoo_spectral}, \ref{tab:2d_r22_sb_spectral}, \ref{tab:2d_sd_spectral}. These improvements suggest that the metrics presented in our composite objective help reduce the low-frequency bias typically introduced by purely pointwise objectives, leading to better recovery of fine-scale structures.

A consistent improvement is also observed for the structure-aware SSIM metric across all datasets and baselines, especially when using SA, followed by GN as shown in Tables \ref{tab:2d_sabw_oooo_structure_aware}, \ref{tab:2d_sabw_ssoo_structure_aware}, \ref{tab:2d_r22_sb_structure_aware}, \ref{tab:2d_sd_structure_aware}, \ref{tab:3d_sabw_structure_aware} and \ref{tab:3d_sdba_structure_aware}. This trend may be partly attributed to the inclusion of SSIM as one of the terms in the composite training objective. However, we emphasize that incorporating a metric into the training objective does not necessarily guarantee improved performance on that metric at inference time; rather, the observed improvement in SSIM is an empirical outcome specific to the settings considered here. The improvement in SSIM metric during inference is also associated with better coherent structure prediction, as illustrated in Figure \ref{fig:ssim_vs_rmse}.

Enstrophy of a flow field is computed as the integral summation of the absolute pointwise vorticities (Refer Table \ref{tab:iqrmse_quantities}). For the domain integrated enstrophy metric (iqRMSE-enstrophy), we observe that there is no significant advantage of using composite losses while training. To a certain extent this holds true for other physics-based metrics presented in Tables \ref{tab:2d_sabw_oooo_physics_informed}, \ref{tab:2d_sabw_ssoo_physics_informed}, \ref{tab:2d_r22_sb_physics_informed}, \ref{tab:2d_sd_physics_informed}, \ref{tab:3d_sabw_physics_informed} and \ref{tab:3d_sdba_physics_informed}. One exception is the ScOT baseline, which benefits substantially from composite training on selected datasets, particularly the 2D R22 bubble-collapse and 2D shock-droplet interaction cases.


The 3D baselines generally perform worse than their 2D counterparts. As shown in the radar charts in Figures~\ref{fig:radar_chart_mse_training}, \ref{fig:radar_chart_sa_training}, and \ref{fig:radar_chart_gn_training}, FFNO outperforms ConvNeXt across the reported metrics. Moreover, FFNO benefits consistently from composite-loss training, showing improvements across all selected metrics. However, these trends should be interpreted with caution. The 3D experiments were constrained by a 24-hour training limit, resulting in substantially fewer training epochs than in the 2D setting. More extensive training is required before drawing firm conclusions about the relative benefits of the composite losses in the 3D setting. The computational resources used to train these baselines are mentioned in section \ref{section:app_computational_resources}.



\subsection{Conclusion}
\label{section:conclustion}
In this work, we introduce six new datasets from the regime of compressible multiphase flows. We provide benchmarks on six baselines for the 2D datasets and two baselines for the 3D datasets. Each baseline is trained with three different loss strategies: MSE alone and composite losses with adaptive loss weighting using SoftAdapt and GradNorm. From the results, one can conclude that no single baseline performs best across all datasets and metrics, and that the choice of this surrogate would clearly depend on the metric of interest. However, the results indicate that using composite losses with adaptive loss-weighting strategies can yield substantial improvements across a range of metrics, particularly those focused on interface quality and spectral content. Among the two adaptive loss-weighting strategies, SoftAdapt yields more consistent improvements across baselines and datasets than GradNorm and is also more computationally efficient. In conclusion, our work highlights the importance of incorporating such complex flow datasets and diverse evaluation metrics that capture different aspects of the data into the SciML community to foster further advances in developing robust data-driven surrogate models.

\textbf{Limitations and Future Work}: There remains a significant gap in the ability of surrogate models to capture fine-scale details in complex flow problems. For example, none of the evaluated surrogate models accurately recovered the small droplets that detach from the primary droplet's surface, although these secondary droplets are a defining characteristic of the SIE regime. Future work could therefore focus on improving the treatment of small-scale interface dynamics and fragmentation processes. Further improvements may also be achieved through more expressive conditioning strategies. In particular, incorporating boundary-condition information during training could improve the generalizability of surrogate models across different simulation setups. The training metrics selected in this work for the composite losses were obtained by analyzing results from a broad list of experiments and selecting only those to include in the composite training loss when a positive correlation was observed on the metric or across a diverse set of metrics during evaluation. In addition, a more detailed field-wise analysis of surrogate predictions would help identify specific failure modes and guide the development of improved architectures and training strategies. Finally, the benchmark could be extended to include more challenging datasets, such as reacting multiphase flows and multiphase flows with phase change, to further push the limits of surrogate modeling in this domain.

\section{Acknowledgments and Disclosures}

This work was funded by the European Union under the ERC Advanced Grant: Project 101094463 — GENUFASD. Furthermore, the authors thank the Gauss Center for Supercomputing eV (GCS) for providing computing time on the supercomputer JUWELS \citep{JUWELS} at the Jülich Supercomputing Center. JUWELS is supported by GCS through funding by the German Federal Ministry of Research, Technology and Space and the Ministry of Culture and Science of the State of North Rhine-Westphalia. Additionally, the authors gratefully acknowledge the Gauss Center for Supercomputing e.V. (www.gauss-centre.eu) for funding this project by providing computing time on the GCS Supercomputer SuperMUC-NG at Leibniz Supercomputing Center (www.lrz.de). The authors declare no competing interests.

\clearpage
{\small
\bibliographystyle{plainnat}
\bibliography{source/neurips_2026_reference}

@article{takamoto2022pdebench,
  title   = {Pdebench: An extensive benchmark for scientific machine learning},
  author  = {Takamoto, Makoto and Praditia, Timothy and Leiteritz, Raphael and MacKinlay, Daniel and Alesiani, Francesco and Pfl{\"u}ger, Dirk and Niepert, Mathias},
  journal = {Advances in neural information processing systems},
  volume  = {35},
  pages   = {1596--1611},
  year    = {2022}
}

@article{gupta2022towards,
  title   = {Towards multi-spatiotemporal-scale generalized pde modeling},
  author  = {Gupta, Jayesh K and Brandstetter, Johannes},
  journal = {arXiv preprint arXiv:2209.15616},
  year    = {2022}
}

@article{hassan2023bubbleml,
  title   = {BubbleML: A multi-physics dataset and benchmarks for machine learning},
  author  = {Hassan, Sheikh Md Shakeel and Feeney, Arthur and Dhruv, Akash and Kim, Jihoon and Suh, Youngjoon and Ryu, Jaiyoung and Won, Yoonjin and Chandramowlishwaran, Aparna},
  journal = {arXiv preprint arXiv:2307.14623},
  year    = {2023}
}

@inproceedings{hassan2025bubbleformer,
  title     = {Bubbleformer: Forecasting Boiling with Transformers},
  author    = {Hassan, Sheikh Md Shakeel and Zou, Xianwei and Dhruv, Akash and Chandramowlishwaran, Aparna},
  booktitle = {The Thirty-ninth Annual Conference on Neural Information Processing Systems Datasets and Benchmarks Track},
  year      = {2025}
}

@article{ohana2024well,
  title   = {The well: a large-scale collection of diverse physics simulations for machine learning},
  author  = {Ohana, Ruben and McCabe, Michael and Meyer, Lucas and Morel, Rudy and Agocs, Fruzsina J and Beneitez, Miguel and Berger, Marsha and Burkhart, Blakesley and Dalziel, Stuart B and Fielding, Drummond B and others},
  journal = {Advances in Neural Information Processing Systems},
  volume  = {37},
  pages   = {44989--45037},
  year    = {2024}
}

@article{chung2023turbulence,
  title   = {Turbulence in focus: Benchmarking scaling behavior of 3D volumetric super-resolution with BLASTNet 2.0 data},
  author  = {Chung, Wai Tong and Akoush, Bassem and Sharma, Pushan and Tamkin, Alex and Jung, Ki Sung and Chen, Jacqueline and Guo, Jack and Brouzet, Davy and Talei, Mohsen and Savard, Bruno and others},
  journal = {Advances in Neural Information Processing Systems},
  volume  = {36},
  pages   = {77430--77484},
  year    = {2023}
}

@article{hu2026realpdebench,
  title   = {RealPDEBench: A Benchmark for Complex Physical Systems with Real-World Data},
  author  = {Hu, Peiyan and Feng, Haodong and Liu, Hongyuan and Yan, Tongtong and Deng, Wenhao and Gao, Tianrun and Zheng, Rong and Zheng, Haoren and Yu, Chenglei and Wang, Chuanrui and others},
  journal = {arXiv preprint arXiv:2601.01829},
  year    = {2026}
}

@article{mao2025benchmarking,
  title   = {Benchmarking neural surrogates on realistic spatiotemporal multiphysics flows},
  author  = {Mao, Runze and Zhang, Rui and Bai, Xuan and Wu, Tianhao and Zhang, Teng and Chen, Zhenyi and Lin, Minqi and Zeng, Bocheng and Xu, Yangchen and Xiang, Yingxuan and others},
  journal = {arXiv preprint arXiv:2512.18595},
  year    = {2025}
}

@article{yu2025climsim,
  title   = {Climsim-online: A large multi-scale dataset and framework for hybrid physics-ml climate emulation},
  author  = {Yu, Sungduk and Hu, Zeyuan and Subramaniam, Akshay and Hannah, Walter and Peng, Liran and Lin, Jerry and Bhouri, Mohamed Aziz and Gupta, Ritwik and L{\"u}tjens, Bj{\"o}rn and Will, Justus C and others},
  journal = {Journal of Machine Learning Research},
  volume  = {26},
  number  = {142},
  pages   = {1--85},
  year    = {2025}
}

@article{koehler2024apebench,
  title   = {Apebench: A benchmark for autoregressive neural emulators of pdes},
  author  = {Koehler, Felix and Niedermayr, Simon and Westermann, R{\"u}diger and Thuerey, Nils},
  journal = {Advances in Neural Information Processing Systems},
  volume  = {37},
  pages   = {120252--120310},
  year    = {2024}
}

@inproceedings{sanchez2020learning,
  title        = {Learning to simulate complex physics with graph networks},
  author       = {Sanchez-Gonzalez, Alvaro and Godwin, Jonathan and Pfaff, Tobias and Ying, Rex and Leskovec, Jure and Battaglia, Peter},
  booktitle    = {International conference on machine learning},
  pages        = {8459--8468},
  year         = {2020},
  organization = {PMLR}
}

@article{brandstetter2022message,
  title   = {Message passing neural PDE solvers},
  author  = {Brandstetter, Johannes and Worrall, Daniel and Welling, Max},
  journal = {arXiv preprint arXiv:2202.03376},
  year    = {2022}
}

@inproceedings{takamoto2023learning,
  title        = {Learning neural pde solvers with parameter-guided channel attention},
  author       = {Takamoto, Makoto and Alesiani, Francesco and Niepert, Mathias},
  booktitle    = {International Conference on Machine Learning},
  pages        = {33448--33467},
  year         = {2023},
  organization = {PMLR}
}

@inproceedings{raonic2023convolutional,
  title     = {Convolutional neural operators},
  author    = {Raonic, Bogdan and Molinaro, Roberto and Rohner, Tobias and Mishra, Siddhartha and de Bezenac, Emmanuel},
  booktitle = {ICLR 2023 workshop on physics for machine learning},
  year      = {2023}
}

@inproceedings{liu2022convnet,
  title     = {A convnet for the 2020s},
  author    = {Liu, Zhuang and Mao, Hanzi and Wu, Chao-Yuan and Feichtenhofer, Christoph and Darrell, Trevor and Xie, Saining},
  booktitle = {Proceedings of the IEEE/CVF conference on computer vision and pattern recognition},
  pages     = {11976--11986},
  year      = {2022}
}

@article{tran2021factorized,
  title   = {Factorized fourier neural operators},
  author  = {Tran, Alasdair and Mathews, Alexander and Xie, Lexing and Ong, Cheng Soon},
  journal = {arXiv preprint arXiv:2111.13802},
  year    = {2021}
}

@article{herde2024poseidon,
  title   = {Poseidon: Efficient foundation models for pdes},
  author  = {Herde, Maximilian and Raonic, Bogdan and Rohner, Tobias and K{\"a}ppeli, Roger and Molinaro, Roberto and de B{\'e}zenac, Emmanuel and Mishra, Siddhartha},
  journal = {Advances in Neural Information Processing Systems},
  volume  = {37},
  pages   = {72525--72624},
  year    = {2024}
}

@inproceedings{liu2021swin,
  title     = {Swin transformer v2: Scaling up capacity and resolution. 2022 IEEE},
  author    = {Liu, Ze and Hu, Han and Lin, Yutong and Yao, Zhuliang and Xie, Zhenda and Wei, Yixuan and Ning, Jia and Cao, Yue and Zhang, Zheng and Dong, Li and others},
  booktitle = {CVF Conference on Computer Vision and Pattern Recognition (CVPR)},
  pages     = {11999--12009},
  year      = {2021}
}

@article{hao2024dpot,
  title   = {Dpot: Auto-regressive denoising operator transformer for large-scale pde pre-training},
  author  = {Hao, Zhongkai and Su, Chang and Liu, Songming and Berner, Julius and Ying, Chengyang and Su, Hang and Anandkumar, Anima and Song, Jian and Zhu, Jun},
  journal = {arXiv preprint arXiv:2403.03542},
  year    = {2024}
}

@techreport{SobolevTrainingWithHigherOrder,
  title       = {Sobolev Training With Higher Order Derivatives},
  author      = {Hartmann, Thomas and Kissel, Matthias and Diepold, Klaus},
  institution = {Chair of Data Processing, Technical University of Munich},
  year        = {2019},
  url         = {https://collab.dvb.bayern/download/attachments/75112352/fp_hartmann.pdf},
  note        = {Accessed: 2026-03-05}
}

@misc{SobolevTrainingOperatorLearning,
  title         = {Sobolev Training for Operator Learning},
  author        = {Namkyeong Cho and Junseung Ryu and Hyung Ju Hwang},
  year          = {2024},
  eprint        = {2402.09084},
  archiveprefix = {arXiv},
  primaryclass  = {cs.LG},
  url           = {https://arxiv.org/abs/2402.09084}
}

@misc{AGeneralizedNovelLoss,
  title         = {A generalised novel loss function for computational fluid dynamics},
  author        = {Zachary Cooper-Baldock and Paulo E. Santos and Russell S. A. Brinkworth and Karl Sammut},
  year          = {2024},
  eprint        = {2411.17059},
  archiveprefix = {arXiv},
  primaryclass  = {cs.LG},
  url           = {https://arxiv.org/abs/2411.17059}
}

@misc{WaveletBasedLoss,
  title         = {Wavelet-based Loss for High-frequency Interface Dynamics},
  author        = {Lukas Prantl and Jan Bender and Tassilo Kugelstadt and Nils Thuerey},
  year          = {2022},
  eprint        = {2209.02316},
  archiveprefix = {arXiv},
  primaryclass  = {cs.LG},
  url           = {https://arxiv.org/abs/2209.02316}
}

@article{nilsson2020understanding,
  title   = {Understanding ssim},
  author  = {Nilsson, Jim and Akenine-M{\"o}ller, Tomas},
  journal = {arXiv preprint arXiv:2006.13846},
  year    = {2020}
}

@article{venkataramanan2021hitchhiker,
  title     = {A hitchhiker’s guide to structural similarity},
  author    = {Venkataramanan, Abhinau K and Wu, Chengyang and Bovik, Alan C and Katsavounidis, Ioannis and Shahid, Zafar},
  journal   = {IEEE Access},
  volume    = {9},
  pages     = {28872--28896},
  year      = {2021},
  publisher = {IEEE}
}

@article{heydari2019softadapt,
  title   = {Softadapt: Techniques for adaptive loss weighting of neural networks with multi-part loss functions},
  author  = {Heydari, A Ali and Thompson, Craig A and Mehmood, Asif},
  journal = {arXiv preprint arXiv:1912.12355},
  year    = {2019}
}

@article{chen2017gradnorm,
  title   = {GradNorm: Gradient Normalization for Adaptive Loss Balancing in Deep Multitask Networks (2018)},
  author  = {Chen, Zhao and Badrinarayanan, Vijay and Lee, Chen-Yu and Rabinovich, Andrew},
  journal = {URL http://arxiv. org/abs/1711.02257. ArXiv},
  volume  = {1711},
  year    = {2017}
}

@article{sankaranwavelit,
  title  = {WaveLiT: A Parameter-Efficient Architecture for Neural PDE Solvers},
  author = {Sankaran, Shyam and Wang, Hanwen and Perdikaris, Paris},
  year   = {2025},
  journal   = {NeurIPS 2025 Workshop: Machine Learning and the Physical Sciences}
}

@article{li2021physics,
  title   = {Physics-informed neural operator for learning partial differential equations. arXiv},
  author  = {Li, Z and Zheng, H and Kovachki, N and Jin, D and Chen, H and Liu, B and Azizzadenesheli, K and Anandkumar, A},
  journal = {arXiv preprint arXiv:2111.03794},
  year    = {2021}
}

@article{cho2024sobolev,
  title   = {Sobolev Training for Operator Learning},
  author  = {Cho, Namkyeong and Ryu, Junseung and Hwang, Hyung Ju},
  journal = {arXiv preprint arXiv:2402.09084},
  year    = {2024}
}

@article{olabiyi2025cronet,
  title     = {CRONet: A convolutional recurrent operator approximator network to accelerate topology optimization},
  author    = {Olabiyi, Ridwan and Yang, Hui and Iquebal, Ashif},
  journal   = {Manufacturing Letters},
  volume    = {44},
  pages     = {1052--1063},
  year      = {2025},
  publisher = {Elsevier}
}

@article{paula2023robust,
  title     = {A robust high-resolution discrete-equations method for compressible multi-phase flow with accurate interface capturing},
  author    = {Paula, Thomas and Adami, Stefan and Adams, Nikolaus A},
  journal   = {Journal of Computational Physics},
  volume    = {491},
  pages     = {112371},
  year      = {2023},
  publisher = {Elsevier}
}

@article{hoppe2022alpaca,
  title     = {ALPACA-a level-set based sharp-interface multiresolution solver for conservation laws},
  author    = {Hoppe, Nils and Winter, Josef M and Adami, Stefan and Adams, Nikolaus A},
  journal   = {Computer Physics Communications},
  volume    = {272},
  pages     = {108246},
  year      = {2022},
  publisher = {Elsevier}
}

@article{anderson1990modern,
  title   = {Modern compressible flow: with historical perspective},
  author  = {Anderson, John David},
  journal = {(No Title)},
  year    = {1990}
}

@article{le2016noble,
  title     = {The Noble-Abel stiffened-gas equation of state},
  author    = {Le M{\'e}tayer, Olivier and Saurel, Richard},
  journal   = {Physics of Fluids},
  volume    = {28},
  number    = {4},
  year      = {2016},
  publisher = {AIP Publishing}
}

@article{theofanous2008physics,
  title     = {On the physics of aerobreakup},
  author    = {Theofanous, TG and Li, GJ},
  journal   = {Physics of fluids},
  volume    = {20},
  number    = {5},
  year      = {2008},
  publisher = {AIP Publishing}
}

@inproceedings{winter2019numerical,
  title     = {Numerical investigation of 3d drop-breakup mechanisms using a sharp interface level-set method},
  author    = {Winter, Josef and Kaiser, Jakob and Adami, Stefan and Adams, Nikolaus},
  booktitle = {11th International Symposium on Turbulence and Shear Flow Phenomena, TSFP 2019},
  year      = {2019}
}

@article{meng2018numerical,
  title     = {Numerical simulation of the aerobreakup of a water droplet},
  author    = {Meng, Jomela C and Colonius, Tim},
  journal   = {Journal of Fluid Mechanics},
  volume    = {835},
  pages     = {1108--1135},
  year      = {2018},
  publisher = {Cambridge University Press}
}

@article{nourgaliev2006adaptive,
  title     = {Adaptive characteristics-based matching for compressible multifluid dynamics},
  author    = {Nourgaliev, Robert R and Dinh, Truc-Nam and Theofanous, Theo G},
  journal   = {Journal of Computational Physics},
  volume    = {213},
  number    = {2},
  pages     = {500--529},
  year      = {2006},
  publisher = {Elsevier}
}

@article{bempedelis2020energy,
  title     = {Energy focusing in shock-collapsed bubble arrays},
  author    = {Bempedelis, N and Ventikos, Y},
  journal   = {Journal of Fluid Mechanics},
  volume    = {900},
  pages     = {A44},
  year      = {2020},
  publisher = {Cambridge University Press}
}

@book{leighton2012acoustic,
  title     = {The acoustic bubble},
  author    = {Leighton, Timothy},
  year      = {2012},
  publisher = {Academic press}
}

@article{wursig2000development,
  title     = {Development of an air bubble curtain to reduce underwater noise of percussive piling},
  author    = {W{\"u}rsig, B and Greene Jr, CR and Jefferson, TA},
  journal   = {Marine environmental research},
  volume    = {49},
  number    = {1},
  pages     = {79--93},
  year      = {2000},
  publisher = {Elsevier}
}

@book{lefebvre2017atomization,
  title     = {Atomization and sprays},
  author    = {Lefebvre, Arthur H and McDonell, Vincent G},
  year      = {2017},
  publisher = {CRC press}
}

@article{ba2016layer,
  title   = {Layer normalization},
  author  = {Ba, Jimmy Lei and Kiros, Jamie Ryan and Hinton, Geoffrey E},
  journal = {arXiv preprint arXiv:1607.06450},
  year    = {2016}
}

@inproceedings{perez2017visual,
  title     = {Visual reasoning with a general conditioning layer, Courville},
  author    = {Perez, E and Strub, F and De Vries, H and Dumoulin, V},
  booktitle = {In Proceedings of the AAAI Conference on Artificial Intelligence},
  year      = {2017}
}

@book{toro2013riemann,
  title     = {Riemann solvers and numerical methods for fluid dynamics: a practical introduction},
  author    = {Toro, Eleuterio F},
  year      = {2013},
  publisher = {Springer Science \& Business Media}
}

@article{fleischmann2020shock,
  title     = {A shock-stable modification of the HLLC Riemann solver with reduced numerical dissipation},
  author    = {Fleischmann, Nico and Adami, Stefan and Adams, Nikolaus A},
  journal   = {Journal of computational physics},
  volume    = {423},
  pages     = {109762},
  year      = {2020},
  publisher = {Elsevier}
}

@article{jiang1996efficient,
  title     = {Efficient implementation of weighted ENO schemes},
  author    = {Jiang, Guang-Shan and Shu, Chi-Wang},
  journal   = {Journal of computational physics},
  volume    = {126},
  number    = {1},
  pages     = {202--228},
  year      = {1996},
  publisher = {Elsevier}
}

@article{gottlieb1998total,
  title   = {Total variation diminishing Runge-Kutta schemes},
  author  = {Gottlieb, Sigal and Shu, Chi-Wang},
  journal = {Mathematics of computation},
  volume  = {67},
  number  = {221},
  pages   = {73--85},
  year    = {1998}
}

@article{JUWELS,
author = {{J\"{u}lich Supercomputing Centre}},
title = {{JUWELS Cluster and Booster: Exascale Pathfinder with Modular Supercomputing Architecture at Juelich Supercomputing Centre}},
journal = {Journal of large-scale research facilities},
number = {A183},
volume = {7},
doi = {10.17815/jlsrf-7-183},
url = {http://dx.doi.org/10.17815/jlsrf-7-183},
year = {2021}
}
}

\clearpage
\appendix
\part{Appendix}
\parttoc
\clearpage
\section{Dataset Details}
The datasets are made publicly available on HuggingFace at \url{https://huggingface.co/FluidVerse}.
\subsection{Dataset Organization}

Each dataset is divided into training and testing sets and stored in HDF5 format using a hierarchical organization. At the top level, groups are defined, with their names encoding the parameters that specify the initial conditions of each trajectory. Within each group, the time-series field data is stored in the shape $ T \times C \times X_{res} \times Y_{res} \times Z_{res}$, where $T, \ C, \ X_{res}, \ Y_{res}, \ Z_{res}$ denote the number of timesteps, channels, and resolution in the X, Y and Z direction respectively. An illustration of this hierarchical structure is provided below:

{\ttfamily\noindent\textcolor{RoyalBlue}{Group}: 005\_Mas1.30\_sb1\_Ax0.0078\_Ay0.0179\_Az0.0145\_Ar0.0022}
\vspace{-2mm}
{%
\ttfamily
\setstretch{1.2} 
\begin{tabbing}
    \hspace{3mm} \textcolor{OliveGreen}{Field}: density \hspace{1.5cm} \= - Shape: (51, 1, 128, 128, 128) \hspace{0.5cm} \= Dtype: float64 \kill
    \hspace{3mm} \textcolor{OliveGreen}{Field}: density \> \textcolor{Orange}{Shape}: (51, 1, 128, 128, 128) \> \textcolor{Purple}{Dtype}: float64 \\
    \hspace{3mm} \textcolor{OliveGreen}{Field}: density\_1 \> \textcolor{Orange}{Shape}: (51, 1, 128, 128, 128) \> \textcolor{Purple}{Dtype}: float64 \\
    \hspace{3mm} \textcolor{OliveGreen}{Field}: density\_2 \> \textcolor{Orange}{Shape}: (51, 1, 128, 128, 128) \> \textcolor{Purple}{Dtype}: float64 \\
    \hspace{3mm} \textcolor{OliveGreen}{Field}: \makecell[l]{diffuse\_volume\vspace{-1mm} \\ \_fraction\_1} \> \textcolor{Orange}{Shape}: (51, 1, 128, 128, 128) \> \textcolor{Purple}{Dtype}: float64 \\
    \hspace{3mm} \textcolor{OliveGreen}{Field}: pressure \> \textcolor{Orange}{Shape}: (51, 1, 128, 128, 128) \> \textcolor{Purple}{Dtype}: float64 \\
    \hspace{3mm} \textcolor{OliveGreen}{Field}: velocityX \> \textcolor{Orange}{Shape}: (51, 1, 128, 128, 128) \> \textcolor{Purple}{Dtype}: float64 \\
    \hspace{3mm} \textcolor{OliveGreen}{Field}: velocityY \> \textcolor{Orange}{Shape}: (51, 1, 128, 128, 128) \> \textcolor{Purple}{Dtype}: float64 \\
    \hspace{3mm} \textcolor{OliveGreen}{Field}: velocityZ \> \textcolor{Orange}{Shape}: (51, 1, 128, 128, 128) \> \textcolor{Purple}{Dtype}: float64 \\
\end{tabbing}
}
\vspace{-10mm}
{\par\bigskip{\scalebox{2}{$\vdots$}}\par\bigskip}
\vspace{-2mm}

{\ttfamily\noindent\textcolor{RoyalBlue}{Group}: 009\_Mas1.30\_sb1\_Ax0.0130\_Ay0.0309\_Az0.0097\_Ar0.0025}
\vspace{-2mm}
{%
\ttfamily
\setstretch{1.2} 
\begin{tabbing}
    \hspace{3mm} \textcolor{OliveGreen}{Field}: density \hspace{1.5cm} \= - Shape: (51, 1, 128, 128, 128) \hspace{0.5cm} \= Dtype: float64 \kill
    \hspace{3mm} \textcolor{OliveGreen}{Field}: density \> \textcolor{Orange}{Shape}: (51, 1, 128, 128, 128) \> \textcolor{Purple}{Dtype}: float64 \\
    \hspace{3mm} \textcolor{OliveGreen}{Field}: density\_1 \> \textcolor{Orange}{Shape}: (51, 1, 128, 128, 128) \> \textcolor{Purple}{Dtype}: float64 \\
    \hspace{3mm} \textcolor{OliveGreen}{Field}: density\_2 \> \textcolor{Orange}{Shape}: (51, 1, 128, 128, 128) \> \textcolor{Purple}{Dtype}: float64 \\
    \hspace{3mm} \textcolor{OliveGreen}{Field}: \makecell[l]{diffuse\_volume\vspace{-1mm} \\ \_fraction\_1} \> \textcolor{Orange}{Shape}: (51, 1, 128, 128, 128) \> \textcolor{Purple}{Dtype}: float64 \\
    \hspace{3mm} \textcolor{OliveGreen}{Field}: pressure \> \textcolor{Orange}{Shape}: (51, 1, 128, 128, 128) \> \textcolor{Purple}{Dtype}: float64 \\
    \hspace{3mm} \textcolor{OliveGreen}{Field}: velocityX \> \textcolor{Orange}{Shape}: (51, 1, 128, 128, 128) \> \textcolor{Purple}{Dtype}: float64 \\
    \hspace{3mm} \textcolor{OliveGreen}{Field}: velocityY \> \textcolor{Orange}{Shape}: (51, 1, 128, 128, 128) \> \textcolor{Purple}{Dtype}: float64 \\
    \hspace{3mm} \textcolor{OliveGreen}{Field}: velocityZ \> \textcolor{Orange}{Shape}: (51, 1, 128, 128, 128) \> \textcolor{Purple}{Dtype}: float64 \\
\end{tabbing}
}

During dataloading, the simulation parameters from the group name are extracted, scaled using min-max normalization and passed to the normalization layer of the baseline architecture as conditioning parameters as described in section \ref{section:app_training_protocol}. 

\subsection{Dataset Specifications}
\label{app_section: dataset_specs}
As mentioned in the main text, the datasets are generated using the high-fidelity finite volume solver ALPACA. Within this framework, the Robust Discrete Equations Method for Interface Capturing (RDEMIC) is employed to produce ground-truth trajectories by solving the two- and three-dimensional compressible Euler equations. The governing equations in vector notation are given as follows:

\begin{equation}
    \label{eq:euler}
    \partial_t \displaystyle \tU_l + \nabla \cdot \displaystyle \tF_l = \displaystyle \tB_l \cdot \nabla \alpha_l 
\end{equation}

where subscript $l$ denotes the index of the phase, $\displaystyle \tU_l$ is the vector of conserved quantities, $\displaystyle \tF_l$ is the flux tensor and $\displaystyle \tB_l$ is the interaction tensor. 

\[
\displaystyle \tU_l =
\begin{bmatrix}
\alpha_l \\
\alpha_l \rho_l \\
\alpha_l \rho_l \mathbf{u}_l \\
\alpha_l E_l
\end{bmatrix},
\quad
\displaystyle \tF_l =
\begin{bmatrix}
0 \\
\alpha_l \rho_l \mathbf{u}_l^{\!T} \\
\alpha_l \rho_l \mathbf{u}_l \otimes \mathbf{u}_l + \alpha_l p_l \mathbf{I} \\
\alpha_l (E_l + p_l)\mathbf{u}_l^{\!T}
\end{bmatrix},
\quad
\displaystyle \tB_l =
\begin{bmatrix}
-\mathbf{u}_{\text{int}}^{\!T} \\
0 \\
p_{\text{int},l} \mathbf{I} \\
p_{\text{int},l} \mathbf{u}_{\text{int}}^{\!T}
\end{bmatrix},
\quad
\]

The quantities $\alpha_l$, $\rho_l$, $\mathbf{u}_l$, $p_l$, and $E_l$ represent the volume fraction, density, velocity vector, pressure, and total energy of phase $l$, respectively. The interface velocity and pressure are denoted by $\mathbf{u}_{\text{int}}$ and $p_{\text{int},l}$. In the absence of surface tension, the interface pressure 
$p_{\text{int},l}$ is identical across all phases. $\mathbf{I}$ represents the identity tensor.

To close the above governing equations, an additional Equation of State (EOS) is required which relates pressure, density, and internal energy. In this work, the stiffened-gas EOS is employed for the generation of all datasets, and is given by

\begin{equation}
\label{eq:eos}
p(\rho, e) = (\gamma - 1)\rho e - \gamma p_{\text{stiff}}
\end{equation}

where $p$ denotes the fluid pressure, $\rho$ the mass density, $e$ the specific internal energy, and $\gamma$ a model constant. The parameter $p_{\text{stiff}}$ represents the contribution from fluid pre-compression.

The simulations use the HLLC Riemann solver \citep{toro2013riemann}. In case of air-bubble collapse in water, the HLLC-LM solver is used, which remedies shock instabilities associated with the low-Mach-number flow in transverse direction to the shock wave \citep{fleischmann2020shock}. A WENO5 spatial reconstruction \citep{jiang1996efficient} is employed to achieve high accuracy and robust shock capturing. Time integration is carried out using a third-order Runge-Kutta scheme \citep{gottlieb1998total}, ensuring stable and accurate evolution of the flow physics.

Table \ref{tab:dataset_metadata} presents the summarized metadata for the generated datasets. The resolution listed here refers to the original output generated directly by the solver; the data is subsequently cropped and downsampled to the target resolution specified in Figure \ref{fig:dataset_overview}. 

\begin{table}[h!]
    \caption{Metadata table for the provided datasets. End-time refers to the total simulation time, CFL refers to Courant-Friedrichs-Lewy criterion, $\Delta t_{\mathrm{solver}}$ is the average solver timestep across trajectories and $\Delta x$ refers to the grid spacing. The reported wall-clock time is obtained using the compute resources described in section \ref{section:app_computational_resources}. }
    \label{tab:dataset_metadata}
    \centering
    \begin{tabularx}{\textwidth}{cccccccc}
    \toprule
    Dataset & Resolution & \makecell{End-time\\{[$\mu s$]}} & CFL & \makecell{$\Delta t_{\mathrm{solver}}$\\{[$\mu s$]}} & \makecell{$\Delta x$\\{[mm]}} & \makecell{\#\\snap-\\shots} & \makecell{Wall-clock\\time\\{[hours]}} \\
    \midrule
    \makecell{2D-SABW\\OOOO}   & [2048 $\times$ 2560]   & 15       & 0.5 & 1.6 $\times 10^{-3}$  & 0.0375 & 101 & 46 \\
    \makecell{2D-SABW\\SSOO}   & [2048 $\times$ 1024]   & 15       & 0.5 & 1.3 $\times 10^{-3}$  & 0.0375 & 101 & 27 \\
    \makecell{3D-SABW\\SSOOSS} & [512 $\times$ 256 $\times$ 256] & 15       & 0.5 & 4.35 $\times 10^{-3}$                & 0.1500 & 51  & 1400  \\
    \makecell{2D-SRBA\\OOOO}   & [2048 $\times$ 2560]   & 70       & 0.5 & 13.0 $\times 10^{-3}$ & 0.0375 & 101 & 21 \\
    \makecell{2D-SDBA\\SSOO}   & [1024 $\times$ 512]    & 91 - 1240 & 0.4 & 13.0 $\times 10^{-3}$ & 0.1030 & 101 & 46 \\
    \makecell{3D-SDBA\\SSOOSS} & [256 $\times$ 128 $\times$ 128] & 91 - 1240         & 0.4 & 36.8 $\times 10^{-3}$                     &  0.4133      & 51  & 620  \\
    \bottomrule
    \end{tabularx}
\end{table}

\subsection{Base Setup}

The datasets capture the time-evolving behavior of bubbles and droplets subjected to an external shock wave. The computational domain is partitioned into pre-shock and post-shock regions. Once the shock wave interacts with the bubble or droplet, the resulting post-shock flow largely governs their deformation and subsequent dynamics. The shock Mach number ($M_s$) serves as an influential non-dimensional parameter characterizing this post-shock regime. For each initial condition, the post-shock density, velocity and pressure are set as Dirichlet boundary conditions in the left boundary of the domain.

\subsubsection{Shock-induced Air Bubble Collapse in Water (SABW)}
\label{section:app_sabw_description}
\begin{figure}[h!]
    \centering
    \includegraphics[width=1.0\textwidth]{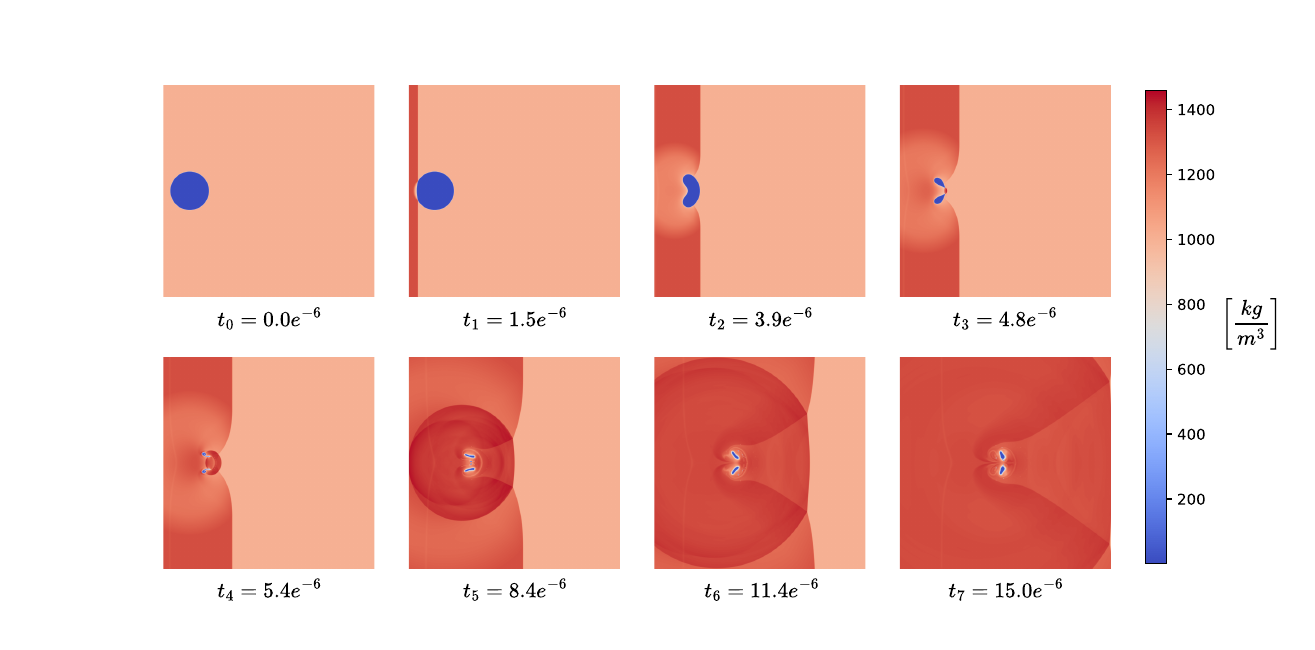}
    \caption{Time evolution of density field for the 2D-SBAW-OOOO dataset.}
    \label{fig:single_bubble_evolving_dynamics}
\end{figure}

In a multiphase flow, one of the key important fields is the evolution of the density. Here, we examine its evolution when an external shock wave interacts with a cylindrical air bubble in water, as shown in Figure \ref{fig:single_bubble_evolving_dynamics}. At $t=1.5 \ \mu s$, the shock wave reaches the bubble, causing the air inside to accelerate due to the transmitted shock. Subsequently, a water jet begins to penetrate the bubble from the upstream side, as observed at $t= 3.9 \ \mu s$. By $t = 4.8 \ \mu s$, this jet impacts the downstream interface of the bubble, generating a localized peak pressure in the domain, commonly referred to as the water hammer effect. This phenomenon produces a radially propagating shock wave and leads to the breakup of the bubble into two fragments in the two-dimensional cylindrical case, or into a toroidal structure in the three-dimensional spherical case. The resulting fragments undergo further breakup under the influence of the radial shock, producing additional pressure waves, as seen at $t = 8.4 \ \mu s$. Eventually, the fragmented bubble is advected downstream with the flow.

The post-shock properties for the SABW configuration are determined using the normal shock relations \citep{le2016noble}, as given in Equation \ref{eq:shock_relations_sabw}. As discussed in the base setup, the domain is divided into pre- and post-shock regions; in the case of SABW, the fluid in the domain is water, and thus the properties of water are used to compute the pre-shock values. In this formulation, $M_s$ denotes the shock Mach number, while $c_{\text{pre}}$ represents the speed of sound in the pre-shock region, calculated using the stiffened gas equation. The pre-shock density ($\rho_{\text{pre}}$) and pressure ($p_{\text{pre}}$) are specified as 1000 kg/m$^3$ and 1.0 × 10$^5$ Pa, respectively. Additionally, for water, the background pressure ($p_\infty$) and the ratio of specific heats ($\gamma$) are taken as 60 × 10$^5$ Pa and 4.4. The density of the bubble is initialized to 1.0 kg/m$^3$ and the pressure inside is initialized to 1.0 × 10$^5$ Pa.

\begin{equation}
    \label{eq:shock_relations_sabw}
    \begin{aligned}
    c_{\text{pre}} &= \sqrt{\gamma\frac{p_{\text{pre}}+p_\infty}{\rho_{\text{pre}}}}\\
    u_s &= M_s*c_{\text{pre}}\\
    p_{\text{post}} &= (p_{\text{pre}}+p_\infty)\left(1+\frac{2\gamma}{(\gamma+1)}(M_s^2-1)\right)-p_\infty\\ 
    \rho_{\text{post}} &= \rho_{\text{pre}}\frac{(\gamma+1)M_s^2}{2+(\gamma-1)M_s^2}\\
    u_{\text{post}} &= u_s\left(1-\frac{\rho_{\text{pre}}}{\rho_{\text{post}}}\right) 
\end{aligned}
\end{equation}

\begin{figure}[h!]
    \centering
    \includegraphics[width=0.5\textwidth]{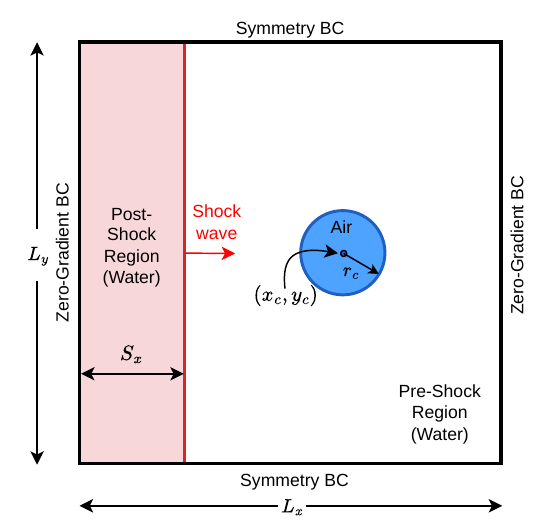}
    \caption{Single air bubble in water setup. The parameter details are presented in Table \ref{tab:single_bubble_setup}. The shock Mach number is 1.72 and the bubble and the surrounding are initialized at atmospheric pressure of 1 bar.}
    \label{fig:single_bubble_setup}
\end{figure}

\begin{figure}[h!]
    \centering
    \includegraphics[width=0.6\textwidth]{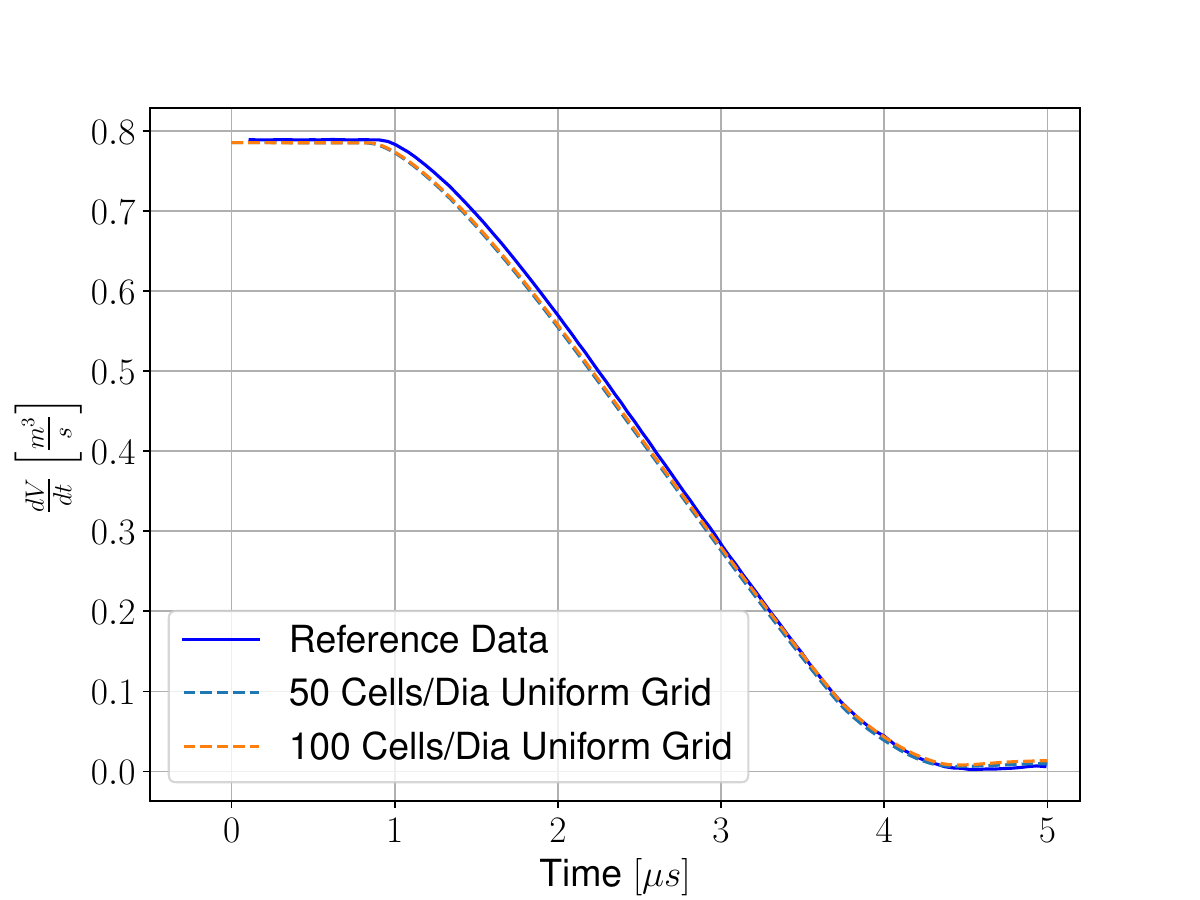}
    \caption{Comparing evolution of rate of change of volume with time for the single air bubble in water with the reference plot shown in \cite{nourgaliev2006adaptive}.}
    \label{fig:single_bubble_comparison}
\end{figure}

The solver was validated on two setups, a single bubble setup and a multi bubble setup. For the single bubble setup, we validate the evolution of the rate of change of volume of air over a given time interval and compare it against the results shown in \cite{nourgaliev2006adaptive}. The setup and the corresponding geometric parameters are shown in Figure \ref{fig:single_bubble_setup} and Table \ref{tab:single_bubble_setup} respectively. The shock Mach number is set to 1.72, the CFL number is 0.4 and the bubble and the surrounding are initialized at atmospheric pressure of 1 bar. The simulation is run for a total duration of 5 $\mu s$. We perform the simulation at two different resolutions of $512\times512$ and $256\times256$ such that there are 100 cells and 50 cells, respectively, in the inital diamter of the bubble. Figure \ref{fig:single_bubble_comparison} shows the rate of change of volume over time for both resolutions and the results are in good agreement with the reference plot sourced from \cite{nourgaliev2006adaptive}.

\begin{table}[h!]
    \caption{Geometric parameters for the single air bubble in water setup.}
    \label{tab:single_bubble_setup}
    \centering
    \begin{tabular}{cc}
        \toprule
        Parameter & Value [m] \\
        \midrule
        $L_x$ & 0.03072 \\
        $L_y$ & 0.03072  \\
        $S_x$ & 0.01040 \\
        $r_c$ & 0.00300   \\
        $(x_c, y_c)$ & (0.01548, 0.001536) \\
        \bottomrule
    \end{tabular}
\end{table}

\begin{figure}[h!]
    \centering
    \includegraphics[width=0.7\textwidth]{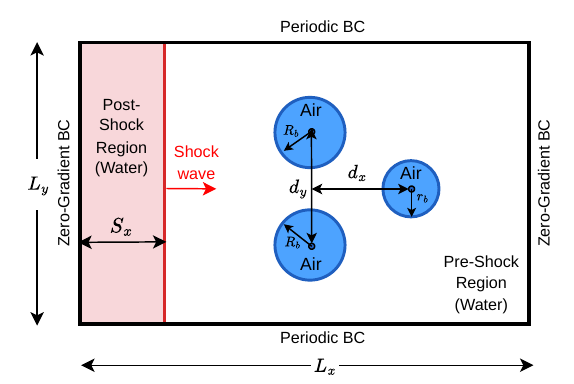}
    \caption{Three staggered air bubble in water setup. The parameter details are presented in Table \ref{tab:3_bubble_setup}. The shock Mach number is set to 1.42 and the bubbles and the surrounding are initialized at atmospheric pressure of 1 bar.}
    \label{fig:3_bubble_setup}
\end{figure}

The next validation case evaluates the solver by comparing the peak pressure in a multi-bubble configuration with a well-established reference \citep{bempedelis2020energy}. As illustrated in Figure \ref{fig:3_bubble_setup}, the setup consists of three staggered air bubbles immersed in water, with the corresponding geometric parameters listed in Table \ref{tab:3_bubble_setup}. The shock Mach number is set to 1.42, the CFL number to 0.5, and both the bubbles and the surrounding medium are initialized at an atmospheric pressure of 1 bar. The simulation is carried out over a total duration of $1.5 \mu s$.

Validation is achieved by comparing the temporal evolution of the peak pressure in the domain with the results reported in \cite{bempedelis2020energy}. The comparison shows good agreement with the reference, with the solver accurately capturing the key events over time. The slight discrepancy observed in the maximum peak pressure throughout the simulation is attributed to differences in interface treatment: the reference employs a level-set method, whereas the present solver uses a diffuse interface approach, which leads to some smearing of peak values.

\begin{table}[h!]
    \caption{Geometric parameters for the staggered three air bubbles in water setup.}
    \label{tab:3_bubble_setup}
    \centering
    \begin{tabular}{cc}
        \toprule
        Parameter & Value [m] \\
        \midrule
        $L_x$ & 0.00350 \\
        $L_y$ & 0.00200  \\
        $S_x$ & 0.00015 \\
        $R_b$ & 0.00050   \\
        $r_b$ & 0.00020  \\
        $(d_x, d_y)$ & (0.00050,0.00150) \\
        \bottomrule
    \end{tabular}
\end{table}

\begin{figure}[h!]
    \centering
    \includegraphics[width=0.6\textwidth]{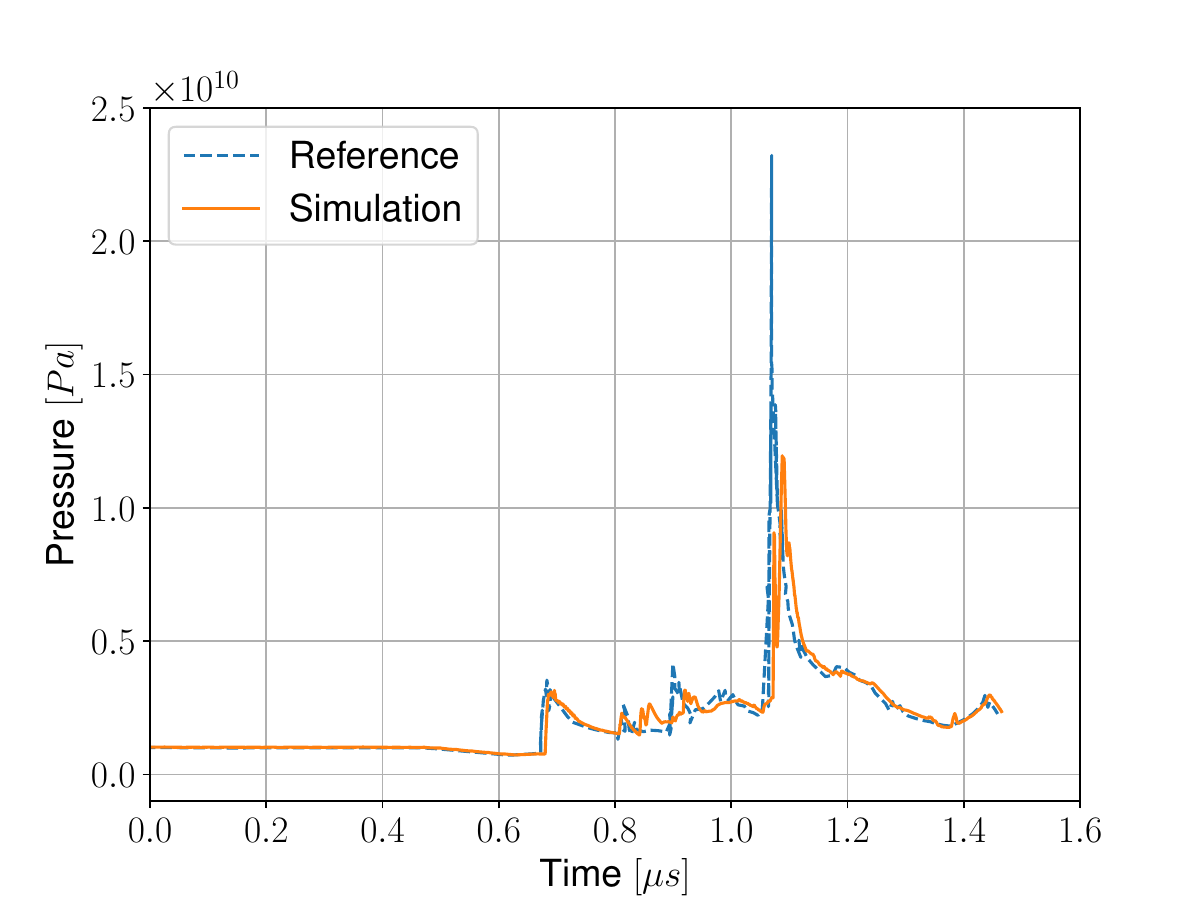}
    \caption{Peak pressure variation with time for the staggered three air bubble in water. The resolution for the simulation is $896\times512$.}
    \label{fig:3_bubble_comparison}
\end{figure}

\begin{figure}[h!]
    \centering
    \includegraphics[width=1.0\textwidth]{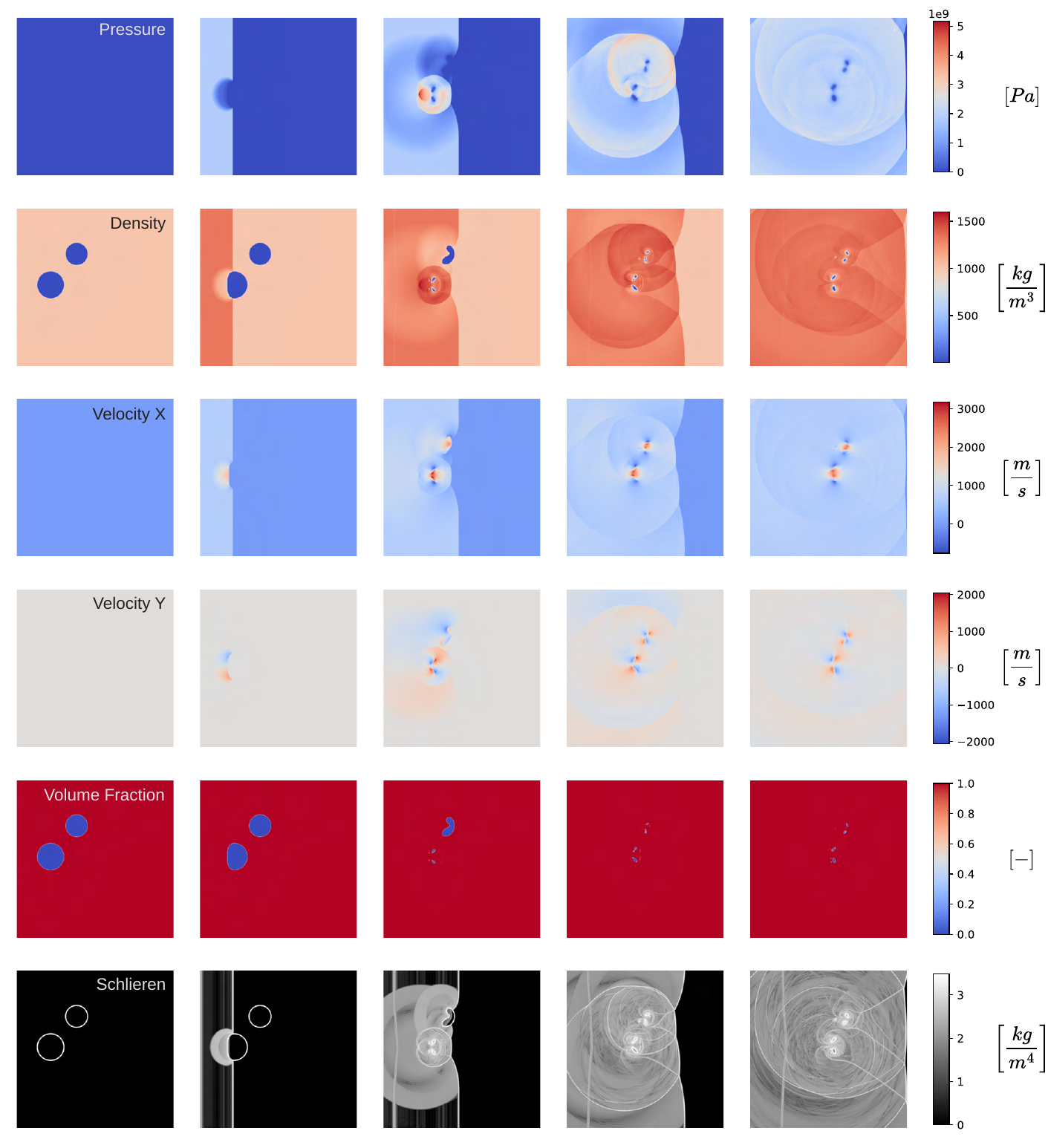}
    \caption{Uniformly spaced time snapshots of Shock-induced Air Bubble collapse in Water (SABW) at shock Mach number 1.70, with open boundary conditions on all sides (OOOO), showing the evolution of pressure, density, x-velocity, y-velocity, volume fraction, and schlieren from $t_0=0.0 \ \mu s$ to $t_{\textit{end}}=15 \ \mu s$.}
    \label{fig:2d_sabw_oooo_6fields}
\end{figure}

\begin{figure}[h!]
    \centering
    \includegraphics[width=1.0\textwidth]{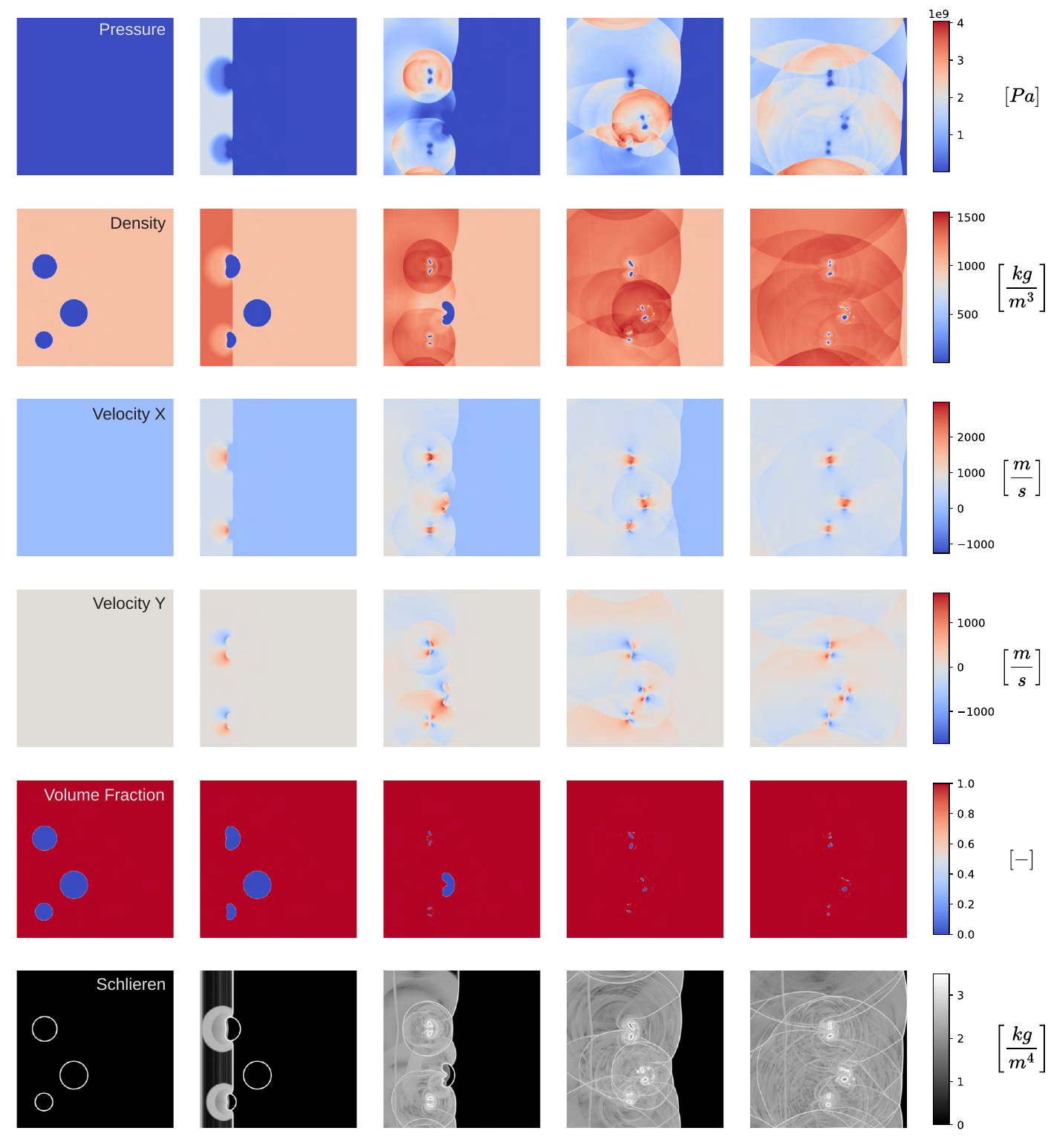}
    \caption{Uniformly spaced time snapshots of Shock-induced Air Bubble collapse in Water (SABW) at shock Mach number 1.70, with North–South symmetric and East–West open boundary conditions (SSOO), showing the evolution of pressure, density, x-velocity, y-velocity, volume fraction, and schlieren from $t_0=0.0 \ \mu s$ to $t_{\textit{end}}=15 \ \mu s$.}
    \label{fig:2d_sabw_ssoo_6fields}
\end{figure}

\begin{figure}[h!]
    \centering
    \includegraphics[width=0.95\textwidth]{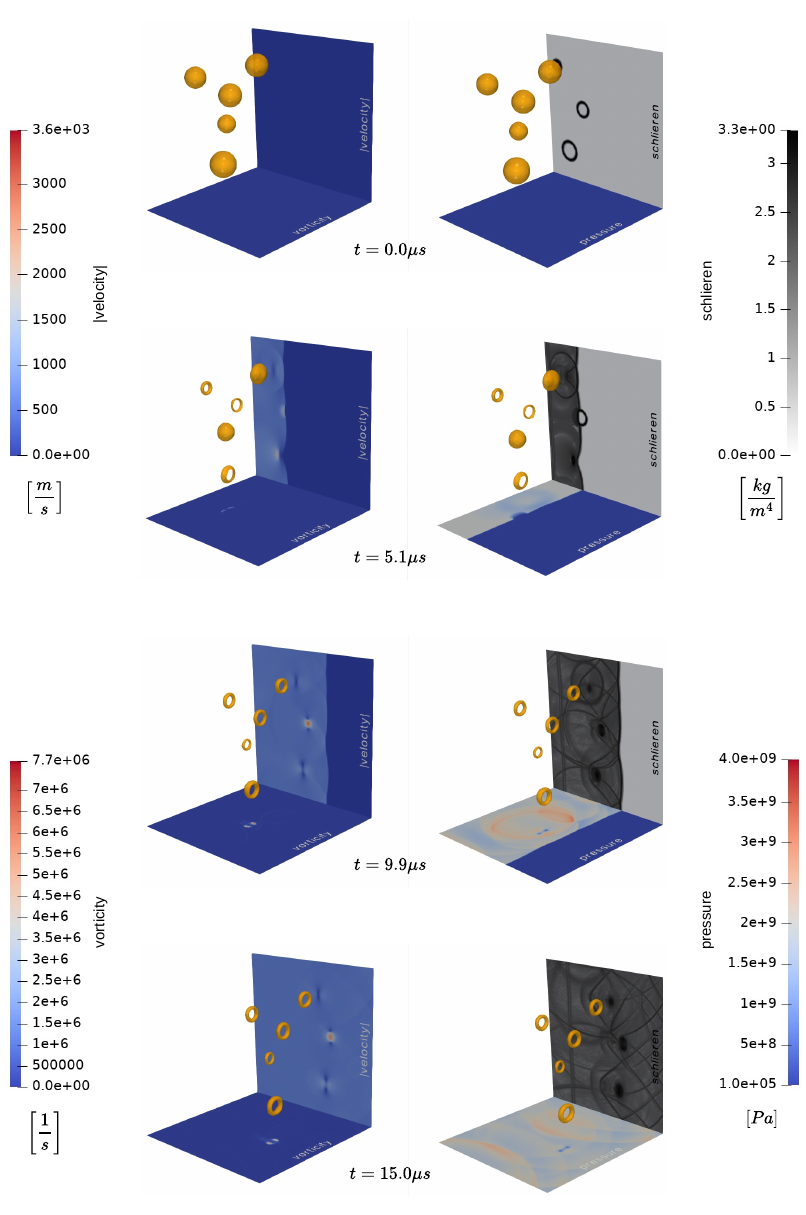}
    \caption{Uniformly spaced time snapshots of a 3D shock-induced air bubble collapse in water (SABW) at a shock Mach number of 1.70. The configuration employs symmetry boundary conditions in the North–South and Top–Bottom directions and open boundaries in the East–West direction (SSOOSS). The figure illustrates the temporal evolution from $t_0 = 0.0 \  \mu s$ to $t_{\textit{end}} = 15 \ \mu s$, showing pressure–vorticity fields on the horizontal plane, schlieren–velocity magnitude on the vertical plane, and a yellow density iso-contour of the bubbles.}
    \label{fig:3d_sabw_ssooss_4fields}
\end{figure}

\clearpage
\subsubsection{Shock-induced Droplet Breakup in Air (SDBA)}
\label{section:app_sdba_description}
The shock induced droplet breakup datasets constitute two distinct breakup modes, namely Rayleigh Taylor Piercing (RTP) and Shear Induced Entrainmet (SIE). The transition between these modes is primarily determined by the Weber number (We), which represents the ratio of aerodynamic forces to surface tension.

\begin{align}
    We=\frac{\rho_{\text{post}} u_{\text{post}}^2D}{\sigma}
\end{align}

In the above equation, $D$ is the droplet diameter and $\sigma$ is the surface tension coefficient. 

When a shock wave impacts a droplet, the initial response—largely independent of the Weber number—is a deformation phase in which the droplet flattens. This deformation arises from a non-uniform pressure distribution around its surface. The subsequent stage is the actual breakup, which is strongly governed by the Weber number. The RTP breakup regime typically occurs for Weber numbers between roughly 10 and 100, whereas the SIE regime dominates at much higher values (We > 1000), with intermediate Weber numbers representing a transition between the two modes \citep{theofanous2008physics}.

In the SIE regime, breakup is driven mainly by strong shear forces acting along the droplet surface. After the droplet has flattened, shear-induced disturbances emerge near the equator; these instabilities originate near the droplet equator after the droplet has flattened out in the first phase and are advected along the droplet surface. As the relative velocity between the droplet and the surrounding gas increases, these disturbances grow due to Kelvin–Helmholtz instability, eventually stripping liquid from the droplet and producing fine droplets downstream. In contrast, in the RTP regime, relatively stronger surface tension suppresses the growth of such shear instabilities, maintaining a smoother interface. As deformation progresses, the upstream side of the droplet becomes concave as the surrounding gas penetrates and pierces the liquid. Unlike RTP, the SIE regime is characterized by a continuous and gradual loss of mass, often resulting in a mist of droplets downstream \citep{winter2019numerical}. 

For determining the post-shock properties for the SDBA setup, we use the shock relations as shown in Equation \ref{eq:shock_relations_sabw} to compute the flow properties behind the shock \citep{anderson1990modern}.

\begin{equation}
    \label{eq:shock_relations_sdba}
    \begin{aligned}
        u_s &= M_s \cdot c_{\text{pre}} \\[3pt]
        u_{\text{pre},\text{rel}} &= -u_s \\[3pt]
        u_{\text{pre}} &= u_{\text{pre},\text{rel}} + u_s \\[3pt]
        T_{\text{post}} &= T_{\text{pre}} \left(1 + \frac{2 \gamma \left(M_s^2 - 1\right)}{\gamma + 1}\right) 
                \left(\frac{2 + (\gamma - 1)M_s^2}{(\gamma + 1)M_s^2}\right) \\[3pt]
        c_{\text{post}} &= \sqrt{\gamma \cdot R \cdot T_{\text{post}}} \\[3pt]
        M_{f,\text{post},\text{rel}} &= \sqrt{\frac{1 + \tfrac{\gamma - 1}{2} M_s^2}{\gamma M_s^2 - \tfrac{\gamma - 1}{2}}} \\[3pt]
        u_{\text{post},\text{rel}} &= M_{f,\text{post},\text{rel}} \cdot c_{\text{post}} \\[3pt]
        u_{\text{post}} &= u_s - u_{\text{post},\text{rel}} \\[3pt]
        \rho_{\text{post}} &= \rho_{\text{pre}} \cdot \frac{(\gamma+1)M_s^2}{2 + (\gamma - 1)M_s^2} \\[3pt]
        p_{\text{post}} &= p_{\text{pre}} \left(1 + \frac{2 \gamma \left(M_s^2 - 1\right)}{\gamma + 1}\right)
    \end{aligned}
\end{equation}

In the above equation set, $M_f$ is the post-shock flow Mach number, $T$ is the temperature, $c$ is the speed of sound, $\gamma = \frac{c_p}{c_v}$ is the ratio of specific heat capacities set to 1.4, and $R$ is the specific gas constant, set to 287 J/kg K. The pre-shock density, $\rho_{\text{pre}}$, = 1.2 kg/m$^3$, the pre-shock pressure, $p_{\text{pre}}$, = 101325 Pa and the pre-shock temperature, $T_{\text{pre}}$ = 300 K. The density of the droplet is 1000 kg/m$^3$ and pressure initialized inside the droplet is 101325 Pa.

The solver was validated for the droplet scenario by comparing the center of mass displacement with the expected trend as shown in \cite{meng2018numerical} when a 2D-axissymmetric spherical droplet whose center is placed on the x-axis interacts with a shockwave of Mach number 1.20.

\begin{figure}[h!]
    \centering
    \includegraphics[width=0.6\textwidth]{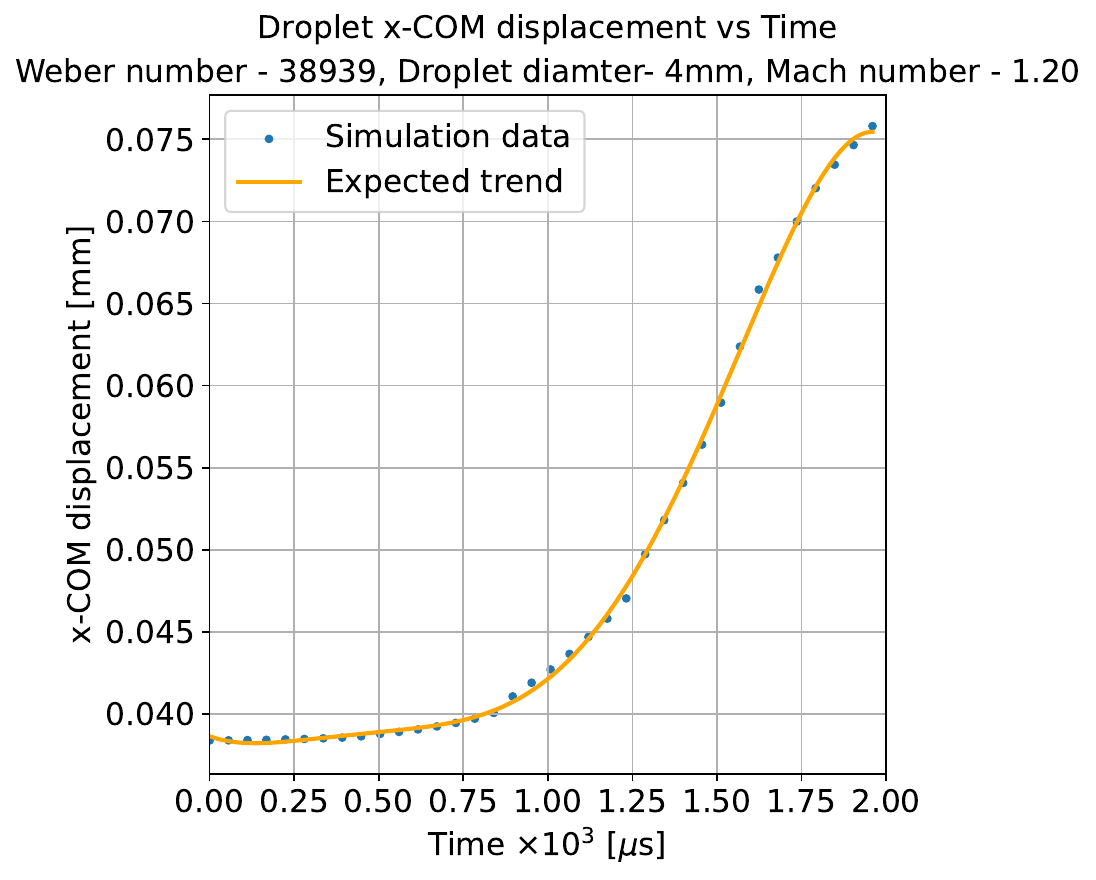}
    \caption{Center-of-mass displacement in x-axis when a 4-mm droplet interacts with a shockwave of Mach 1.2}
    \label{fig:droplet_com_comparison}
\end{figure}

\begin{figure}[h!]
    \centering
    \includegraphics[width=1.0\textwidth]{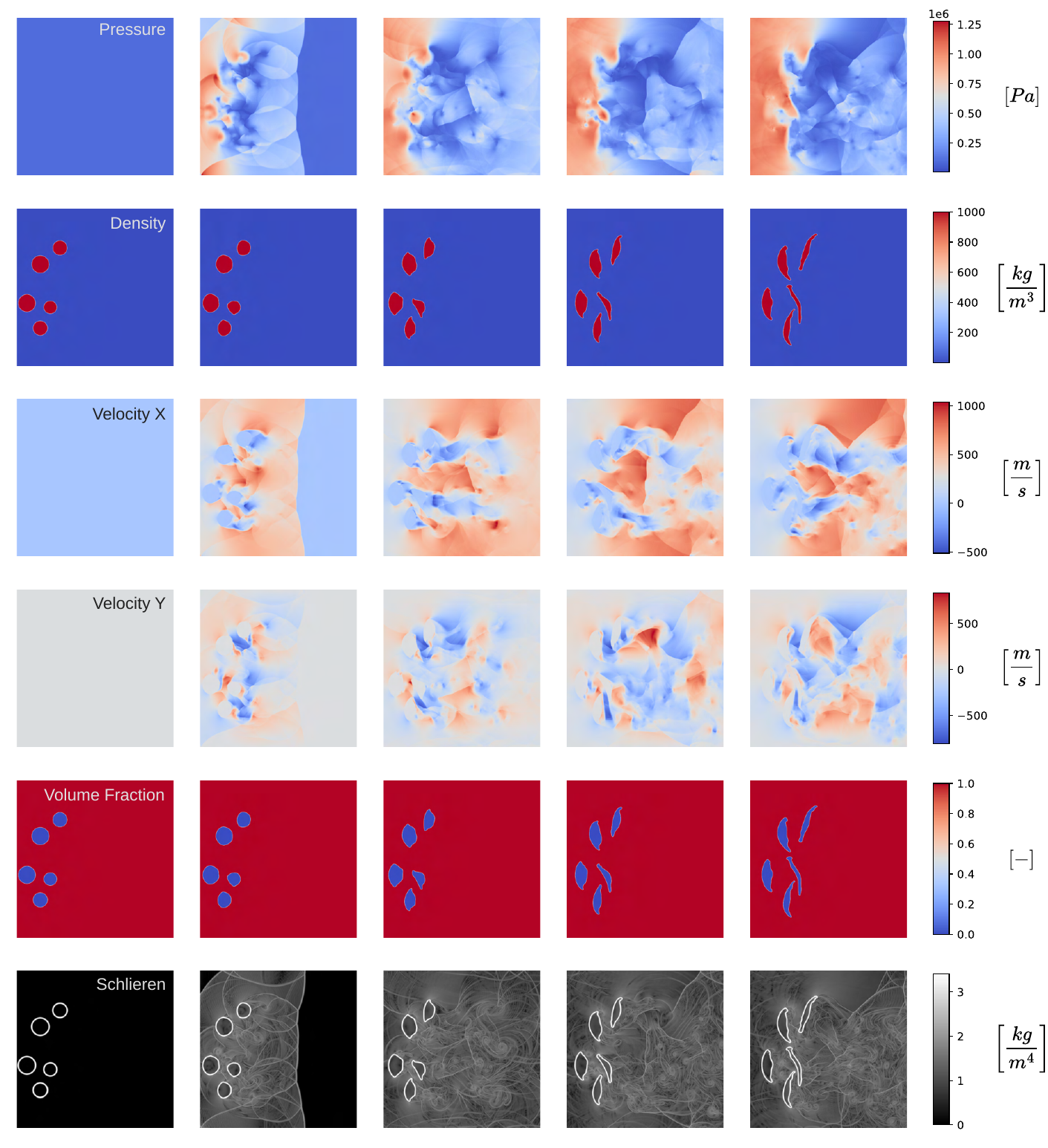}
    \caption{Uniformly spaced time snapshots of Shock-induced Droplet Breakup in Air (SDBA) at shock Mach number 2.0, all droplets in the RTP regime (Weber < 100) having a surface tension coefficient $\sigma=60.514$, under North–South symmetric and East–West open boundary conditions (SSOO), showing the evolution of pressure, density, x-velocity, y-velocity, volume fraction, and schlieren from $t_0=0.0 \ \mu s$ to $t_{\textit{end}}= 215.5 \ \mu s$.}
    \label{fig:2d_sdba_RTP_6fields}
\end{figure}

\begin{figure}[h!]
    \centering
    \includegraphics[width=1.0\textwidth]{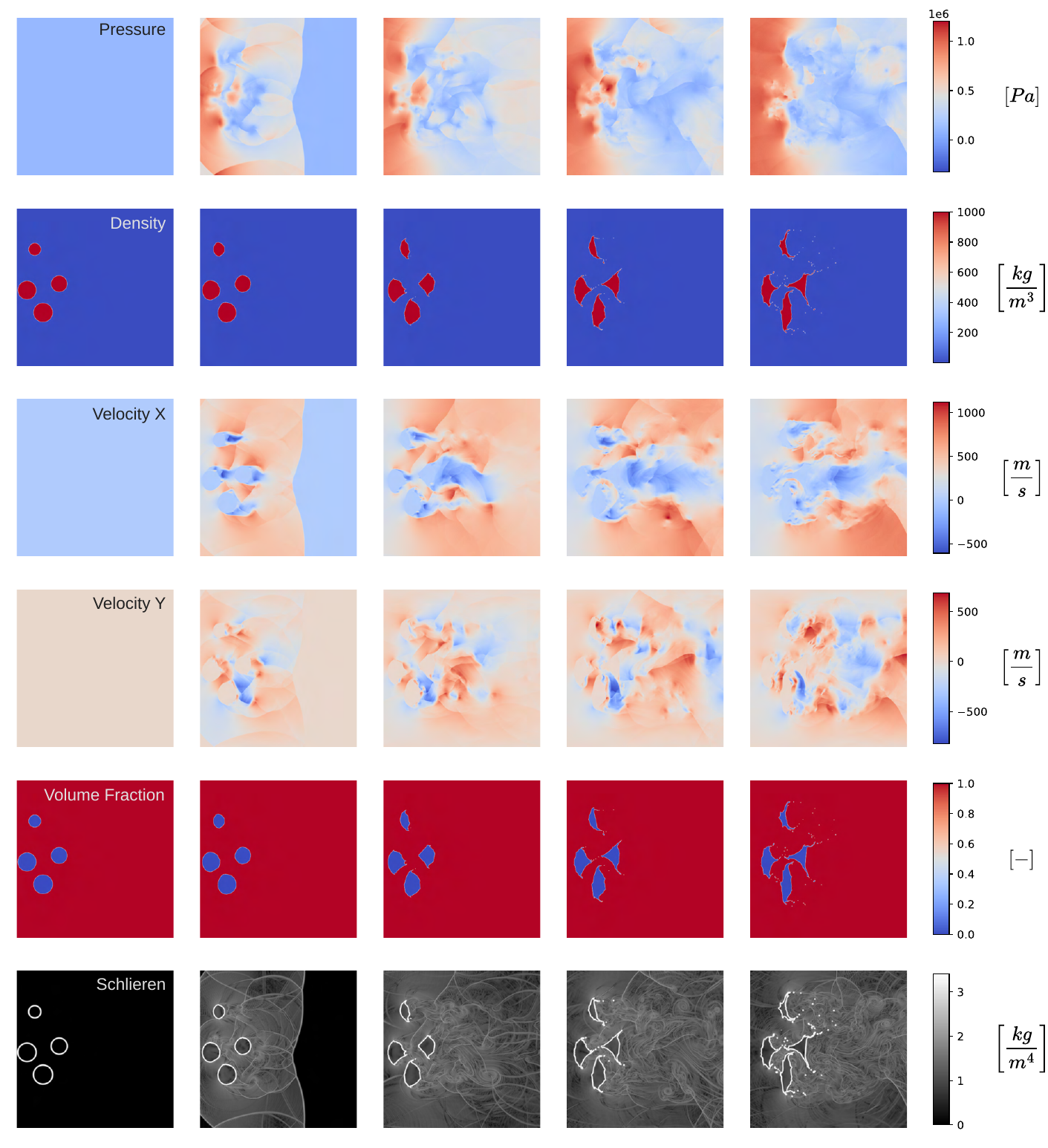}
    \caption{Uniformly spaced time snapshots of Shock-induced Droplet Breakup in Air (SDBA) at shock Mach number 2.0, with all droplets in the SIE regime (Weber > 1000) having a surface tension coefficient $\sigma=0.0719$, under North–South symmetric and East–West open boundary conditions (SSOO), showing the evolution of pressure, density, x-velocity, y-velocity, volume fraction, and schlieren from $t_0=0.0 \ \mu s$ to $t_{\textit{end}}= 215.5 \ \mu s$.}
    \label{fig:2d_sdba_SIE_6fields}
\end{figure}

\begin{figure}[h!]
    \centering
    \includegraphics[width=0.95\textwidth]{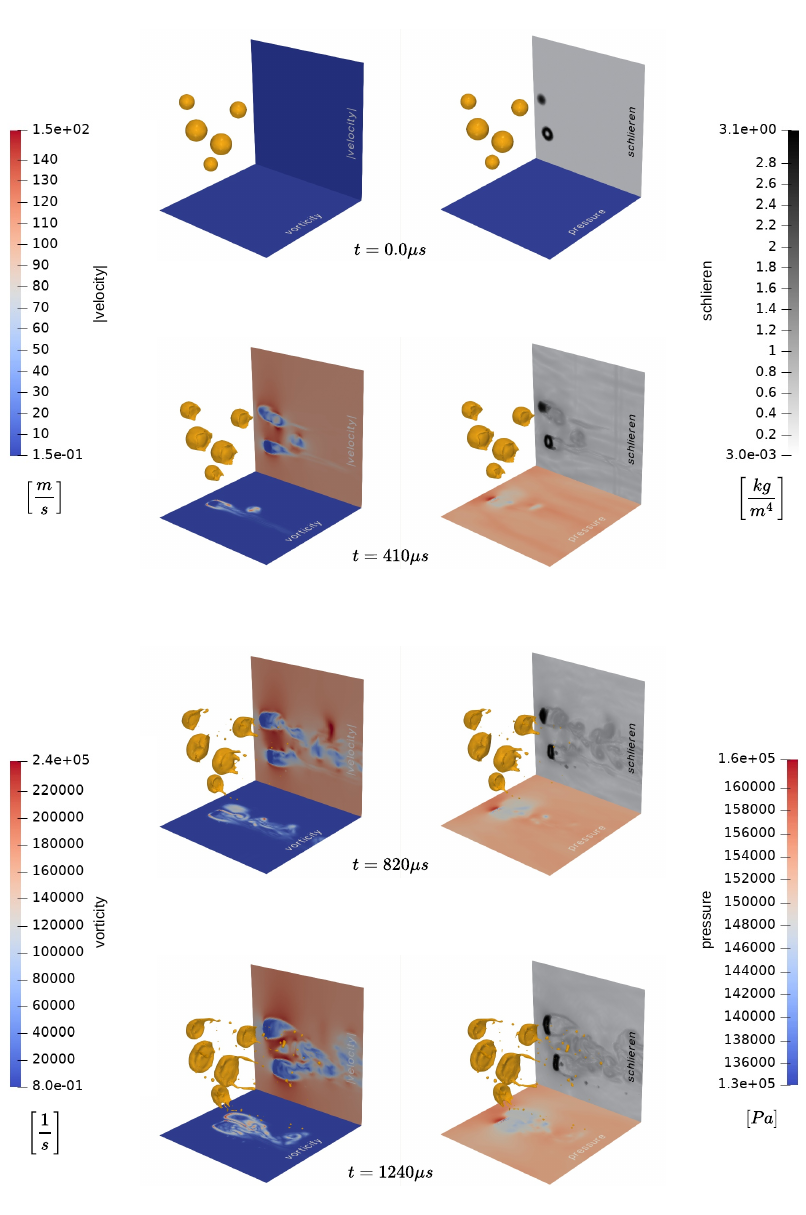}
    \caption{Uniformly spaced time snapshots of a 3D shock-induced droplet breakup in air (SDBA) at a shock Mach number of 1.20, with the breakup mode in the SIE regime. The configuration employs symmetry boundary conditions in the North–South and Top–Bottom directions and open boundaries in the East–West direction (SSOOSS). The figure illustrates the temporal evolution from $t_0 = 0.0\mu s$ to $t_{\textit{end}} = 1240 \ \mu s$, showing pressure–vorticity fields on the horizontal plane, schlieren–velocity magnitude on the vertical plane, and a yellow density iso-contour of the droplets.}
    \label{fig:3d_sdba_sie_ssooss_4fields}
\end{figure}

\subsubsection{Shock-induced R22 Bubble Breakup in Air (SRBA)}
\label{section:app_srba_description}
\begin{figure}[h!]
    \centering
    \includegraphics[width=\textwidth]{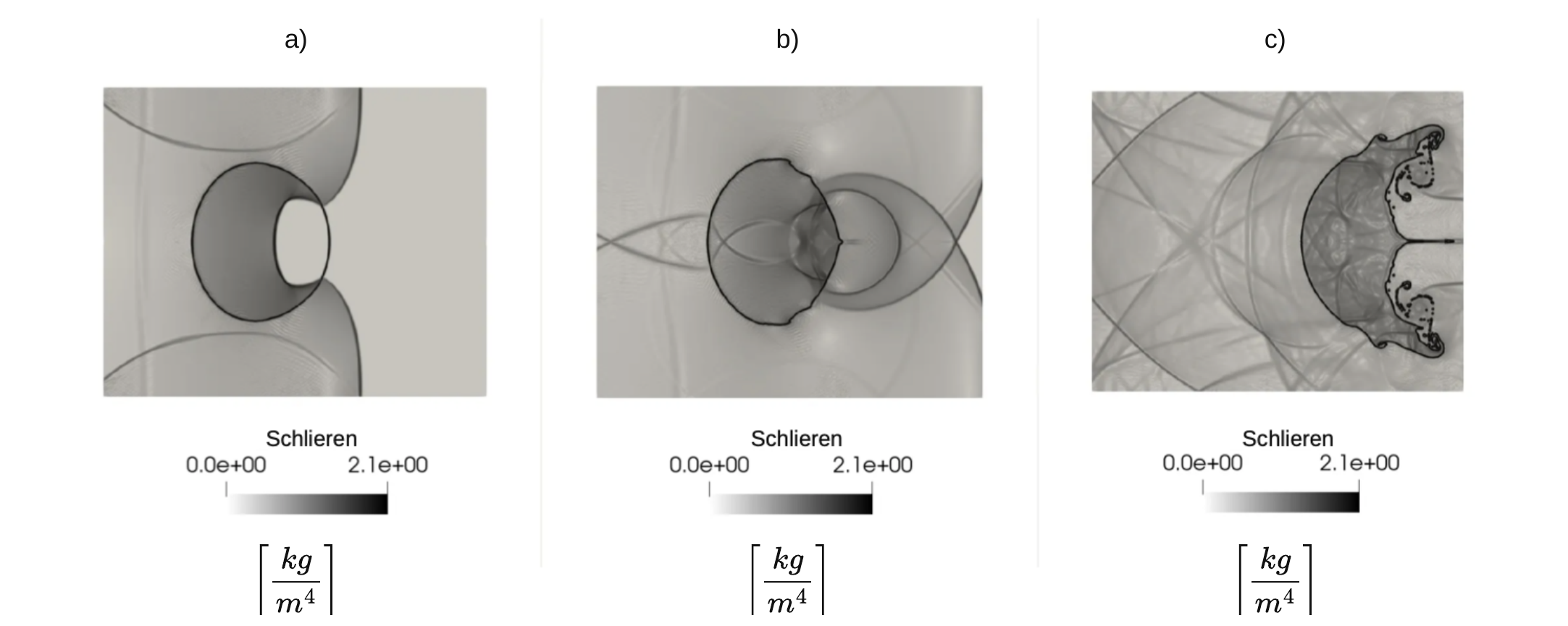}
    \caption{Schlieren images showing the evolution of the interaction of shock with R22 bubble in air at (a)$t = 141.1 \ \mu s$, (b) $ t = 240.5 \ \mu s$ and (c) $t = 640.9 \ \mu s$.}
    \label{fig:r22_schlieren_evolution}
\end{figure}

The dynamics for the SRBA case is notably different from the SABW scenario. Since the R22 bubble is denser than the surrounding air, the transmitted shock within the bubble propagates more slowly than the incident shock, as illustrated in Figure \ref{fig:r22_schlieren_evolution}(a). As the transmitted shock converges toward the downstream side of the bubble, a small cusp forms, as seen in Figure \ref{fig:r22_schlieren_evolution}(b), which eventually evolves into a thin R22 jet, shown in Figure \ref{fig:r22_schlieren_evolution}(c). During the same time interval, small vortices develop along the interface and are subsequently amplified as they move downstream of the bubble.

For determining the post-shock properties for SRBA setup, we again use the normal shock relations given in Equation \ref{eq:shock_relations_sabw}. In this case, the medium in the domain is air, and thus the properties of air are used to compute the pre-shock values. The pre-shock density ($\rho_{\text{pre}}$) and pressure ($p_{\text{pre}}$) are specified as 1.2041 kg/m$^3$ and 1.01 $\times$ 10$^5$ Pa, respectively. Furthermore, for air, the background pressure ($p_\infty$) and the ratio of specific heats ($\gamma$) are taken as 0.00 Pa and 1.249. The R22 bubble is initialized with a density of 3.7975 kg/m$^3$.

\begin{figure}[h!]
    \centering
    \includegraphics[width=1.0\textwidth]{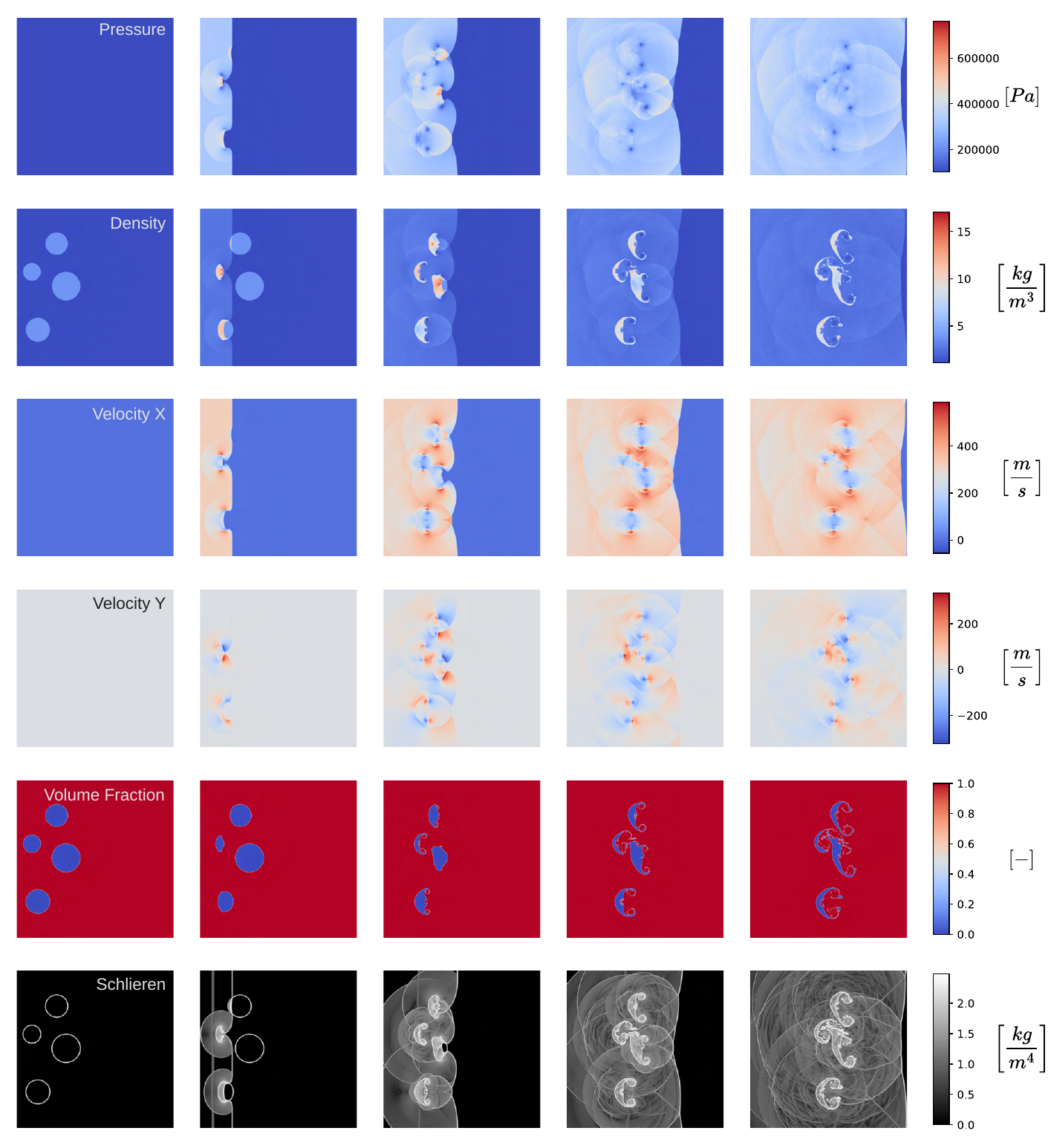}
    \caption{Uniformly spaced time snapshots of Shock-induced R22 Bubble collapse in Water (SABW) at shock Mach number 1.70, with open boundary conditions on all sides (OOOO), showing the evolution of pressure, density, x-velocity, y-velocity, volume fraction, and schlieren from  $t_0=0.0 \ \mu s$ to $t_{\textit{end}}=7.0 \ \mu s$.}
    \label{fig:2d_srba_6fields}
\end{figure}

\clearpage
\subsection{Governing Parameter Coverage}
\label{section:governing_parameter_coverage}
In this section, we provide a detailed description of the curated datasets. To ensure sufficient diversity in the out-of-distribution (OOD) trajectories, we focus on generating various scenarios for four selected datasets: 2D-SABW (OOOO) and 2D-SDBA (SSOO) in two dimensions, and 3D-SABW (SSOOSS) and 3D-SDBA (SSOOSS) in three dimensions.

\subsubsection{2D-SRBA (OOOO) Dataset}

The dataset describes shock-induced interactions between cylindrical
R22 bubbles and the surrounding air, with an R22-to-air
density ratio of approximately $3:1$. All boundaries in this dataset are open. The bubble radii are sampled using a Sobol sequence between [2.0 mm, 3.5 mm].
Table~\ref{tab:srba_dataset} summarizes the dataset
configuration and the parameter distributions used for
training and evaluation.

\begin{table}[h!]
    \centering
    \caption{Dataset configuration and parameter distributions
    for 2D-SRBA (OOOO).}
    \label{tab:srba_dataset}

    \renewcommand{\arraystretch}{1.2}
    \setlength{\tabcolsep}{8pt}

    \begin{tabular}{lccc}
        \toprule
        \textbf{Split} &
        \textbf{Mach number} &
        \textbf{Bubble count} &
        \textbf{Trajectories} \\
        \midrule

        Training
        & $\{1.3, 1.7, 2.1\}$
        & $\{1,2,3,4,5\}$
        & 240 \\

        Testing
        & $\{1.3, 1.7, 2.1\}$
        & $\{1,2,3,4,5\}$
        & 60 \\

        \bottomrule
    \end{tabular}
\end{table}

\subsubsection{2D-SABW (SSOO) Dataset}

The dataset describes shock-induced interactions between cylindrical
air bubbles and the surrounding water, with an air-to-water
density ratio of approximately $1:1000$. The top and bottom boundaries are reflective, while the left and right boundaries are open. The bubble radii are sampled using a Sobol sequence between [2.0 mm, 3.5 mm].
Table~\ref{tab:sabw_ssoo_dataset} summarizes the dataset
configuration and the parameter distributions used for
training and evaluation.

\begin{table}[h!]
    \centering
    \caption{Dataset configuration and parameter distributions
    for 2D-SABW (SSOO).}
    \label{tab:sabw_ssoo_dataset}

    \renewcommand{\arraystretch}{1.2}
    \setlength{\tabcolsep}{8pt}

    \begin{tabular}{lccc}
        \toprule
        \textbf{Split} &
        \textbf{Mach number} &
        \textbf{Bubble count} &
        \textbf{Trajectories} \\
        \midrule

        Training
        & $\{1.3, 1.7, 2.1\}$
        & $\{1,2,3,4,5\}$
        & 240 \\

        Testing
        & $\{1.3, 1.7, 2.1\}$
        & $\{1,2,3,4,5\}$
        & 60 \\

        \bottomrule
    \end{tabular}
\end{table}

\subsubsection{2D-SABW (OOOO) Dataset}

The dataset describes shock-induced interactions between cylindrical
air bubbles and the surrounding water, with an air-to-water
density ratio of approximately $1:1000$. All boundaries in this dataset are open. The bubble radii are sampled using a Sobol sequence between [2.0 mm, 3.5 mm].
Table~\ref{tab:sabw_oooo_dataset} summarizes the dataset
configuration and the parameter distributions used for
training and evaluation.

\begin{table}[h!]
    \centering
    \caption{Dataset configuration and parameter distributions
    for 2D-SABW (OOOO). Curly braces \(\{\}\) denote discrete values, while square brackets \([ ]\) indicate continuous ranges.
    The OOD test sets evaluate generalization to unseen
    parameter values.}
    \label{tab:sabw_oooo_dataset}

    \renewcommand{\arraystretch}{1.2}
    \setlength{\tabcolsep}{8pt}

    \begin{tabular}{lccc}
        \toprule
        \textbf{Split} &
        \textbf{Mach number} &
        \textbf{Bubble count} &
        \textbf{Trajectories} \\
        \midrule

        Training
        & $\{1.3, 1.7, 2.1\}$
        & $\{1,2,3,4,5\}$
        & 240 \\

        Testing
        & $\{1.3, 1.7, 2.1\}$
        & $\{1,2,3,4,5\}$
        & 60 \\

        \midrule

        Interpolation-Mach
        & $[1.3, 2.1 ]$
        & $\{1,2,3,4,5\}$
        & 60 \\

        Extrapolation-Mach
        & $\{1.1, 2.3\}$
        & $\{1,2,3,4,5\}$
        & 20 \\

        \makecell[l]{Extrapolation-Mach\\and Bubble count}
        & $\{1.1, 2.3\}$
        & $\{6,7\}$
        & 20 \\

        \bottomrule
    \end{tabular}
\end{table}

\subsubsection{2D-SDBA (SSOO) Dataset}

The dataset captures shock-induced interactions between cylindrical water droplets and the surrounding air, characterized by an air-to-water density ratio of approximately 1:1000. Reflective boundary conditions are imposed at the top and bottom boundaries, while open boundary conditions are applied at the left and right boundaries. The droplet radii are sampled using a Sobol sequence over the range of 2.0–3.5 mm. To capture both the Rayleigh–Taylor Piercing (RTP) and Shear-Induced Entrainment (SIE) breakup regimes, the Weber numbers are sampled from two distinct ranges, [20, 60] and [20,000, 60,000], respectively. Sobol sampling is also employed for the Weber numbers to ensure a systematic coverage of the parameter space.
Table~\ref{tab:sdba_ssoo_dataset} summarizes the dataset
configuration and the parameter distributions used for
training and evaluation.

\begin{table*}[h!]
    \centering
    \caption{Dataset configuration and parameter distributions
    for 2D-SDBA (SSOO). Curly braces \(\{\}\) denote discrete values, while square brackets \([ ]\) indicate continuous ranges.
    The OOD test sets evaluate generalization to unseen
    parameter values.}
    \label{tab:sdba_ssoo_dataset}

    \renewcommand{\arraystretch}{1.2}
    \setlength{\tabcolsep}{8pt}

    \begin{tabular}{lccc}
        \toprule
        \textbf{Split} &
        \textbf{Mach number} &
        \textbf{Droplet count} &
        \textbf{Trajectories} \\
        \midrule

        Training
        & \multirow{2}{*}{\makecell[c]{
            $\{1.2, 1.6, 2.0,$ \\
            $2.4, 2.8, 3.2\}$
        }}
        & $\{1,2,3,4,5\}$
        & 240 \\

        Testing
        &
        & $\{1,2,3,4,5\}$
        & 60 \\

        \midrule

        Interpolation-Mach
        & $[1.2, 3.2]$
        & $\{1,2,3,4,5\}$
        & 60 \\

        Extrapolation-Mach
        & $\{1.05, 3.35\}$
        & $\{1,2,3,4,5\}$
        & 20 \\

        \bottomrule
    \end{tabular}
\end{table*}

\subsubsection{3D-SABW (SSOOSS) Dataset}

The dataset describes shock-induced interactions between
spherical air bubbles and the surrounding water, with an air-to-water
density ratio of approximately $1:1000$. The left and right boundaries are open, while the other boundaries are reflecting. The bubble radii are sampled using a Sobol sequence between [2.0 mm, 3.5 mm].
Table~\ref{tab:3d_sabw_ssooss_dataset} summarizes the dataset
configuration and the parameter distributions used for
training and evaluation.

\begin{table}[h!]
    \centering
    \caption{Dataset configuration and parameter distributions
    for 3D-SABW (SSOOSS). Curly braces \(\{\}\) denote discrete values, while square brackets \([ ]\) indicate continuous ranges.
    The OOD test sets evaluate generalization to unseen
    parameter values.}
    \label{tab:3d_sabw_ssooss_dataset}

    \renewcommand{\arraystretch}{1.2}
    \setlength{\tabcolsep}{8pt}

    \begin{tabular}{lccc}
        \toprule
        \textbf{Split} &
        \textbf{Mach number} &
        \textbf{Bubble count} &
        \textbf{Trajectories} \\
        \midrule

        Training
        & $\{1.3, 1.7, 2.1\}$
        & $\{1, 3, 5\}$
        & 144 \\

        Testing
        & $\{1.3, 1.7, 2.1\}$
        & $\{1, 3, 5\}$
        & 36 \\

        \midrule

        Interpolation-Mach
        & $[1.3, 2.1 ]$
        & $\{1,2,3,4,5\}$
        & 20 \\

        Extrapolation-Mach
        & $\{1.1, 2.3\}$
        & $\{1,2,3,4,5\}$
        & 20 \\

        \bottomrule
    \end{tabular}
\end{table}

\subsubsection{3D-SDBA (SSOOSS) Dataset}

The dataset describes shock-induced interactions between
spherical droplets and the surrounding air, with an air-to-water
density ratio of approximately $1:1000$. The left and right boundaries are open, while the other boundaries are reflecting. The droplet radii are sampled using a Sobol sequence between [2.0 mm, 3.5 mm]. To capture both the RTP and SIE breakup regimes, Weber numbers are sampled from two distinct intervals: [20, 60] and [20,000, 60,000]. Sobol sampling is employed within each interval to ensure systematic and well-distributed coverage of the parameter space.
Table~\ref{tab:3d_sdba_ssooss_dataset} summarizes the dataset
configuration and the parameter distributions used for
training and evaluation.

\begin{table*}[h!]
    \centering
    \caption{Dataset configuration and parameter distributions
    for 3D-SDBA (SSOOSS). Curly braces \(\{\}\) denote discrete values, while square brackets \([ ]\) indicate continuous ranges.
    The OOD test sets evaluate generalization to unseen
    parameter values.}
    \label{tab:3d_sdba_ssooss_dataset}

    \renewcommand{\arraystretch}{1.2}
    \setlength{\tabcolsep}{8pt}

    \begin{tabular}{lccc}
        \toprule
        \textbf{Split} &
        \textbf{Mach number} &
        \textbf{Droplet count} &
        \textbf{Trajectories} \\
        \midrule

        Training
        & \multirow{2}{*}{\makecell[c]{
            $\{1.2, 1.6, 2.0,$ \\
            $2.4, 2.8, 3.2\}$
        }}
        & $\{1,3,5\}$
        & 144 \\

        Testing
        &
        & $\{1,3,5\}$
        & 36 \\

        \midrule

        Interpolation-Mach
        & $[1.2, 3.2]$
        & $\{1,2,3,4,5\}$
        & 20 \\

        Extrapolation-Mach
        & $\{1.05, 3.35\}$
        & $\{1,2,3,4,5\}$
        & 16 \\

        \bottomrule
    \end{tabular}
\end{table*}

\subsection{Downsampling of Datasets}
Direct simulations on coarse grids are often of limited value because key phenomena, such as boundary layers or interface deformation, are poorly resolved, while training surrogate models on high-resolution data remains computationally challenging. A common workaround is to generate datasets at high resolution and then downsample them to a coarser grid by averaging primitive variables, such as velocity or pressure over neighboring cells. However, this naive averaging is not appropriate for compressible flows, since it does not preserve key conserved quantities such as momentum and energy. In this section, we present a conservative downsampling strategy that ensures conservation of total mass, momentum and energy when downsampling the dataset.

\subsubsection{Methodology}

When generating higher-resolution datasets with ALPACA, we also store the densities of each individual phase. Our methodology is designed to be general and applicable to any multiphase solver, which should, in principle, provide access to the densities of each phase. As all datasets in this work involve two phases, we save the phasic densities in addition to the mixed density field. The use of these phasic densities in the downsampling process is described in the following paragraphs.

Another key property of multiphase simulation is the volume fraction, $\alpha$, defining the share of each phase in a given cell. Here, $\alpha_1$ defines the volume fraction of phase-1 and $\alpha_2$ for phase-2 can be computed as $\alpha_2 = 1 - \alpha_1$. In cells where only one phase is present, $\alpha$ is either 0 or 1. However, $\alpha$ $\in$ (0, 1) in cells intersected by the interface.

When downsampling the domain, these three equations need to be satisfied. The mass conservation states that the total mass in the domain must remain conserved during downsampling. Hence, the total mass in the fine grid cells must equal the total mass of the coarse grid cell, leading to Equation \ref{eq:mass_conservation} for phase-1 ($\Phi_1$), a similar equation is applicable for phase-2 ($\Phi_2$) as well.
\begin{equation}
    \label{eq:mass_conservation}
    \sum_{i=1}^{\delta^2} \rho_{i,\Phi_1} V_i \alpha_i \overset{!}{=} \hat{\rho}_{I,\Phi_1} \sum_{i=1}^{\delta^2} V_i \alpha_i
\end{equation}

The parameter $\delta$ defines the factor by which the grid is downsampled in each spatial direction and $i$ represents the index of the cells at the finer resolution.  Equation \ref{eq:mass_conservation} can be rearranged to compute the downsampled density of phase-1 in the coarse cell $I$ as shown in Equation \ref{eq:downsampled_density}. Here $\hat{\alpha}_I$ is the downsampled volume fraction of phase-1 in the coarse cell $I$ and is computed as shown in Equation \ref{eq:downsampled_volume_fraction}.

\begin{equation}
    \label{eq:downsampled_density}
    \hat{\rho}_{I,\Phi_1} = \frac{\sum_{i=1}^{\delta^2} \rho_{i,\Phi_1} V_i \alpha_i}{\sum_{i=1}^{\delta^2} V_i \alpha_i} =\frac{1}{\delta^2\hat{\alpha}_I}\sum_{i=1}^{\delta^2} \rho_{i,\Phi_1} \alpha_i
\end{equation}

\begin{equation}
    \label{eq:downsampled_volume_fraction}
    \hat{\alpha}_I = \frac{1}{\delta^2} \sum_{i=1}^{\delta^2} \alpha_i
\end{equation}

To get the mixed density in the coarse cell, we combine the densities of both phases weighted by their volume fractions as shown in Equation \ref{eq:mixed_density}.
\begin{equation}\label{eq:mixed_density}
    \hat{\rho}_I = \hat{\rho}_{I,\Phi_1} \hat{\alpha}_I + \hat{\rho}_{I,\Phi_2} (1 - \hat{\alpha}_I)
\end{equation}

Here we want to emphasize that both conservative and average downsampling lead to the same result for volume fraction and density fields. This is made evident by inserting Equation \ref{eq:downsampled_density} into Equation \ref{eq:mixed_density}. 
\begin{equation}
    \hat{\rho}_I = \hat{\rho}_{I,\Phi_1} \hat{\alpha}_I + \hat{\rho}_{I,\Phi_2} (1 - \hat{\alpha}_I) = \frac{1}{\delta^2} \sum_{i=1}^{\delta^2} \rho_{i,\Phi_1} \alpha_i + \frac{1}{\delta^2} \sum_{i=1}^{\delta^2} \rho_{i,\Phi_2} (1 - \alpha_i) = \frac{1}{\delta^2} \sum_{i=1}^{\delta^2} \rho_i
\end{equation}

To get the downsampled x-velocity field $u_x$, we use the momentum conservation in the x-direction. Similar to mass conservation, the total momentum in the fine grid cells must equal the total momentum in the coarse grid cells as shown in Equation \ref{eq:momentum_conservation} for phase-1, and a similar equation applies for phase-2 and in every Cartesian direction.

\begin{equation}
    \label{eq:momentum_conservation}
    \sum_{i=1}^{\delta^2} \rho_{i,\Phi_1} u_{x,i, \Phi_1} V_i \alpha_i \overset{!}{=} \hat{\rho}_{I,\Phi_1} \hat{u}_{x,I,\Phi_1} \sum_{i=1}^{\delta^2} V_i \alpha_i 
\end{equation}

In the above Equation, $u_{x,i,\Phi_1}$ is the x-component of the velocity of phase-1 for the fine grid cell $i$. At the interface cells, we assume this quantity is equal for both phases and can therefore be written as $u_{x,i}$. Outside the interface, $u_{x,i}$ = $u_{x,i, \Phi_1}  \alpha_i$ +  $u_{x,i, \Phi_2} (1 - \alpha_i)$ = $u_{x,i, \Phi_1}$  for $\alpha_i = 1$. Thus Equation \ref{eq:momentum_conservation} can be rewritten as shown in Equation \ref{eq:momentum_conservation_2}.

\begin{equation}
    \label{eq:momentum_conservation_2}
    \sum_{i=1}^{\delta^2} \rho_{i,\Phi_1} u_{x,i} V_i \alpha_i \overset{!}{=} \hat{\rho}_{I,\Phi_1} \hat{u}_{x,I,\Phi_1} \sum_{i=1}^{\delta^2} V_i \alpha_i 
\end{equation}

From Equation \ref{eq:momentum_conservation_2}, we get the downsampled x-velocity for phase-1 for the coarse grid cell $I$.

\begin{equation}\label{eq:downsampled_velocity}
    \hat{u}_{x,I,\Phi_1} = \frac{\sum_{i=1}^{\delta^2} \rho_{i,\Phi_1} u_{x,i} V_i \alpha_i}{\hat{\rho}_{I,\Phi_1} \sum_{i=1}^{\delta^2} V_i \alpha_i} = \frac{1}{\delta^2 \hat{\rho}_{I,\Phi_1} \hat{\alpha}_I} \sum_{i=1}^{\delta^2} \rho_{i,\Phi_1} u_{x,i} \alpha_i
\end{equation}

To get the mixed momentum of the coarse cell, we combine the momentum of both phases weighted by their volume fractions as shown in Equation \ref{eq:mixed_velocity}. Dividing by the mixed density $\hat{\rho}_I$ gives us the downsampled mixed velocity in x-direction.
\begin{equation}\label{eq:mixed_velocity}
    \hat{u}_{x,I} = \frac{\hat{\rho}_{I,\Phi_1} \hat{u}_{x,I,\Phi_1} \hat{\alpha}_I + \hat{\rho}_{I,\Phi_2} \hat{u}_{x,I,\Phi_2} (1 - \hat{\alpha}_I)}{\hat{\rho}_I}
\end{equation}

To compute the pressure field $\hat{p}$ and the total energy field $\hat{E}$ of the downsampled flow, we need to use the stiffened gas EOS as shown in Equation \ref{eq:eos}, which is used by ALPACA for the generation of the datasets.

By rearranging this equation we can compute the internal energy for phase-1 as shown in Equation \ref{eq:internal_energy}.
\begin{equation}\label{eq:internal_energy}
    e_{i,\Phi_1} = \frac{p_i + \gamma_{\Phi_1} p_{\infty,\Phi_1}}{(\gamma_{\Phi_1} - 1) \rho_{i,\Phi_1}}
\end{equation}

Similar to the assumptions for no slip for the interfacial velocity, we assume that the interface is in mechanical equilibrium and thus pressure is equal for both phases at the interface cells such that $p_{i,\Phi_1}$ = $p_{i,\Phi_2}$ = $p_i$ for these cells. Outside the interface, $p_{i}$ = $p_{i, \Phi_1}  \alpha_i$ +  $p_{i, \Phi_2} (1 - \alpha_i)$ = $p_{i, \Phi_1}$  for $\alpha_i = 1$.
Finally the total energy per unit mass for cell $i$ in phase-1 can be computed by combining the internal and kinetic energy as shown in Equation \ref{eq:total_energy}.
\begin{equation}\label{eq:total_energy}
    E_{i,\Phi_1} = e_{i,\Phi_1} + \frac{1}{2} (u_{x,i}^2 + u_{y,i}^2)
\end{equation}

Here we use the mixed velocities $u_{x,i}$ and $u_{y,i} $, because the velocities per phase are equal as described above. With Equation \ref{eq:total_energy}, we can now downsample the total energy field using the total energy conservation principle: The total energy in the fine grid cells must equal the total energy of the coarse grid cells as shown in Equation \ref{eq:energy_conservation} for phase-1, and a similar equation applies for phase-2 as well.
\begin{equation}\label{eq:energy_conservation}
    \sum_{i=1}^{\delta^2} \rho_{i,\Phi_1} E_{i,\Phi_1} V_i \alpha_i \overset{!}{=} \hat{\rho}_{I,\Phi_1} \hat{E}_{I,\Phi_1} \sum_{i=1}^{\delta^2} V_i \alpha_i
\end{equation}
From this we get the downsampled total energy for phase-1 in coarse cell $I$ as shown in Equation \ref{eq:downsampled_total_energy}.
\begin{equation}\label{eq:downsampled_total_energy}
    \hat{E}_{I,\Phi_1} = \frac{\sum_{i=1}^{\delta^2} \rho_{i,\Phi_1} E_{i,\Phi_1} V_i \alpha_i}{\hat{\rho}_{I,\Phi_1} \sum_{i=1}^{\delta^2} V_i \alpha_i} = \frac{1}{\delta^2 \hat{\rho}_{I,\Phi_1} \hat{\alpha}_I} \sum_{i=1}^{\delta^2} \rho_{i,\Phi_1} E_{i,\Phi_1} \alpha_i
\end{equation}

Similarly to velocity, the mixed total energy can be computed as:
\begin{equation}\label{eq:mixed_total_energy}
    \hat{E}_I = \frac{\hat{\rho}_{I,\Phi_1} \hat{E}_{I,\Phi_1} \hat{\alpha}_I + \hat{\rho}_{I,\Phi_2} \hat{E}_{I,\Phi_2} (1 - \hat{\alpha}_I)}{\hat{\rho}_I}
\end{equation}

We compute the internal energy per phase by rearranging Equation \ref{eq:total_energy} and adapting it to the downsampled values obtained from Equations \ref{eq:downsampled_velocity} and \ref{eq:downsampled_total_energy}.
\begin{equation}\label{eq:downsampled_internal_energy}
    \hat{e}_{I,\Phi_1} = \hat{E}_{I,\Phi_1} - \frac{1}{2} (\hat{u}_{x,I,\Phi_1}^2 + \hat{u}_{y,I,\Phi_1}^2)
\end{equation}

We refer to the EOS to compute the downsampled pressure for cell $I$ in phase-1 as shown in Equation \ref{eq:downsampled_pressure}.
\begin{equation}\label{eq:downsampled_pressure}
    \hat{p}_{I,\Phi_1} = (\gamma_{\Phi_1} - 1) \hat{\rho}_{I,\Phi_1} \hat{e}_{I,\Phi_1} - \gamma_{\Phi_1} p_{\infty,\Phi_1}
\end{equation}
To get the final downsampled pressure $\hat{p}_I$ for the coarse cell $I$, we combine the pressures of both phases weighted by their volume fractions as shown in Equation \ref{eq:mixed_pressure}.
\begin{equation}\label{eq:mixed_pressure}
    \hat{p}_I = \hat{p}_{I,\Phi_1} \hat{\alpha}_I + \hat{p}_{I,\Phi_2} (1 - \hat{\alpha}_I)
\end{equation}

From these downsampled primitive fields we can further compute derived fields such as schlieren or vorticity.

\subsubsection{Results}

To compare the primitive and conservative downsampling methods, we compute the total mass, momentum and energy for the finer-resolution dataset at a particular timestep before  downsampling in the entire domain as shown in Equation \ref{eq:total_mass_momentum_energy_before_ds}. Here $N$ is the total number of cells in the domain.
\begin{equation}
    \label{eq:total_mass_momentum_energy_before_ds}
    M_{tot} = \sum_{i=1}^{N} \rho_i V_i \qquad\qquad P_{tot, x/y/z} = \sum_{i=1}^{N} \rho_i u_{x/y/z, i} V_i \qquad\qquad E_{tot} = \sum_{i=1}^{N} \rho_i E_i V_i
\end{equation}

After applying the average and conservative downsampling methods, we compute the total mass, momentum and energy again as shown in Equation \ref{eq:total_mass_momentum_energy_after_ds}.
\begin{equation}
    \label{eq:total_mass_momentum_energy_after_ds}
    M_{tot, ds} = \sum_{I=1}^{M} \hat{\rho}_I \hat{V}_I \qquad\quad P_{tot, ds, x/y/z} = \sum_{I=1}^{M} \hat{\rho}_I \hat u_{x/y/z, I} \hat{V}_I \qquad\quad E_{tot, ds} = \sum_{I=1}^{M} \hat{\rho}_I \hat{E}_I \hat{V}_I
\end{equation}
In the above equation, $M (= N / \delta^2)$ is the total number of coarse cells in the 2D downsampled domain.

\begin{figure}[h!]
    \centering
    \includegraphics[width=1.0\textwidth]{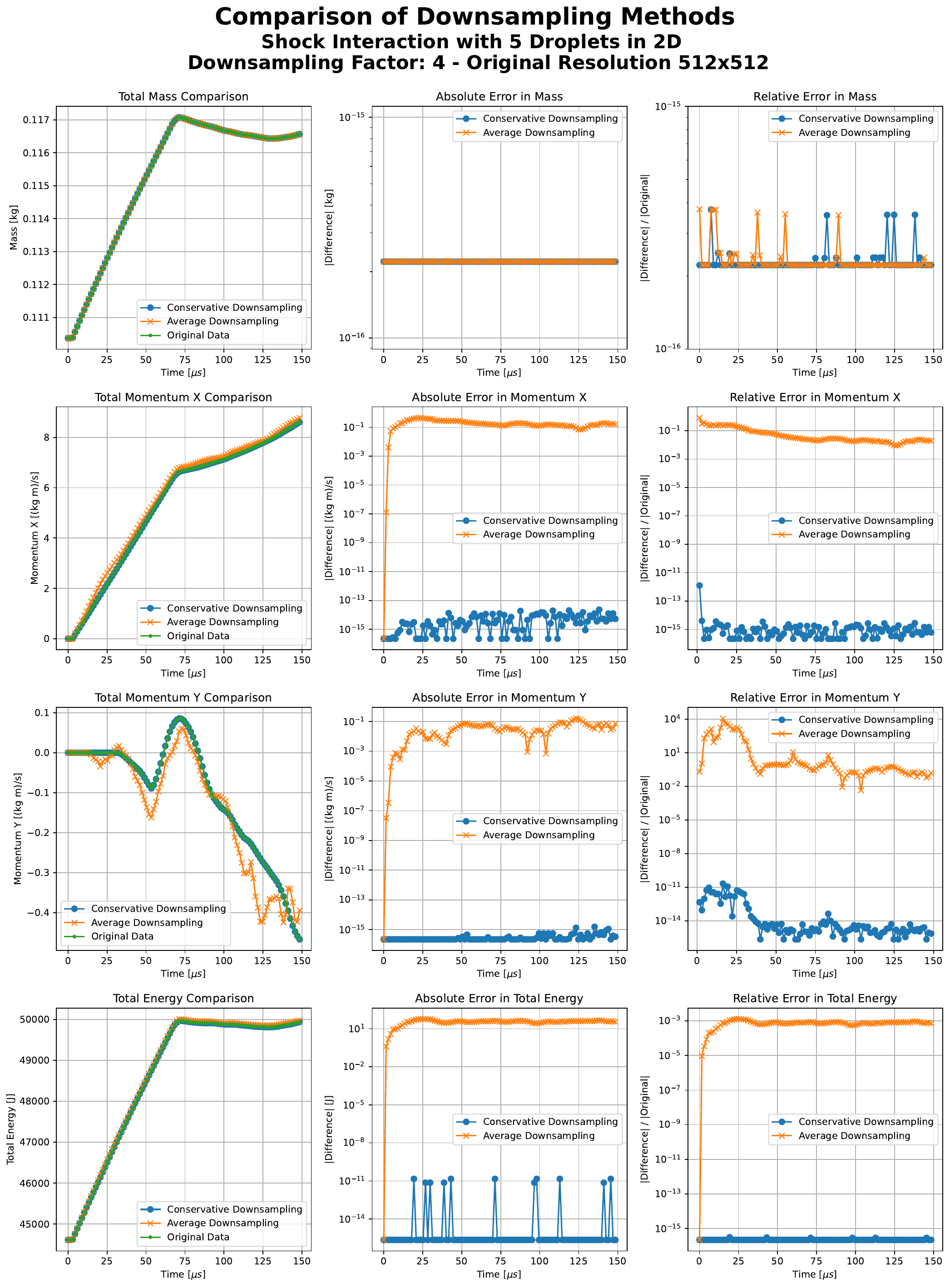}
    \caption{Comparison of mass, momentum and energy in the original grid [512x512] and downsampled grid [128x128] over time for the Shock-induced Droplet Breakup in Air (SDBA) case at shock Mach number 2.40 with 5 droplets. As observed in the y-momentum plot, downsampling by direct averaging leads to significant deviation from the reference, in contrast to the conservative method.}
    \label{fig:downsampling_plot}
\end{figure}

We choose a random trajectory from the 2D-shock-induced droplet breakup in air (SDBA) dataset and compare the results obtained from both downsampling methods. The average method downsamples each primitive field variable independently (e.g. density, velocity) by averaging the neighboring cells. The mass, momentum and energy of the averaged fields are computed after downsampling as shown in Figure \ref{fig:downsampling_plot}. In contrast, the conservative method operates on the conserved quantities such as the momentum ($\rho u V$) and energy ($\rho E V$), ensuring conservation in the downsampled resolution and subsequently extracting the primitive variables of interest such as velocity.
The absolute and relative errors in the conserved quantities obtained from conservative downsampling are in the order of machine precision. On the other hand,  significant discrepancies are observed in the y-momentum, obtained from averaging the y-velocities on the fine grid - with relative errors ranging from $\mathcal{O}(10^{-2})$ - $\mathcal{O}(10)$, highlighting the need for a conservative downsampling approach. Although the relative error in the y-momentum during the initial timesteps is in the order of $\mathcal{O}(10^4)$, this is primarily because the corresponding values on the original grid are nearly zero and division by such small numbers results in large relative errors.

\begin{figure}[h!]
    \centering
    \includegraphics[width=\textwidth]{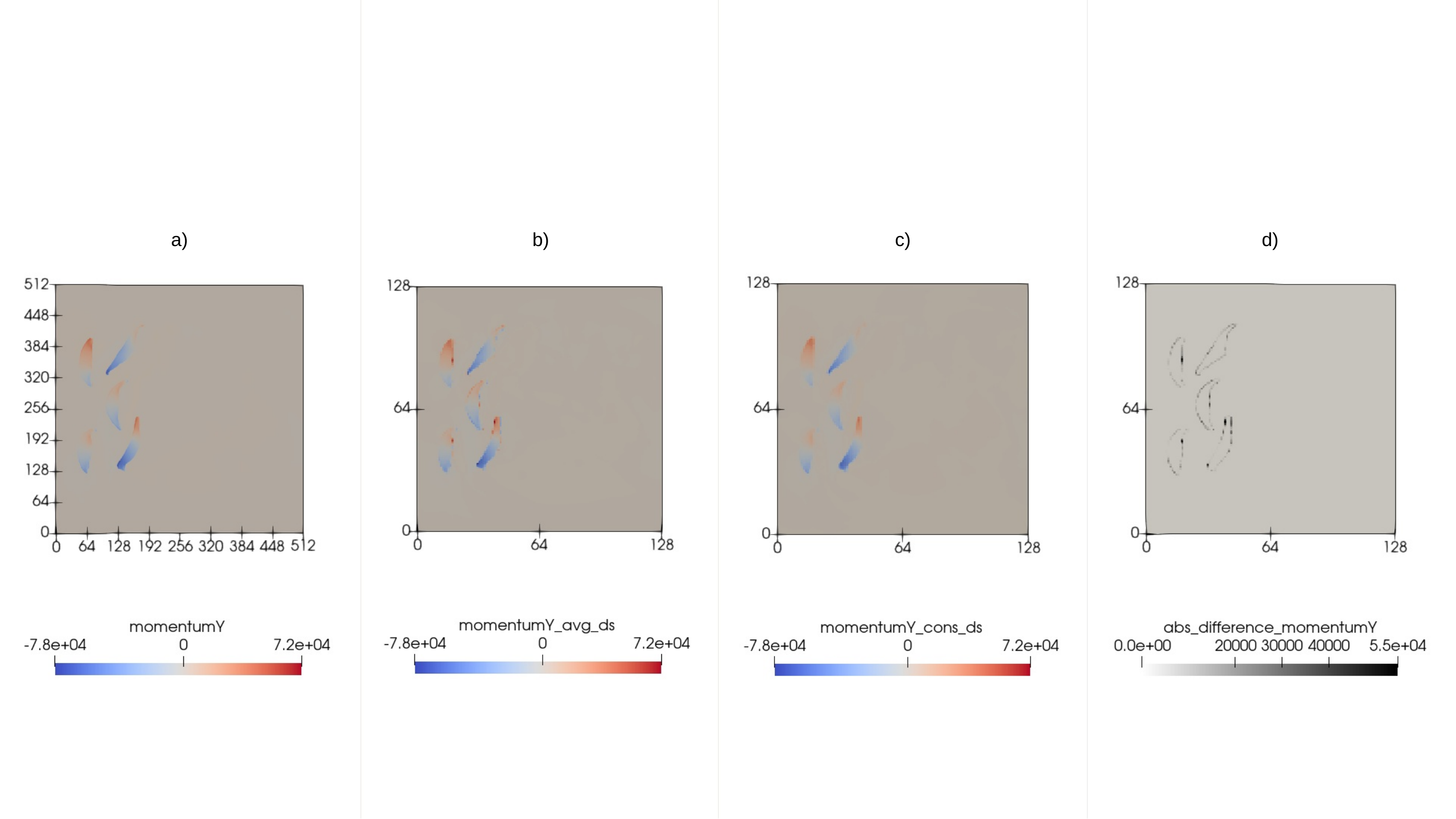}
    \caption{(a) y-momentum at the original resolution [512x512], (b) y-momentum at the downsampled resolution [128x128] obtained by averaging the y-velocities, and (c) y-momentum at the downsampled resolution [128x128] obtained by the conservative downsampling method. (d) Absolute difference between (b) and (c). The snapshot is taken at time $t= 148 \ \mu s$. Artifacts are observed at the interface when using the average downsampling method as seen in (b).}
    \label{fig:downsampling_momentumY_comparison}
\end{figure}

To gain insights into the impact of both downsampling methods, we visualize in Figure \ref{fig:downsampling_momentumY_comparison} the y-momentum field on the original resolution - (a) and the downsampled resolution - (b) and (c). When comparing the downsampled field to the original grid, the values obtained by averaging the y-velocities shows significant errors at the interface, whereas these errors are not present when using the conservative downsampling method.

\clearpage
\section{Benchmarking Details}

\subsection{Baseline Models}
\label{section:app_baseline_models}
We present additional details on the baseline models used in our benchmarking experiments. Our evaluation covers a diverse set of state-of-the-art architectures trained from scratch with random initialization, including convolutional models (CNeXt, CNO), spectral models (FFNO), and transformer-based architectures (ScOT). 
In addition, we fine-tune two pretrained models: Poseidon-B and DPOT-M. Details on the chosen hyperparameters for the baselines are as follows:

\begin{enumerate}
    \item \textbf{Convolutional Neural Operator (CNO): }
    A CNO, much like a U-Net, processes an input function through a sequence of encoder layers that progressively decrease spatial resolution while increasing channel depth. This encoded representation is then passed to decoder layers, which reverse the process by restoring spatial resolution and reducing the number of channels. At corresponding spatial scales (or spectral bands), encoder and decoder features are connected via ResNet-style skip connections.

    The central idea of the architecture is its upsampling and downsampling scheme, 
    where low-pass filtering is applied to avoid introducing artificial high-frequency components during resampling. This approach aligns with the Shannon sampling theorem, ensuring that discrete representations remain consistent with the underlying continuous functions.

    \begin{table}[h!]
        \caption{CNO hyperparameters.}
        \label{tab:cno_hyperparameters}
        \centering
        \begin{tabular}{lc}
            \toprule
            Hyperparameter  & 50M \\
            \midrule
            Latent channels                & 256 \\
            Depth                          & 24 \\
            \# blocks (except at neck)     & 8 \\
            \# blocks (at neck)            & 14 \\
            Channel multiplier             & 24 \\
            Activation                     & custom LeakyReLU \\
            \bottomrule
        \end{tabular}
    \end{table}
    
    \item \textbf{ConvNeXt (CNeXt): }
    A ConvNeXt is a modern convolutional neural network architecture introduced 
    to bridge the performance gap between traditional CNNs and vision transformers. 
    It builds on the classical ResNet design but incorporates several transformer-inspired updates, such as large-kernel depthwise convolutions, Layer Normalization, GELU activations, and a patch-based stem. Organized as a hierarchical model with multiple stages that progressively reduce spatial resolution while increasing channel capacity, ConvNeXt serves as a powerful and efficient feature extractor for a wide range of vision tasks. Despite remaining fully convolutional, its design achieves performance comparable to transformer-based models. The 3D datasets are trained on the 10M variant and the 2D datasets are trained on the 50M variant.

    \begin{table}[h!]
        \caption{CNeXt hyperparameters.}
        \label{tab:cnext_hyperparameters}
        \centering
        \begin{tabular}{lcc}
            \toprule
            Hyperparameter & 10M & 50M \\
            \midrule
            Latent channels                       & 32     & 80 \\
            \# stages                             & 4      & 4 \\
            \# blocks per stage (except at neck)  & 1      & 1 \\
            \# blocks (at neck)                   & 1      & 1 \\
            Channel multiplier                    & 2      & 2 \\
            Activation                            & GELU   & GELU \\
            \bottomrule
        \end{tabular}
    \end{table}

    \item \textbf{Factorized Fourrier Neural Operator(FFNO): }
    Like FNO, it operates in the Fourier domain to capture global interactions, but instead of learning a full multi-dimensional spectral kernel, FFNO factorizes the operator into a sequence of one-dimensional Fourier transformations applied along each spatial dimension. This separable formulation significantly reduces the number of parameters and computational cost while maintaining strong expressive power. Similar to ConvNeXt, the 3D datasets are trained using the 10M variant, while the 2D datasets are trained using the 50M variant.

    \begin{table}[h!]
        \caption{FFNO hyperparameters.}
        \label{tab:ffno_hyperparameters}
        \centering
        \begin{tabular}{lcc}
            \toprule
            Hyperparameter & 10M & 50M \\
            \midrule
            Latent channels                    & 96  & 240 \\
            FNO layers                         & 8  & 12 \\
            Modes                              & 20  & 16 \\
            MLP expansion factor               & 4  & 4 \\
            \# MLP layers                      & 2  & 2 \\
            Activation in MLP                  & ReLU   & ReLU \\
            Activation in input and output heads & GELU & GELU \\
            \bottomrule
        \end{tabular}
    \end{table}

    \item \textbf{Scalable Operator Transformer (ScOT) and Poseidon: }
    The ScOT model is built upon the Poseidon framework \citep{herde2024poseidon}, where Poseidon provides the foundation-model framework for learning generalizable solution operators, and ScOT serves as its core architectural backbone. At its core, ScOT employs a hierarchical transformer design inspired by vision transformers, using a window-based mechanism in which the input domain is divided into a uniform grid of non-overlapping patches, with added support for non-square inputs. Each patch is averaged via a shared spatial weighting scheme and then linearly projected into a latent embedding space defined by the number of latent channels, yielding a piecewise-constant latent representation that reduces the cost of global attention while retaining essential local structure. This embedded representation is then processed through multiple stages of SwinV2 Transformer blocks \citep{liu2021swin} arranged in a UNet-like hierarchy, where feature maps are progressively downsampled and later upsampled, with the number of blocks per stage governed by the “depths” hyperparameter. Within each stage, attention is computed using windowed multi-head self-attention (MHSA), restricting interactions to local regions for efficiency, while shifted windows between layers enable cross-window communication and capture global context. The encoder incorporates patch merging to decrease spatial resolution and increase feature dimensionality for learning global patterns, whereas the decoder restores resolution via patch expansion, with skip connections implemented through ConvNeXt blocks \citep{liu2022convnet}—controlled by the “skip-connections” hyperparameter—linking corresponding encoder and decoder stages to preserve fine-grained information.

    \begin{table}[h!]
        \centering
        \begin{minipage}[t]{0.48\textwidth}
            \centering
            \caption{ScOT hyperparameters.}
            \label{tab:scot_hyperparameters}
            \begin{tabular}{lc}
                \toprule
                Hyperparameter & 50M \\
                \midrule
                Latent channels     & 150 \\
                Patch size          & 4 \\
                Depths              & [4, 4, 4] \\
                \# attention heads  & [6, 12, 24] \\
                Skip connections    & [3, 3, 0] \\
                Window size         & 16 \\
                MLP ratio           & 4.0 \\
                Activation          & GELU \\
                \bottomrule
            \end{tabular}
        \end{minipage}
        \hfill
        \begin{minipage}[t]{0.48\textwidth}
            \centering
            \caption{Poseidon-B hyperparameters.}
            \label{tab:poseidon-b_hyperparameters}
            \begin{tabular}{lc}
                \toprule
                Hyperparameter & 158M \\
                \midrule
                Latent channels     & 96 \\
                Patch size          & 4 \\
                Depths              & [8, 8, 8, 8] \\
                \# attention heads  & [3, 6, 12, 24] \\
                Skip connections    & [2, 2, 2, 0] \\
                Window size         & 16 \\
                MLP ratio           & 4.0 \\
                Activation          & GELU \\
                \bottomrule
            \end{tabular}
        \end{minipage}
    \end{table}

    \item \textbf{Denoising Pre-trained Operator Transformer (DPOT): } DPOT is a large-scale neural operator architecture designed to learn solution mappings of PDEs. It is built around a Fourier transformer backbone, where attention is performed in the frequency domain to efficiently capture global spatial dependencies, enabling the model to approximate integral operators over function spaces. The architecture combines temporal aggregation layers to encode information from multiple time steps with multi-head Fourier attention layers, which apply learnable transformations in Fourier space to model complex multi-scale dynamics. Furthermore, DPOT is paired with an auto-regressive denoising pre-training strategy, where the model predicts future states from noise-corrupted inputs to improve robustness and generalization. This design allows DPOT to scale to very large models and datasets, functioning as a foundation model for PDEs that can be fine-tuned across diverse downstream tasks.
    
    \begin{table}[h!]
        \caption{DPOT-M hyperparameters.}
        \label{tab:dpot-m_hyperparameters}
        \centering
        \begin{tabular}{lc}
            \toprule
            Hyperparameter & 122M \\
            \midrule
            Latent channels      & 1024 \\
            Patch size           & 8 \\
            \# blocks            & 8 \\
            MLP ratio            & 4.0 \\
            Depth                & 12 \\
            Modes                & 32 \\
            Mixing Type          & AFNO \\
            Time Aggregation     & Exponential MLP \\
            Activation           & GELU \\
            \bottomrule
        \end{tabular}
    \end{table}
\end{enumerate}

\subsection{Training Protocol}

\label{section:app_training_protocol}

As mentioned in the main text, we employ a many-to-one training strategy for the models intialized with random weights, where 4 historic snapshots are passed to the model as input and the next snapshot is predicted and compared with the target. We use this setup for benchmarking the baselines mentioned in the previous section, but the benchmarking repository is capabale of handling many-to-many scenarios as well. For the pre-trained models, we adhere to the original input–output configurations specified in their respective works.

\begin{figure}[h!]
    \centering
    \includegraphics[width=8cm]{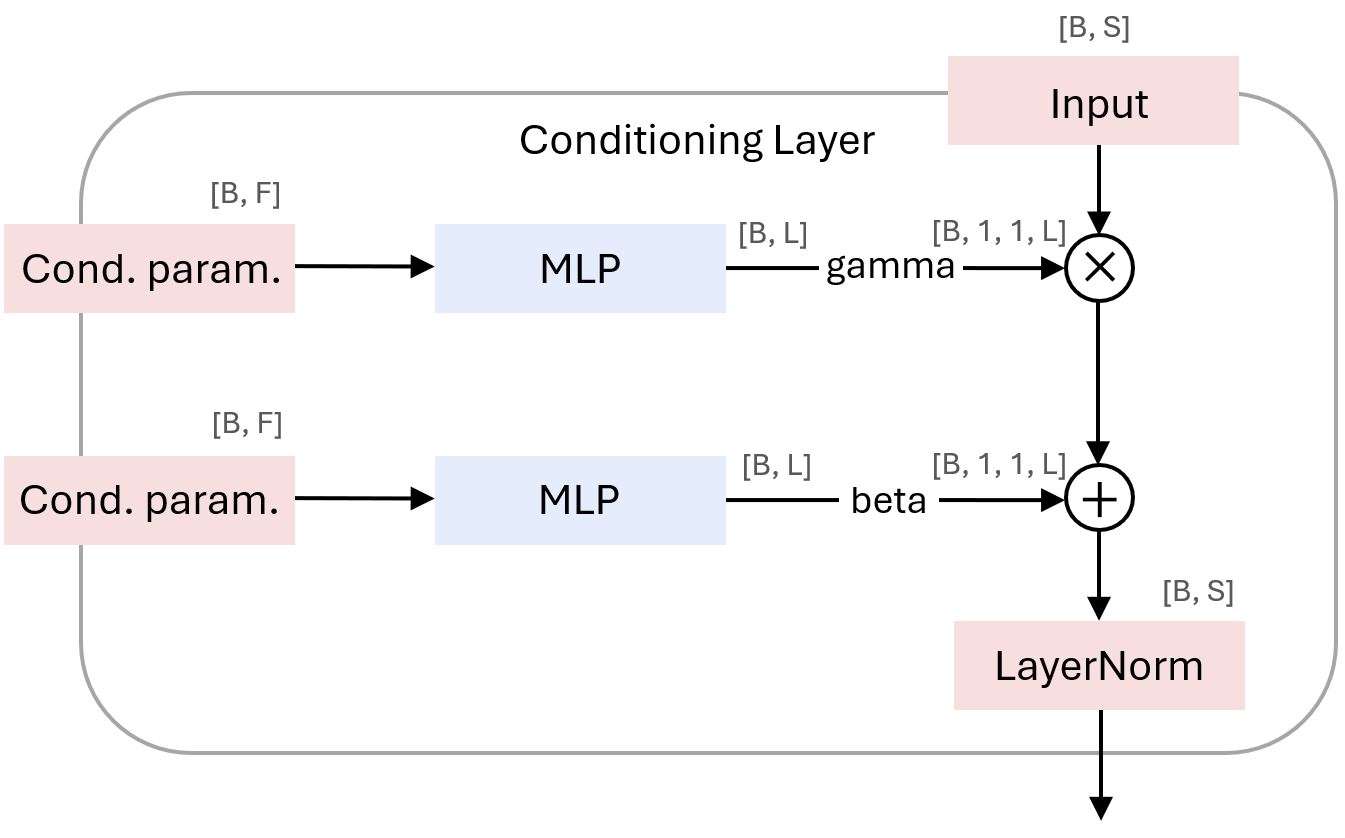}
    \caption{Illustration of the AdaNorm conditioning module, applicable to all baselines. $F$ denotes the number of conditioning parameters, and $L$ the latent dimension.}
    \label{fig:conditioning_module}
\end{figure}

A key central feature of our benchmarking repository is the ability to integrate conditioning parameters into all baseline architectures via custom normalization layers (Figure \ref{fig:conditioning_module}) where intermediate feature statistics are modulated through learned scale and shift terms derived from the conditioning vector. Consider an input to the custom norm, $x \in \mathbb{R}^{B \times S \times L}$, where $B$ denotes the batch size, $S$ represent the spatial resolution, and $L$ is the latent dimension. The custom norm is similar to a standard LayerNorm \citep{ba2016layer}, with the modification that the affine parameters ($\gamma_{\theta}$ and $\beta_{\theta}$) are made functions of the conditioning parameter vector $c$ \citep{perez2017visual}, where $c \in \mathbb{R}^{B \times F}$, and $F$ is the number of conditioning features.

\begin{equation}
    \begin{aligned}
    y(c) &= \gamma_{\theta}(c)\cdot\frac{x - \mu}{\sigma} + \beta_{\theta}(c) \\
    \mu &= \frac{1}{BSL} \sum_{b=1}^{B} \sum_{s=1}^{S} \sum_{\ell=1}^{L} x \\
    \sigma &= \sqrt{ \frac{1}{BSL} \sum_{b=1}^{B} \sum_{s=1}^{S} \sum_{\ell=1}^{L} \left(x - \mu\right)^2 + \epsilon }.
    \end{aligned}
\end{equation}

In our benchmarking repository, we have several other conditioning methods implemented, but in this work we focus on Adaptive Normalization (AdaNorm), which is a simple and lightweight mapping of the conditioning feature vector $c$, through an MLP to the desired latent dimension $L$. 

The hyperparameters used for training the 2D and 3D baselines are shown in Table \ref{tab:training-pipeline}.

\begin{table}[H]
    \centering
    \caption{Training hyperparameters used across experiments.}
    \label{tab:training-pipeline}
    \begin{tabular}{lll}
    \toprule
    Hyperparameter & 2D & 3D \\
    \midrule
    Train time         & \texttt{time-restricted to 24 hours} & \texttt{time-restricted to 24 hours} \\
    & \texttt{on a 4 NVIDIA-A100} & \texttt{on a single NVIDIA-H100} \\
    Checkpointing metric & \multirow{2}{*}{\texttt{RMSE}} & \multirow{2}{*}{\texttt{RMSE}} \\
    for baselines       &                                &                                \\
    Batch size           & 16 & 1 \\
    Validation split ratio     & 0.2 & 0.2 \\
    Optimizer            & \texttt{AdamW} & \texttt{AdamW} \\
    Learning rate (LR)       & \texttt{1e-3} & \texttt{1e-4} \\
    LR scheduler         & \texttt{cosine} & \texttt{cosine} \\
    Weight decay         & \texttt{1e-6} & \texttt{1e-6} \\
    \bottomrule
    \end{tabular}
\end{table}

Table \ref{tab:training_epochs} summarizes the number of training epochs completed by the baseline models across different datasets and training strategies. For ConvNeXt and FFNO, the 50M-parameter variants are employed for the 2D datasets, while the 10M-parameter variants are used for the 3D datasets. For 2D experiments, training is limited to a maximum of 128 epochs or 24 hours of wall-clock time, whichever is reached first. For 3D experiments, the maximum number of epochs is reduced to 50, while retaining the same 24-hour wall-clock limit.

As shown in Table \ref{tab:training_epochs}, most baseline models complete the full 128 epochs for the 2D experiments. Validation is performed every four epochs, during which the models are evaluated through autoregressive rollouts up to the maximum permissible rollout horizon. Model checkpoints are selected based on the lowest validation RMSE achieved throughout training. Furthermore, we verify that the validation loss has converged by the end of training, ensuring that the reported results are obtained from sufficiently trained models.

\begin{table}[H]
    \centering
    \scriptsize
    \caption{Number of training epochs for each baseline, dataset, and training loss strategy. A dash (--) indicates that the baseline was not trained on that dataset. The models }
    \label{tab:training_epochs}
    \begin{tabular}{llcccccc}
    \toprule
    Training loss & Dataset
    & \makecell{\texttt{CNO}\\\texttt{(50M)}}
    & \makecell{\texttt{ConvNeXt}\\\texttt{(50M/10M)}}
    & \makecell{\texttt{DPOT-M}\\\texttt{(122M)}}
    & \makecell{\texttt{FFNO}\\\texttt{(50M/10M)}}
    & \makecell{\texttt{Poseidon-B}\\\texttt{(158M)}}
    & \makecell{\texttt{ScOT}\\\texttt{(50M)}} \\
    \midrule

    \multirow{6}{*}{\texttt{MSE loss}}
    & 2D-SRBA (OOOO)   & 128 & 128 & 116           & 112 & 128 & 128 \\
    & 2D-SABW (OOOO)   & 128 & 128 & 116           & 112 & 128 & 128 \\
    & 2D-SABW (SSOO)   & 128 & 128 & 116           & 112 & 128 & 128 \\
    & 2D-SDBA (SSOO)   & 128 & 128 & 20$^{\ast}$   & 112 & 128 & 128 \\
    & 3D-SABW (SSOOSS) & --  & 50  & --            & 50  & --  & --  \\
    & 3D-SDBA (SSOOSS) & --  & 50  & --            & 50  & --  & --  \\
    \addlinespace

    \multirow{6}{*}{\texttt{Comp.loss (SA)}}
    & 2D-SRBA (OOOO)   & 128 & 128 & 116 & 112 & 128 & 128 \\
    & 2D-SABW (OOOO)   & 128 & 128 & 116 & 112 & 128 & 128 \\
    & 2D-SABW (SSOO)   & 128 & 128 & 116 & 112 & 128 & 128 \\
    & 2D-SDBA (SSOO)   & 128 & 128 & 116 & 112 & 128 & 128 \\
    & 3D-SABW (SSOOSS) & --  & 50  & --  & 50  & --  & --  \\
    & 3D-SDBA (SSOOSS) & --  & 50  & --  & 50  & --  & --  \\
    \addlinespace

    \multirow{6}{*}{\texttt{Comp.loss (GN)}}
    & 2D-SRBA (OOOO)   & 128 & 128 & 100 & 112 & 128 & 128 \\
    & 2D-SABW (OOOO)   & 128 & 128 & 100 & 100 & 128 & 128 \\
    & 2D-SABW (SSOO)   & 128 & 128 & 100 & 108 & 128 & 128 \\
    & 2D-SDBA (SSOO)   & 128 & 128 & 100 & 100 & 128 & 128 \\
    & 3D-SABW (SSOOSS) & --  & 50  & --  & 50  & --  & --  \\
    & 3D-SDBA (SSOOSS) & --  & 50  & --  & 50  & --  & --  \\
    \bottomrule
    \end{tabular}

    \vspace{2pt}
    \parbox{\linewidth}{\scriptsize $^{\ast}$Validation loss diverged; the best model checkpoint was retained at epoch 20.}
\end{table}

\subsection{Evaluation Protocol}
\label{section:app_evaluation_protocol}
Model performance is evaluated on a held-out test set. During inference, each model is applied autoregressively, starting from the initial condition and rolling out for the full length of a test trajectory.

Inference produces prediction and target tensors of shape $(N, R, T, C, S)$, where $N$ is the number of test trajectories, $R$ the number of rollout steps, and $T$ the number of predicted timesteps per rollout. Each metric is reduced over the spatial dimensions only, yielding an error tensor of shape $E \in \mathbb{R}^{N \times (R T) \times C}$. This preserves the dependence on trajectory, prediction time, and channel.

The resulting tensor enables evaluation at multiple levels. First, the mean and standard deviation over the trajectory dimension are computed as $\mu,\sigma \in \mathbb{R}^{(RT)\times C}$, which describe the temporal evolution of the error and the variability across test trajectories, reported separately for each channel.

A scalar overall loss for each metric $\mu_{\mathrm{overall}}$ across all trajectories of the test dataset is obtained by aggregating over the remaining dimensions of $\mu$, i.e.\ over prediction time and channels, in accordance with the formulation used by that metric. Tables \ref{tab:2d_sabw_oooo_pointwise} through \ref{tab:3d_sdba_physics_informed} showcase this scalar overall loss for all the metrics in this work for different datasets.



\subsection{Metrics}
\label{section:app_metrics}
As mentioned in the main text, we perform two sets of ablations from a metric perspective. In the first set of experiments, we perform training of the baselines with only MSE loss and in the second set of experiments, we train the baselines with a composite loss which is a combination of several metrics: MSE, H1, SSIM and IRMSE. The initial weights for these metrics is specified in Table \ref{tab:initial_loss_coefficients}. During inference, we evaluate the performance of the trained baselines on a broader set of metrics, as mentioned in Table \ref{tab:loss_metrics_list}. 

\begin{table}[H]
    \centering
    \caption{Initial loss coefficients for the composite loss.}
    \label{tab:initial_loss_coefficients}
    \begin{tabular}{ll}
    \toprule
    Initial loss weights & \texttt{default}\\
    \midrule
    $\lambda_{\text{MSE}}$ & \texttt{1.0}     \\
    $\lambda_{\text{H1}}$  & \texttt{2.0}  \\
    $\lambda_{\text{SSIM}}$               & \texttt{0.5}           \\
    $\lambda_{\text{IRMSE}}$            & \texttt{1.0}           \\
    \bottomrule
    \end{tabular}
\end{table}

\begin{table}[h!]
    \caption{Overview of implemented evaluation metrics. Metrics are grouped by conceptual type.}
    \label{tab:loss_metrics_list}
    \centering
    \begin{tabularx}{\textwidth}{p{5.4cm}p{2.3cm}p{2.6cm}p{2.3cm}}
        \toprule
        Name & Acronym & Type & Source \\
        \midrule
        Mean squared error & MSE & Pointwise &  \\[3pt]

        Mean absolute error & MAE & Pointwise &  \\ [3pt]

        Variance-normalized RMSE & VRMSE & Pointwise & \cite{ohana2024well} \\ [3pt]

        H$^1$-seminorm & H1 & Pointwise & \cite{SobolevTrainingWithHigherOrder, SobolevTrainingOperatorLearning} \\ [3pt]

        H$^2$-seminorm & H2 & Pointwise & \cite{SobolevTrainingWithHigherOrder, SobolevTrainingOperatorLearning} \\ [3pt]

        Interface-focused RMSE & IRMSE & Feature-focused & \cite{hassan2023bubbleml, hassan2025bubbleformer} \\ [3pt]
        
        Shock-focused RMSE & SRMSE & Feature-focused & \cite{AGeneralizedNovelLoss} \\ [3pt]

        Multilevel wavelet & MLW & Spectral & \cite{WaveletBasedLoss} \\ [3pt]

        Wavelet frequency-binned RMSE & wfRMSE & Spectral & \cite{takamoto2022pdebench, WaveletBasedLoss} \\ [3pt]

        Structural similarity index & SSIM & Structure-aware & \cite{nilsson2020understanding} \\ [3pt]
 
        Pearson correlation coefficient & PCC & Structure-aware & \cite{hu2026realpdebench, mao2025benchmarking} \\ [3pt]

        Integral quantities of interest: & iqRMSE & Physics-informed & \cite{takamoto2022pdebench} \\ [-10pt]
        \makecell[l]{Mass, Momentum, Kinetic energy,\\ Enstrophy, Center of mass displacement} &  &  &  \\ 

        \bottomrule
    \end{tabularx}
\end{table}

The inputs to each loss function are the prediction ($y$) and target ($\hat{y}$) tensors, where 
\[  
\hat{y}, y \in \mathbb{R}^{B \times T \times C \times S}
\]  
B, T, C, S correspond to the batch, time, channel and spatial dimensions respectively. Spatial sums are written over the full discrete grid, and metrics are computed per sample before averaging over the batch dimension. For readability, the batch index is omitted in the formulas described in this section. 

\subsubsection{Pointwise metrics}
The \textbf{Mean Square Error (MSE)} and its varient the \textbf{Root Mean Square Error (RMSE)} are defined as shown in Equations \ref{eq:mse} and \ref{eq:rmse} respectively.
\begin{equation}
    \label{eq:mse}
    \mathcal{L}_{\text{MSE}}
    =
    \frac{1}{N_{T, C, S}}
    \sum_{T, C, S}
    (\hat{y} - y)^2
\end{equation}

\begin{equation}
\label{eq:rmse}
\mathcal{L}_{\text{RMSE}}
=
\sqrt{
\frac{1}{N_{T, C, S}}
\sum_{T, C, S}
(\hat{y} - y)^2
}
\end{equation}
The \textbf{Varience scaled Root Mean Square Error (VRMSE)} metric divides the mean squared error by the variance of the target field as shown in Equation \ref{eq:vrmse}.
\begin{equation}
\label{eq:vrmse}
\mathcal{L}_{\text{VRMSE}}
=
\sqrt{
\frac{1}{N_{T, C}}\sum_{T, C}\left(\frac{
\sum_{S}(\hat{y} - y)^2
}{
\sum_{S}(y - \bar{y})^2 + \epsilon
}\right)
}
\end{equation}
where $\bar{y}$ is the spatial mean of the target field and $\epsilon$ is $10^{-8}$. 

Another useful variant of the MSE family is the \textbf{Normalized Root Mean Square Error (nRMSE)}, which normalizes the RMSE by the mean squared magnitude of the target field. The nRMSE instead normalizes by the mean squared magnitude of the reference field as shown in Equation \ref{eq:nrmse}.

\begin{equation}
    \label{eq:nrmse}
    \mathcal{L}_{\text{nRMSE}}
=
\sqrt{
\frac{1}{N_{T, C}}\sum_{T, C}\left(\frac{
\sum_{S}(\hat{y} - y)^2
}{
\sum_{S}y ^2
}\right)
},
\end{equation}

The Sobolev loss of order $K$ is defined as
\begin{equation}
\mathcal{L}_{\text{Sob},K} =
\lambda_0 \frac{1}{N_{T,C,S}} \sum_{T,C,S} (\hat y - y)^2
+
\sum_{k=1}^{K} \lambda_k
\frac{1}{N_{T,C,S}} \sum_{T,C,S}
\left\| \nabla_S^{(k)} \hat y - \nabla_S^{(k)} y \right\|^2,
\end{equation}

where $\nabla^{(k)}$ denotes the $k$-th order spatial derivative, $\lambda_k$ are weighting coefficients, and $\|\cdot\|^2$ is the pointwise squared Euclidean norm of the derivative vector/tensor \cite{SobolevTrainingWithHigherOrder, SobolevTrainingOperatorLearning}. In this work, the derivative terms of the Sobolev loss are each defined as separate metrics, corresponding to the $H^1$ and $H^2$ components of the Sobolev space. 

\begin{figure}[ht]
    \centering
    \includegraphics[width=10cm]{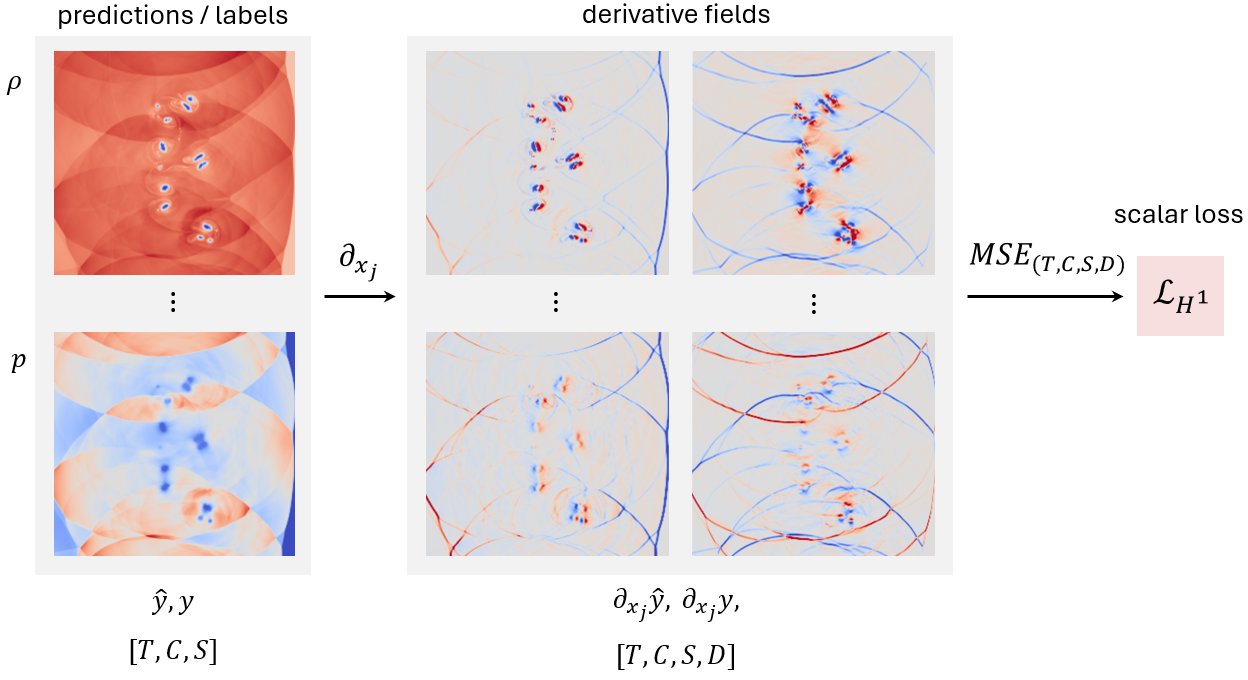}
    \caption{$H^1$ semi-norm metric.}
    \label{fig:H1}
\end{figure}

The\textbf{\boldmath H$^1$ semi-norm} metric, as shown in Figure \ref{fig:H1} penalizes mismatches in first-order spatial derivatives as defined in Equation \ref{eq:H1}:
\begin{equation}
\label{eq:H1}
\mathcal{L}_{\text{H}^1}
=
\frac{1}{N_{T,C,S}}
\sum_{T,C,S}\sum_{j=1}^d
\left(\partial_{x_j}\hat y-\partial_{x_j}y\right)^2.
\end{equation}
The \textbf{\boldmath H$^2$ semi-norm} penalizes discrepancies in second-order derivatives as shown in Equation \ref{eq:H2}.

\begin{equation}
\label{eq:H2}
\mathcal{L}_{\text{H}^2}
=
\frac{1}{N_{T,C,S}}
\sum_{T,C,S}\sum_{i=1}^d\sum_{j=1}^d
\left(\partial_{x_i x_j}^2 \hat y-\partial_{x_i x_j}^2 y\right)^2.
\end{equation}

Spatial derivatives are computed on regular grids using fixed convolution stencils, supporting both standard central-difference and Sobel kernels, the latter combining differentiation with mild local smoothing.

\subsubsection{Feature-focused metrics}

\begin{figure}[h!]
    \centering
    \includegraphics[width=10cm]{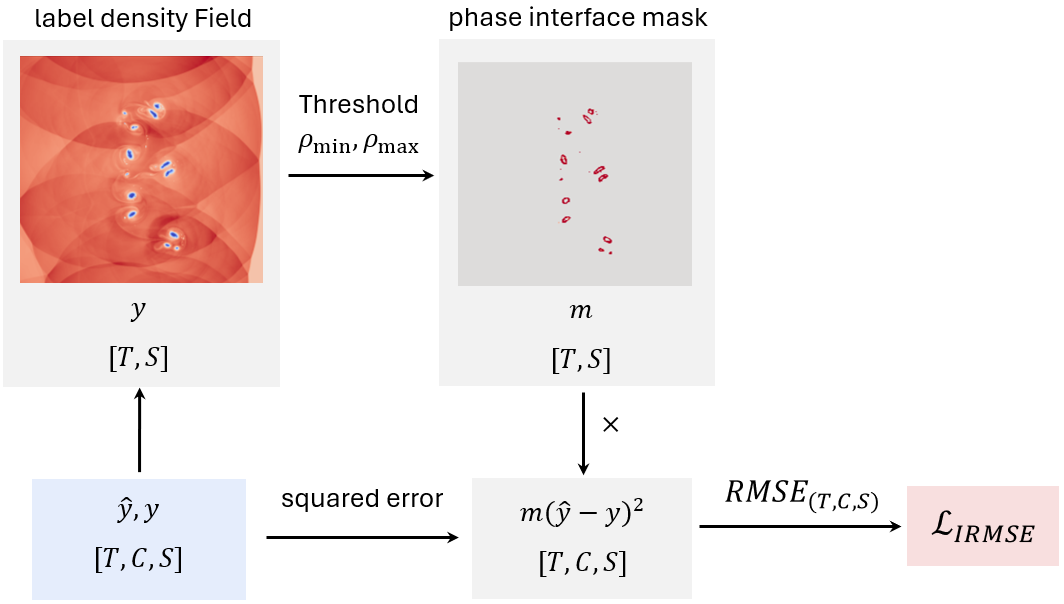}
    \caption{IRMSE metric.}
    \label{fig:IRMSE}
\end{figure}

The \textbf{Interface-focused Root Mean Square Error (IRMSE)} restricts error evaluation to the vicinity of the phase boundary (Figure \ref{fig:IRMSE}) as defined in Equation \ref{eq:irmse}.

\begin{equation}
\label{eq:irmse}
\mathcal{L}_{\text{IRMSE}}
=
\sqrt{
\frac{1}{N_{TC}}
\sum_{TC}
\frac{
\sum_{S} m
\left(\hat{y}-y\right)^2
}{
\sum_{S} m + \varepsilon
}
}.
\end{equation}

Here $m$ denotes the mask of spatial locations corresponding to the interface region.

In the original formulation, $m$ is defined as the zero level-set of a signed-distance field $\phi(x)$. In this work, the interface is identified using a threshold on the density field $\rho$. Since a hard threshold would introduce discontinuities in the loss, a soft interface mask is constructed using sigmoid transitions around the density bounds $\rho_{\min}$ and $\rho_{\max}$, which maintains differentiability:
\begin{equation}
m
=
\sigma\!\left(\frac{\rho-\rho_{\min}}{s}\right)
\sigma\!\left(\frac{\rho_{\max}-\rho}{s}\right),
\end{equation}

where $s$ controls the softness of the transition.

\begin{table}[H]
    \centering
    \caption{Configuration settings for interface-focused RMSE.}
    \label{tab:metric-config-feature}
    \begin{tabular}{llll}
    \toprule
    Hyperparamter &  \texttt{SABW} & \texttt{SDBA} & \texttt{SRBA} \\
    \midrule
    Threshold field    & \texttt{density}    & \texttt{density}  & \texttt{density}  \\
    Threshold range    & \texttt{(350, 550)} & \texttt{(350, 550)} & \texttt{(4.5, 8.5)} \\
    Blur sigma         & \texttt{3.0}        & \texttt{3.0} & \texttt{3.0}  \\
    Threshold softness & \texttt{0.05}       & \texttt{0.05} & \texttt{0.05} \\
    \bottomrule
    \end{tabular}
\end{table}

The GRMSE emphasizes regions where the target field exhibits strong spatial variation as described in \cite{AGeneralizedNovelLoss}. We define a member of this family called shock-RMSE with a special focus in capturing strong shocks from the pressure field. A spatial weighting function $w(y)$ is constructed from the magnitude of the gradient of the target pressure field:
\begin{equation}
w
\propto
\left\|
\nabla_S y
\right\|_2,
\end{equation}

The \textbf{Shock-focused Root Mean Square Error (SRMSE)} is then defined as shown in Equation \ref{eq:shock_rmse}.
\begin{equation}
    \label{eq:shock_rmse}
    \mathcal{L}_{\text{SRMSE}}
=
\sqrt{
\frac{1}{N_{T, C}}
\sum_{T,C}
\frac{
\sum_{S} w
\left(\hat{y}-y\right)^2
}{
\sum_{S} w + \varepsilon
}
}.
\end{equation}

\begin{figure}[ht]
    \centering
    \includegraphics[width=10cm]{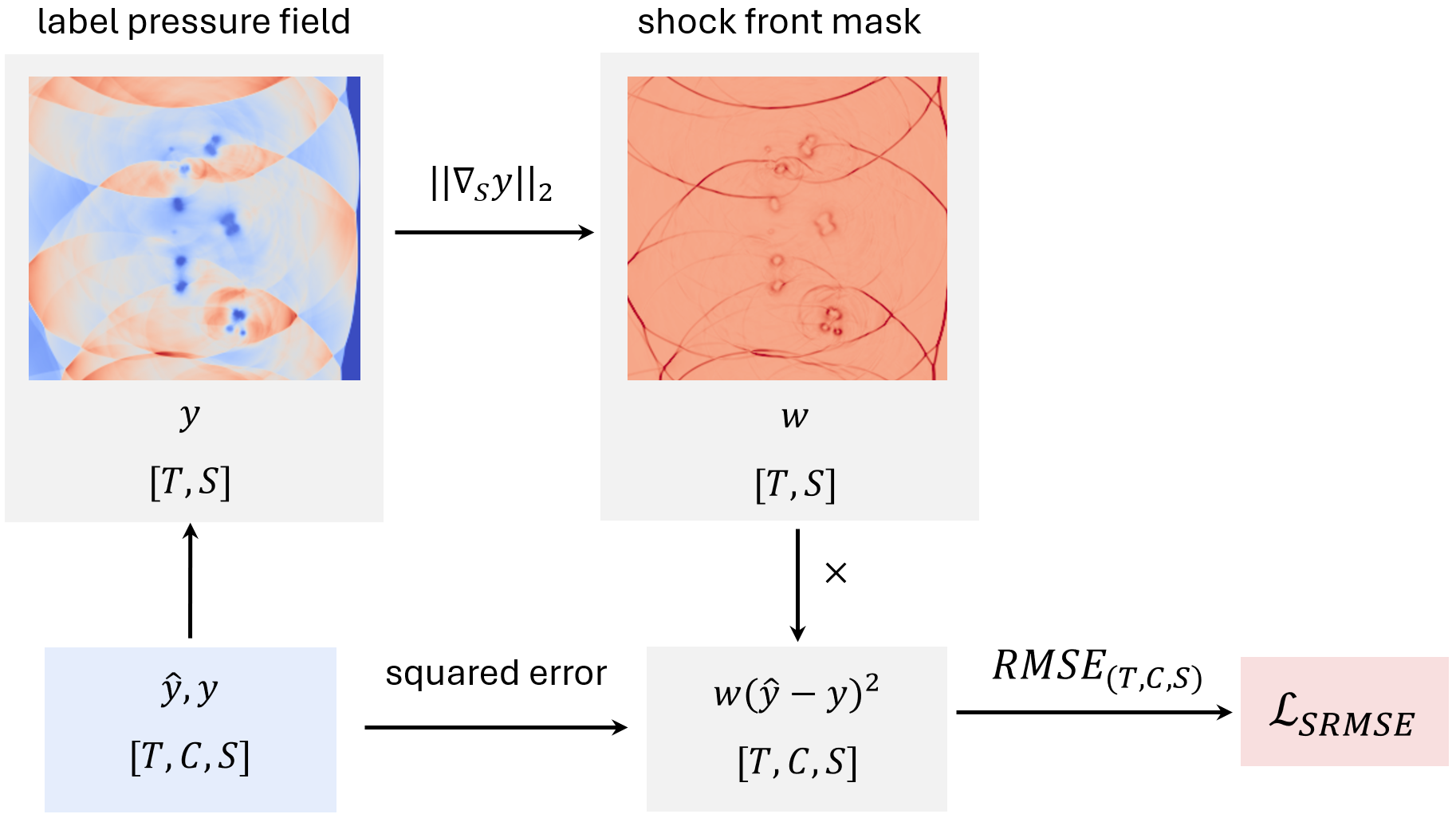}
    \caption{SRMSE metric.}
    \label{fig:GRMSE}
\end{figure}

In this formulation, spatial locations with large gradients receive higher weights. For this implementation, we define $w$ using the spatial gradients of the pressure field (Figure \ref{fig:GRMSE}). This effectively produces a mask which emphasizes shock fronts, providing a complementary metric to IRMSE.

\begin{table}[H]
    \centering
    \caption{Configuration settings for shock-focused RMSE.}
    \label{tab:metric-config-shock}
    \begin{tabular}{ll}
    \toprule
    Hyperparameter & \texttt{default}\\
    \midrule
    Gradient threshold field & \texttt{pressure}     \\
    Gradient threshold range & \texttt{(0.3, 4.0)}  \\
    Threshold softness               & \texttt{0.03}           \\
    Blur sigma               & \texttt{1.5}           \\
    Derivative stencil       & \texttt{Central Difference}  \\
    \bottomrule
    \end{tabular}
\end{table}

Both IRMSE and SRMSE optionally include Gaussian smoothing, as proposed in the original GRMSE formulation \cite{AGeneralizedNovelLoss}. A Gaussian filter with standard deviation $\sigma$ is applied to the weighting masks before computing the loss. This is implemented as separable convolution, where a one-dimensional Gaussian kernel is applied sequentially along each spatial axis. The smoothing reduces high-frequency noise in the weighting fields and leads to more stable gradients during training.

\subsubsection{Spectral metrics}
In this work, spectral decomposition is implemented using the multi-level DWT. The key idea is to represent a field in terms of its spatially localized frequency content across different spatial scales. At each decomposition level \(i\), the input field is split into a low-frequency component and a high-frequency component,

\begin{equation}
LF_i(y), \quad HF_i(y),
\end{equation}
and analogously for the prediction \(\hat y\),
\begin{equation}
LF_i(\hat y), \quad HF_i(\hat y).
\end{equation}

\begin{figure}[ht]
    \centering
    \includegraphics[width=9cm]{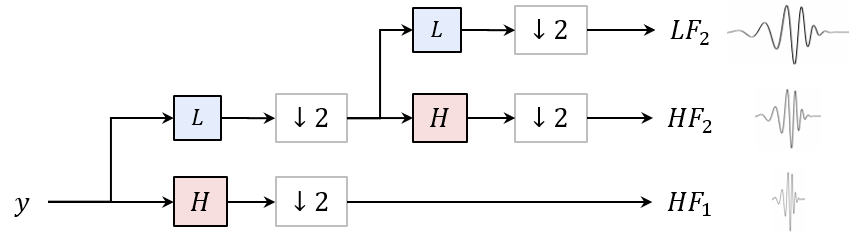}
    \caption{Conceptual illustration of the DWT, recursively decomposing an input $y$ into frequency-band contributions.}
    \label{fig:DWT}
\end{figure}

The low-frequency component \(LF_i\) contains the coarse, slowly varying structure of the field, while the high-frequency component \(HF_i\) captures finer-scale detail. The transform is then applied recursively only to the low-frequency component. Repeating this procedure over \(I\) levels yields one final coarse approximation \(LF_I\) together with a hierarchy of detail coefficients,

\begin{equation}
HF_1, HF_2, \ldots, HF_I,
\end{equation}
where \(HF_1\) corresponds to the finest details and higher levels represent progressively coarser spatial scales (Figure ~\ref{fig:DWT}).

For multidimensional fields, the transform is applied independently along each spatial dimension. This produces directional detail subbands (e.g., horizontal, vertical, and diagonal features in 2D). Collectively, these directional components form the set of high-frequency coefficients $HF_i$ at level $i$. In this work, orthogonal Daubechies wavelets are used, following their usage in related studies \cite{WaveletBasedLoss}.

\begin{figure}[ht]
    \centering
    \includegraphics[width=13.5cm]{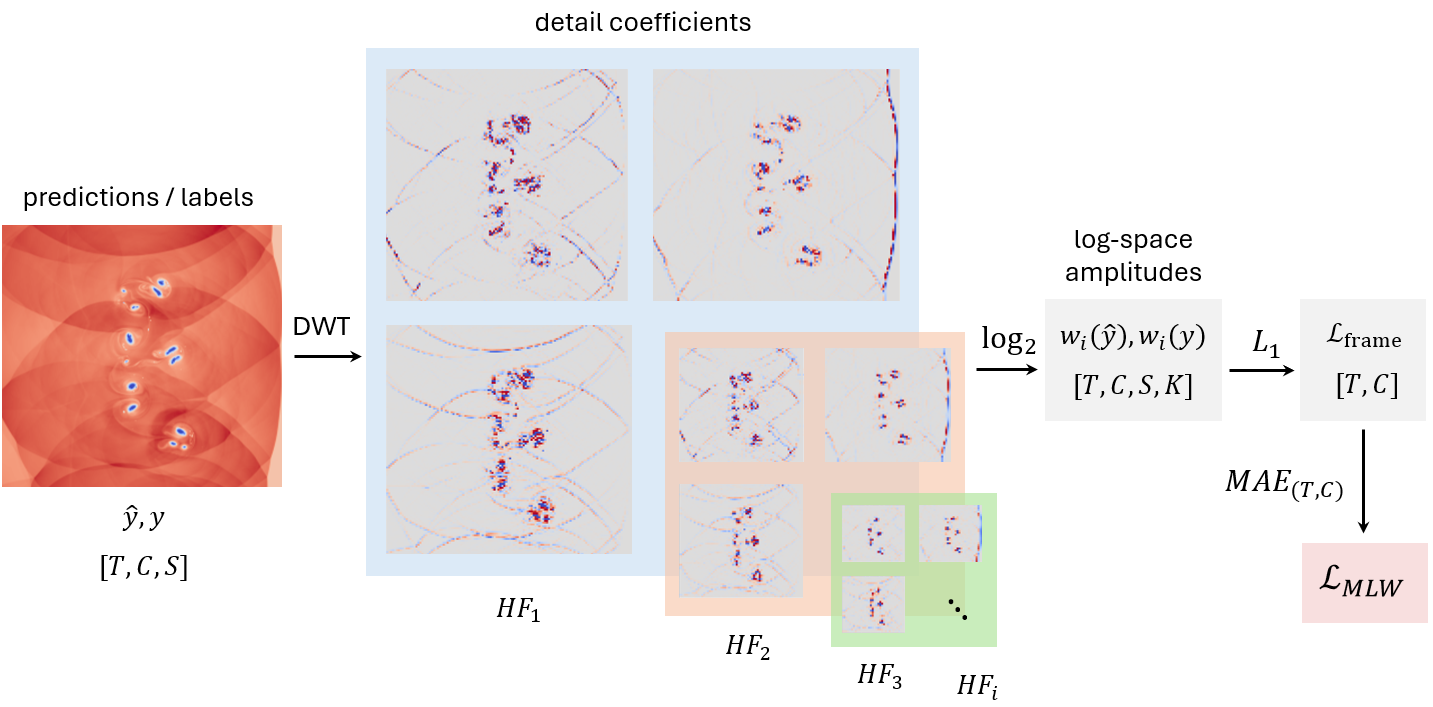}
    \caption{Multilevel wavelet loss (MLW) metric.}
    \label{fig:MLW}
\end{figure}

For the \textbf{Multilevel Wavelet (MLW) } metric, the DWT is first computed over the spatial dimensions of $y$ and $\hat{y}$, yielding the detail coefficients $HF_i$ at each level $i$ as depicted in Figure \ref{fig:MLW}. The coefficients are then passed through a logarithmic transform:
\begin{equation}
w_i(y) = \log_2\bigl(|HF_i(y)| + \varepsilon\bigr), \qquad
w_i(\hat{y}) = \log_2\bigl(|HF_i(\hat{y})| + \varepsilon\bigr),
\end{equation}
where $\varepsilon$ is a small constant for numerical stability.

The wavelet spectral loss is then defined as:
\begin{equation}
\mathcal{L}_{\text{w}} =
\sum_i \left\| w_i(y) - w_i(\hat{y}) \right\|_1,
\end{equation}
with the sum taken over all decomposition levels and orientations.

The logarithmic scaling makes the more loss sensitive to discrepancies in fine-scale structure (i.e., thin interfaces and small-scale oscillations). When applied to trajectories with multiple timesteps, the loss optionally includes a separate 1D wavelet transform over the time dimension, applied independently at each spatial location. In this case, the loss is constructed as the sum of spatial and temporal contributions, where $\alpha$ and $\beta$ are the respective weighting coefficients:

\begin{equation}
    \mathcal{L}_{\text{MLW}}=\alpha\mathcal{L}_{ws} + \beta\mathcal{L}_{wt}
\end{equation}

The settings used for $\alpha$ and $\beta$ are provided in Table \ref{tab:metric-config-mlw}.

\begin{table}[h!]
    \caption{Configuration settings for \texttt{MultilevelWaveletLoss}.}
    \label{tab:metric-config-mlw}
    \centering
    \begin{tabular}{ll}
        \toprule
        Hyperparameter & Default \\
        \midrule
        Wavelet  & \texttt{db2} \\
        Alpha    & \texttt{1} \\
        Beta     & \texttt{0.1} \\
        DWT mode & \texttt{reflect} \\
        \bottomrule
    \end{tabular}
\end{table}

The \textbf{Wavelet frequency-binned Root Mean Square Error (wfRMSE)} metric measures RMSE across wavelet frequency bands. This metric is inspired by the frequency-binned Fourier RMSE used in \cite{takamoto2022pdebench}, but replaces global Fourier modes with localized wavelet bands to avoid periodicity assumptions.

For each wavelet level $i$, the wfRMSE of the detail coefficients is computed as
\begin{equation}
\text{RMSE}_i =
\sqrt{ \bigl(HF_i(\hat{y}) - HF_i(y)\bigr)^2  }.
\end{equation}

The total wavelet frequency-binned RMSE is defined as a weighted sum over scales,
\begin{equation}
\mathcal{L}_{\text{wfRMSE}} =
\sum_i \alpha_i \, \text{RMSE}_i,
\end{equation}
where the weights $\alpha_i$ control the relative contribution of different frequency bands. By adjusting $\alpha_i$, the metric can be tuned to prioritize specific ranges of dynamically important scales.

\begin{table}[h!]
    \centering
    \caption{Configuration settings for \texttt{wfRMSE}.}
    \label{tab:metric-config-wavelet-binned}
    \begin{tabular}{llll}
    \toprule
    Hyperparameter & \texttt{high} & \texttt{mid} & \texttt{low} \\
    \midrule
    Wavelet       & \texttt{db2}     & \texttt{db2}     & \texttt{db2} \\
    Levels        & \texttt{3}       & \texttt{3}       & \texttt{3} \\
    Level weights & \texttt{(1,0,0)} & \texttt{(0,1,0)} & \texttt{(0,0,1)} \\
    DWT mode      & \texttt{reflect} & \texttt{reflect} & \texttt{reflect} \\
    \bottomrule
    \end{tabular}
\end{table}

\subsubsection{Structure-aware metrics}

The \textbf{Pearson Correlation Coefficient (PCC)} metric measures the linear correlation between two signals and is defined as:
\begin{equation}
    r = \frac{1}{N_{T,C}} \sum_{T,C}
    \frac{\sum_S (y - \bar{y})(\hat{y} - \bar{\hat{y}})}
    {\sqrt{\sum_S (y - \bar{y})^2 \vphantom{\sum_S (\hat{y} - \bar{\hat{y}})^2}}
     \sqrt{\sum_S (\hat{y} - \bar{\hat{y}})^2}} \, .
\end{equation}

Here $\bar{\hat{y}}$ and $\bar{y}$ denote the means of the predictions and labels. The corresponding loss is defined as:
\begin{equation}
  \mathcal{L}_{\text{PCC}} = 1 - r .  
\end{equation}

PCC is invariant under affine transformations of the inputs, making it insensitive to uniform shifts or global scaling differences between fields. Minimizing $1-r$ therefore encourages the prediction to reproduce the overall covariance structure of the target, promoting spatial alignment of features.

The \textbf{Structural Similarity Index (SSIM)} for two given fields $y,\hat{y}$ is evaluated over sliding spatial windows and decomposes similarity into three components: luminance, contrast, and structure as described in \cite{nilsson2020understanding} and \cite{venkataramanan2021hitchhiker},
\begin{equation}
\text{SSIM}(y,\hat{y}) = l(y,\hat{y})\, c(y,\hat{y})\, s(y,\hat{y}),
\end{equation}

These components are defined as:
\begin{equation}
l(y,\hat{y}) =
\frac{2\mu_1 \mu_2 + C_1}{\mu_1^2 + \mu_2^2 + C_1},
\qquad
c(y,\hat{y}) =
\frac{2\sigma_1 \sigma_2 + C_2}{\sigma_1^2 + \sigma_2^2 + C_2},
\qquad
s(y,\hat{y}) =
\frac{\sigma_{12} + C_3}{\sigma_1 \sigma_2 + C_3},
\end{equation}
where $\mu_i$, $\sigma_i^2$, and $\sigma_{12}$ denote the local means, variances, and covariance of the two fields computed within a spatial window. $C_1$, $C_2$, and $C_3$ are small constants that stabilize division (typically $C_3 = C_2 / 2$).

The global SSIM score is obtained by spatially averaging local scores:
\begin{equation}
\text{MSSIM}(y,\hat{y}) =
\frac{1}{N_W}
\sum_{w\in\mathcal{W}}
\text{SSIM}(y,\hat{y}; w).
\end{equation}

\begin{figure}[ht]
    \centering
    \includegraphics[width=10cm]{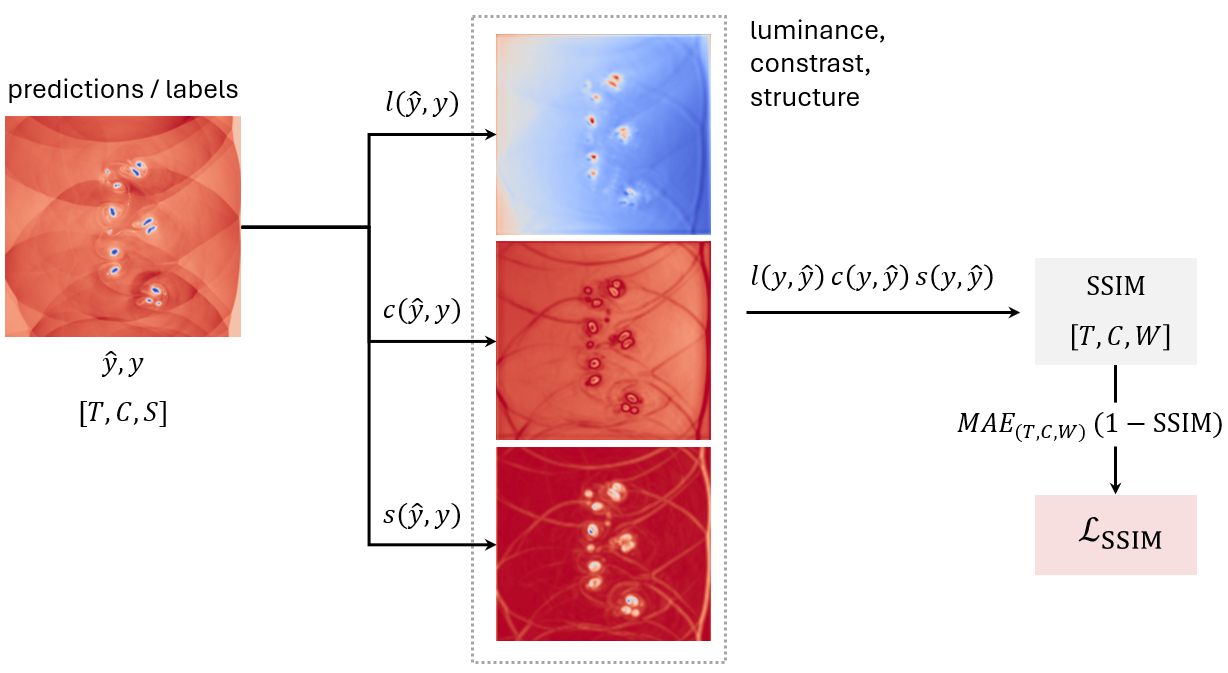}
    \caption{SSIM metric.}
    \label{fig:SSIM}
\end{figure}

The resulting values range from $-1$ to $1$, with $1$ denoting perfect similarity. When used as a training objective, this metric is written in the form (Figure \ref{fig:SSIM}):
\begin{equation}
\mathcal{L}_{\text{SSIM}} = 1 - \text{MSSIM}.
\end{equation}

The hyperparameters used for the SSIM metric are provided in Table \ref{tab:metric-config-ssim}.
\begin{table}[h!]
    \centering
    \caption{Configuration settings for \texttt{SSIM}.}
    \label{tab:metric-config-ssim}
    \begin{tabular}{ll}
    \toprule
    Hyperparameter & \texttt{default} \\
    \midrule
    Window size & \texttt{11} \\
    Sigma       & \texttt{1.5} \\
    K1          & \texttt{0.01} \\
    K2          & \texttt{0.03} \\
    L           & \texttt{10} \\
    \bottomrule
    \end{tabular}
\end{table}

\subsubsection{Physics-informed metrics}

Let $c(y)$ denote a scalar quantity of interest. On a discrete grid, we approximate the integral of this quantity over $\Omega$ by a sum over grid cells.

\begin{equation}
Q(y)=\sum_S c(y) \ \Delta V
\end{equation}

 where $\Delta V$ is the cell volume or quadrature weight. 
 
 The \textbf{Domain-integrated Quantity of Interest RMSE (iqRMSE)} is defined as:

\begin{equation}
\mathrm{iqRMSE}
=
\sqrt{
\frac{1}{N_{T}}
\sum_{T}
\left(
\frac{
Q(\hat{y}) - Q(y)
}{
Q(y) + \varepsilon
}
\right)^2
}.
\end{equation}

\begin{figure}[ht]
    \centering
    \includegraphics[width=13.5cm]{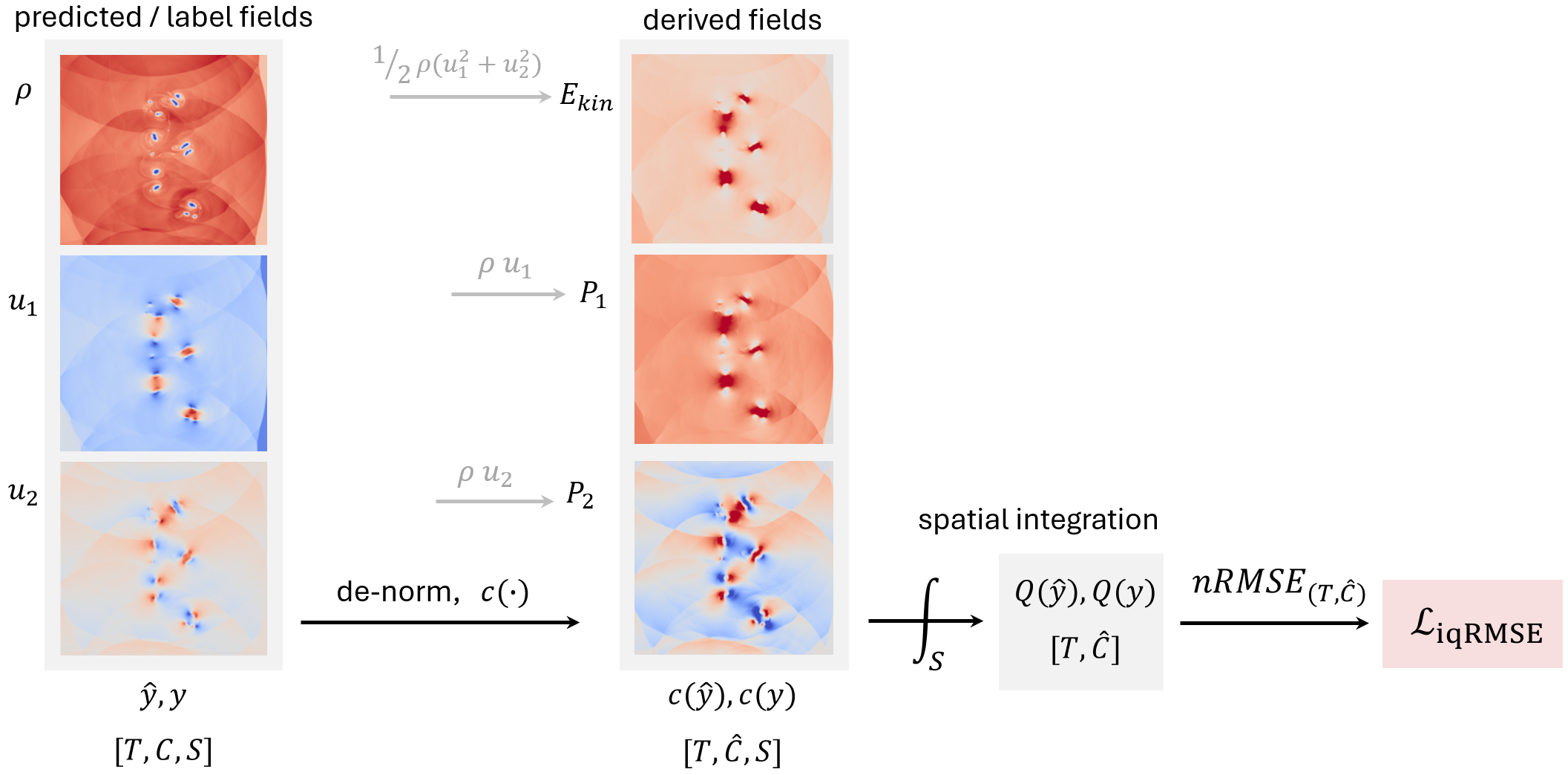}
    \caption{iqRMSE metric.}
    \label{fig:cRMSE}
\end{figure}

This metric measures discrepancies in the total amount of the conserved quantity, independent of how errors are distributed spatially within the domain (Figure \ref{fig:cRMSE}). The implemented quantities are summarized in Table \ref{tab:iqrmse_quantities}.

In the implementation, all predicted and target fields are first converted back to physical units before computing the metric. To keep the resulting values well-scaled, the conservation error is normalized by the corresponding target quantity, resulting in an nRMSE-style normalization that yields loss values of order $O(1)$.

\begin{table}[t]
\centering
\fontsize{10}{12}\selectfont
\caption{Quantities included in the iqRMSE metric.}
\label{tab:iqrmse_quantities}
\begin{tabular}{ll}
\toprule
Quantity & Expression \\
\midrule
Mass & $\int_\Omega \rho \, d\Omega$ \\ [3pt]

Momentum along axis $k$ & $\int_\Omega \rho u_k \, d\Omega$ \\ [3pt]

Kinetic energy & $\int_\Omega \tfrac{1}{2}\rho |\mathbf{u}|^2 \, d\Omega$ \\ [3pt]


Enstrophy & $\int_\Omega \tfrac{1}{2}|\nabla \times \mathbf{u}|^2 \, d\Omega$ \\ [3pt]

Center of mass along axis $k$ & $\frac{\int_\Omega \rho x_k \, d\Omega}{\int_\Omega \rho \, d\Omega}$ \\


\bottomrule
\end{tabular}
\end{table}


\subsection{Adaptive loss balancing framework}
\label{section:app_loss_balancing}
Given a set of $M$ loss components: $\{\mathcal{L}_i\}_{i=1}^M$, the scalar training objective is written as
\begin{equation}\label{eq:scalarloss}
\mathcal{J}(\theta)
=
\sum_{i=1}^{M} \lambda_i \mathcal{L}_i(\theta),
\qquad \lambda_i > 0,
\end{equation}
where $\lambda_i$ denotes the weight assigned to loss component $i$.

\begin{figure}[h!]
    \centering
    \includegraphics[width=12cm]{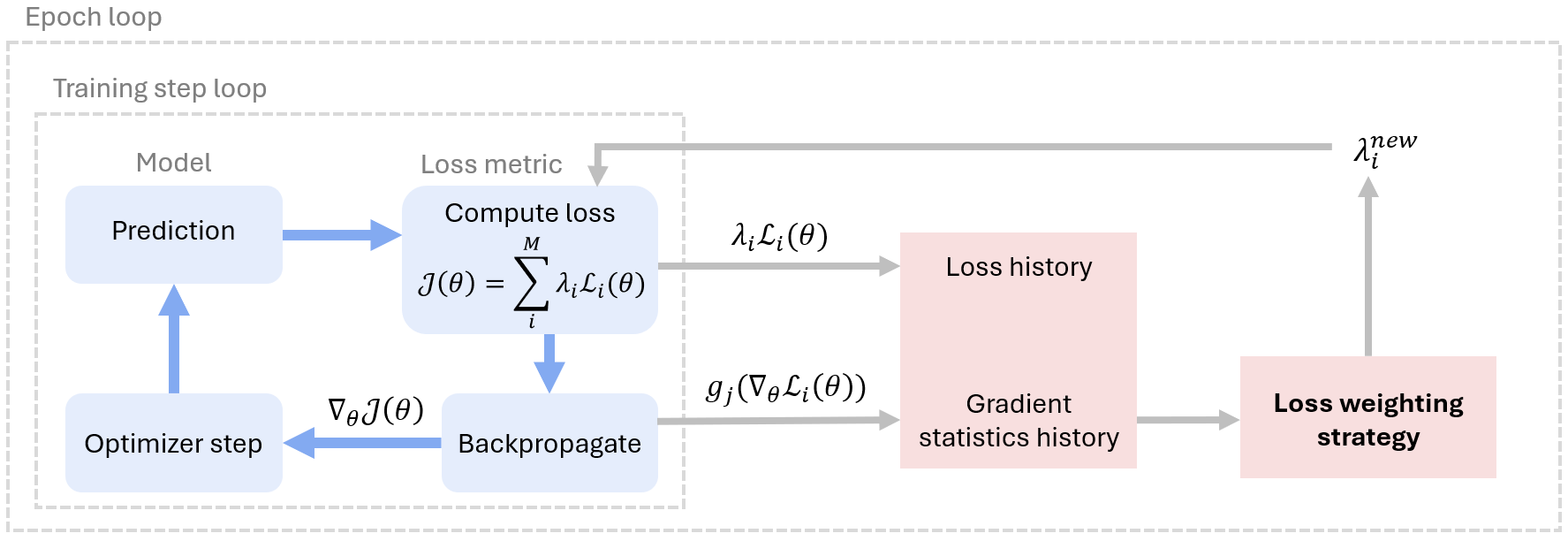}
    \caption{Conceptual illustration of the loss balancing framework. }
    \label{fig:lossbalancing}
\end{figure}

A generalized loss-balancing framework is illustrated in Figure \ref{fig:lossbalancing}. The scalarized training loss $\mathcal{J}(\theta)$ is computed via Equation \ref{eq:scalarloss}, and the constituent loss components are individually sampled and stored. The same is optionally performed for component gradients, whereby only aggregated statistics across parameters are stored (indicated here by $g_j$) as opposed to full gradient tensors, in order to reduce memory overhead. These histories form the input to the \emph{loss weighting strategy}.  The sampling frequency and update frequency are configurable independently. By default, loss values or gradients are sampled at every training step, while the loss weights are updated once per epoch. This makes it possible to trade off responsiveness against computational overhead. In this work, we employ two loss balancing strategies: SoftAdapt  as shown in algorithm \ref{alg:app_softadapt} and the Gradient Normalization (GradNorm) as shown in algorithm \ref{alg:app_grad_norm}. 

\subsubsection{SoftAdapt}

SoftAdapt \cite{heydari2019softadapt} adjusts the weights of loss components according to their recent rate of change. The key idea is to prioritize objectives whose loss has decreased more slowly (or increased), indicating that the model is currently struggling to optimize them. To estimate this behavior, SoftAdapt approximates the recent slope of each loss component using a short history of past values.

Let $\mathcal{L}_i^{(t)}$ denote the value of loss component $i$ at training step $t$. The recent rate of change is approximated as $s_i^{(t)} = \mathcal{L}_i^{(t)} - \mathcal{L}_i^{(t-1)}$. These slopes are converted into loss component weights through a softmax transformation

\begin{equation}
    \lambda_i^{(t)} =
    \frac{\exp(\beta s_i^{(t)})}
         {\sum_{j=1}^{T} \exp(\beta s_j^{(t)})},
    \label{eq:softadapt}
\end{equation}

where $\beta$ is a tunable temperature parameter that controls how strongly the weighting emphasizes poorly improving losses. For $\beta>0$, losses with larger positive slopes receive higher weights, while $\beta<0$ favors losses that are decreasing fastest. Setting $\beta=0$ results in uniform weighting. The hyperparameter value chosen for this work is presented in Table \ref{tab:softadapt_config}.

\begin{table}[h!]
    \centering
    \caption{Configuration settings for \texttt{SoftAdapt}.}
    \label{tab:softadapt_config}
    \begin{tabular}{ll}
    \toprule
    Hyperparameter & \texttt{default} \\
    \midrule
    Temperature ($\beta$) & \texttt{1.0} \\
    \bottomrule
    \end{tabular}
\end{table}

\begin{algorithm}[h!]
\caption{Training with SoftAdapt}
\label{alg:app_softadapt}
\begin{algorithmic}[1]
\State Choose temperature parameter $\beta$
\State Initialize weights $\lambda_i \leftarrow \frac{1}{M} \ \forall i$
\State Store initial losses $\mathcal{L}_i^{(0)}$
\For{each epoch}
    \For{each training step}
        \State Train network with weighted loss $\mathcal{J} = \sum_{i=1}^{M} \lambda_i \mathcal{L}_i$
        \State Record current losses $\mathcal{L}_i^{(t)}$
    \EndFor

    \State Compute slope estimates $s_i^{(t)} = \mathcal{L}_i^{(t)} - \mathcal{L}_i^{(t-1)}$
    \State Compute weights $\lambda_i^{(t)}$ (Equation \ref{eq:softadapt})
    \State Normalize weights such that $\sum_{i=1}^{M} \lambda_i = 1$
\EndFor
\end{algorithmic}
\end{algorithm}

\subsubsection{Gradient Normalization (GradNorm)}

GradNorm \citep{chen2017gradnorm} updates the loss component weights such that the gradient norm of each weighted loss component matches a target determined by the component’s relative training speed. In practice, the method is applied to a shared parameter subset $\theta_{\text{sh}}\subset\theta$, typically the last shared layer. For task $i$, the gradient norm is defined as $G_\theta^{(i)} = \|\nabla_\theta (\lambda_i \mathcal{L}_i)\|_2$. The target gradient norm $\hat{G}_{\theta_{\text{sh}}}^{(i)} = G_{\theta_{\text{sh}}} \,[r_i]^\alpha$ is based on the normalized loss ratio $\tilde{\mathcal{L}}_i$ and the relative inverse training rate $r_i$,
\begin{equation}
    r_i = \frac{\tilde{\mathcal{L}}_i}{\frac{1}{M}\sum_j \tilde{\mathcal{L}}_j}, \qquad \tilde{\mathcal{L}}_i = \frac{\mathcal{L}_i}{\mathcal{L}_i(0)},
    \label{eq:gradnorm}
\end{equation}

and the asymmetry hyperparameter $\alpha$. The hyperparameter values selected for this work are presented in Table \ref{tab:gradnorm_config}.

\begin{algorithm}[h!]
\caption{Training with GradNorm}
\label{alg:app_grad_norm}
\begin{algorithmic}[1]
\State Initialize task weights $\lambda_i \leftarrow \lambda_i(0) \ \forall i$ 
\State Store initial task losses $\mathcal{L}_i(0)$
\For{each epoch}
    \For{each training step}
        \State Train network for with weighted loss $\mathcal{J} = \sum_{i=1}^{M} \lambda_i \mathcal{L}_i$
        \State Compute gradient norms on shared parameters $\theta_{\text{sh}}$ as $G_{\theta_{\text{sh}}}^{(i)} = \|\nabla_{\theta_{\text{sh}}} (\lambda_i \mathcal{L}_i)\|_2$
    \EndFor
    \State Compute average gradient norm across the $M$ loss components: $G_{\theta_{\text{sh}}} = \frac{1}{M}\sum_{i=1}^{M} G_{\theta_{\text{sh}}}^{(i)}$
    \State Compute normalized losses $\tilde{\mathcal{L}}_i$ and inverse training rates $r_i$ (Equation \ref{eq:gradnorm})
    \State Define target gradient norms $\hat{G}_{\theta_{\text{sh}}}^{(i)} = G_{\theta_{\text{sh}}} \,[r_i]^\alpha$
    \State Compute GradNorm objective $\mathcal{L}_{\text{grad}} = \sum_{i=1}^{M} \left| G_{\theta_{\text{sh}}}^{(i)} - \hat{G}_{\theta_{\text{sh}}}^{(i)} \right|$
    \State Update task weights using $\nabla_{\lambda_i} \mathcal{L}_{\text{grad}}$ 
          (targets $\hat{G}_{\theta_{\text{sh}}}^{(i)}$ treated as constants)
    \State Renormalize weights such that $\sum_{i=1}^{M} \lambda_i = M$
\EndFor
\end{algorithmic}
\end{algorithm}

\begin{table}[h!]
    \centering
    \caption{Configuration settings for \texttt{GradNorm}.}
    \label{tab:gradnorm_config}
    \begin{tabular}{ll}
    \toprule
    Hyperparameter & \texttt{default} \\
    \midrule
    Asymmetry parameter ($\alpha$) & \texttt{1.0} \\
    Loss weight learning rate & \texttt{0.001} \\
    \bottomrule
    \end{tabular}
\end{table}

\subsection{Computational Resources}
\label{section:app_computational_resources}

Each trajectory of the high fidelity 2D-datasets and the 3D-SDBA (SSOOSS) datasets was generated on an HPC-cluster using a single Intel-Xeon(R) Platinum 8480+ node with 112 CPU-cores, while a single 3D-SABW (SSOOSS) trajectory was generated on a single node of Intel Skylake Xeon Platinum 8174 with 24 CPU-cores.

For training surrogate models, we employ a Data Distributed Parallel strategy on 4 NVIDIA-A100 GPUs each with 40GB of VRAM for the 2D baselines, while the 3D baselines were trained on 4 NVIDIA-H100 GPUs, each with 90GB VRAM.
\clearpage
\section{Detailed results}
\label{section:app_extended_results}

\subsection{Metric performance comparison}
\label{section:app_metric_performance_comparison}

\begin{figure}[ht]
    \centering
    \includegraphics[width=0.6\textwidth]{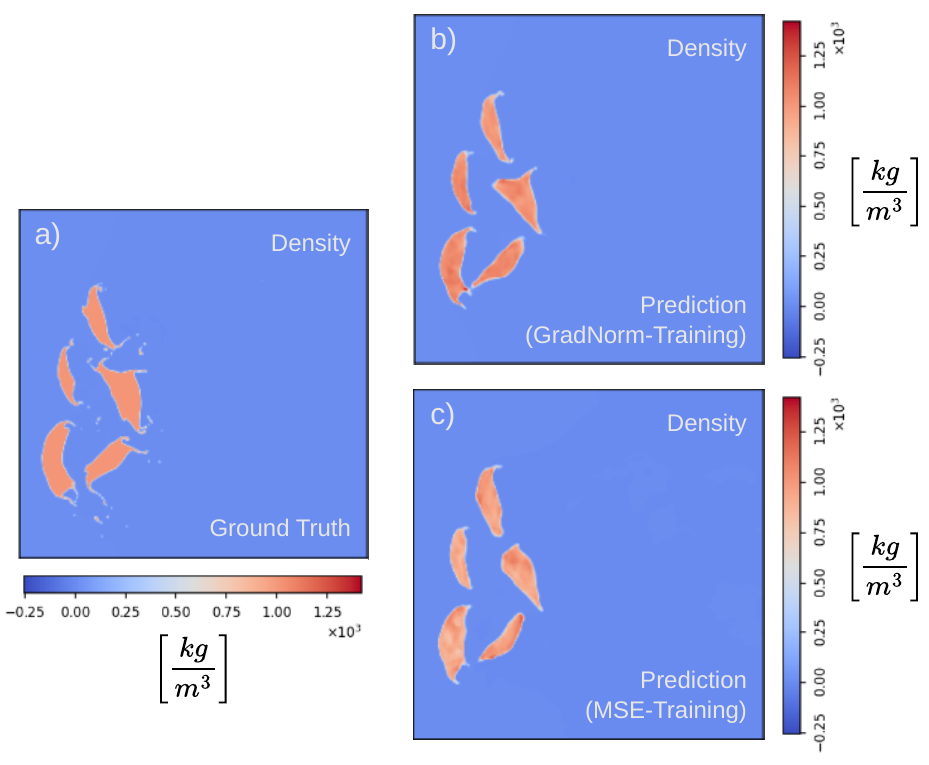}
    \caption{(a) Reference density snapshot for a randomly chosen 2D-SDBA dataset at t=70 s. (b) Corresponding FFNO (50M) prediction trained with composite loss and further using GradNorm adaptive loss weighting strategy. RMSE = 0.46 and SSIM = 0.28.(c) Corresponding FFNO prediction trained with only MSE loss. RMSE = 0.46 and SSIM = 0.32}
    \label{fig:ssim_vs_rmse}
\end{figure}

Figure \ref{fig:ssim_vs_rmse} shows the difference in the prediction quality for the last timestep of a 2D-SDBA trajectory using FFNO with two different training strategies and also comparison with the ground truth. It can be observed that the predictions having better SSIM values have more structural coherence with the ground truth even though RMSE indicates equally good predictions for both. 

\begin{figure}[h!]
    \centering
    \includegraphics[width=0.8\textwidth]{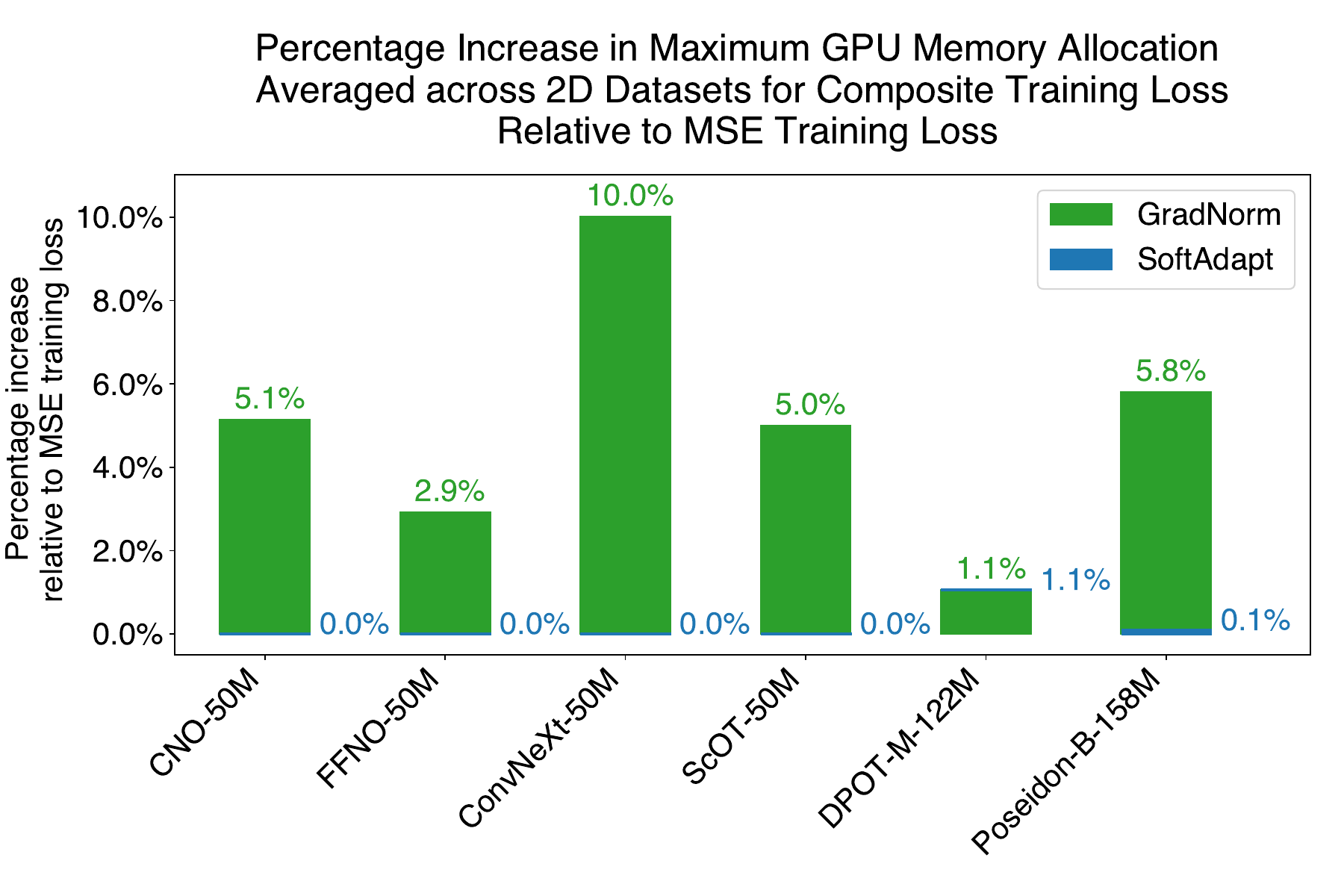}
    \caption{Peak GPU memory allocation during training}
    \label{fig:app_peak_gpu_memory}
\end{figure}

\begin{figure}[h!]
    \centering
    \includegraphics[width=0.8\textwidth]{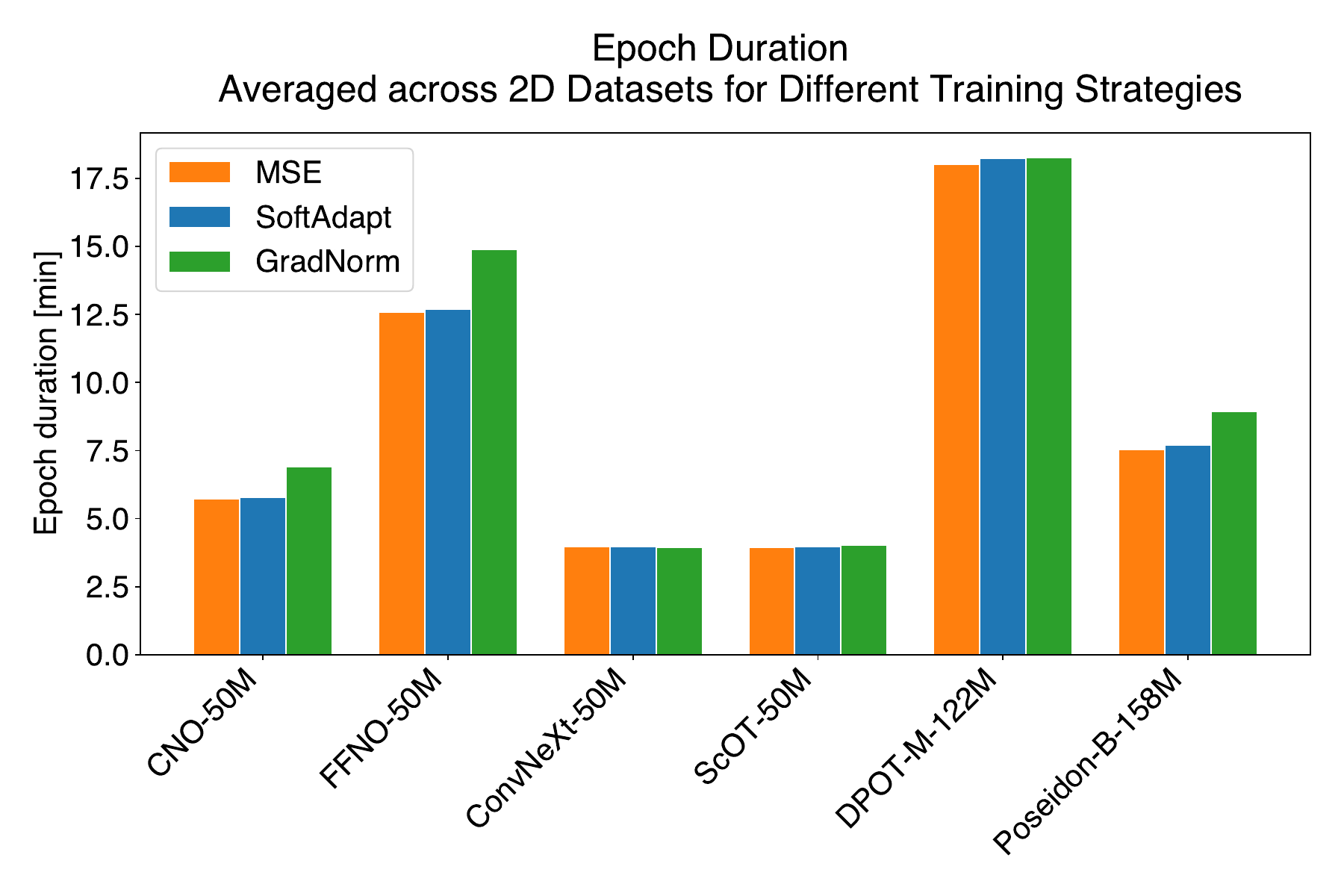}
    \caption{Epoch duration}
    \label{fig:app_epoch_duration}
\end{figure}

As the results in the main text indicate clear advantages of using composite losses with adaptive loss-weighting strategies, a natural question concerns the additional computational cost incurred by these methods. We examine this trade-off from two perspectives: the peak GPU memory requirement, shown in Figure~\ref{fig:app_peak_gpu_memory}, and the training time required to complete one epoch, shown in Figure~\ref{fig:app_epoch_duration}. Since GradNorm requires gradient information to adapt the loss weights, it is expected to introduce additional memory and computational overhead. In contrast, SoftAdapt does not require gradient storage and therefore introduces negligible overhead in both memory usage and training time per epoch.

\subsection{Inference metric evolution over rollout steps}
\label{section:app_rollout_metrics}
This section presents the evolution of inference metrics across rollout steps for in-distribution test-dataset trajectories. The metrics are averaged across all trajectories in the test dataset at each rollout step and the corresponding mean and standard deviation are presented in the following figures. The figures are organized by dataset, with each figure corresponding to a specific dataset and containing the results for all baselines and training strategies evaluated on that dataset. The metrics are categorized into pointwise, spectral, feature-focused, structure-aware, and physics-informed. Poseidon-M has a different starting point in all the plots shown in this section, since the first prediction timestep depends on the number of input timesteps required by the baseline and the stride between those timesteps.
\clearpage
\subsubsection{2D shock-induced air bubble collapse in water with open boundaries [2D-SABW (OOOO)]}
\begin{figure}[h!]
    \centering
    \includegraphics[width=0.8\textwidth]{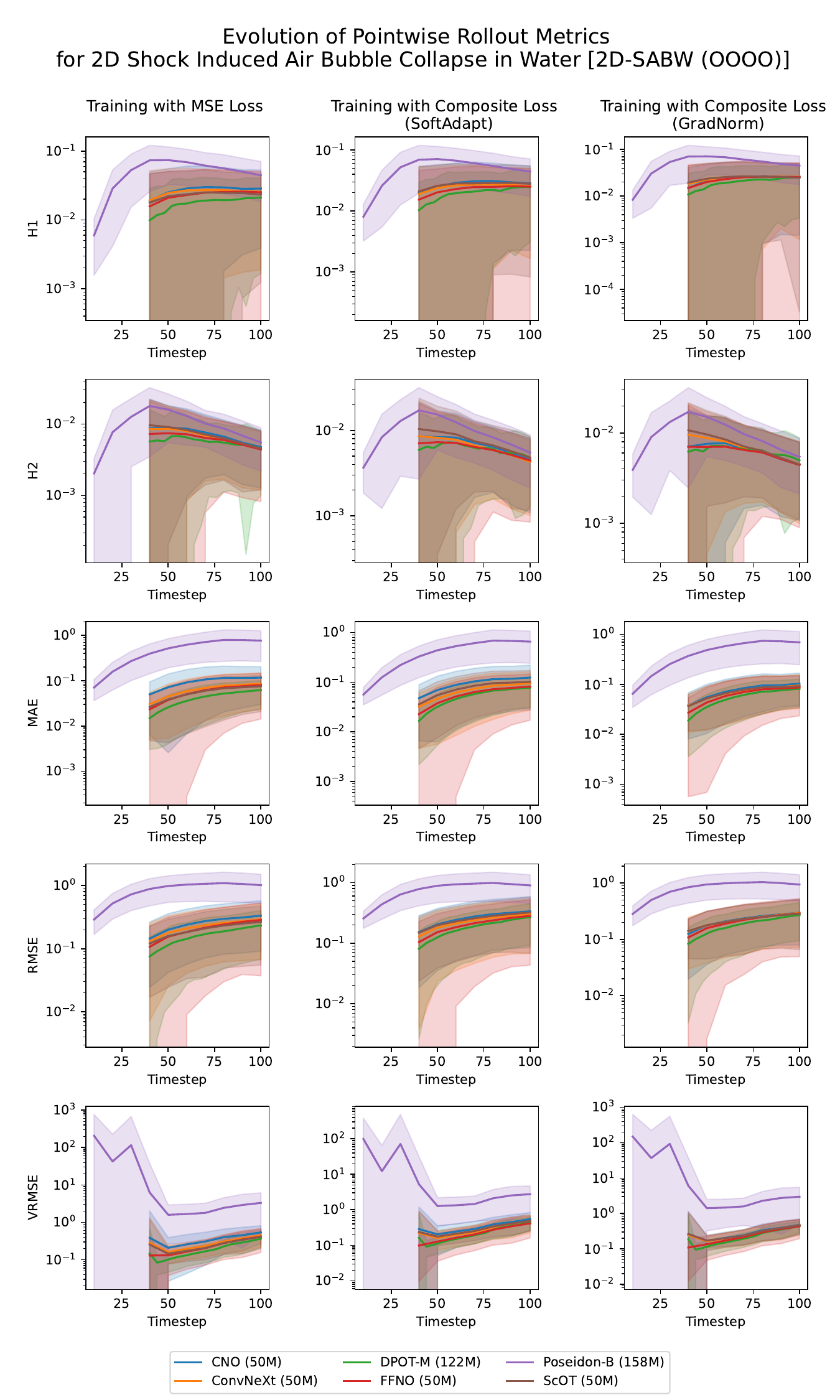}
    \caption{ Pointwise evaluation of baselines on the 2D-SABW (OOOO) test dataset, reporting errors
in pointwise field values (MAE, RMSE, VRMSE) and the field derivatives (H1, H2) averaged across
all fields for each timestep.}
    \label{fig:2d_sabw_oooo_pointwise_rollout_evolution}
\end{figure}
\begin{figure}[h!]
    \centering
    \includegraphics[width=0.8\textwidth]{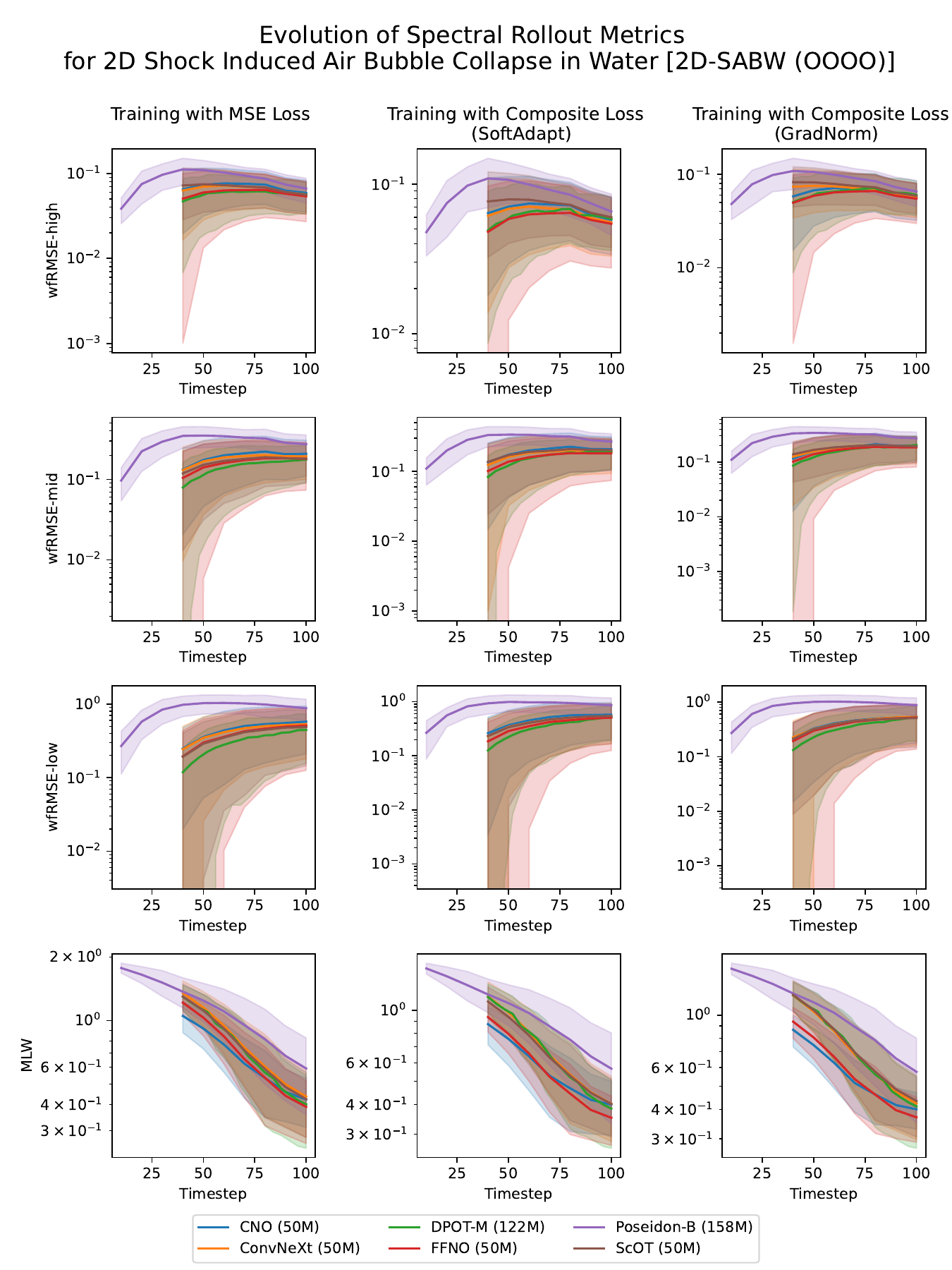}
    \caption{Wavelet based spectral metrics for the 2D-SABW (OOOO) test dataset, averaged across all fields for each timestep.}
    \label{fig:2d_sabw_oooo_spectral_rollout_evolution}
\end{figure}
\begin{figure}[h!]
    \centering
    \includegraphics[width=0.8\textwidth]{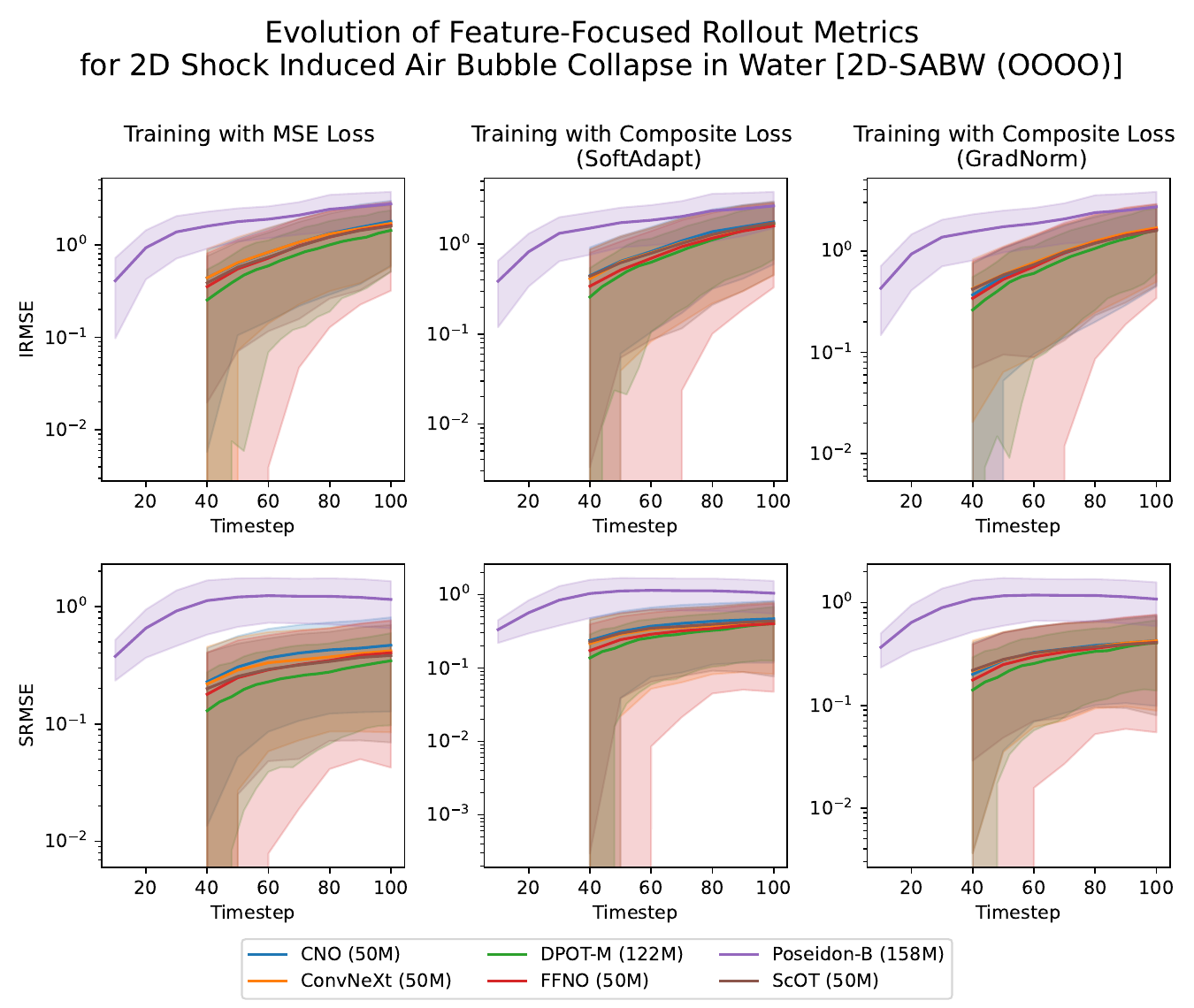}
    \caption{Feature focused metrics for capturing interface (IRMSE) and shocks (SRMSE) on the 2D-SABW (OOOO) test dataset, averaged across all fields for each timestep.}
    \label{fig:2d_sabw_oooo_feature_focused_rollout_evolution}
\end{figure}
\begin{figure}[h!]
    \centering
    \includegraphics[width=0.8\textwidth]{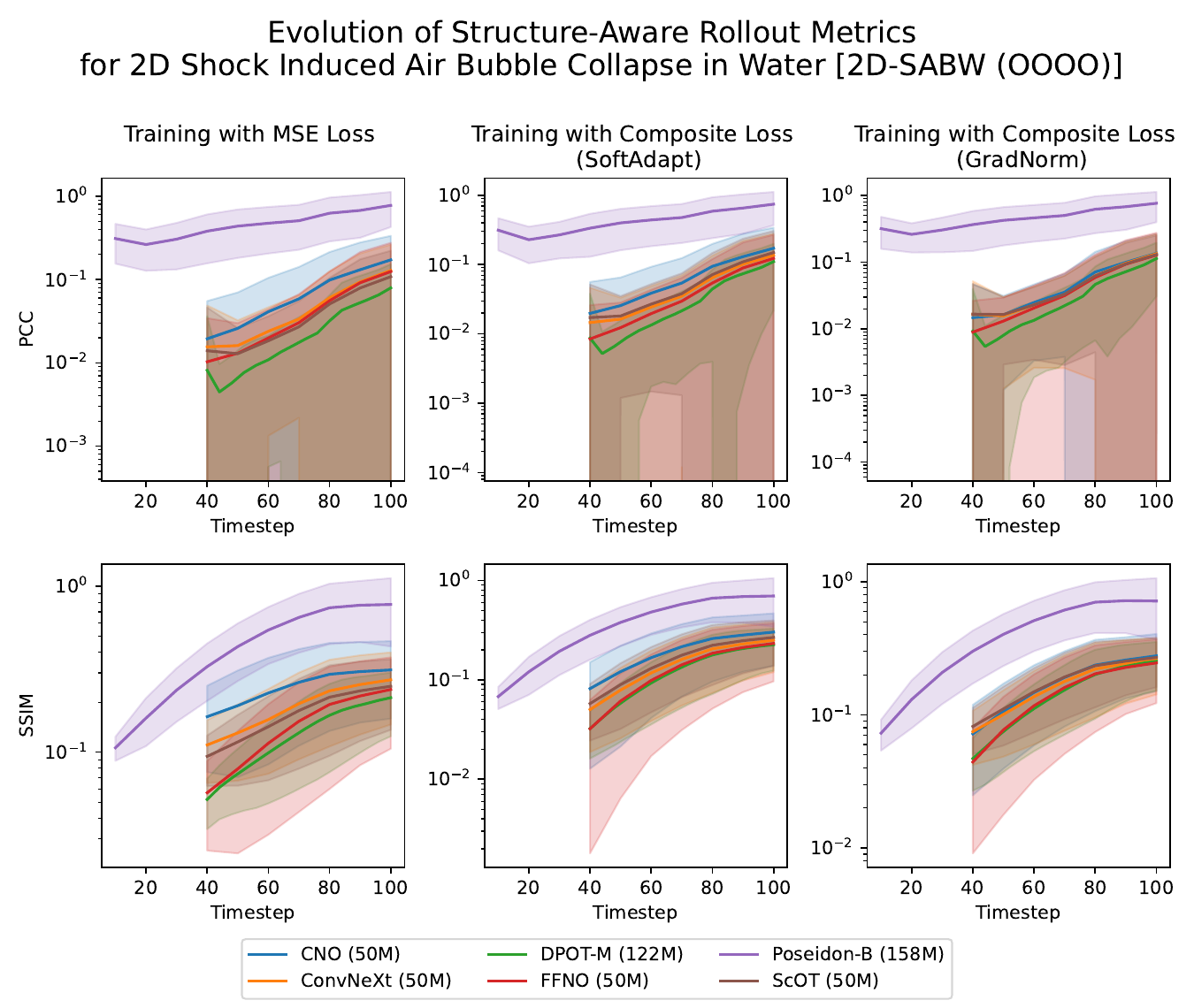}
    \caption{Structure-aware metrics for the 2D-SABW (OOOO) test dataset, averaged across all fields for each timestep.}
    \label{fig:2d_sabw_oooo_structure_aware_rollout_evolution}
\end{figure}
\begin{figure}[h!]
    \centering
    \includegraphics[width=0.8\textwidth]{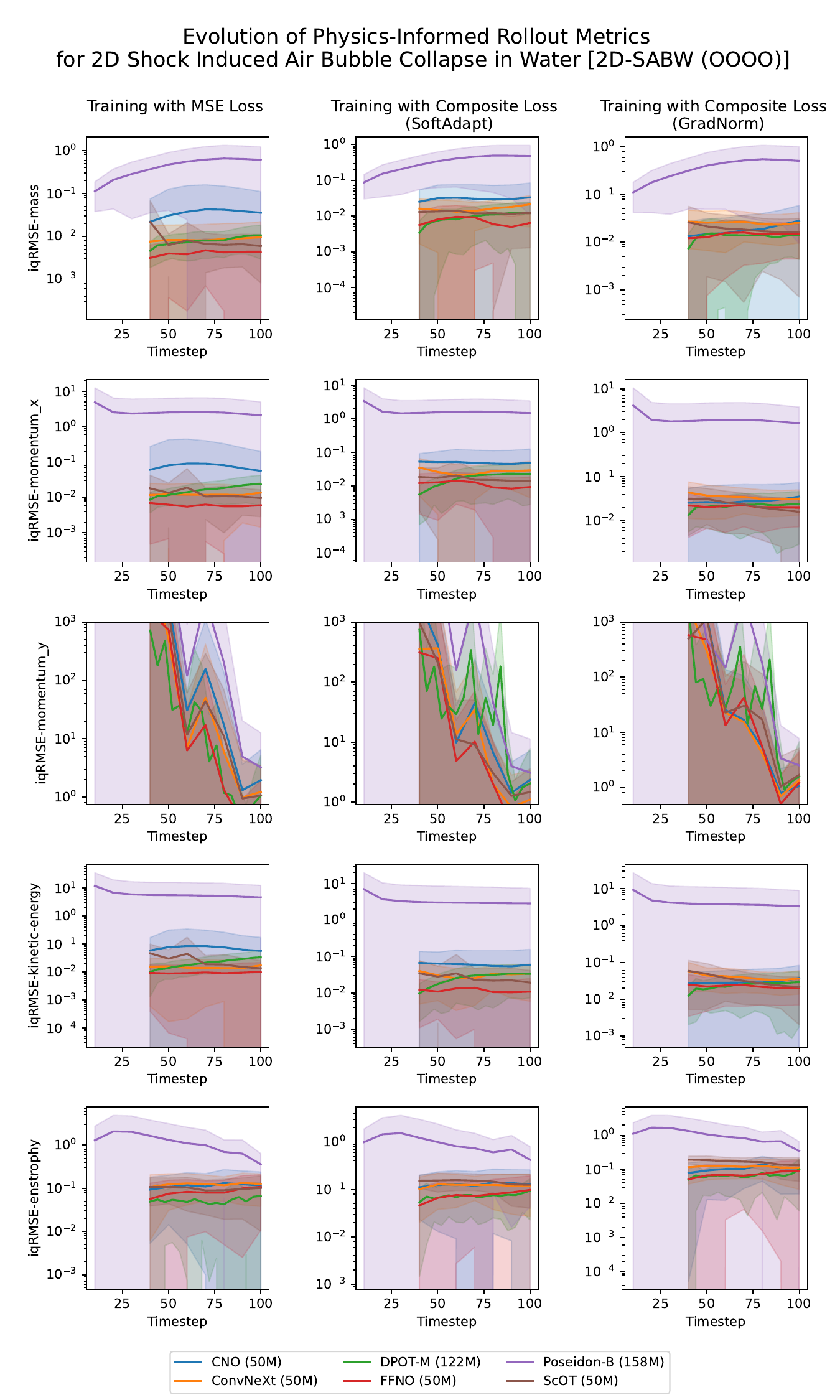}
    \caption{Physics-informed metrics reporting the nRMSE of integral quantities of interest on the 2D-SABW (OOOO) test dataset, averaged across all fields for each timestep.}
    \label{fig:2d_sabw_oooo_physics_informed_rollout_evolution}
\end{figure}

\clearpage
\subsubsection{2D shock-induced air bubble collapse in water with symmetry boundaries [2D-SABW (SSOO)]}
\begin{figure}[h!]
    \centering
    \includegraphics[width=0.8\textwidth]{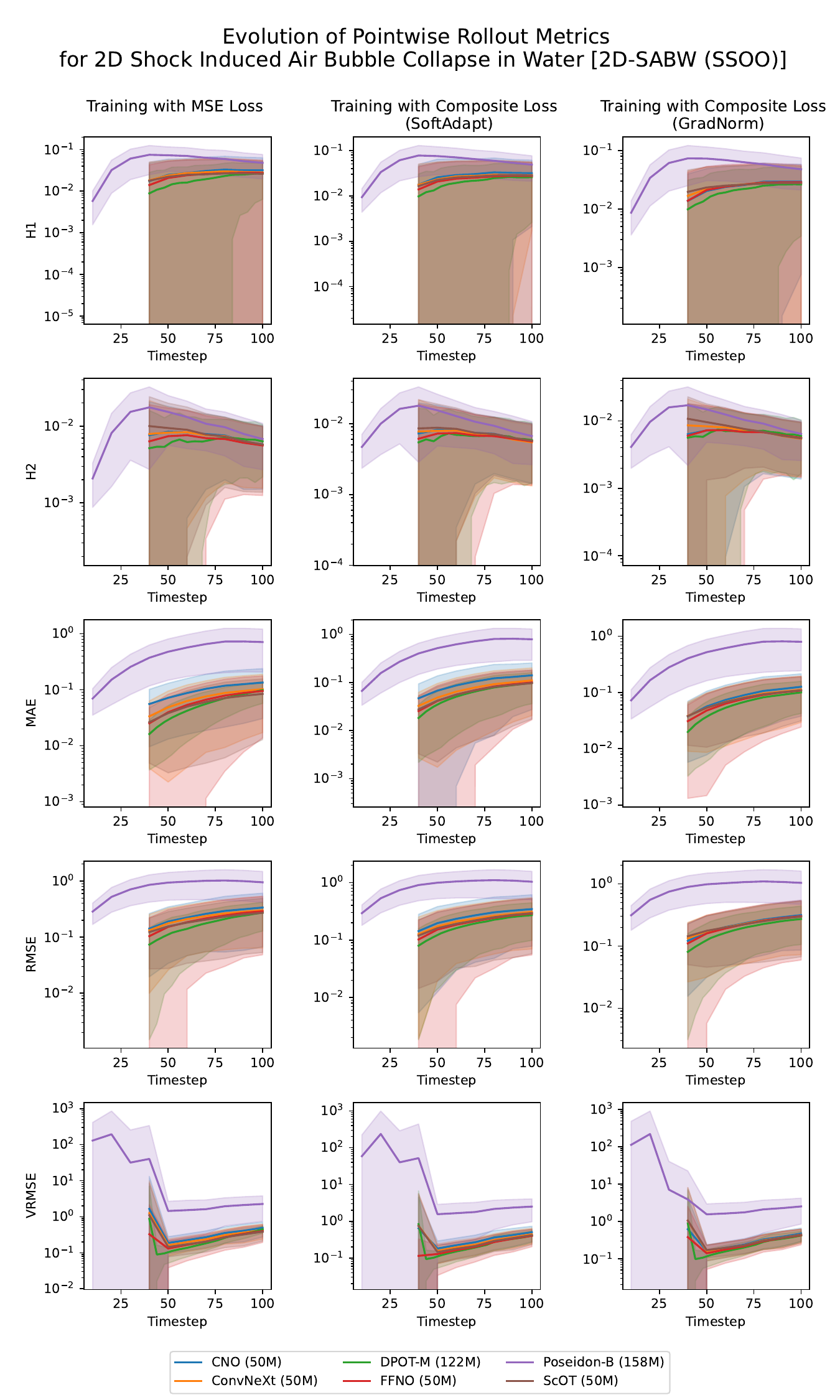}
    \caption{ Pointwise evaluation of baselines on the 2D-SABW (SSOO) test dataset, reporting errors
in pointwise field values (MAE, RMSE, VRMSE) and the field derivatives (H1, H2) averaged across
all fields for each timestep.}
    \label{fig:2d_sabw_ssoo_pointwise_rollout_evolution}
\end{figure}
\begin{figure}[h!]
    \centering
    \includegraphics[width=0.8\textwidth]{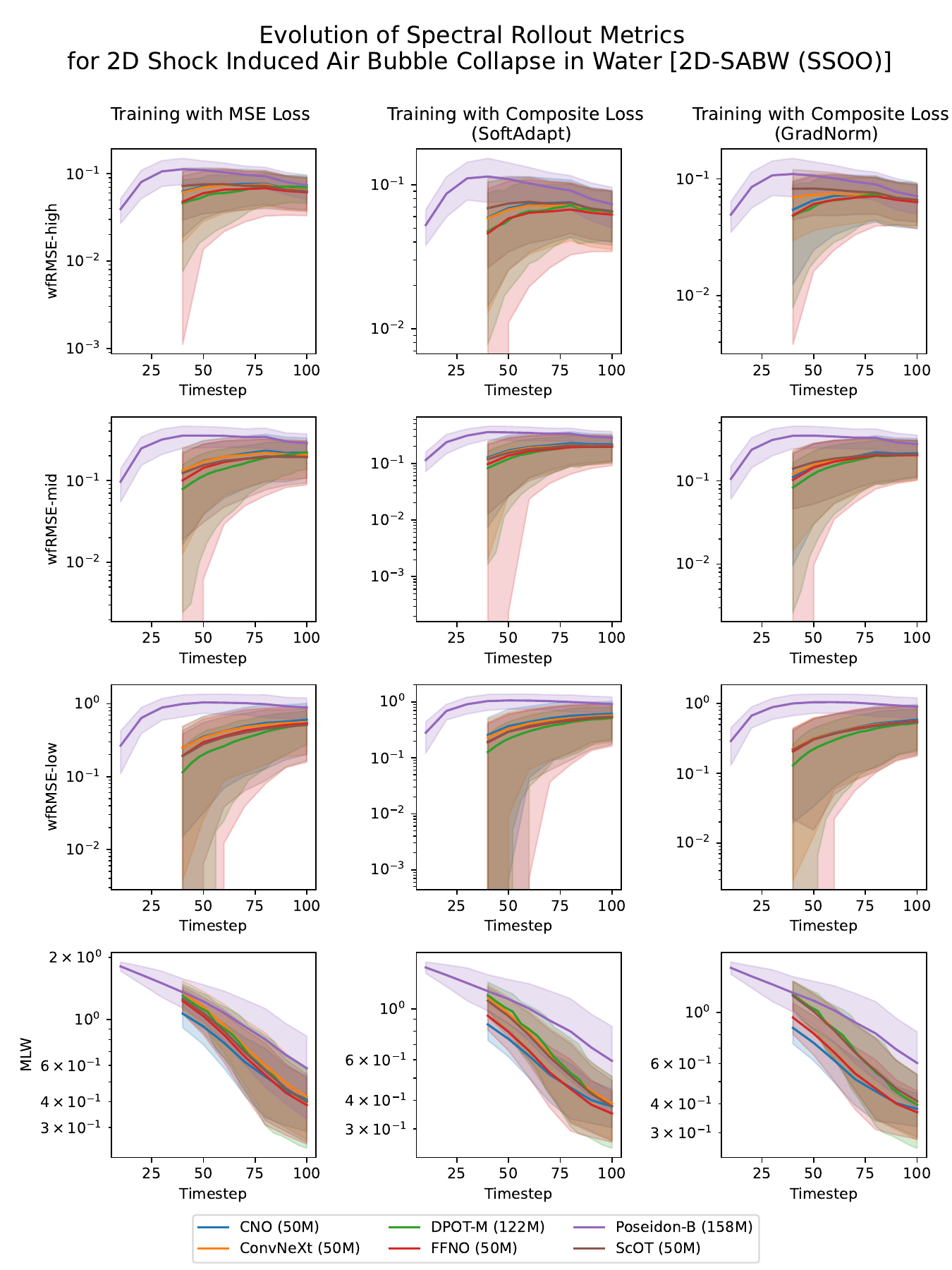}
    \caption{Wavelet based spectral metrics for the 2D-SABW (SSOO) test dataset, averaged across all fields for each timestep.}
    \label{fig:2d_sabw_ssoo_spectral_rollout_evolution}
\end{figure}
\begin{figure}[h!]
    \centering
    \includegraphics[width=0.8\textwidth]{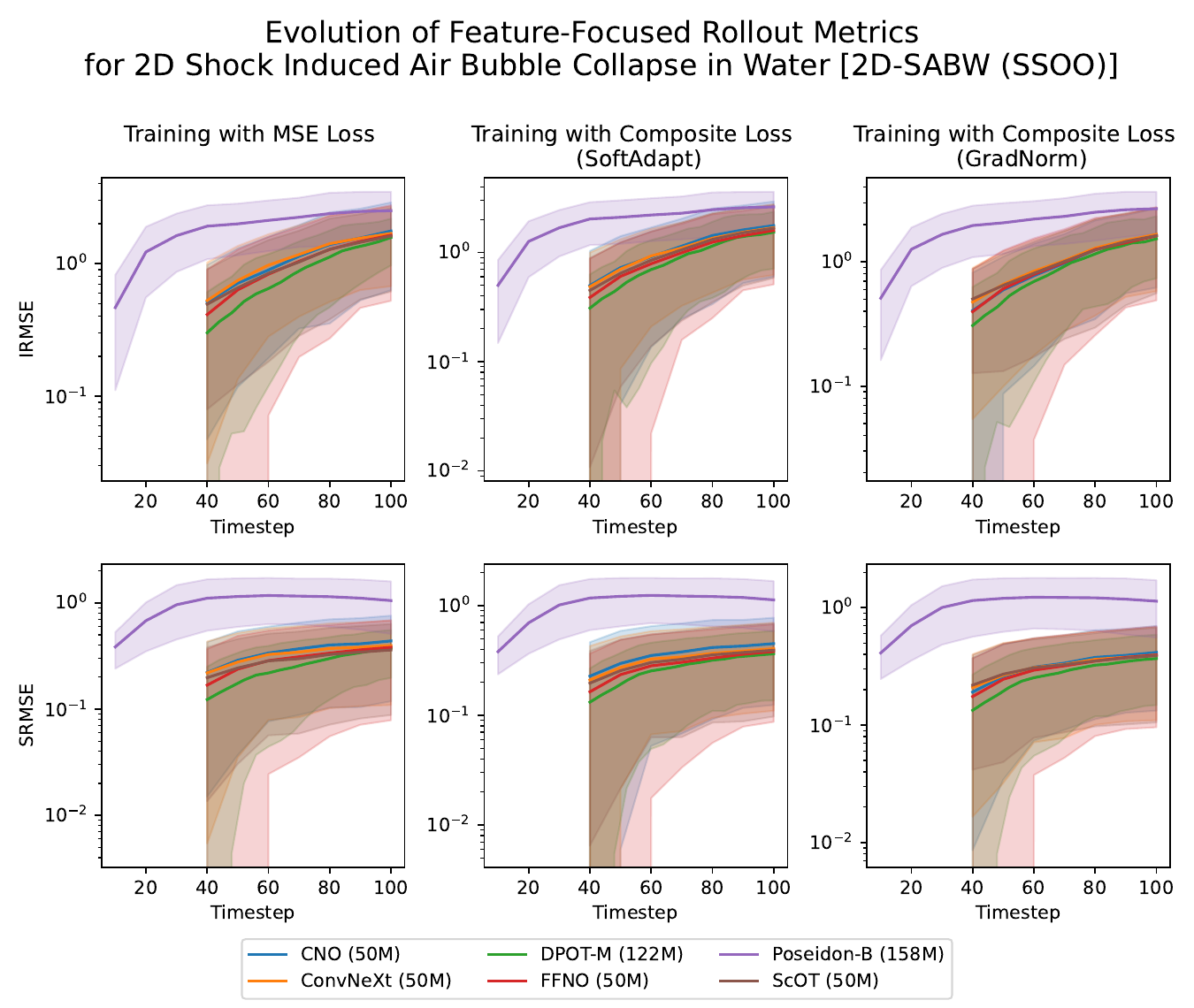}
    \caption{Feature focused metrics for capturing interface (IRMSE) and shocks (SRMSE) on the 2D-SABW (SSOO) test dataset, averaged across all fields for each timestep.}
    \label{fig:2d_sabw_ssoo_feature_focused_rollout_evolution}
\end{figure}
\begin{figure}[h!]
    \centering
    \includegraphics[width=0.8\textwidth]{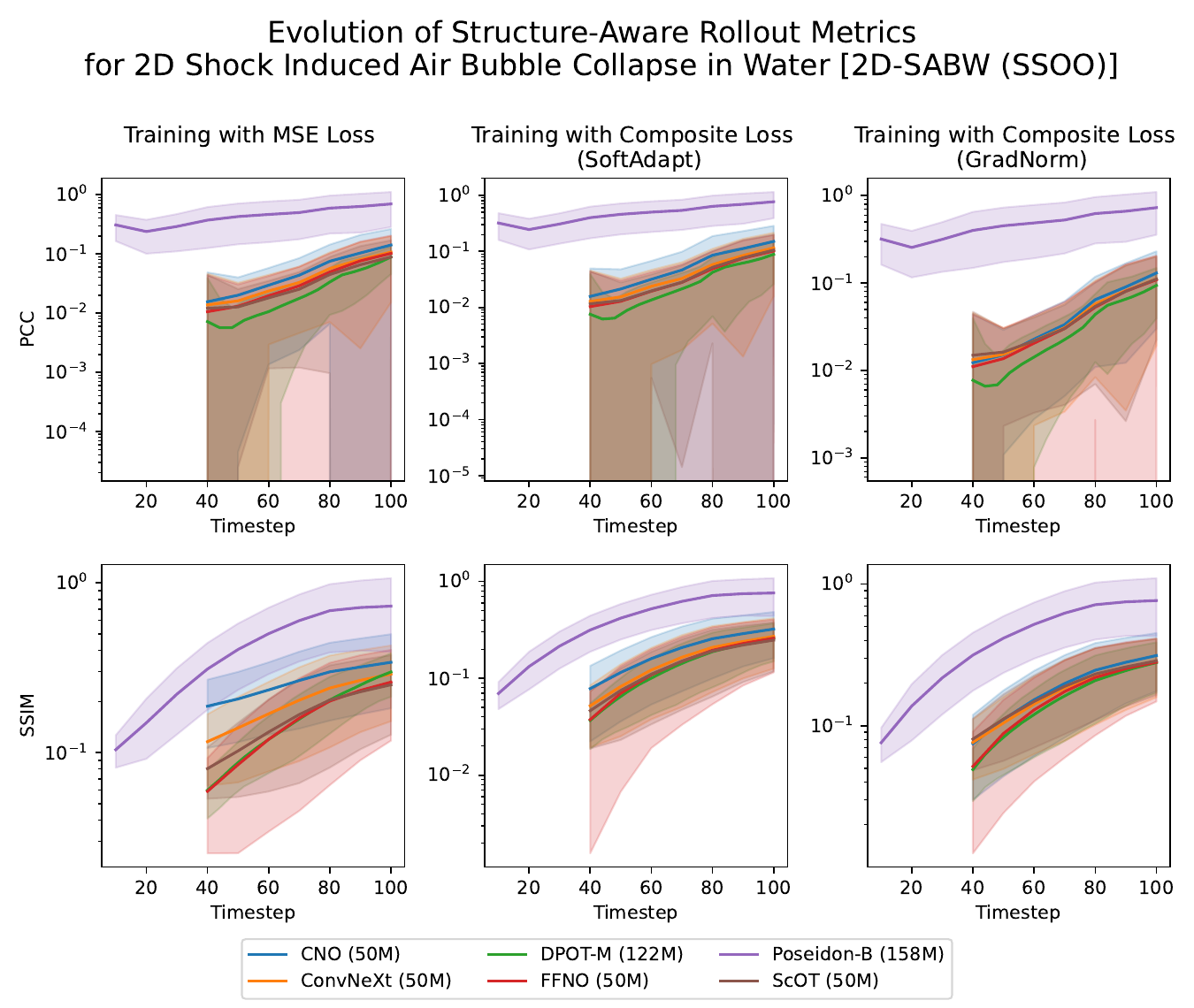}
    \caption{Structure-aware metrics for the 2D-SABW (SSOO) test dataset, averaged across all fields for each timestep.}
    \label{fig:2d_sabw_ssoo_structure_aware_rollout_evolution}
\end{figure}
\begin{figure}[h!]
    \centering
    \includegraphics[width=0.8\textwidth]{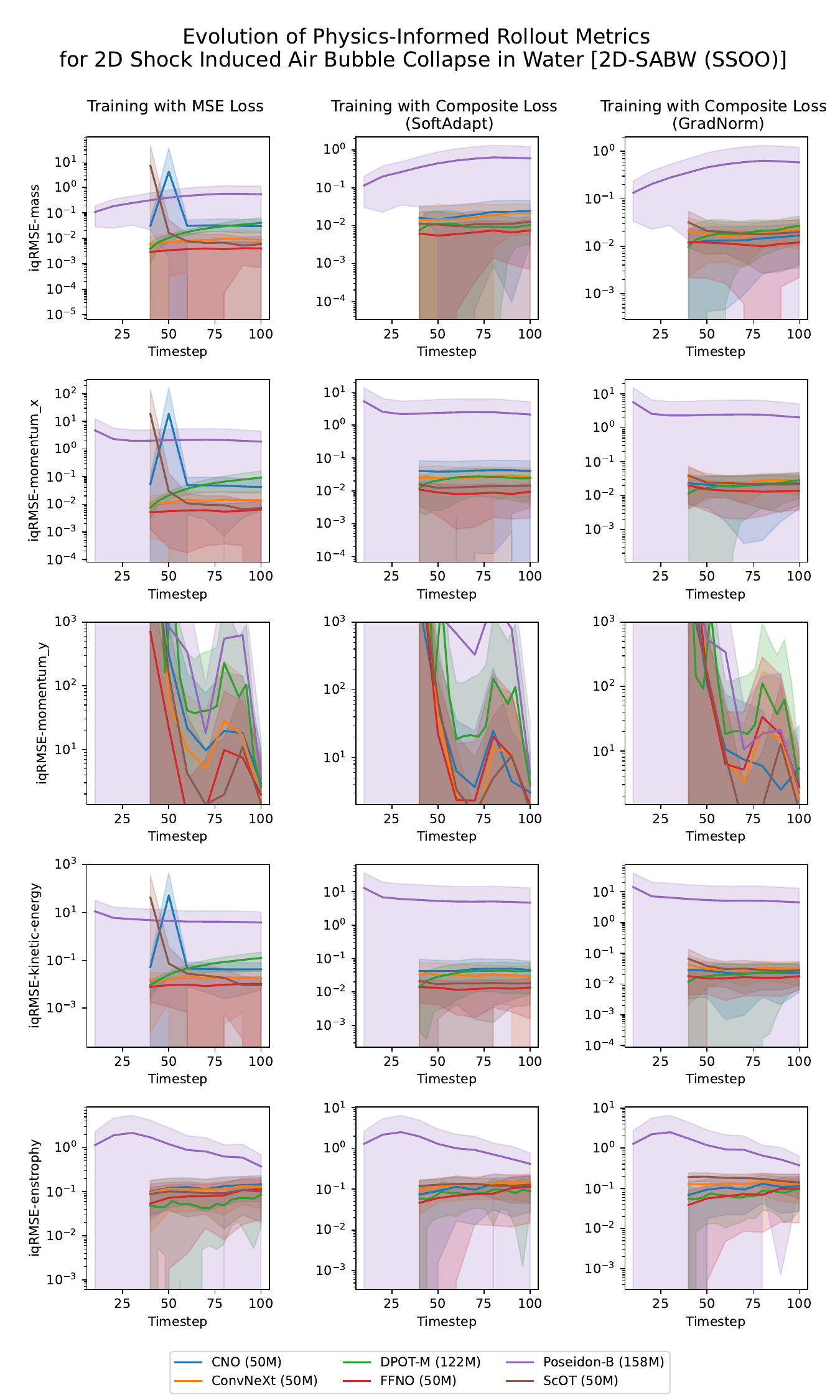}
    \caption{Physics-informed metrics reporting the nRMSE of integral quantities of interest on the 2D-SABW (SSOO) test dataset, averaged across all fields for each timestep.}
    \label{fig:2d_sabw_ssoo_physics_informed_rollout_evolution}
\end{figure}

\clearpage
\subsubsection{2D shock-induced R22 bubble collapse in air with open boundaries [2D-SRBA (OOOO)]}
\begin{figure}[h!]
    \centering
    \includegraphics[width=0.8\textwidth]{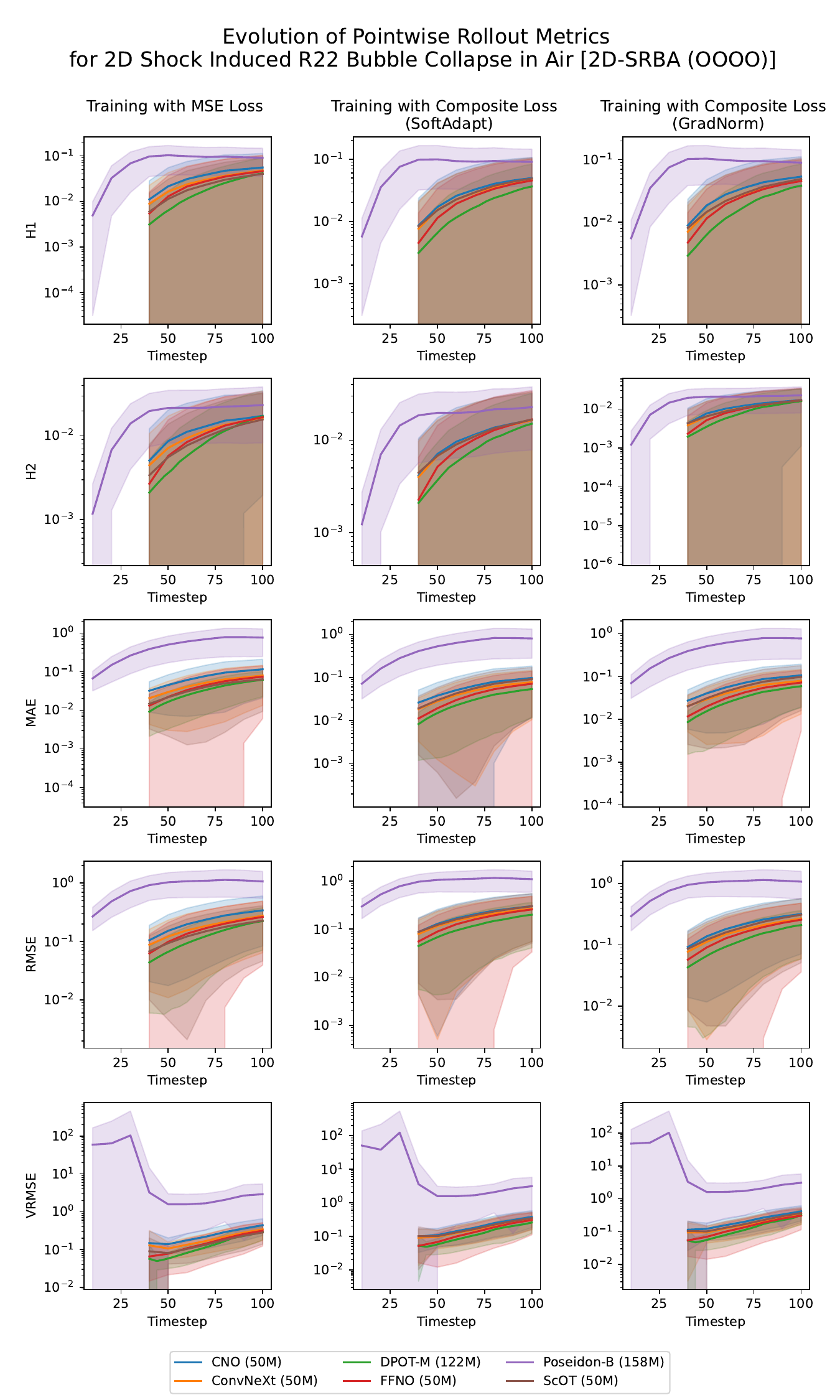}
    \caption{ Pointwise evaluation of baselines on the 2D-SRBA (OOOO) test dataset, reporting errors
in pointwise field values (MAE, RMSE, VRMSE) and the field derivatives (H1, H2) averaged across
all fields for each timestep.}
    \label{fig:2d_srba_oooo_pointwise_rollout_evolution}
\end{figure}
\begin{figure}[h!]
    \centering
    \includegraphics[width=0.8\textwidth]{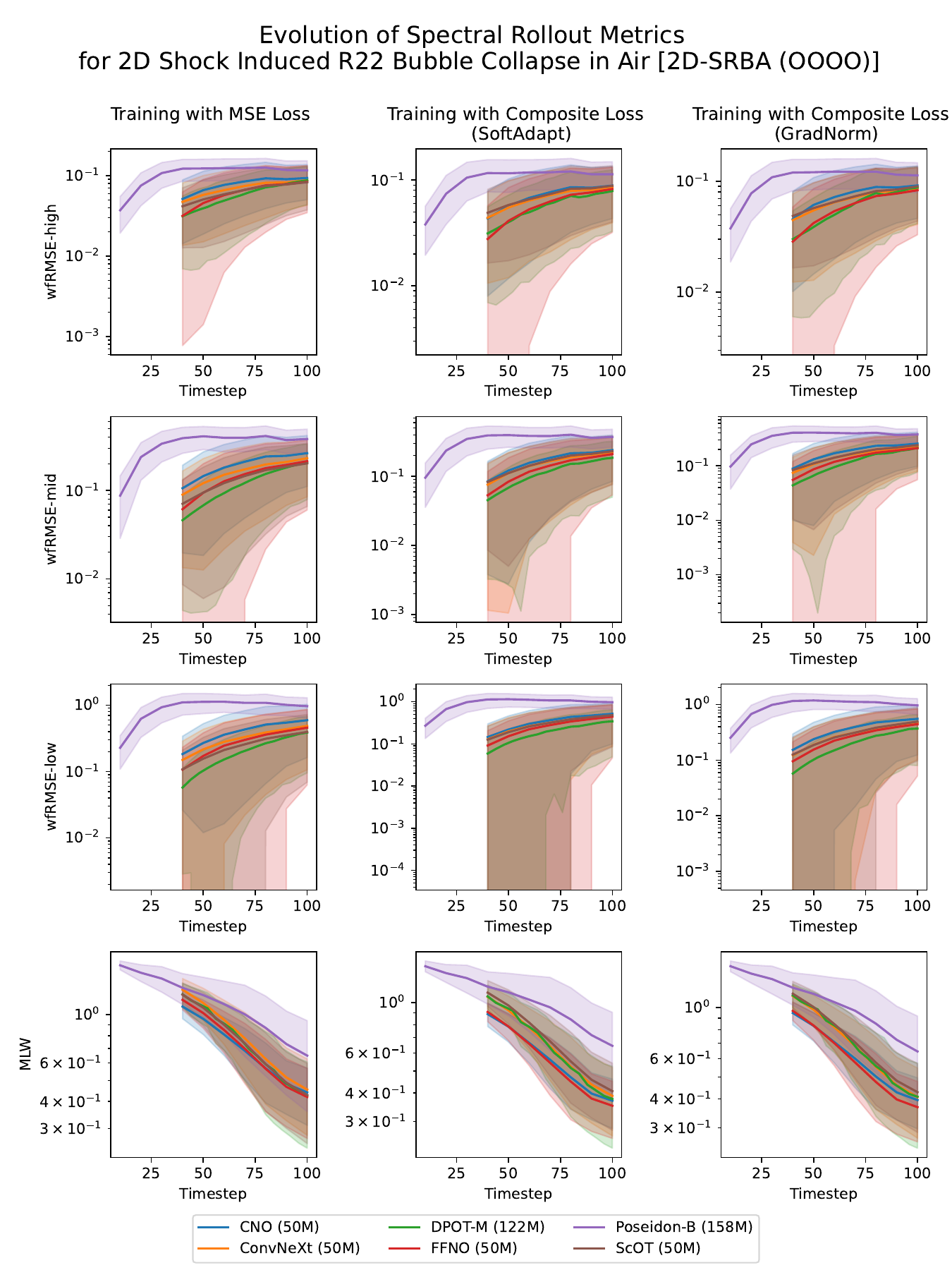}
    \caption{Wavelet based spectral metrics for the 2D-SRBA (OOOO) test dataset, averaged across all fields for each timestep.}
    \label{fig:2d_srba_oooo_spectral_rollout_evolution}
\end{figure}
\begin{figure}[h!]
    \centering
    \includegraphics[width=0.8\textwidth]{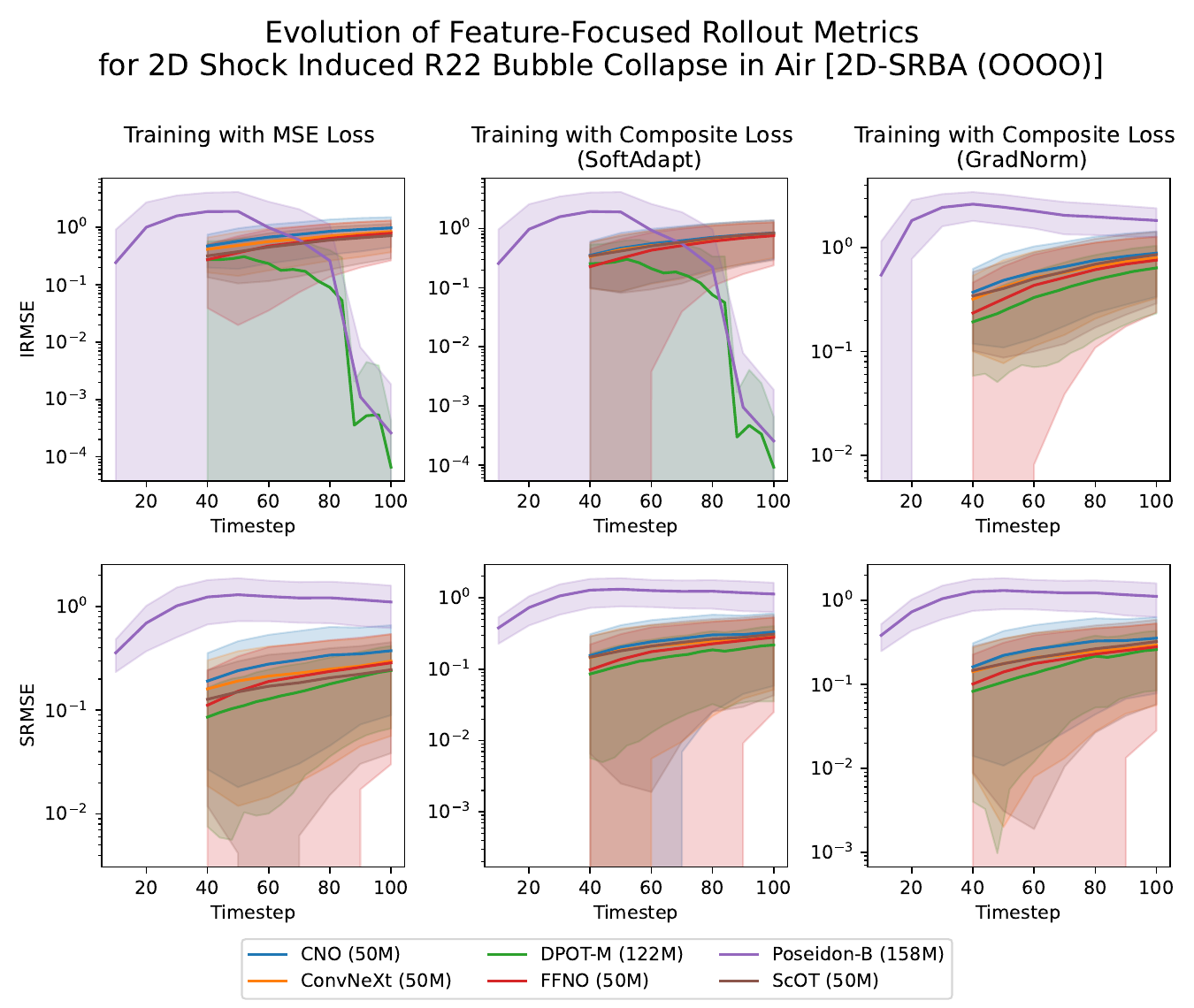}
    \caption{Feature focused metrics for capturing interface (IRMSE) and shocks (SRMSE) on the 2D-SRBA (OOOO) test dataset, averaged across all fields for each timestep.}
    \label{fig:2d_srba_oooo_feature_focused_rollout_evolution}
\end{figure}
\begin{figure}[h!]
    \centering
    \includegraphics[width=0.8\textwidth]{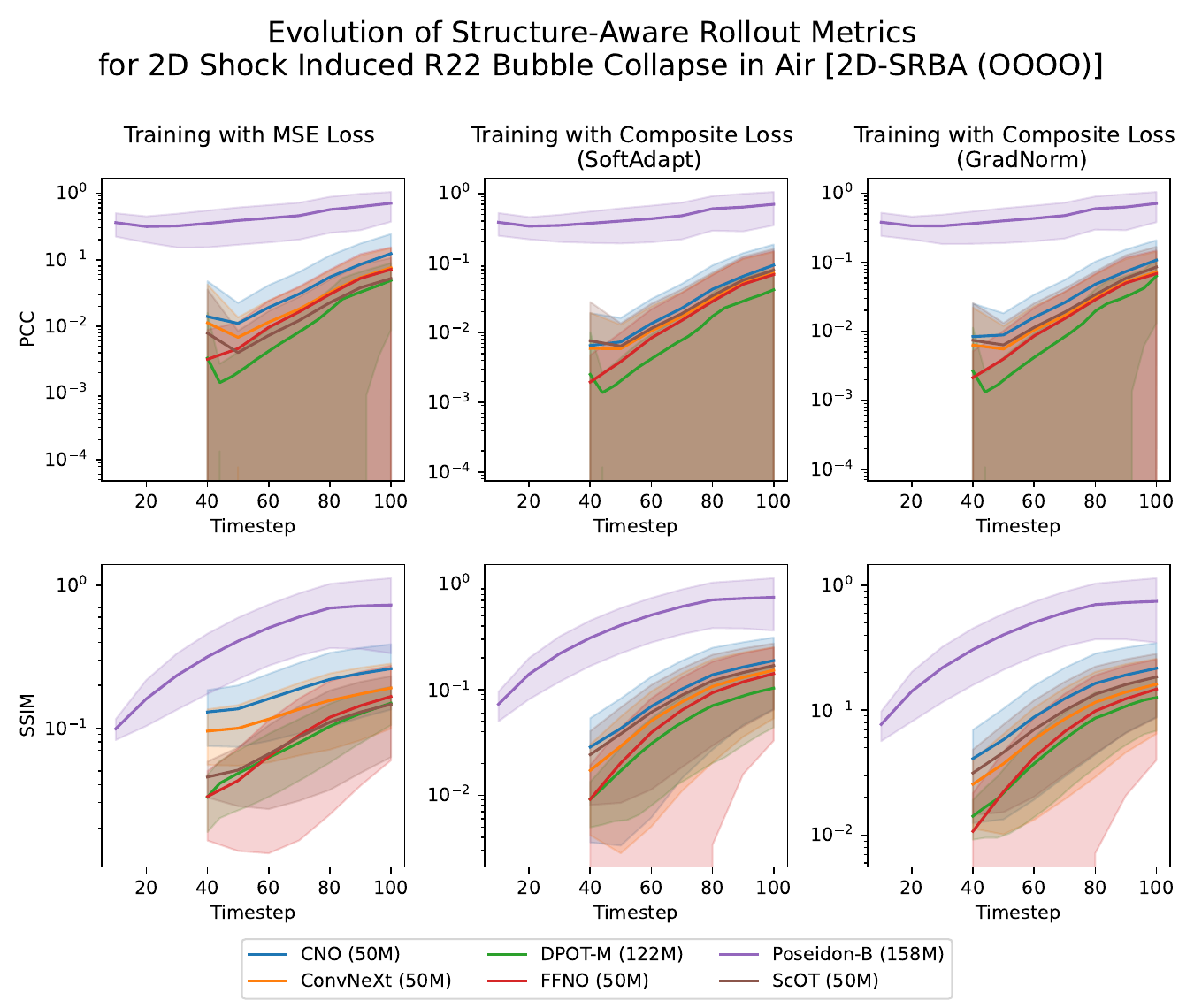}
    \caption{Structure-aware metrics for the 2D-SRBA (OOOO) test dataset, averaged across all fields for each timestep.}
    \label{fig:2d_srba_oooo_structure_aware_rollout_evolution}
\end{figure}
\begin{figure}[h!]
    \centering
    \includegraphics[width=0.8\textwidth]{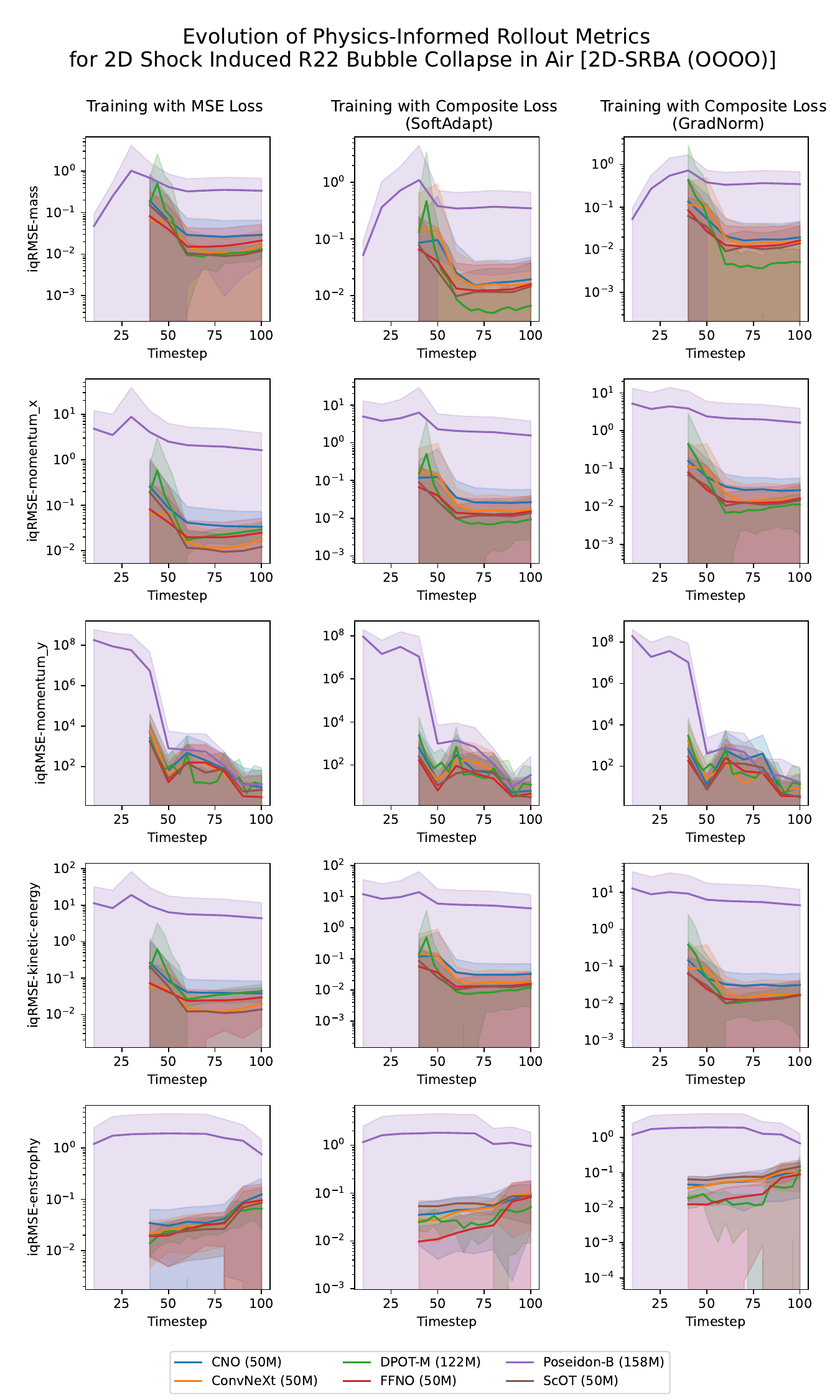}
    \caption{Physics-informed metrics reporting the nRMSE of integral quantities of interest on the 2D-SRBA (OOOO) test dataset, averaged across all fields for each timestep.}
    \label{fig:2d_srba_oooo_physics_informed_rollout_evolution}
\end{figure}

\clearpage
\subsubsection{2D shock-induced droplet breakup in air with symmetry boundaries [2D-SDBA (SSOO)]}
\begin{figure}[h!]
    \centering
    \includegraphics[width=0.8\textwidth]{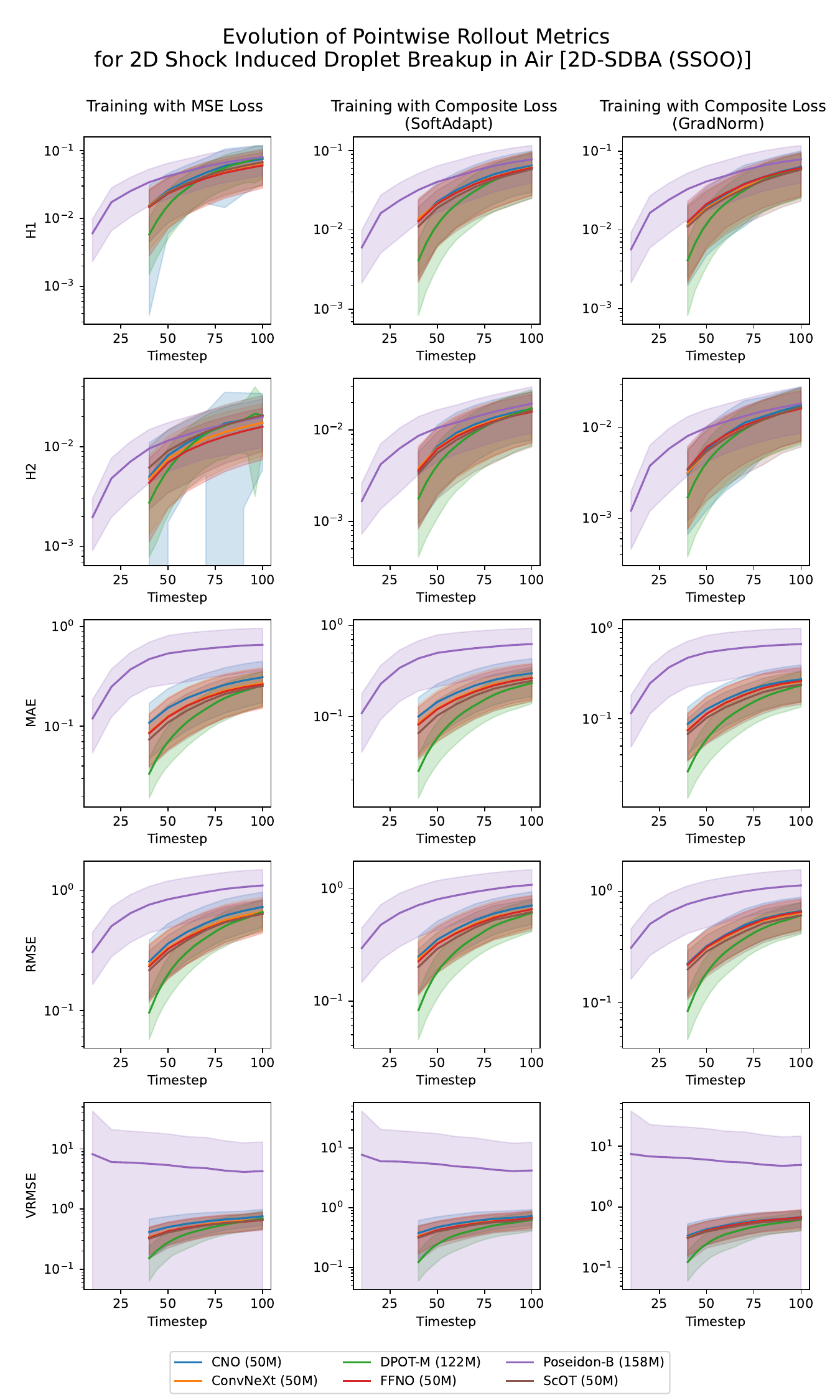}
    \caption{ Pointwise evaluation of baselines on the 2D-SDBA (SSOO) test dataset, reporting errors
in pointwise field values (MAE, RMSE, VRMSE) and the field derivatives (H1, H2) averaged across
all fields for each timestep.}
    \label{fig:2d_sdba_ssoo_pointwise_rollout_evolution}
\end{figure}
\begin{figure}[h!]
    \centering
    \includegraphics[width=0.8\textwidth]{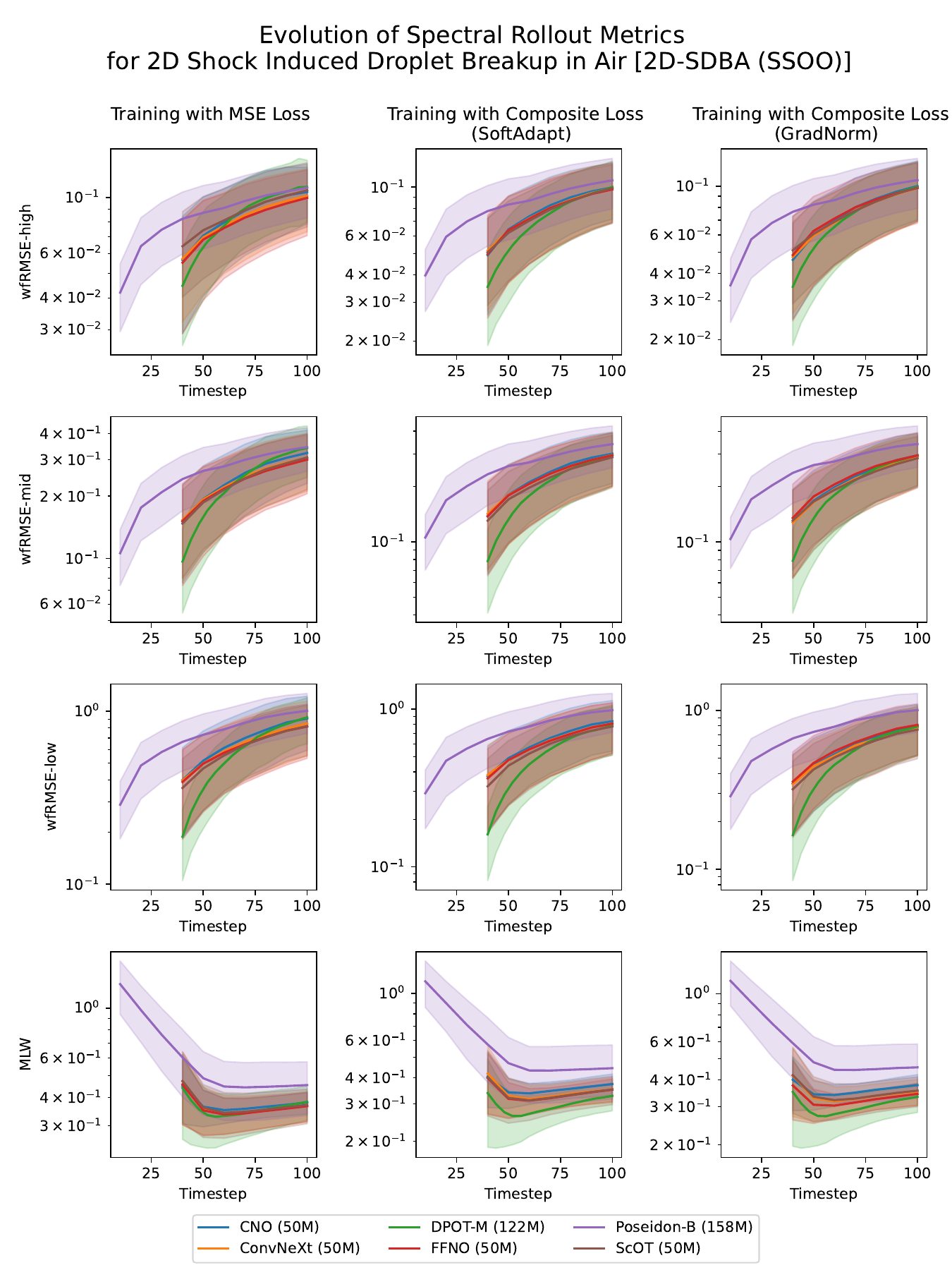}
    \caption{Wavelet based spectral metrics for the 2D-SDBA (SSOO) test dataset, averaged across all fields for each timestep.}
    \label{fig:2d_sdba_ssoo_spectral_rollout_evolution}
\end{figure}
\begin{figure}[h!]
    \centering
    \includegraphics[width=0.8\textwidth]{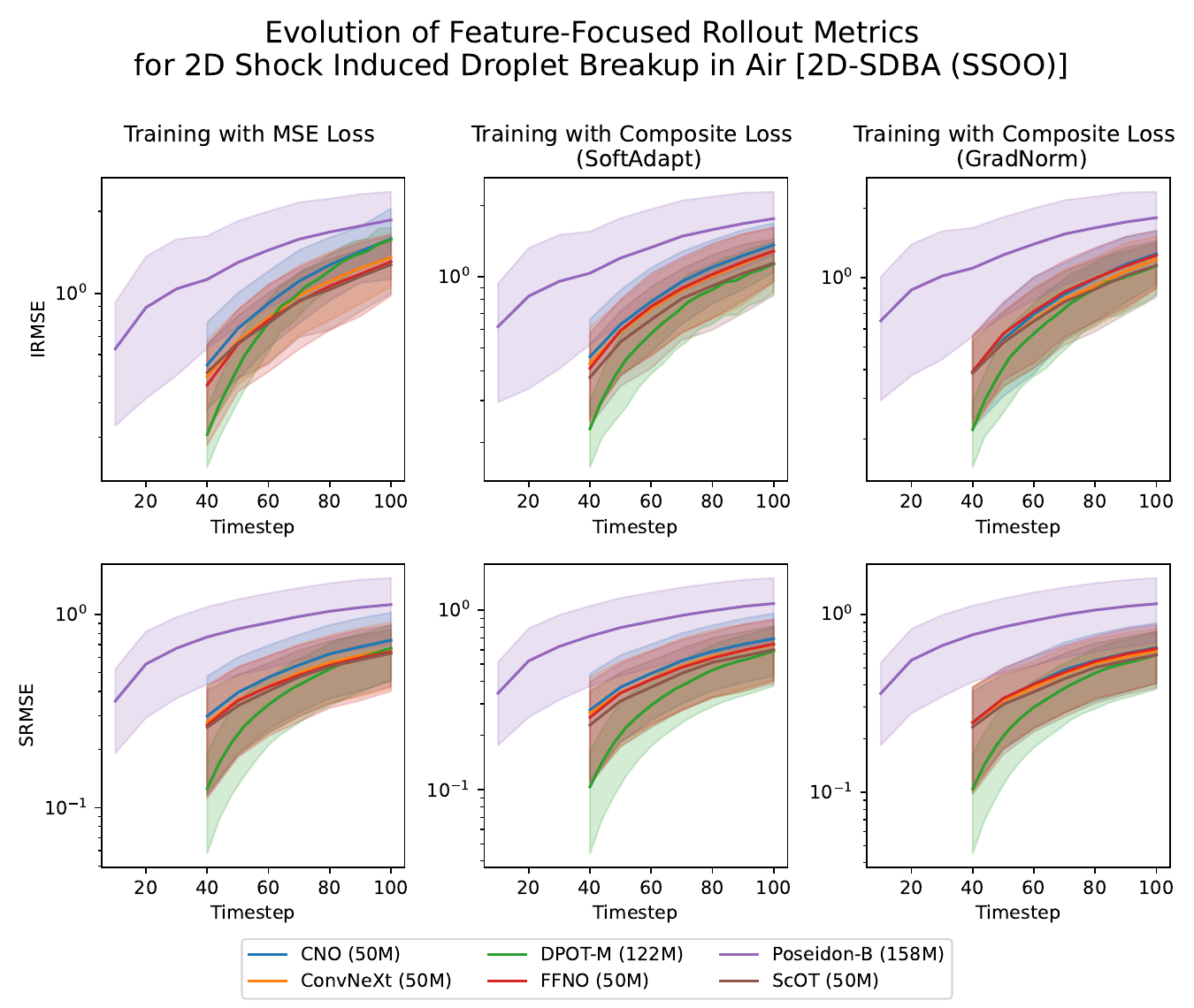}
    \caption{Feature focused metrics for capturing interface (IRMSE) and shocks (SRMSE) on the 2D-SDBA (SSOO) test dataset, averaged across all fields for each timestep.}
    \label{fig:2d_sdba_ssoo_feature_focused_rollout_evolution}
\end{figure}
\begin{figure}[h!]
    \centering
    \includegraphics[width=0.8\textwidth]{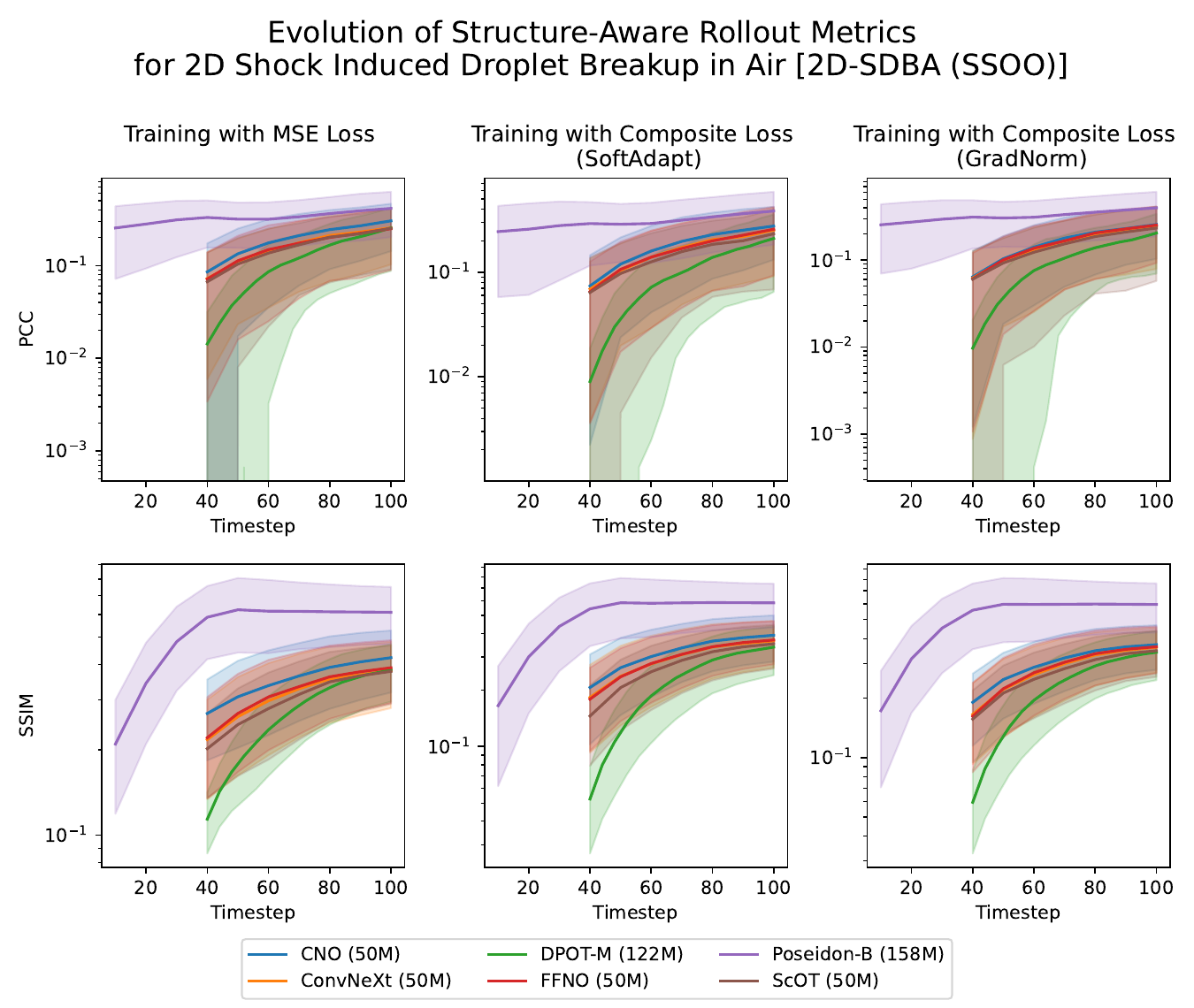}
    \caption{Structure-aware metrics for the 2D-SDBA (SSOO) test dataset, averaged across all fields for each timestep.}
    \label{fig:2d_sdba_ssoo_structure_aware_rollout_evolution}
\end{figure}
\begin{figure}[h!]
    \centering
    \includegraphics[width=0.65\textwidth]{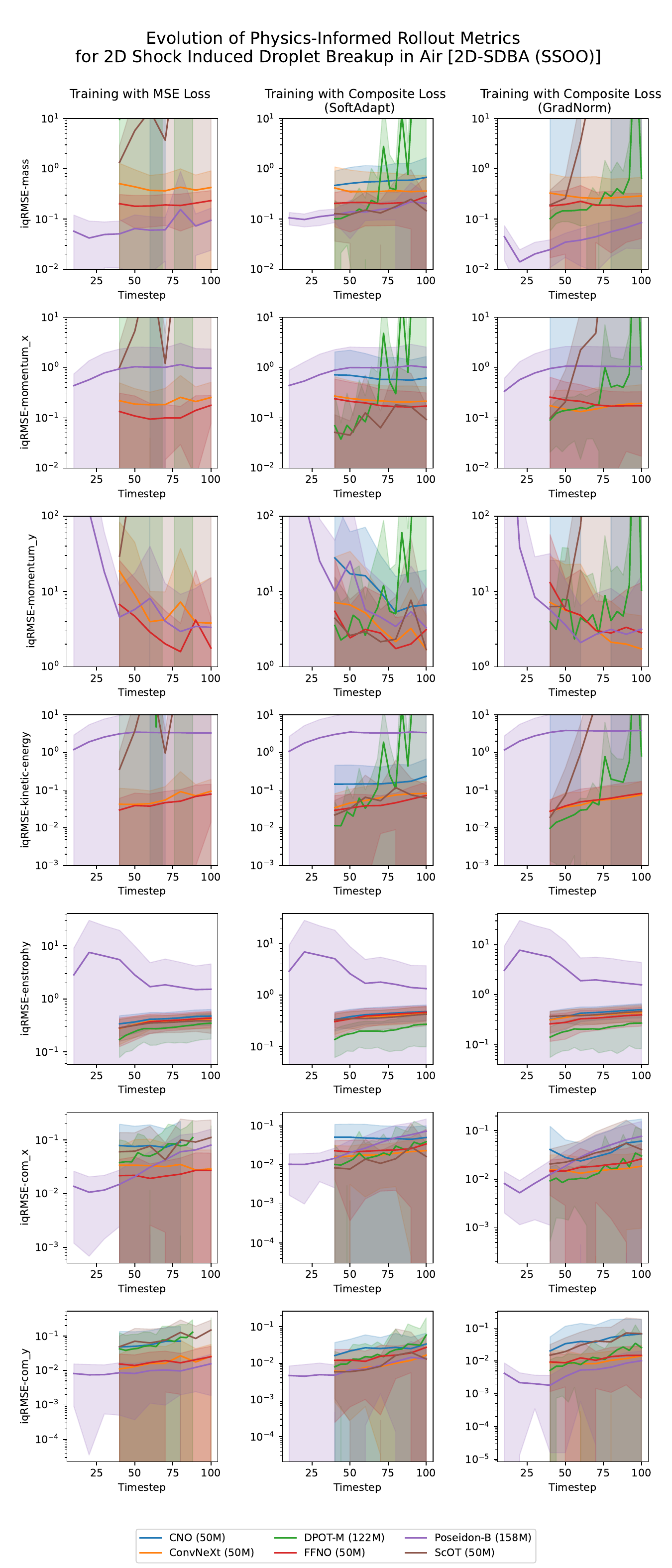}
    \caption{Physics-informed metrics reporting the nRMSE of integral quantities of interest on the 2D-SDBA (SSOO) test dataset, averaged across all fields for each timestep.}
    \label{fig:2d_sdba_ssoo_physics_informed_rollout_evolution}
\end{figure}

\clearpage
\subsubsection{3D shock-induced air bubble collapse in water with symmetry boundaries [3D-SABW (SSOOSS)]}
\begin{figure}[h!]
    \centering
    \includegraphics[width=0.8\textwidth]{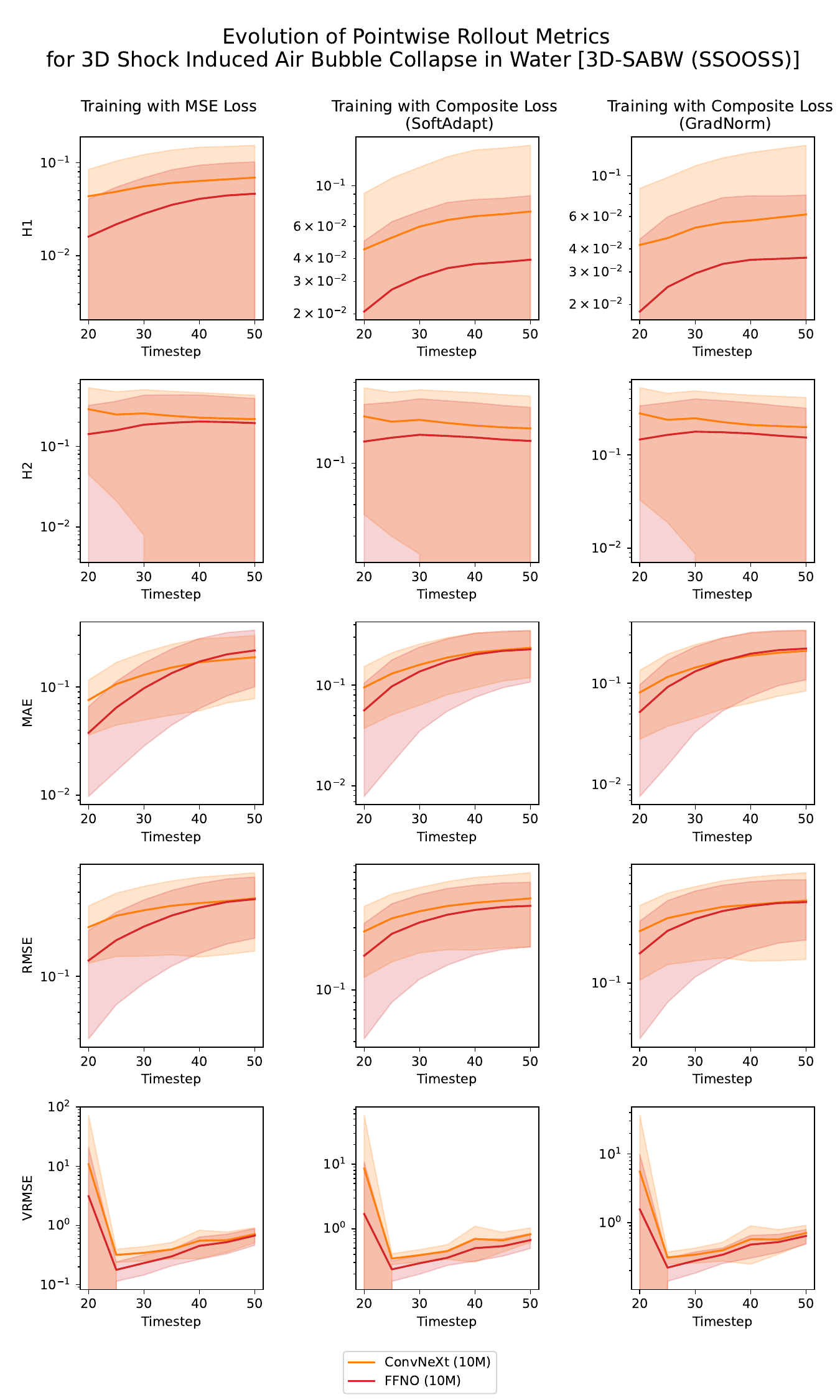}
    \caption{ Pointwise evaluation of baselines on the 3D-SABW (SSOOSS) test dataset, reporting errors in pointwise field values (MAE, RMSE, VRMSE) and the field derivatives (H1, H2) averaged across all fields for each timestep.}
    \label{fig:3d_sabw_ssooss_pointwise_rollout_evolution}
\end{figure}
\begin{figure}[h!]
    \centering
    \includegraphics[width=0.8\textwidth]{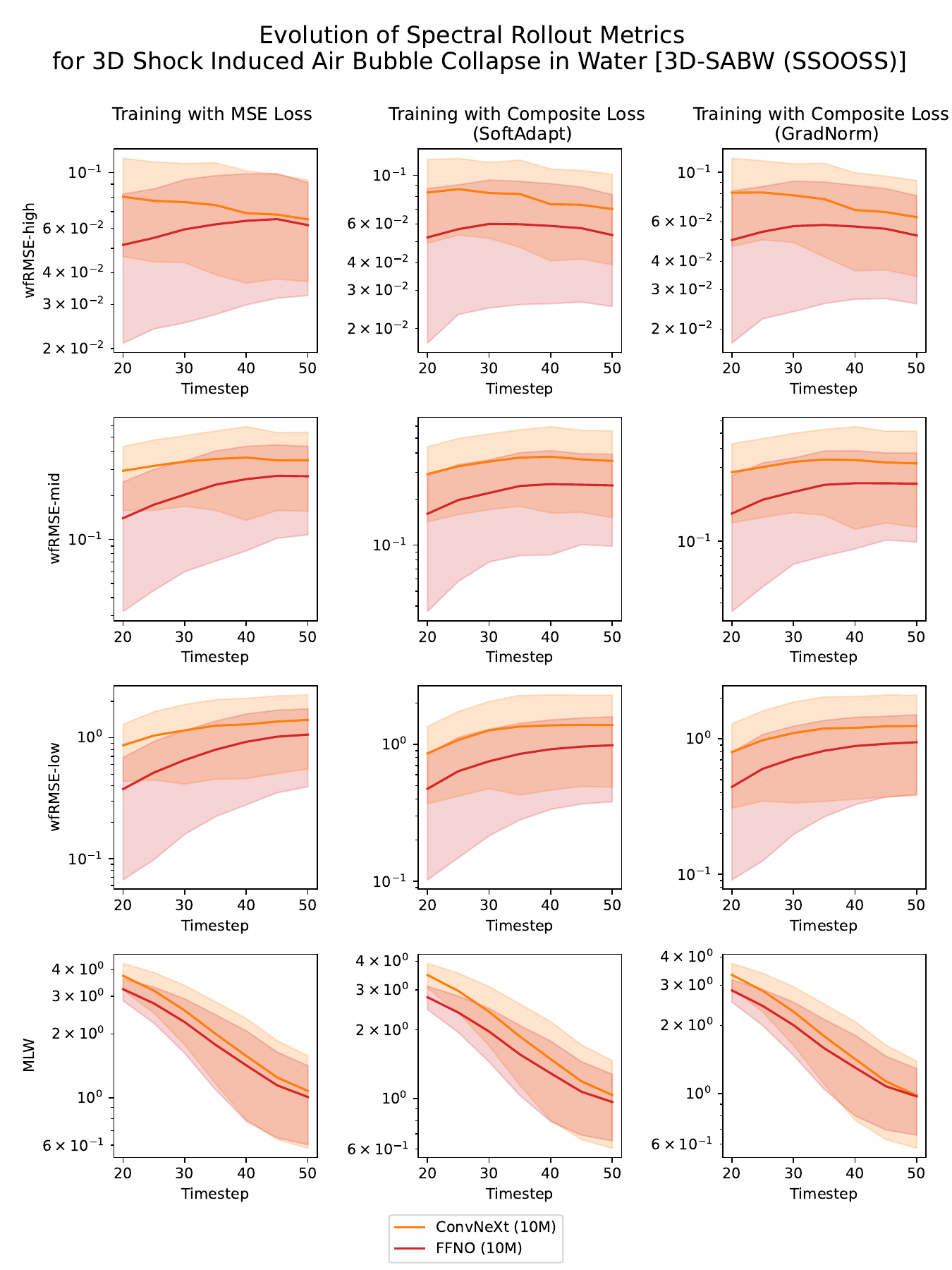}
    \caption{Wavelet based spectral metrics for the 3D-SABW (SSOOSS) test dataset, averaged across all fields for each timestep.}
    \label{fig:3d_sabw_ssooss_spectral_rollout_evolution}
\end{figure}
\begin{figure}[h!]
    \centering
    \includegraphics[width=0.8\textwidth]{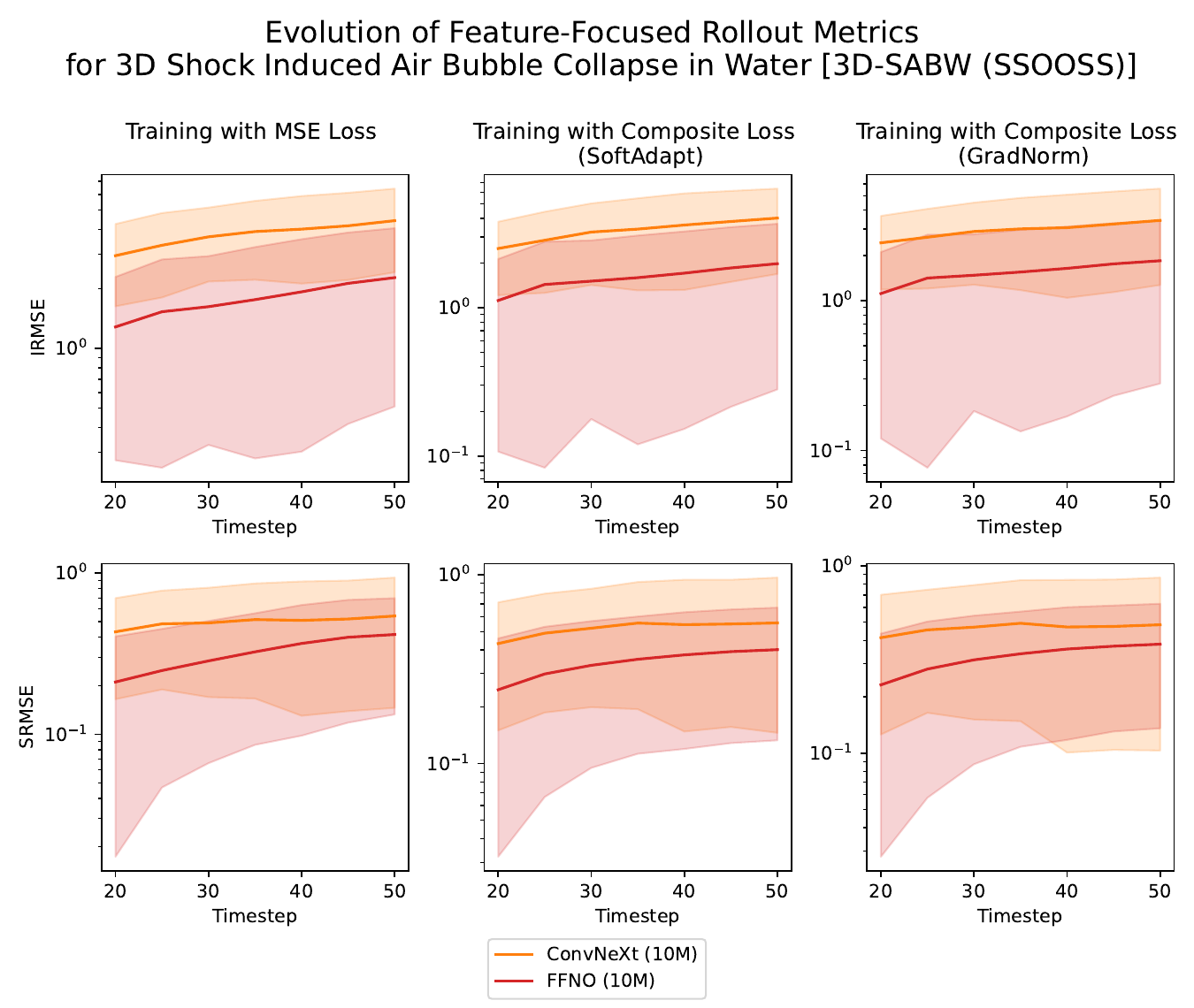}
    \caption{Feature focused metrics for capturing interface (IRMSE) and shocks (SRMSE) on the 3D-SABW (SSOOSS) test dataset, averaged across all fields for each timestep.}
    \label{fig:3d_sabw_ssooss_feature_focused_rollout_evolution}
\end{figure}
\begin{figure}[h!]
    \centering
    \includegraphics[width=0.8\textwidth]{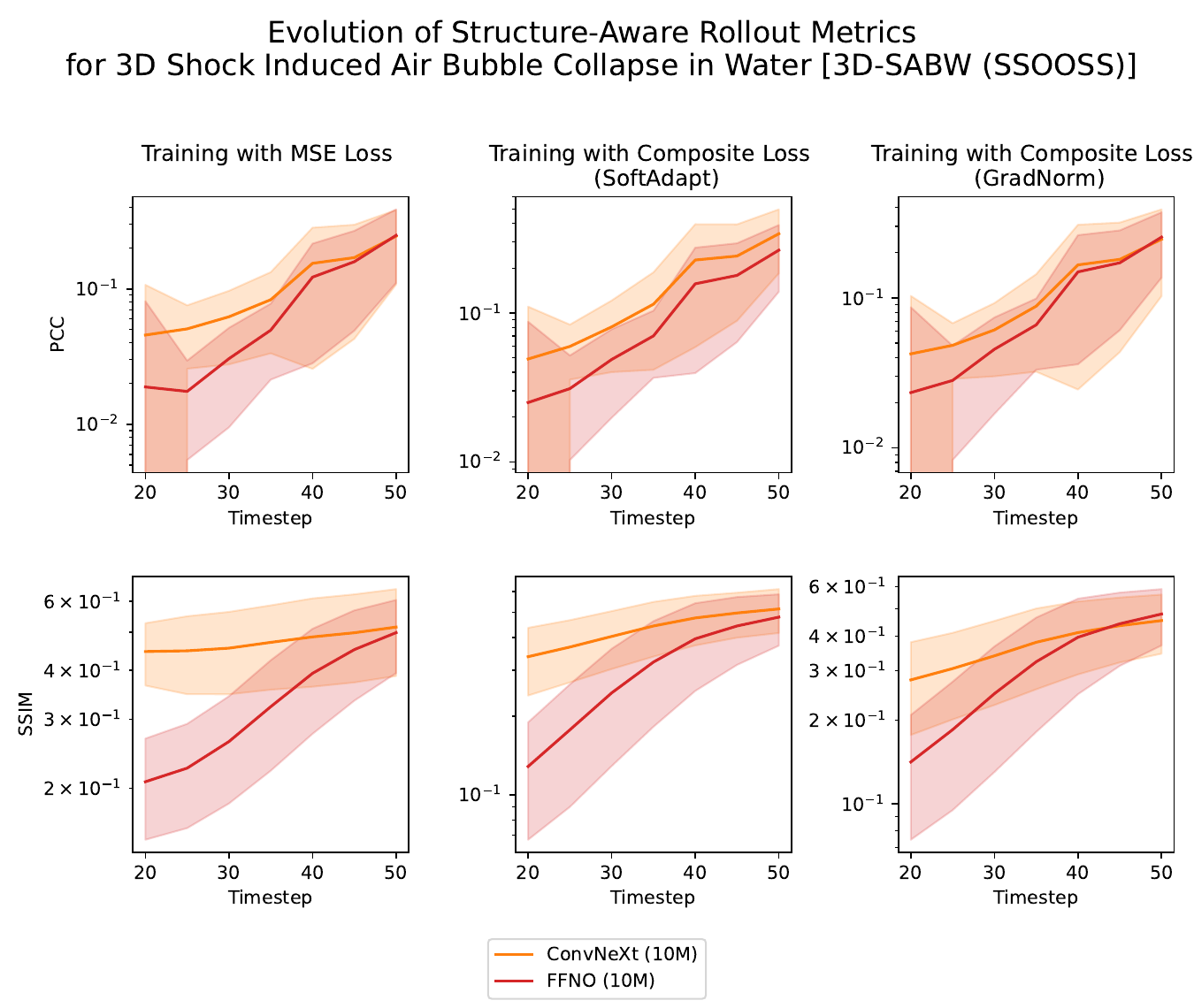}
    \caption{Structure-aware metrics for the 3D-SABW (SSOOSS) test dataset, averaged across all fields for each timestep.}
    \label{fig:3d_sabw_ssooss_structure_aware_rollout_evolution}
\end{figure}
\begin{figure}[h!]
    \centering
    \includegraphics[width=0.8\textwidth]{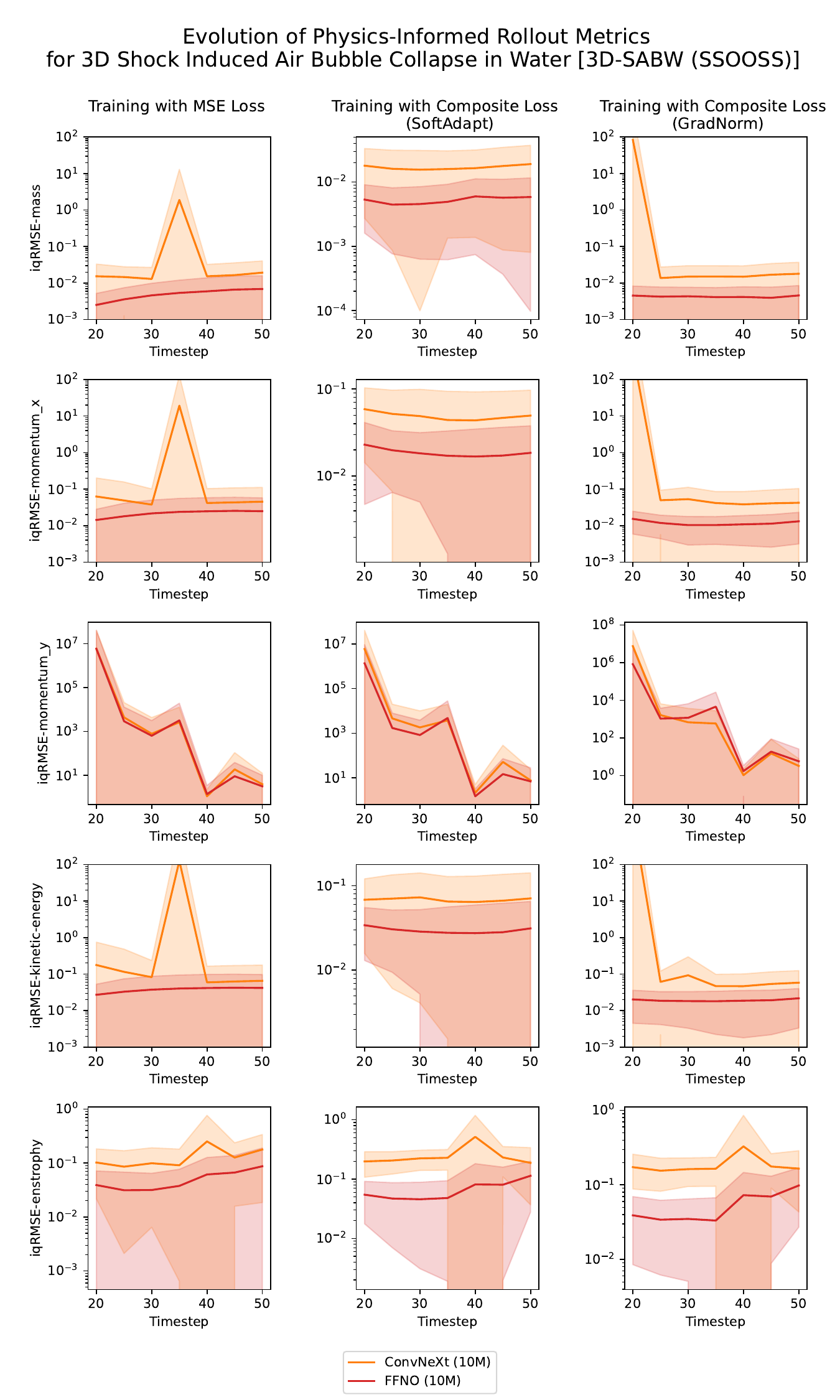}
    \caption{Physics-informed metrics reporting the nRMSE of integral quantities of interest on the 3D-SABW (SSOOSS) test dataset, averaged across all fields for each timestep.}
    \label{fig:3d_sabw_ssooss_physics_informed_rollout_evolution}
\end{figure}

\clearpage
\subsubsection{3D shock-induced droplet-breakup in air with symmetry boundaries [3D-SDBA (SSOOSS)]}
\begin{figure}[h!]
    \centering
    \includegraphics[width=0.8\textwidth]{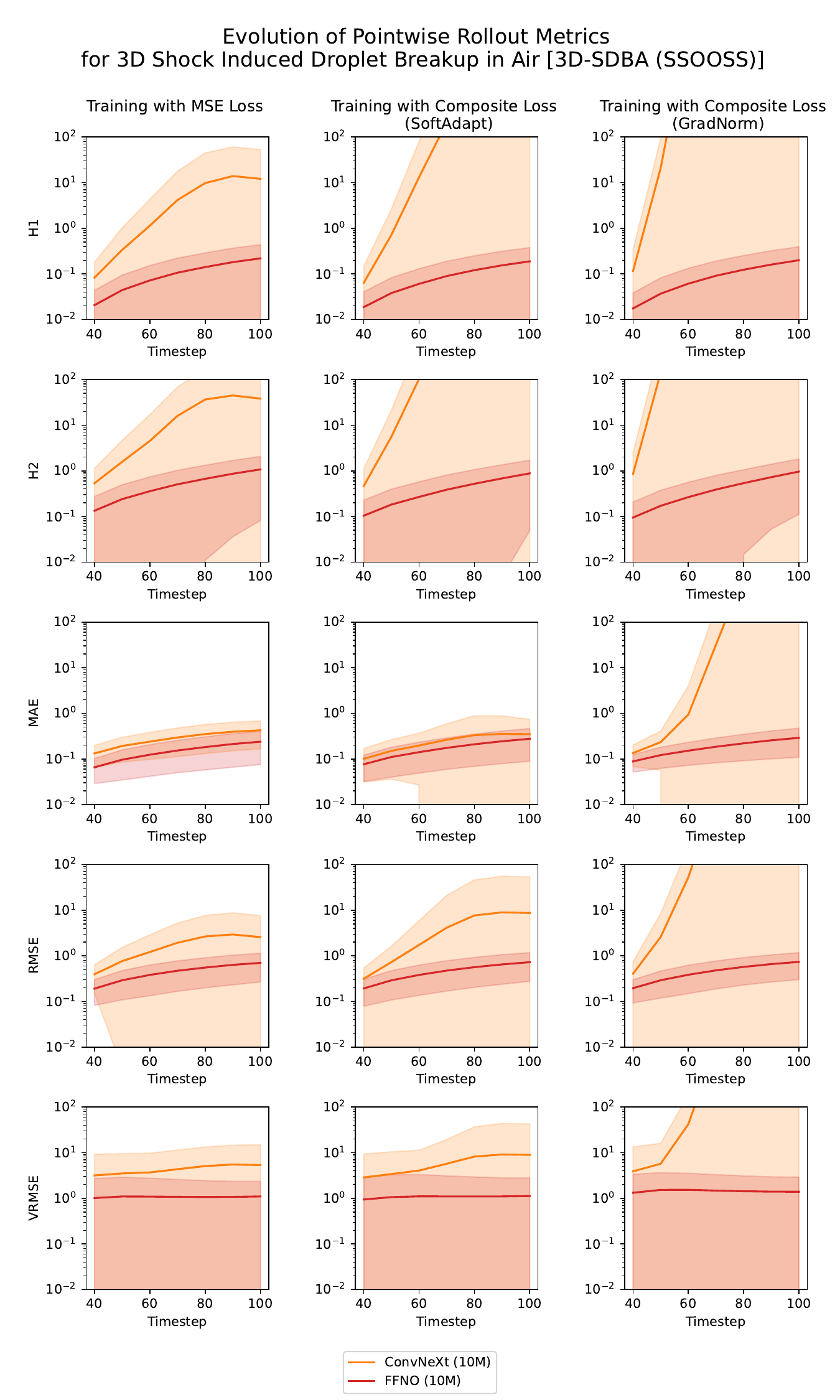}
    \caption{ Pointwise evaluation of baselines on the 3D-SDBA (SSOOSS) test dataset, reporting errors in pointwise field values (MAE, RMSE, VRMSE) and the field derivatives (H1, H2) averaged across all fields for each timestep.}
    \label{fig:3d_sdba_ssooss_pointwise_rollout_evolution}
\end{figure}
\begin{figure}[h!]
    \centering
    \includegraphics[width=0.8\textwidth]{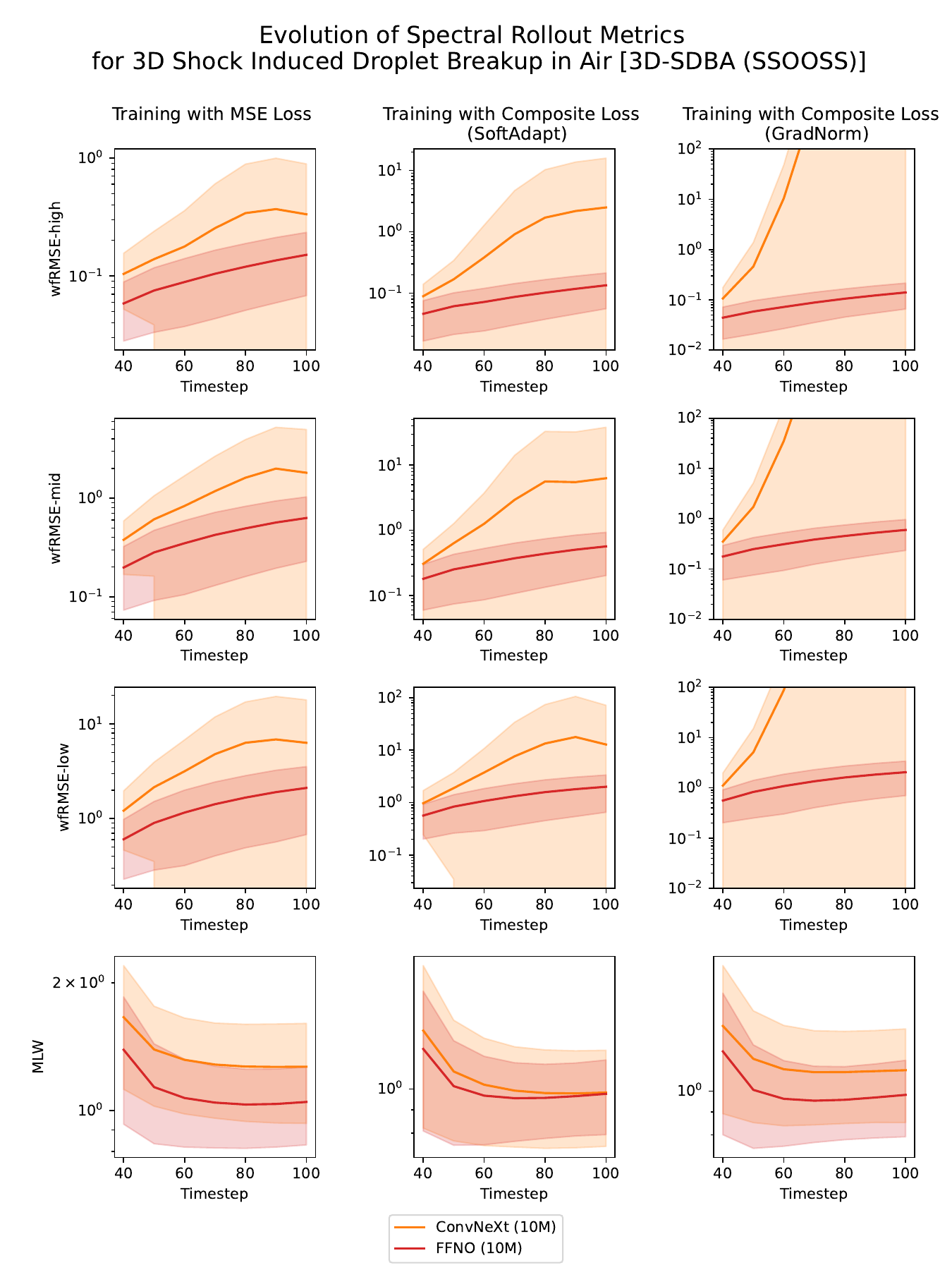}
    \caption{Wavelet based spectral metrics for the 3D-SDBA (SSOOSS) test dataset, averaged across all fields for each timestep.}
    \label{fig:3d_sdba_ssooss_spectral_rollout_evolution}
\end{figure}
\begin{figure}[h!]
    \centering
    \includegraphics[width=0.8\textwidth]{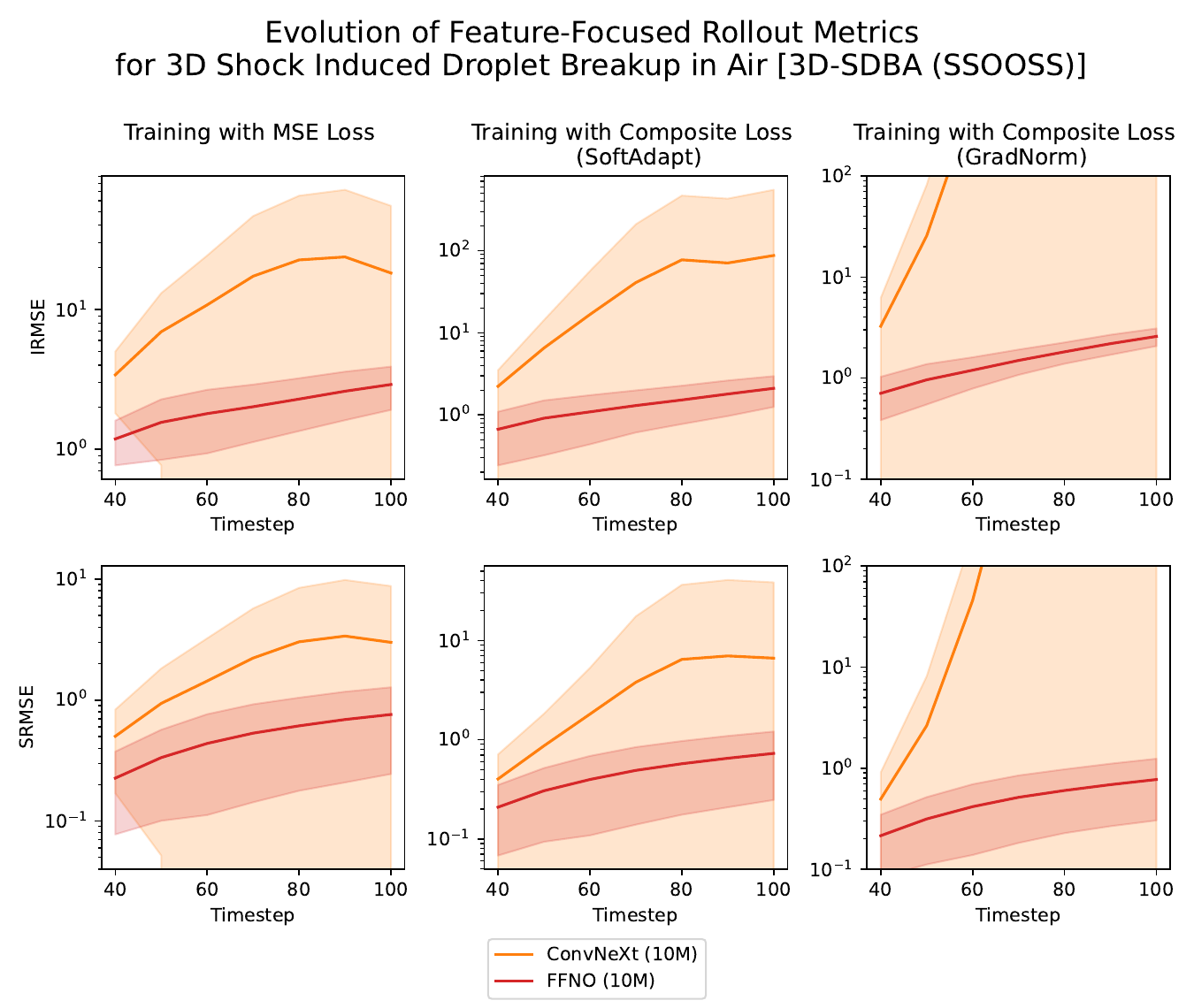}
    \caption{Feature focused metrics for capturing interface (IRMSE) and shocks (SRMSE) on the 3D-SDBA (SSOOSS) test dataset, averaged across all fields for each timestep.}
    \label{fig:3d_sdba_ssooss_feature_focused_rollout_evolution}
\end{figure}
\begin{figure}[h!]
    \centering
    \includegraphics[width=0.8\textwidth]{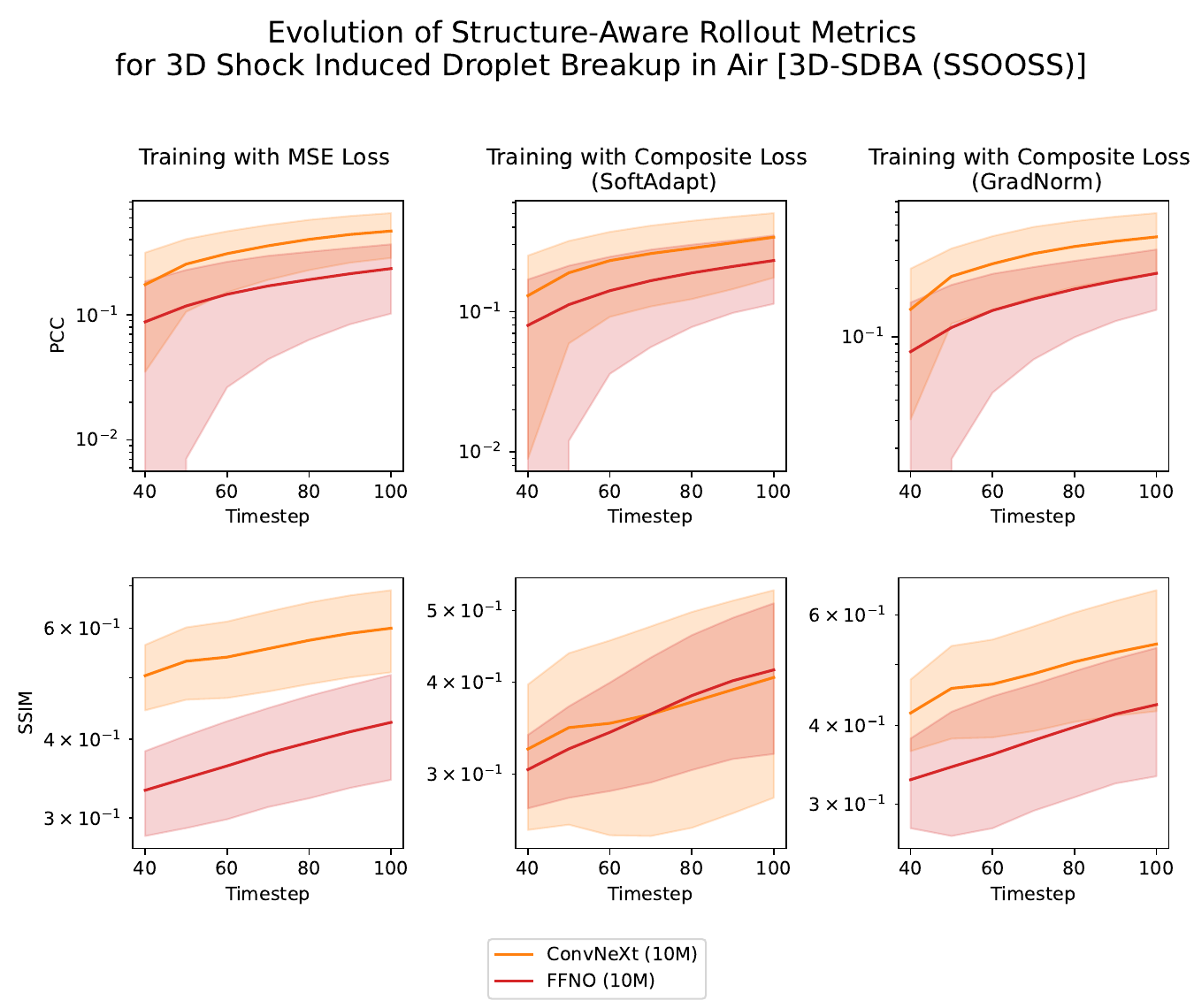}
    \caption{Structure-aware metrics for the 3D-SDBA (SSOOSS) test dataset, averaged across all fields for each timestep.}
    \label{fig:3d_sdba_ssooss_structure_aware_rollout_evolution}
\end{figure}
\begin{figure}[h!]
    \centering
    \includegraphics[width=0.65\textwidth]{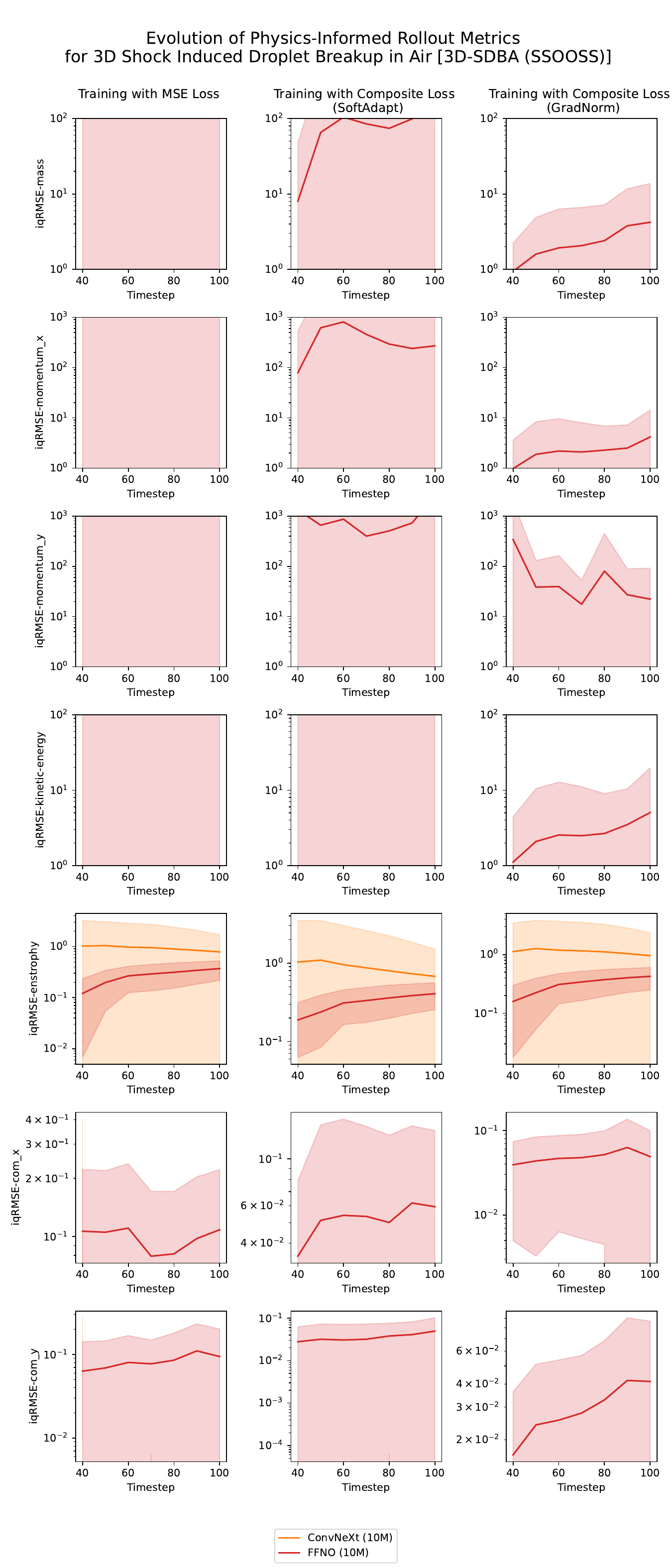}
    \caption{Physics-informed metrics reporting the nRMSE of integral quantities of interest on the 3D-SDBA (SSOOSS) test dataset, averaged across all fields for each timestep.}
    \label{fig:3d_sdba_ssooss_physics_informed_rollout_evolution}
\end{figure}

\clearpage
\subsection{Overall inference metrics}
\label{section:app_overall_inference_metrics}
The following tables report the long-term rollout inference metrics. For each trajectory in the in-distribution test dataset, the trained baseline model is applied autoregressively, starting from the initial condition and continuing until the trajectory ends. The evolution of the metrics, averaged across all the trajectories, for each rollout for the test dataset is presented in section \ref{section:app_rollout_metrics}. The final metric presented in this section is obtained by averaging the metric values across all trajectories and rollout steps. The tables are organized by dataset, with each table corresponding to a specific dataset and containing the results for all baselines and training strategies evaluated on that dataset. The metrics are categorized into pointwise, spectral, feature-focused, structure-aware, and physics-informed.

\begin{table}[h!]
    \centering
    \scriptsize
    \caption{Pointwise evaluation of baselines on the 2D-SABW (OOOO) test dataset, reporting errors in pointwise field values (MAE, RMSE, VRMSE) and the field derivatives (H1, H2) averaged across all fields and all rollout steps. Lower values are better. Darker shade indicates lower value for each baseline within a group of training loss strategies. Bold values denote the lowest among all baselines for a given metric and the underlined values denote the second-lowest values.}
    \label{tab:2d_sabw_oooo_pointwise}
\begin{tabular}{llccccc}
    \toprule
    Model & Training loss & H1 & H2 & MAE & RMSE & VRMSE \\
    \midrule

    \multirow{3}{*}{\makecell[l]{\texttt{ConvNeXt}\\\texttt{(50M)}}}
    & \texttt{MSE loss}
    & \sabw{0}{0.0252} & \sabw{50}{0.0068} & \sabw{100}{0.0640} & \sabw{50}{0.2255} & \sabw{50}{0.2858} \\
    & \texttt{Comp.loss (SA)}
    & \sabw{50}{0.0251} & \sabw{100}{0.0067} & \sabw{50}{0.0691} & \sabw{0}{0.2299} & \sabw{100}{0.2841} \\
    & \texttt{Comp.loss (GN)}
    & \sabw{100}{0.0244} & \sabw{0}{0.0070} & \sabw{0}{0.0719} & \sabw{100}{0.2254} & \sabw{0}{0.2875} \\
    \addlinespace

    \multirow{3}{*}{\makecell[l]{\texttt{CNO}\\\texttt{(50M)}}}
    & \texttt{MSE loss}
    & \sabw{50}{0.0270} & \sabw{0}{0.0073} & \sabw{0}{0.0951} & \sabw{50}{0.2575} & \sabw{0}{0.3655} \\
    & \texttt{Comp.loss (SA)}
    & \sabw{0}{0.0277} & \sabw{50}{0.0069} & \sabw{0}{0.0951} & \sabw{0}{0.2618} & \sabw{50}{0.3447} \\
    & \texttt{Comp.loss (GN)}
    & \sabw{100}{0.0237} & \sabw{100}{0.0064} & \sabw{100}{0.0777} & \sabw{100}{0.2265} & \sabw{100}{0.2967} \\
    \addlinespace

    \multirow{3}{*}{\makecell[l]{\texttt{FFNO}\\\texttt{(50M)}}}
    & \texttt{MSE loss}
    & \sabw{50}{0.0230} & \sabw{0}{0.0063} & \sabw{100}{0.0571} & \sabw{50}{0.2081} & \sabw{50}{0.2389} \\
\bottomrule
    \end{tabular}
\end{table}

\begin{table}[h!]
    \centering
    \scriptsize
    \caption{Wavelet based spectral metrics for the 2D-SABW (OOOO) test dataset. Lower values are better. Darker shade indicates lower value for each baseline within a group of training loss strategies. Bold values denote the lowest among all baselines for a given metric and the underlined values denote the second-lowest values.}
    \label{tab:2d_sabw_oooo_spectral}
    \begin{tabular}{llcccc}
    \toprule
        Model & Training loss & wfRMSE- & wfRMSE- & wfRMSE- & MLW \\
            &               & high & mid & low &  \\
        \midrule

    \multirow{3}{*}{\makecell[l]{\texttt{ConvNeXt}\\\texttt{(50M)}}}
    & \texttt{MSE loss}
    & \sabw{50}{0.0657} & \sabw{0}{0.1813} & \sabw{0}{0.4326} & \sabw{0}{0.8191} \\
    & \texttt{Comp.loss (SA)}
    & \sabw{100}{0.0644} & \sabw{50}{0.1784} & \sabw{50}{0.4306} & \sabw{100}{0.7089} \\
    & \texttt{Comp.loss (GN)}
    & \sabw{0}{0.0689} & \sabw{100}{0.1764} & \sabw{100}{0.4192} & \sabw{50}{0.7489} \\
    \addlinespace

    \multirow{3}{*}{\makecell[l]{\texttt{CNO}\\\texttt{(50M)}}}
    & \texttt{MSE loss}
    & \sabw{0}{0.0695} & \sabw{50}{0.1934} & \sabw{50}{0.4578} & \sabw{0}{0.6822} \\
    & \texttt{Comp.loss (SA)}
    & \sabw{50}{0.0677} & \sabw{0}{0.1947} & \sabw{0}{0.4722} & \sabw{50}{\underline{0.5831}} \\
    & \texttt{Comp.loss (GN)}
    & \sabw{100}{0.0647} & \sabw{100}{0.1795} & \sabw{100}{0.4233} & \sabw{100}{\textbf{0.5763}} \\
\bottomrule
    \end{tabular}
\end{table}

\begin{table}[h!]
    \centering
    \scriptsize
    \caption{Feature focused metrics for capturing interface (IRMSE) and shocks (SRMSE) on the 2D-SABW (OOOO) test dataset. Lower values are better. Darker shade indicates lower value for each baseline within a group of training loss strategies. Bold values denote the lowest among all baselines for a given metric and the underlined values denote the second-lowest values.}
    \label{tab:2d_sabw_oooo_feature_focused}
    \begin{tabular}{llcc}
    \toprule
    Model & Training loss & IRMSE & SRMSE \\
    \midrule

    \multirow{3}{*}{\makecell[l]{\texttt{ConvNeXt}\\\texttt{(50M)}}}
    & \texttt{MSE loss}
    & \sabw{0}{1.0677} & \sabw{50}{0.3405} \\
    & \texttt{Comp.loss (SA)}
    & \sabw{50}{1.0435} & \sabw{0}{0.3442} \\
    & \texttt{Comp.loss (GN)}
    & \sabw{100}{1.0279} & \sabw{100}{0.3385} \\
    \addlinespace

    \multirow{3}{*}{\makecell[l]{\texttt{CNO}\\\texttt{(50M)}}}
    & \texttt{MSE loss}
    & \sabw{50}{1.0874} & \sabw{50}{0.3775} \\
    & \texttt{Comp.loss (SA)}
    & \sabw{0}{1.1015} & \sabw{0}{0.3827} \\
    & \texttt{Comp.loss (GN)}
    & \sabw{100}{0.9846} & \sabw{100}{0.3392} \\
    \addlinespace

    \multirow{3}{*}{\makecell[l]{\texttt{FFNO}\\\texttt{(50M)}}}
    & \texttt{MSE loss}
    & \sabw{0}{0.9791} & \sabw{50}{0.3104} \\
\bottomrule
    \end{tabular}
\end{table}

\begin{table}[h!]
    \centering
    \scriptsize
    \caption{Structure-aware metrics for the 2D-SABW (OOOO) test dataset. Lower values are better. Darker shade indicates lower value for each baseline within a group of training loss strategies. Bold values denote the lowest among all baselines for a given metric and the underlined values denote the second-lowest values.}
    \label{tab:2d_sabw_oooo_structure_aware}
    \begin{tabular}{llcc}
    \toprule
    Model & Training loss & PCC & SSIM \\
    \midrule

    \multirow{3}{*}{\makecell[l]{\texttt{ConvNeXt}\\\texttt{(50M)}}}
    & \texttt{MSE loss}
    & \sabw{100}{0.0531} & \sabw{0}{0.1938} \\
    & \texttt{Comp.loss (SA)}
    & \sabw{0}{0.0560} & \sabw{100}{0.1560} \\
    & \texttt{Comp.loss (GN)}
    & \sabw{50}{0.0543} & \sabw{50}{0.1736} \\
    \addlinespace

    \multirow{3}{*}{\makecell[l]{\texttt{CNO}\\\texttt{(50M)}}}
    & \texttt{MSE loss}
    & \sabw{0}{0.0782} & \sabw{0}{0.2512} \\
    & \texttt{Comp.loss (SA)}
    & \sabw{50}{0.0763} & \sabw{50}{0.2042} \\
    & \texttt{Comp.loss (GN)}
    & \sabw{100}{0.0568} & \sabw{100}{0.1841} \\
    \addlinespace

    \multirow{3}{*}{\makecell[l]{\texttt{FFNO}\\\texttt{(50M)}}}
    & \texttt{MSE loss}
    & \sabw{50}{0.0493} & \sabw{50}{0.1505} \\
\bottomrule
    \end{tabular}
\end{table}

\begin{table}[h!]
    \centering
    \scriptsize
    \setlength{\tabcolsep}{4pt}
    \caption{Physics-informed metrics reporting the nRMSE of integral quantities of interest on the 2D-SABW (OOOO) test dataset. Lower values are better. Darker shade indicates lower value for each baseline within a group of training loss strategies. Bold values denote the lowest among all baselines for a given metric and the underlined values denote the second-lowest values.}
    \label{tab:2d_sabw_oooo_physics_informed}
    \begin{tabular}{@{}p{1.7cm}p{2.3cm}ccccc@{}}
    \toprule
    Model & Training loss & iqRMSE- & iqRMSE- & iqRMSE- & iqRMSE- & iqRMSE- \\
          &               & mass & momentum & momentum & kinetic-energy & enstrophy \\
          &               &  & \_x & \_y &  &  \\
    \midrule

    \multirow{3}{*}{\makecell[l]{\texttt{ConvNeXt}\\\texttt{(50M)}}}
    & \texttt{MSE loss}
    & \sabw{100}{0.0084} & \sabw{100}{0.0122} & \sabw{0}{\texttt{>10}} & \sabw{100}{0.0149} & \sabw{0}{0.1232} \\
    & \texttt{Comp.loss (SA)}
    & \sabw{50}{0.0162} & \sabw{50}{0.0269} & \sabw{0}{\texttt{>10}} & \sabw{50}{0.0317} & \sabw{50}{0.1205} \\
    & \texttt{Comp.loss (GN)}
    & \sabw{0}{0.0255} & \sabw{0}{0.0350} & \sabw{0}{\texttt{>10}} & \sabw{0}{0.0408} & \sabw{100}{0.1198} \\
    \addlinespace

    \multirow{3}{*}{\makecell[l]{\texttt{CNO}\\\texttt{(50M)}}}
    & \texttt{MSE loss}
    & \sabw{0}{0.0355} & \sabw{0}{0.0753} & \sabw{0}{\texttt{>10}} & \sabw{0}{0.0713} & \sabw{50}{0.1142} \\
    & \texttt{Comp.loss (SA)}
    & \sabw{50}{0.0303} & \sabw{50}{0.0491} & \sabw{0}{\texttt{>10}} & \sabw{50}{0.0601} & \sabw{0}{0.1271} \\
\bottomrule
    \end{tabular}
\end{table}

\begin{table}[h!]
    \centering
    \scriptsize
    \caption{Pointwise evaluation of baselines on the 2D-SABW (SSOO) test dataset, reporting errors in pointwise field values (MAE, RMSE, VRMSE) and the field derivatives (H1, H2) averaged across all fields and all rollout steps. Lower values are better. Darker shade indicates lower value for each baseline within a group of training loss strategies. Bold values denote the lowest among all baselines for a given metric and the underlined values denote the second-lowest values.}
    \label{tab:2d_sabw_ssoo_pointwise}
    \begin{tabular}{llccccc}
    \toprule
    Model & Training loss & H1 & H2 & MAE & RMSE & VRMSE \\
    \midrule

    \multirow{3}{*}{\makecell[l]{\texttt{ConvNeXt}\\\texttt{(50M)}}}
    & \texttt{MSE loss}
    & \ssoo{0}{0.0265} & \ssoo{0}{0.0074} & \ssoo{100}{0.0726} & \ssoo{0}{0.2295} & \ssoo{0}{0.4207} \\
    & \texttt{Comp.loss (SA)}
    & \ssoo{50}{0.0260} & \ssoo{100}{0.0070} & \ssoo{50}{0.0753} & \ssoo{50}{0.2289} & \ssoo{100}{0.3504} \\
    & \texttt{Comp.loss (GN)}
    & \ssoo{100}{0.0248} & \ssoo{50}{0.0072} & \ssoo{0}{0.0789} & \ssoo{100}{0.2250} & \ssoo{50}{0.3923} \\
    \addlinespace

    \multirow{3}{*}{\makecell[l]{\texttt{CNO}\\\texttt{(50M)}}}
    & \texttt{MSE loss}
    & \ssoo{50}{0.0281} & \ssoo{0}{0.0074} & \ssoo{0}{0.0999} & \ssoo{50}{0.2517} & \ssoo{0}{0.5123} \\
    & \texttt{Comp.loss (SA)}
    & \ssoo{0}{0.0284} & \ssoo{50}{0.0071} & \ssoo{50}{0.0998} & \ssoo{0}{0.2593} & \ssoo{50}{0.3719} \\
    & \texttt{Comp.loss (GN)}
    & \ssoo{100}{0.0245} & \ssoo{100}{0.0066} & \ssoo{100}{0.0865} & \ssoo{100}{0.2286} & \ssoo{100}{0.3434} \\
    \addlinespace

    \multirow{3}{*}{\makecell[l]{\texttt{FFNO}\\\texttt{(50M)}}}
    & \texttt{MSE loss}
    & \ssoo{50}{0.0240} & \ssoo{0}{0.0067} & \ssoo{100}{0.0640} & \ssoo{50}{0.2089} & \ssoo{50}{\underline{0.2620}} \\
\bottomrule
    \end{tabular}
\end{table}

\begin{table}[h!]
    \centering
    \scriptsize
    \caption{Wavelet based spectral metrics for the 2D-SABW (SSOO) test dataset. Lower values are better. Darker shade indicates lower value for each baseline within a group of training loss strategies. Bold values denote the lowest among all baselines for a given metric and the underlined values denote the second-lowest values.}
    \label{tab:2d_sabw_ssoo_spectral}
    \begin{tabular}{llcccc}
    \toprule
        Model & Training loss & wfRMSE- & wfRMSE- & wfRMSE- & MLW \\
            &               & high & mid & low &  \\
        \midrule

    \multirow{3}{*}{\makecell[l]{\texttt{ConvNeXt}\\\texttt{(50M)}}}
    & \texttt{MSE loss}
    & \ssoo{50}{0.0694} & \ssoo{0}{0.1904} & \ssoo{0}{0.4452} & \ssoo{0}{0.8141} \\
    & \texttt{Comp.loss (SA)}
    & \ssoo{100}{0.0667} & \ssoo{50}{0.1842} & \ssoo{50}{0.4388} & \ssoo{100}{0.6925} \\
    & \texttt{Comp.loss (GN)}
    & \ssoo{0}{0.0705} & \ssoo{100}{0.1811} & \ssoo{100}{0.4211} & \ssoo{50}{0.7396} \\
    \addlinespace

    \multirow{3}{*}{\makecell[l]{\texttt{CNO}\\\texttt{(50M)}}}
    & \texttt{MSE loss}
    & \ssoo{0}{0.0709} & \ssoo{0}{0.1965} & \ssoo{50}{0.4642} & \ssoo{0}{0.6835} \\
    & \texttt{Comp.loss (SA)}
    & \ssoo{50}{0.0691} & \ssoo{50}{0.1952} & \ssoo{0}{0.4766} & \ssoo{50}{\underline{0.5666}} \\
    & \texttt{Comp.loss (GN)}
    & \ssoo{100}{0.0670} & \ssoo{100}{0.1841} & \ssoo{100}{0.4321} & \ssoo{100}{\textbf{0.5659}} \\
\bottomrule
    \end{tabular}
\end{table}

\begin{table}[h!]
    \centering
    \scriptsize
    \caption{Feature focused metrics for capturing interface (IRMSE) and shocks (SRMSE) on the 2D-SABW (SSOO) test dataset. Lower values are better. Darker shade indicates lower value for each baseline within a group of training loss strategies. Bold values denote the lowest among all baselines for a given metric and the underlined values denote the second-lowest values.}
    \label{tab:2d_sabw_ssoo_feature_focused}
    \begin{tabular}{llcc}
    \toprule
    Model & Training loss & IRMSE & SRMSE \\
    \midrule

    \multirow{3}{*}{\makecell[l]{\texttt{ConvNeXt}\\\texttt{(50M)}}}
    & \texttt{MSE loss}
    & \ssoo{0}{1.1529} & \ssoo{0}{0.3310} \\
    & \texttt{Comp.loss (SA)}
    & \ssoo{50}{1.1136} & \ssoo{50}{0.3287} \\
    & \texttt{Comp.loss (GN)}
    & \ssoo{100}{1.0645} & \ssoo{100}{0.3216} \\
    \addlinespace

    \multirow{3}{*}{\makecell[l]{\texttt{CNO}\\\texttt{(50M)}}}
    & \texttt{MSE loss}
    & \ssoo{50}{1.1379} & \ssoo{50}{0.3523} \\
    & \texttt{Comp.loss (SA)}
    & \ssoo{0}{1.1555} & \ssoo{0}{0.3631} \\
    & \texttt{Comp.loss (GN)}
    & \ssoo{100}{1.0193} & \ssoo{100}{0.3266} \\
    \addlinespace

    \multirow{3}{*}{\makecell[l]{\texttt{FFNO}\\\texttt{(50M)}}}
    & \texttt{MSE loss}
    & \ssoo{0}{1.0427} & \ssoo{50}{0.2973} \\
\bottomrule
    \end{tabular}
\end{table}

\begin{table}[h!]
    \centering
    \scriptsize
    \caption{Structure-aware metrics for the 2D-SABW (SSOO) test dataset. Lower values are better. Darker shade indicates lower value for each baseline within a group of training loss strategies. Bold values denote the lowest among all baselines for a given metric and the underlined values denote the second-lowest values.}
    \label{tab:2d_sabw_ssoo_structure_aware}
    \begin{tabular}{llcc}
    \toprule
    Model & Training loss & PCC & SSIM \\
    \midrule

    \multirow{3}{*}{\makecell[l]{\texttt{ConvNeXt}\\\texttt{(50M)}}}
    & \texttt{MSE loss}
    & \ssoo{100}{0.0472} & \ssoo{0}{0.2037} \\
    & \texttt{Comp.loss (SA)}
    & \ssoo{0}{0.0490} & \ssoo{100}{0.1630} \\
    & \texttt{Comp.loss (GN)}
    & \ssoo{50}{0.0474} & \ssoo{50}{0.1822} \\
    \addlinespace

    \multirow{3}{*}{\makecell[l]{\texttt{CNO}\\\texttt{(50M)}}}
    & \texttt{MSE loss}
    & \ssoo{50}{0.0612} & \ssoo{0}{0.2649} \\
    & \texttt{Comp.loss (SA)}
    & \ssoo{0}{0.0663} & \ssoo{50}{0.2041} \\
    & \texttt{Comp.loss (GN)}
    & \ssoo{100}{0.0527} & \ssoo{100}{0.1967} \\
    \addlinespace

    \multirow{3}{*}{\makecell[l]{\texttt{FFNO}\\\texttt{(50M)}}}
    & \texttt{MSE loss}
    & \ssoo{50}{0.0428} & \ssoo{50}{0.1596} \\
\bottomrule
    \end{tabular}
\end{table}

\begin{table}[h!]
    \centering
    \scriptsize
    \setlength{\tabcolsep}{4pt}
    \caption{Physics-informed metrics reporting the nRMSE of integral quantities of interest on the 2D-SABW (SSOO) test dataset. Lower values are better. Darker shade indicates lower value for each baseline within a group of training loss strategies. Bold values denote the lowest among all baselines for a given metric and the underlined values denote the second-lowest values.}
    \label{tab:2d_sabw_ssoo_physics_informed}
    \begin{tabular}{@{}p{2.2cm}p{2.3cm}ccccc@{}}
    \toprule
    Model & Training loss & iqRMSE- & iqRMSE- & iqRMSE- & iqRMSE- & iqRMSE- \\
          &               & mass & momentum & momentum & kinetic-energy & enstrophy \\
          &               &  & \_x & \_y &  &  \\
    \midrule

    \multirow{3}{*}{\makecell[l]{\texttt{ConvNeXt}\\\texttt{(50M)}}}
    & \texttt{MSE loss}
    & \ssoo{100}{0.0082} & \ssoo{100}{0.0133} & \ssoo{0}{\texttt{>10}} & \ssoo{100}{0.0180} & \ssoo{100}{0.1171} \\
    & \texttt{Comp.loss (SA)}
    & \ssoo{50}{0.0172} & \ssoo{50}{0.0263} & \ssoo{0}{\texttt{>10}} & \ssoo{50}{0.0317} & \ssoo{50}{0.1267} \\
    & \texttt{Comp.loss (GN)}
    & \ssoo{0}{0.0194} & \ssoo{0}{0.0277} & \ssoo{0}{\texttt{>10}} & \ssoo{0}{0.0328} & \ssoo{0}{0.1330} \\
    \addlinespace

    \multirow{3}{*}{\makecell[l]{\texttt{CNO}\\\texttt{(50M)}}}
    & \texttt{MSE loss}
    & \ssoo{0}{0.6255} & \ssoo{0}{2.7342} & \ssoo{0}{\texttt{>10}} & \ssoo{0}{7.4899} & \ssoo{0}{0.1267} \\
    & \texttt{Comp.loss (SA)}
    & \ssoo{50}{0.0196} & \ssoo{50}{0.0405} & \ssoo{0}{\texttt{>10}} & \ssoo{50}{0.0458} & \ssoo{50}{0.1075} \\
\bottomrule
    \end{tabular}
\end{table}

\begin{table}[h!]
    \centering
    \scriptsize
    \caption{Pointwise evaluation of baselines on the 2D-SRBA (OOOO) test dataset, reporting errors in pointwise field values (MAE, RMSE, VRMSE) and the field derivatives (H1, H2) averaged across all fields and all rollout steps. Lower values are better. Darker shade indicates lower value for each baseline within a group of training loss strategies. Bold values denote the lowest among all baselines for a given metric and the underlined values denote the second-lowest values.}
    \label{tab:2d_r22_sb_pointwise}
    \begin{tabular}{llccccc}
    \toprule
    Model & Training loss & H1 & H2 & MAE & RMSE & VRMSE \\
    \midrule

    \multirow{3}{*}{\makecell[l]{\texttt{ConvNeXt}\\\texttt{(50M)}}}
    & \texttt{MSE loss}
    & \cw{0}{0.0294} & \cw{0}{0.0111} & \cw{50}{0.0509} & \cw{0}{0.1847} & \cw{0}{0.1979} \\
    & \texttt{Comp.loss (SA)}
    & \cw{50}{0.0283} & \cw{50}{0.0107} & \cw{100}{0.0493} & \cw{50}{0.1777} & \cw{100}{0.1833} \\
    & \texttt{Comp.loss (GN)}
    & \cw{100}{0.0277} & \cw{100}{0.0106} & \cw{0}{0.0510} & \cw{100}{0.1773} & \cw{50}{0.1849} \\
    \addlinespace

    \multirow{3}{*}{\makecell[l]{\texttt{CNO}\\\texttt{(50M)}}}
    & \texttt{MSE loss}
    & \cw{0}{0.0365} & \cw{0}{0.0124} & \cw{0}{0.0756} & \cw{0}{0.2329} & \cw{0}{0.2538} \\
    & \texttt{Comp.loss (SA)}
    & \cw{100}{0.0314} & \cw{100}{0.0111} & \cw{100}{0.0635} & \cw{100}{0.2008} & \cw{100}{0.2091} \\
    & \texttt{Comp.loss (GN)}
    & \cw{50}{0.0338} & \cw{50}{0.0116} & \cw{50}{0.0696} & \cw{50}{0.2148} & \cw{50}{0.2293} \\
    \addlinespace

    \multirow{3}{*}{\makecell[l]{\texttt{FFNO}\\\texttt{(50M)}}}
    & \texttt{MSE loss}
    & \cw{0}{0.0270} & \cw{0}{0.0104} & \cw{0}{0.0443} & \cw{0}{0.1674} & \cw{0}{0.1667} \\
\bottomrule
    \end{tabular}
\end{table}

\begin{table}[h!]
    \centering
    \scriptsize
    \caption{Wavelet based spectral metrics for the 2D-SRBA (OOOO) test dataset. Lower values are better. Darker shade indicates lower value for each baseline within a group of training loss strategies. Bold values denote the lowest among all baselines for a given metric and the underlined values denote the second-lowest values.}
    \label{tab:2d_r22_sb_spectral}
    \begin{tabular}{llcccc}
    \toprule
    Model & Training loss & \makecell{wfRMSE-\\high} & \makecell{wfRMSE-\\mid} & \makecell{wfRMSE-\\low} & MLW \\
    \midrule

    \multirow{3}{*}{\makecell[l]{\texttt{ConvNeXt}\\\texttt{(50M)}}}
    & \texttt{MSE loss}
    & \cw{0}{0.0709} & \cw{0}{0.1673} & \cw{0}{0.3275} & \cw{0}{0.8222} \\
    & \texttt{Comp.loss (SA)}
    & \cw{100}{0.0687} & \cw{50}{0.1592} & \cw{50}{0.3106} & \cw{100}{0.6784} \\
    & \texttt{Comp.loss (GN)}
    & \cw{50}{0.0689} & \cw{100}{0.1584} & \cw{100}{0.3083} & \cw{50}{0.7178} \\
    \addlinespace

    \multirow{3}{*}{\makecell[l]{\texttt{CNO}\\\texttt{(50M)}}}
    & \texttt{MSE loss}
    & \cw{0}{0.0792} & \cw{0}{0.1993} & \cw{0}{0.4141} & \cw{0}{0.7187} \\
    & \texttt{Comp.loss (SA)}
    & \cw{100}{0.0721} & \cw{100}{0.1751} & \cw{100}{0.3511} & \cw{100}{\underline{0.5906}} \\
    & \texttt{Comp.loss (GN)}
    & \cw{50}{0.0754} & \cw{50}{0.1881} & \cw{50}{0.3794} & \cw{50}{0.6305} \\
    \addlinespace

    \multirow{3}{*}{\makecell[l]{\texttt{FFNO}\\\texttt{(50M)}}}
    & \texttt{MSE loss}
    & \cw{0}{0.0626} & \cw{0}{0.1461} & \cw{0}{0.2932} & \cw{0}{0.7409} \\
\bottomrule
    \end{tabular}
\end{table}

\begin{table}[h!]
    \centering
    \scriptsize
    \caption{Feature focused metrics for capturing interface (IRMSE) and shocks (SRMSE) on the 2D-SRBA (OOOO) test dataset. Lower values are better. Darker shade indicates lower value for each baseline within a group of training loss strategies. Bold values denote the lowest among all baselines for a given metric and the underlined values denote the second-lowest values.}
    \label{tab:2d_r22_sb_feature_focused}
    \begin{tabular}{llcc}
    \toprule
    Model & Training loss & IRMSE & SRMSE \\
    \midrule

    \multirow{3}{*}{\makecell[l]{\texttt{ConvNeXt}\\\texttt{(50M)}}}
    & \texttt{MSE loss}
    & \cw{0}{0.6297} & \cw{0}{0.2299} \\
    & \texttt{Comp.loss (SA)}
    & \cw{50}{0.5821} & \cw{50}{0.2231} \\
    & \texttt{Comp.loss (GN)}
    & \cw{100}{0.5763} & \cw{100}{0.2230} \\
    \addlinespace

    \multirow{3}{*}{\makecell[l]{\texttt{CNO}\\\texttt{(50M)}}}
    & \texttt{MSE loss}
    & \cw{0}{0.7471} & \cw{0}{0.2975} \\
    & \texttt{Comp.loss (SA)}
    & \cw{100}{0.6207} & \cw{100}{0.2582} \\
    & \texttt{Comp.loss (GN)}
    & \cw{50}{0.6536} & \cw{50}{0.2795} \\
    \addlinespace

    \multirow{3}{*}{\makecell[l]{\texttt{FFNO}\\\texttt{(50M)}}}
    & \texttt{MSE loss}
    & \cw{0}{0.5446} & \cw{0}{0.2070} \\
    & \texttt{Comp.loss (SA)}
    & \cw{100}{0.5078} & \cw{100}{0.1951} \\
    & \texttt{Comp.loss (GN)}
    & \cw{50}{0.5114} & \cw{50}{0.1975} \\
    \addlinespace

    \multirow{3}{*}{\makecell[l]{\texttt{ScOT}\\\texttt{(50M)}}}
    & \texttt{MSE loss}
    & \cw{100}{0.5205} & \cw{100}{0.1868} \\
    & \texttt{Comp.loss (SA)}
    & \cw{50}{0.5956} & \cw{50}{0.2323} \\
    & \texttt{Comp.loss (GN)}
    & \cw{0}{0.5966} & \cw{0}{0.2345} \\
    \addlinespace

    \multirow{3}{*}{\makecell[l]{\texttt{DPOT-M}\\\texttt{(122M)}}}
    & \texttt{MSE loss}
    & \cw{50}{\underline{0.1522}} & \cw{50}{\underline{0.1571}} \\
    & \texttt{Comp.loss (SA)}
    & \cw{100}{\textbf{0.1462}} & \cw{100}{\textbf{0.1554}} \\
    & \texttt{Comp.loss (GN)}
    & \cw{0}{0.4103} & \cw{0}{0.1711} \\
    \addlinespace

    \multirow{3}{*}{\makecell[l]{\texttt{Poseidon-B}\\\texttt{(158M)}}}
    & \texttt{MSE loss}
    & \cw{50}{0.8480} & \cw{100}{1.0545} \\
    & \texttt{Comp.loss (SA)}
    & \cw{100}{0.8398} & \cw{0}{1.0778} \\
    & \texttt{Comp.loss (GN)}
    & \cw{0}{1.9987} & \cw{50}{1.0770} \\

    \bottomrule
    \end{tabular}
\end{table}

\begin{table}[h!]
    \centering
    \scriptsize
    \caption{Structure-aware metrics for the 2D-SRBA (OOOO) test dataset. Lower values are better. Darker shade indicates lower value for each baseline within a group of training loss strategies. Bold values denote the lowest among all baselines for a given metric and the underlined values denote the second-lowest values.}
    \label{tab:2d_r22_sb_structure_aware}
    \begin{tabular}{llcc}
    \toprule
    Model & Training loss & PCC & SSIM \\
    \midrule

    \multirow{3}{*}{\makecell[l]{\texttt{ConvNeXt}\\\texttt{(50M)}}}
    & \texttt{MSE loss}
    & \cw{0}{0.0300} & \cw{0}{0.1382} \\
    & \texttt{Comp.loss (SA)}
    & \cw{100}{0.0277} & \cw{100}{0.0806} \\
    & \texttt{Comp.loss (GN)}
    & \cw{100}{0.0277} & \cw{50}{0.0891} \\
    \addlinespace

    \multirow{3}{*}{\makecell[l]{\texttt{CNO}\\\texttt{(50M)}}}
    & \texttt{MSE loss}
    & \cw{0}{0.0483} & \cw{0}{0.1911} \\
    & \texttt{Comp.loss (SA)}
    & \cw{100}{0.0357} & \cw{100}{0.1047} \\
    & \texttt{Comp.loss (GN)}
    & \cw{50}{0.0411} & \cw{50}{0.1262} \\
    \addlinespace

    \multirow{3}{*}{\makecell[l]{\texttt{FFNO}\\\texttt{(50M)}}}
    & \texttt{MSE loss}
    & \cw{0}{0.0269} & \cw{0}{0.0940} \\
    & \texttt{Comp.loss (SA)}
    & \cw{100}{0.0250} & \cw{100}{0.0696} \\
    & \texttt{Comp.loss (GN)}
    & \cw{50}{0.0252} & \cw{50}{0.0732} \\
    \addlinespace

    \multirow{3}{*}{\makecell[l]{\texttt{ScOT}\\\texttt{(50M)}}}
    & \texttt{MSE loss}
    & \cw{100}{0.0206} & \cw{100}{0.0908} \\
    & \texttt{Comp.loss (SA)}
    & \cw{50}{0.0305} & \cw{50}{0.0926} \\
    & \texttt{Comp.loss (GN)}
    & \cw{0}{0.0316} & \cw{0}{0.1039} \\
    \addlinespace

    \multirow{3}{*}{\makecell[l]{\texttt{DPOT-M}\\\texttt{(122M)}}}
    & \texttt{MSE loss}
    & \cw{50}{\underline{0.0156}} & \cw{0}{0.0854} \\
    & \texttt{Comp.loss (SA)}
    & \cw{100}{\textbf{0.0138}} & \cw{100}{\textbf{0.0518}} \\
    & \texttt{Comp.loss (GN)}
    & \cw{0}{0.0165} & \cw{50}{\underline{0.0640}} \\
    \addlinespace

    \multirow{3}{*}{\makecell[l]{\texttt{Poseidon-B}\\\texttt{(158M)}}}
    & \texttt{MSE loss}
    & \cw{100}{0.4512} & \cw{0}{0.4469} \\
    & \texttt{Comp.loss (SA)}
    & \cw{0}{0.4684} & \cw{50}{0.4464} \\
    & \texttt{Comp.loss (GN)}
    & \cw{50}{0.4659} & \cw{100}{0.4424} \\

    \bottomrule
    \end{tabular}
\end{table}
\begin{table}[h!]
    \centering
    \scriptsize
    \caption{Physics-informed metrics reporting the nRMSE of integral quantities of interest on the 2D-SRBA (OOOO) test dataset. Lower values are better. Darker shade indicates lower value for each baseline within a group of training loss strategies. Bold values denote the lowest among all baselines for a given metric and the underlined values denote the second-lowest values.}
    \label{tab:2d_r22_sb_physics_informed}
    \setlength{\tabcolsep}{4pt}
    \begin{tabular}{@{}p{2.2cm}p{2.3cm}ccccc@{}}
    \toprule
    Model & Training loss & iqRMSE- & iqRMSE- & iqRMSE- & iqRMSE- & iqRMSE- \\
          &               & mass & \makecell{momentum\\\_x} & \makecell{momentum\\\_y} & \makecell{kinetic-\\energy} & enstrophy \\
    \midrule

    \multirow{3}{*}{\makecell[l]{\texttt{ConvNeXt}\\\texttt{(50M)}}}
    & \texttt{MSE loss}
    & \cw{100}{0.0277} & \cw{100}{0.0276} & \cw{0}{\texttt{>10}} & \cw{100}{0.0266} & \cw{100}{0.0446} \\
    & \texttt{Comp.loss (SA)}
    & \cw{0}{0.0541} & \cw{0}{0.0565} & \cw{0}{\texttt{>10}} & \cw{0}{0.0521} & \cw{50}{0.0537} \\
    & \texttt{Comp.loss (GN)}
    & \cw{50}{0.0420} & \cw{50}{0.0419} & \cw{0}{\texttt{>10}} & \cw{50}{0.0376} & \cw{0}{0.0666} \\
    \addlinespace

    \multirow{3}{*}{\makecell[l]{\texttt{CNO}\\\texttt{(50M)}}}
    & \texttt{MSE loss}
    & \cw{0}{0.0566} & \cw{0}{0.0747} & \cw{0}{\texttt{>10}} & \cw{0}{0.0773} & \cw{50}{0.0556} \\
    & \texttt{Comp.loss (SA)}
    & \cw{100}{0.0396} & \cw{50}{0.0546} & \cw{0}{\texttt{>10}} & \cw{50}{0.0586} & \cw{100}{0.0551} \\
    & \texttt{Comp.loss (GN)}
    & \cw{50}{0.0403} & \cw{100}{0.0517} & \cw{0}{\texttt{>10}} & \cw{100}{0.0502} & \cw{0}{0.0657} \\
\bottomrule
    \end{tabular}
\end{table}

\begin{table}[h!]
    \centering
    \scriptsize
    \caption{Pointwise evaluation of baselines on the 2D-SDBA (SSOO) test dataset, reporting errors in pointwise field values (MAE, RMSE, VRMSE) and the field derivatives (H1, H2) averaged across all fields and all rollout steps. Lower values are better. Darker shade indicates lower value for each baseline within a group of training loss strategies. Bold values denote the lowest among all baselines for a given metric and the underlined values denote the second-lowest values.}
    \label{tab:2d_sd_pointwise}
    \begin{tabular}{llccccc}
    \toprule
    Model & Training loss & H1 & H2 & MAE & RMSE & VRMSE \\
    \midrule

    \multirow{3}{*}{\makecell[l]{\texttt{ConvNeXt}\\\texttt{(50M)}}}
    & \texttt{MSE loss}
    & \sdcw{0}{0.0411} & \sdcw{0}{0.0116} & \sdcw{0}{0.1891} & \sdcw{0}{0.4764} & \sdcw{0}{0.5366} \\
    & \texttt{Comp.loss (SA)}
    & \sdcw{50}{0.0380} & \sdcw{50}{0.0106} & \sdcw{50}{0.1854} & \sdcw{50}{0.4696} & \sdcw{50}{0.5242} \\
    & \texttt{Comp.loss (GN)}
    & \sdcw{100}{0.0352} & \sdcw{100}{\underline{0.0100}} & \sdcw{100}{0.1771} & \sdcw{100}{0.4552} & \sdcw{100}{0.5168} \\
    \addlinespace

    \multirow{3}{*}{\makecell[l]{\texttt{CNO}\\\texttt{(50M)}}}
    & \texttt{MSE loss}
    & \sdcw{0}{0.0468} & \sdcw{0}{0.0134} & \sdcw{0}{0.2199} & \sdcw{0}{0.5207} & \sdcw{0}{0.6074} \\
    & \texttt{Comp.loss (SA)}
    & \sdcw{50}{0.0403} & \sdcw{50}{0.0110} & \sdcw{50}{0.2111} & \sdcw{50}{0.5057} & \sdcw{50}{0.5755} \\
    & \texttt{Comp.loss (GN)}
    & \sdcw{100}{0.0376} & \sdcw{100}{0.0107} & \sdcw{100}{0.1924} & \sdcw{100}{0.4736} & \sdcw{100}{0.5357} \\
    \addlinespace

    \multirow{3}{*}{\makecell[l]{\texttt{FFNO}\\\texttt{(50M)}}}
    & \texttt{MSE loss}
    & \sdcw{0}{0.0385} & \sdcw{0}{0.0106} & \sdcw{0}{0.1847} & \sdcw{100}{0.4615} & \sdcw{0}{0.5230} \\
\bottomrule
    \end{tabular}
\end{table}

\begin{table}[h!]
    \centering
    \scriptsize
    \caption{Wavelet based spectral metrics for the 2D-SDBA (SSOO) test dataset. Lower values are better. Darker shade indicates lower value for each baseline within a group of training loss strategies. Bold values denote the lowest among all baselines for a given metric and the underlined values denote the second-lowest values.}
    \label{tab:2d_sd_spectral}
    \begin{tabular}{llcccc}
    \toprule
        Model & Training loss & wfRMSE- & wfRMSE- & wfRMSE- & MLW \\
            &               & high & mid & low &  \\
    \midrule

    \multirow{3}{*}{\makecell[l]{\texttt{ConvNeXt}\\\texttt{(50M)}}}
    & \texttt{MSE loss}
    & \sdcw{0}{0.0827} & \sdcw{0}{0.2403} & \sdcw{0}{0.6461} & \sdcw{0}{0.3711} \\
    & \texttt{Comp.loss (SA)}
    & \sdcw{50}{0.0783} & \sdcw{50}{0.2277} & \sdcw{50}{0.6173} & \sdcw{100}{0.3456} \\
    & \texttt{Comp.loss (GN)}
    & \sdcw{100}{0.0754} & \sdcw{100}{0.2162} & \sdcw{100}{0.5839} & \sdcw{50}{0.3465} \\
    \addlinespace

    \multirow{3}{*}{\makecell[l]{\texttt{CNO}\\\texttt{(50M)}}}
    & \texttt{MSE loss}
    & \sdcw{0}{0.0855} & \sdcw{0}{0.2488} & \sdcw{0}{0.6829} & \sdcw{0}{0.3773} \\
    & \texttt{Comp.loss (SA)}
    & \sdcw{50}{0.0792} & \sdcw{50}{0.2318} & \sdcw{50}{0.6359} & \sdcw{100}{0.3599} \\
    & \texttt{Comp.loss (GN)}
    & \sdcw{100}{0.0768} & \sdcw{100}{0.2215} & \sdcw{100}{0.5972} & \sdcw{50}{0.3624} \\
\bottomrule
    \end{tabular}
\end{table}

\begin{table}[h!]
    \centering
    \scriptsize
    \caption{Feature focused metrics for capturing interface (IRMSE) and shocks (SRMSE) on the 2D-SDBA (SSOO) test dataset. Lower values are better. Darker shade indicates lower value for each baseline within a group of training loss strategies. Bold values denote the lowest among all baselines for a given metric and the underlined values denote the second-lowest values.}
    \label{tab:2d_sd_feature_focused}
    \begin{tabular}{llcc}
    \toprule
    Model & Training loss & IRMSE & SRMSE \\
    \midrule

    \multirow{3}{*}{\makecell[l]{\texttt{ConvNeXt}\\\texttt{(50M)}}}
    & \texttt{MSE loss}
    & \sdcw{0}{0.9550} & \sdcw{0}{0.4865} \\
    & \texttt{Comp.loss (SA)}
    & \sdcw{50}{0.8639} & \sdcw{50}{0.4703} \\
    & \texttt{Comp.loss (GN)}
    & \sdcw{100}{0.7957} & \sdcw{100}{0.4505} \\
    \addlinespace

    \multirow{3}{*}{\makecell[l]{\texttt{CNO}\\\texttt{(50M)}}}
    & \texttt{MSE loss}
    & \sdcw{0}{1.0893} & \sdcw{0}{0.5368} \\
    & \texttt{Comp.loss (SA)}
    & \sdcw{50}{0.9306} & \sdcw{50}{0.5056} \\
    & \texttt{Comp.loss (GN)}
    & \sdcw{100}{0.8388} & \sdcw{100}{0.4680} \\
    \addlinespace

    \multirow{3}{*}{\makecell[l]{\texttt{FFNO}\\\texttt{(50M)}}}
    & \texttt{MSE loss}
    & \sdcw{0}{0.9170} & \sdcw{0}{0.4739} \\
\bottomrule
    \end{tabular}
\end{table}

\begin{table}[h!]
    \centering
    \scriptsize
    \caption{Structure-aware metrics for the 2D-SDBA (SSOO) test dataset. Lower values are better. Darker shade indicates lower value for each baseline within a group of training loss strategies. Bold values denote the lowest among all baselines for a given metric and the underlined values denote the second-lowest values.}
    \label{tab:2d_sd_structure_aware}
    \begin{tabular}{llcc}
    \toprule
    Model & Training loss & PCC & SSIM \\
    \midrule

    \multirow{3}{*}{\makecell[l]{\texttt{ConvNeXt}\\\texttt{(50M)}}}
    & \texttt{MSE loss}
    & \sdcw{0}{0.1713} & \sdcw{0}{0.3158} \\
    & \texttt{Comp.loss (SA)}
    & \sdcw{50}{0.1656} & \sdcw{50}{0.2949} \\
    & \texttt{Comp.loss (GN)}
    & \sdcw{100}{0.1604} & \sdcw{100}{0.2846} \\
    \addlinespace

    \multirow{3}{*}{\makecell[l]{\texttt{CNO}\\\texttt{(50M)}}}
    & \texttt{MSE loss}
    & \sdcw{0}{0.2034} & \sdcw{0}{0.3572} \\
    & \texttt{Comp.loss (SA)}
    & \sdcw{50}{0.1864} & \sdcw{50}{0.3205} \\
    & \texttt{Comp.loss (GN)}
    & \sdcw{100}{0.1692} & \sdcw{100}{0.3043} \\
    \addlinespace

    \multirow{3}{*}{\makecell[l]{\texttt{FFNO}\\\texttt{(50M)}}}
    & \texttt{MSE loss}
    & \sdcw{0}{0.1690} & \sdcw{0}{0.3228} \\
\bottomrule
    \end{tabular}
\end{table}

\begin{table}[h!]
    \centering
    \scriptsize
    \setlength{\tabcolsep}{4pt}
    \caption{Physics-informed metrics reporting the nRMSE of integral quantities of interest on the 2D-SDBA (SSOO) test dataset. Lower values are better. Darker shade indicates lower value for each baseline within a group of training loss strategies. Bold values denote the lowest among all baselines for a given metric and the underlined values denote the second-lowest values.}
    \label{tab:2d_sd_physics_informed}
    \begin{tabular}{@{}p{1.6cm}p{2.3cm}cccccc@{}}
    \toprule
    Model & Training loss & iqRMSE- & iqRMSE- & iqRMSE- & iqRMSE- & iqRMSE- & iqRMSE- \\
          &               & mass & momentum & momentum & com\_x & com\_y & enstrophy \\
          &               &  & \_x & \_y &  &  \\
    \midrule

    \multirow{3}{*}{\makecell[l]{\texttt{ConvNeXt}\\\texttt{(50M)}}}
    & \texttt{MSE loss}
    & \sdcw{0}{0.4169} & \sdcw{50}{0.2133} & \sdcw{0}{7.2569} & \sdcw{0}{0.0319} & \sdcw{0}{0.0180} & \sdcw{100}{0.3690} \\
    & \texttt{Comp.loss (SA)}
    & \sdcw{50}{0.3653} & \sdcw{0}{0.2270} & \sdcw{50}{4.1769} & \sdcw{50}{0.0190} & \sdcw{100}{0.0095} & \sdcw{50}{0.3804} \\
    & \texttt{Comp.loss (GN)}
    & \sdcw{100}{0.2826} & \sdcw{100}{0.1657} & \sdcw{100}{3.7943} & \sdcw{100}{\underline{0.0157}} & \sdcw{50}{0.0102} & \sdcw{0}{0.3865} \\
    \addlinespace

    \multirow{3}{*}{\makecell[l]{\texttt{CNO}\\\texttt{(50M)}}}
    & \texttt{MSE loss}
    & \sdcw{0}{\texttt{>10}} & \sdcw{0}{\texttt{>10}} & \sdcw{0}{\texttt{>10}} & \sdcw{0}{\texttt{>10}} & \sdcw{0}{\texttt{>10}} & \sdcw{100}{0.4177} \\
    & \texttt{Comp.loss (SA)}
    & \sdcw{100}{0.5645} & \sdcw{100}{0.6296} & \sdcw{0}{\texttt{>10}} & \sdcw{50}{0.0488} & \sdcw{100}{0.0250} & \sdcw{50}{0.4205} \\
\bottomrule
    \end{tabular}
\end{table}

\begin{table}[h!]
    \centering
    \scriptsize
    \caption{Pointwise evaluation of baselines on the 3D-SABW (SSOOSS) test dataset, reporting errors in pointwise field values (MAE, RMSE, VRMSE) and the field derivatives (H1, H2) averaged across all fields and all rollout steps. Lower values are better. Darker shade indicates lower value for each baseline within a group of training loss strategies. Bold values denote the lowest among all baselines for a given metric and the underlined values denote the second-lowest values.}
    \label{tab:3d_sabw_pointwise}
    \begin{tabular}{llccccc}
    \toprule
    Model & Training loss & H1 & H2 & MAE & RMSE & VRMSE \\
    \midrule

    \multirow{3}{*}{\makecell[l]{\texttt{ConvNeXt}\\\texttt{(10M)}}}
    & \texttt{MSE loss}
    & \cellcolor[HTML]{5ACCDA}0.0583 & \cellcolor[HTML]{5ACCDA}0.2428 & \cellcolor[HTML]{17BECF}\underline{0.1426} & \cellcolor[HTML]{17BECF}0.3697 & \cellcolor[HTML]{9EDAE5}1.9591 \\
    & \texttt{Comp.loss (SA)}
    & \cellcolor[HTML]{9EDAE5}0.0618 & \cellcolor[HTML]{9EDAE5}0.2434 & \cellcolor[HTML]{9EDAE5}0.1779 & \cellcolor[HTML]{9EDAE5}0.4190 & \cellcolor[HTML]{5ACCDA}1.6982 \\
    & \texttt{Comp.loss (GN)}
    & \cellcolor[HTML]{17BECF}0.0534 & \cellcolor[HTML]{17BECF}0.2271 & \cellcolor[HTML]{5ACCDA}0.1576 & \cellcolor[HTML]{5ACCDA}0.3723 & \cellcolor[HTML]{17BECF}1.2033 \\
    \addlinespace

    \multirow{3}{*}{\makecell[l]{\texttt{FFNO}\\\texttt{(10M)}}}
    & \texttt{MSE loss}
    & \cellcolor[HTML]{9EDAE5}0.0333 & \cellcolor[HTML]{9EDAE5}0.1831 & \cellcolor[HTML]{17BECF}\textbf{0.1318} & \cellcolor[HTML]{17BECF}\textbf{0.3056} & \cellcolor[HTML]{9EDAE5}0.7823 \\
    & \texttt{Comp.loss (SA)}
    & \cellcolor[HTML]{5ACCDA}\underline{0.0329} & \cellcolor[HTML]{5ACCDA}\underline{0.1739} & \cellcolor[HTML]{9EDAE5}0.1590 & \cellcolor[HTML]{9EDAE5}0.3493 & \cellcolor[HTML]{5ACCDA}\underline{0.6104} \\
    & \texttt{Comp.loss (GN)}
    & \cellcolor[HTML]{17BECF}\textbf{0.0303} & \cellcolor[HTML]{17BECF}\textbf{0.1631} & \cellcolor[HTML]{5ACCDA}0.1527 & \cellcolor[HTML]{5ACCDA}\underline{0.3373} & \cellcolor[HTML]{17BECF}\textbf{0.5767} \\

    \bottomrule
    \end{tabular}
\end{table}

\begin{table}[h!]
    \centering
    \scriptsize
    \caption{Wavelet based spectral metrics for the 3D-SABW (SSOOSS) test dataset. Lower values are better. Darker shade indicates lower value for each baseline within a group of training loss strategies. Bold values denote the lowest among all baselines for a given metric and the underlined values denote the second-lowest values.}
    \label{tab:3d_sabw_spectral}
    \begin{tabular}{llcccc}
    \toprule
    Model & Training loss & wfRMSE-high & wfRMSE-mid & wfRMSE-low & MLW \\
    \midrule

    \multirow{3}{*}{\makecell[l]{\texttt{ConvNeXt}\\\texttt{(10M)}}}
    & \texttt{MSE loss}
    & \cellcolor[HTML]{17BECF}0.0729 & \cellcolor[HTML]{5ACCDA}0.3385 & \cellcolor[HTML]{5ACCDA}1.1939 & \cellcolor[HTML]{9EDAE5}2.1995 \\
    & \texttt{Comp.loss (SA)}
    & \cellcolor[HTML]{9EDAE5}0.0790 & \cellcolor[HTML]{9EDAE5}0.3488 & \cellcolor[HTML]{9EDAE5}1.2437 & \cellcolor[HTML]{5ACCDA}2.0639 \\
    & \texttt{Comp.loss (GN)}
    & \cellcolor[HTML]{5ACCDA}0.0734 & \cellcolor[HTML]{17BECF}0.3181 & \cellcolor[HTML]{17BECF}1.1102 & \cellcolor[HTML]{17BECF}1.9722 \\
    \addlinespace

    \multirow{3}{*}{\makecell[l]{\texttt{FFNO}\\\texttt{(10M)}}}
    & \texttt{MSE loss}
    & \cellcolor[HTML]{9EDAE5}0.0600 & \cellcolor[HTML]{5ACCDA}\underline{0.2220} & \cellcolor[HTML]{5ACCDA}\underline{0.7633} & \cellcolor[HTML]{9EDAE5}1.9471 \\
    & \texttt{Comp.loss (SA)}
    & \cellcolor[HTML]{5ACCDA}\underline{0.0570} & \cellcolor[HTML]{9EDAE5}0.2236 & \cellcolor[HTML]{9EDAE5}0.7982 & \cellcolor[HTML]{17BECF}\textbf{1.7184} \\
    & \texttt{Comp.loss (GN)}
    & \cellcolor[HTML]{17BECF}\textbf{0.0550} & \cellcolor[HTML]{17BECF}\textbf{0.2128} & \cellcolor[HTML]{17BECF}\textbf{0.7617} & \cellcolor[HTML]{5ACCDA}\underline{1.7481} \\

    \bottomrule
    \end{tabular}
\end{table}

\begin{table}[h!]
    \centering
    \scriptsize
    \caption{Feature focused metrics for capturing interface (IRMSE) and shocks (SRMSE) on the 3D-SABW (SSOOSS) test dataset. Lower values are better. Darker shade indicates lower value for each baseline within a group of training loss strategies. Bold values denote the lowest among all baselines for a given metric and the underlined values denote the second-lowest values.}
    \label{tab:3d_sabw_feature_focused}
    \begin{tabular}{llcc}
    \toprule
    Model & Training loss & IRMSE & SRMSE \\
    \midrule

    \multirow{3}{*}{\makecell[l]{\texttt{ConvNeXt}\\\texttt{(10M)}}}
    & \texttt{MSE loss}
    & \cellcolor[HTML]{9EDAE5}3.7761 & \cellcolor[HTML]{5ACCDA}0.4985 \\
    & \texttt{Comp.loss (SA)}
    & \cellcolor[HTML]{5ACCDA}3.3504 & \cellcolor[HTML]{9EDAE5}0.5203 \\
    & \texttt{Comp.loss (GN)}
    & \cellcolor[HTML]{17BECF}2.9574 & \cellcolor[HTML]{17BECF}0.4668 \\
    \addlinespace

    \multirow{3}{*}{\makecell[l]{\texttt{FFNO}\\\texttt{(10M)}}}
    & \texttt{MSE loss}
    & \cellcolor[HTML]{9EDAE5}1.7955 & \cellcolor[HTML]{17BECF}\textbf{0.3214} \\
    & \texttt{Comp.loss (SA)}
    & \cellcolor[HTML]{5ACCDA}\underline{1.6019} & \cellcolor[HTML]{9EDAE5}0.3427 \\
    & \texttt{Comp.loss (GN)}
    & \cellcolor[HTML]{17BECF}\textbf{1.5421} & \cellcolor[HTML]{5ACCDA}\underline{0.3260} \\

    \bottomrule
    \end{tabular}
\end{table}

\begin{table}[h!]
    \centering
    \scriptsize
    \caption{Structure-aware metrics for the 3D-SABW (SSOOSS) test dataset. Lower values are better. Darker shade indicates lower value for each baseline within a group of training loss strategies. Bold values denote the lowest among all baselines for a given metric and the underlined values denote the second-lowest values.}
    \label{tab:3d_sabw_structure_aware}
    \begin{tabular}{llcc}
    \toprule
    Model & Training loss & PCC & SSIM \\
    \midrule

    \multirow{3}{*}{\makecell[l]{\texttt{ConvNeXt}\\\texttt{(10M)}}}
    & \texttt{MSE loss}
    & \cellcolor[HTML]{17BECF}0.1160 & \cellcolor[HTML]{9EDAE5}0.4745 \\
    & \texttt{Comp.loss (SA)}
    & \cellcolor[HTML]{9EDAE5}0.1589 & \cellcolor[HTML]{5ACCDA}0.4342 \\
    & \texttt{Comp.loss (GN)}
    & \cellcolor[HTML]{5ACCDA}0.1189 & \cellcolor[HTML]{17BECF}0.3721 \\
    \addlinespace

    \multirow{3}{*}{\makecell[l]{\texttt{FFNO}\\\texttt{(10M)}}}
    & \texttt{MSE loss}
    & \cellcolor[HTML]{17BECF}\textbf{0.0923} & \cellcolor[HTML]{9EDAE5}0.3374 \\
    & \texttt{Comp.loss (SA)}
    & \cellcolor[HTML]{9EDAE5}0.1107 & \cellcolor[HTML]{17BECF}\textbf{0.3130} \\
    & \texttt{Comp.loss (GN)}
    & \cellcolor[HTML]{5ACCDA}\underline{0.1053} & \cellcolor[HTML]{5ACCDA}\underline{0.3165} \\

    \bottomrule
    \end{tabular}
\end{table}

\begin{table}[h!]
    \centering
    \scriptsize
    \setlength{\tabcolsep}{4pt}
    \caption{Physics-informed metrics reporting the nRMSE of integral quantities of interest on the 3D-SABW (SSOOSS) test dataset. Lower values are better. Darker shade indicates lower value for each baseline within a group of training loss strategies. Bold values denote the lowest among all baselines for a given metric and the underlined values denote the second-lowest values.}
    \label{tab:3d_sabw_physics_informed}
    \begin{tabular}{@{}p{2.2cm}p{2.3cm}ccccc@{}}
    \toprule
    Model & Training loss & iqRMSE- & iqRMSE- & iqRMSE- & iqRMSE- & iqRMSE- \\
          &               & mass & \makecell{momentum\\\_x} & \makecell{momentum\\\_y} & \makecell{kinetic-\\energy} & enstrophy \\
    \midrule

    \multirow{3}{*}{\makecell[l]{\texttt{ConvNeXt}\\\texttt{(10M)}}}
    & \texttt{MSE loss}
    & \cellcolor[HTML]{5ACCDA}0.2839 & \cellcolor[HTML]{5ACCDA}2.7939 & \cellcolor[HTML]{9EDAE5}\texttt{>10} & \cellcolor[HTML]{9EDAE5}\texttt{>10} & \cellcolor[HTML]{17BECF}0.1337 \\
    & \texttt{Comp.loss (SA)}
    & \cellcolor[HTML]{17BECF}0.0170 & \cellcolor[HTML]{17BECF}0.0490 & \cellcolor[HTML]{9EDAE5}\texttt{>10} & \cellcolor[HTML]{17BECF}0.0682 & \cellcolor[HTML]{9EDAE5}0.2560 \\
    & \texttt{Comp.loss (GN)}
    & \cellcolor[HTML]{9EDAE5}\texttt{>10} & \cellcolor[HTML]{9EDAE5}\texttt{>10} & \cellcolor[HTML]{9EDAE5}\texttt{>10} & \cellcolor[HTML]{9EDAE5}\texttt{>10} & \cellcolor[HTML]{5ACCDA}0.1894 \\
    \addlinespace

    \multirow{3}{*}{\makecell[l]{\texttt{FFNO}\\\texttt{(10M)}}}
    & \texttt{MSE loss}
    & \cellcolor[HTML]{5ACCDA}\underline{0.0051} & \cellcolor[HTML]{9EDAE5}0.0218 & \cellcolor[HTML]{9EDAE5}\texttt{>10} & \cellcolor[HTML]{9EDAE5}0.0376 & \cellcolor[HTML]{17BECF}\textbf{0.0508} \\
    & \texttt{Comp.loss (SA)}
    & \cellcolor[HTML]{9EDAE5}0.0053 & \cellcolor[HTML]{5ACCDA}\underline{0.0187} & \cellcolor[HTML]{9EDAE5}\texttt{>10} & \cellcolor[HTML]{5ACCDA}\underline{0.0297} & \cellcolor[HTML]{9EDAE5}0.0673 \\
    & \texttt{Comp.loss (GN)}
    & \cellcolor[HTML]{17BECF}\textbf{0.0043} & \cellcolor[HTML]{17BECF}\textbf{0.0118} & \cellcolor[HTML]{9EDAE5}\texttt{>10} & \cellcolor[HTML]{17BECF}\textbf{0.0192} & \cellcolor[HTML]{5ACCDA}\underline{0.0546} \\

    \bottomrule
    \end{tabular}
\end{table}

\begin{table}[h!]
    \centering
    \scriptsize
    \caption{Pointwise evaluation of baselines on the 3D-SDBA (SSOOSS) test dataset, reporting errors in pointwise field values (MAE, RMSE, VRMSE) and the field derivatives (H1, H2) averaged across all fields and all rollout steps. Lower values are better. Darker shade indicates lower value for each baseline within a group of training loss strategies. Bold values denote the lowest among all baselines for a given metric and the underlined values denote the second-lowest values.}
    \label{tab:3d_sdba_pointwise}
    \begin{tabular}{llccccc}
    \toprule
    Model & Training loss & H1 & H2 & MAE & RMSE & VRMSE \\
    \midrule

    \multirow{3}{*}{\makecell[l]{\texttt{ConvNeXt}\\\texttt{(10M)}}}
    & \texttt{MSE loss}
    & \cellcolor[HTML]{BCBD22}5.9460 & \cellcolor[HTML]{DBDB8D}\texttt{>10} & \cellcolor[HTML]{CCCC58}0.2900 & \cellcolor[HTML]{BCBD22}1.7821 & \cellcolor[HTML]{BCBD22}4.3839 \\
    & \texttt{Comp.loss (SA)}
    & \cellcolor[HTML]{DBDB8D}\texttt{>10} & \cellcolor[HTML]{DBDB8D}\texttt{>10} & \cellcolor[HTML]{BCBD22}0.2495 & \cellcolor[HTML]{CCCC58}4.6078 & \cellcolor[HTML]{CCCC58}6.0536 \\
    & \texttt{Comp.loss (GN)}
    & \cellcolor[HTML]{DBDB8D}\texttt{>10} & \cellcolor[HTML]{DBDB8D}\texttt{>10} & \cellcolor[HTML]{DBDB8D}\texttt{>10} & \cellcolor[HTML]{DBDB8D}\texttt{>10} & \cellcolor[HTML]{DBDB8D}\texttt{>10} \\
    \addlinespace

    \multirow{3}{*}{\makecell[l]{\texttt{FFNO}\\\texttt{(10M)}}}
    & \texttt{MSE loss}
    & \cellcolor[HTML]{DBDB8D}0.1122 & \cellcolor[HTML]{DBDB8D}0.5487 & \cellcolor[HTML]{BCBD22}\textbf{0.1532} & \cellcolor[HTML]{BCBD22}\textbf{0.4605} & \cellcolor[HTML]{BCBD22}\textbf{1.0754} \\
    & \texttt{Comp.loss (SA)}
    & \cellcolor[HTML]{BCBD22}\textbf{0.0959} & \cellcolor[HTML]{BCBD22}\textbf{0.4316} & \cellcolor[HTML]{CCCC58}\underline{0.1761} & \cellcolor[HTML]{CCCC58}\underline{0.4683} & \cellcolor[HTML]{CCCC58}\underline{1.0765} \\
    & \texttt{Comp.loss (GN)}
    & \cellcolor[HTML]{CCCC58}\underline{0.0987} & \cellcolor[HTML]{CCCC58}\underline{0.4491} & \cellcolor[HTML]{DBDB8D}0.1876 & \cellcolor[HTML]{DBDB8D}0.4744 & \cellcolor[HTML]{DBDB8D}1.4433 \\

    \bottomrule
    \end{tabular}
\end{table}

\begin{table}[h!]
    \centering
    \scriptsize
    \caption{Wavelet based spectral metrics for the 3D-SDBA (SSOOSS) test dataset. Lower values are better. Darker shade indicates lower value for each baseline within a group of training loss strategies. Bold values denote the lowest among all baselines for a given metric and the underlined values denote the second-lowest values.}
    \label{tab:3d_sdba_spectral}
    \begin{tabular}{llcccc}
    \toprule
    Model & Training loss & wfRMSE-high & wfRMSE-mid & wfRMSE-low & MLW \\
    \midrule

    \multirow{3}{*}{\makecell[l]{\texttt{ConvNeXt}\\\texttt{(10M)}}}
    & \texttt{MSE loss}
    & \cellcolor[HTML]{BCBD22}0.2453 & \cellcolor[HTML]{BCBD22}1.2022 & \cellcolor[HTML]{BCBD22}4.4012 & \cellcolor[HTML]{DBDB8D}1.3508 \\
    & \texttt{Comp.loss (SA)}
    & \cellcolor[HTML]{CCCC58}1.1313 & \cellcolor[HTML]{CCCC58}3.2171 & \cellcolor[HTML]{CCCC58}8.3366 & \cellcolor[HTML]{BCBD22}1.0554 \\
    & \texttt{Comp.loss (GN)}
    & \cellcolor[HTML]{DBDB8D}\texttt{>10} & \cellcolor[HTML]{DBDB8D}\texttt{>10} & \cellcolor[HTML]{DBDB8D}\texttt{>10} & \cellcolor[HTML]{CCCC58}1.1599 \\
    \addlinespace

    \multirow{3}{*}{\makecell[l]{\texttt{FFNO}\\\texttt{(10M)}}}
    & \texttt{MSE loss}
    & \cellcolor[HTML]{DBDB8D}0.1047 & \cellcolor[HTML]{DBDB8D}0.4203 & \cellcolor[HTML]{DBDB8D}1.3961 & \cellcolor[HTML]{DBDB8D}1.1078 \\
    & \texttt{Comp.loss (SA)}
    & \cellcolor[HTML]{BCBD22}\textbf{0.0888} & \cellcolor[HTML]{BCBD22}\textbf{0.3741} & \cellcolor[HTML]{BCBD22}\textbf{1.3179} & \cellcolor[HTML]{CCCC58}\underline{1.0079} \\
    & \texttt{Comp.loss (GN)}
    & \cellcolor[HTML]{CCCC58}\underline{0.0903} & \cellcolor[HTML]{CCCC58}\underline{0.3858} & \cellcolor[HTML]{CCCC58}\underline{1.3299} & \cellcolor[HTML]{BCBD22}\textbf{1.0078} \\

    \bottomrule
    \end{tabular}
\end{table}

\begin{table}[h!]
    \centering
    \scriptsize
    \caption{Feature focused metrics for capturing interface (IRMSE) and shocks (SRMSE) on the 3D-SDBA (SSOOSS) test dataset. Lower values are better. Darker shade indicates lower value for each baseline within a group of training loss strategies. Bold values denote the lowest among all baselines for a given metric and the underlined values denote the second-lowest values.}
    \label{tab:3d_sdba_feature_focused}
    \begin{tabular}{llcc}
    \toprule
    Model & Training loss & IRMSE & SRMSE \\
    \midrule

    \multirow{3}{*}{\makecell[l]{\texttt{ConvNeXt}\\\texttt{(10M)}}}
    & \texttt{MSE loss}
    & \cellcolor[HTML]{DBDB8D}\texttt{>10} & \cellcolor[HTML]{BCBD22}2.0698 \\
    & \texttt{Comp.loss (SA)}
    & \cellcolor[HTML]{DBDB8D}\texttt{>10} & \cellcolor[HTML]{CCCC58}3.8508 \\
    & \texttt{Comp.loss (GN)}
    & \cellcolor[HTML]{DBDB8D}\texttt{>10} & \cellcolor[HTML]{DBDB8D}\texttt{>10} \\
    \addlinespace

    \multirow{3}{*}{\makecell[l]{\texttt{FFNO}\\\texttt{(10M)}}}
    & \texttt{MSE loss}
    & \cellcolor[HTML]{DBDB8D}2.0507 & \cellcolor[HTML]{CCCC58}0.5131 \\
    & \texttt{Comp.loss (SA)}
    & \cellcolor[HTML]{BCBD22}\textbf{1.3402} & \cellcolor[HTML]{BCBD22}\textbf{0.4788} \\
    & \texttt{Comp.loss (GN)}
    & \cellcolor[HTML]{CCCC58}\underline{1.5661} & \cellcolor[HTML]{CCCC58}\underline{0.5041} \\

    \bottomrule
    \end{tabular}
\end{table}

\begin{table}[h!]
    \centering
    \scriptsize
    \caption{Structure-aware metrics for the 3D-SDBA (SSOOSS) test dataset. Lower values are better. Darker shade indicates lower value for each baseline within a group of training loss strategies. Bold values denote the lowest among all baselines for a given metric and the underlined values denote the second-lowest values.}
    \label{tab:3d_sdba_structure_aware}
    \begin{tabular}{llcc}
    \toprule
    Model & Training loss & PCC & SSIM \\
    \midrule

    \multirow{3}{*}{\makecell[l]{\texttt{ConvNeXt}\\\texttt{(10M)}}}
    & \texttt{MSE loss}
    & \cellcolor[HTML]{DBDB8D}0.3439 & \cellcolor[HTML]{DBDB8D}0.5562 \\
    & \texttt{Comp.loss (SA)}
    & \cellcolor[HTML]{BCBD22}0.2486 & \cellcolor[HTML]{BCBD22}\underline{0.3653} \\
    & \texttt{Comp.loss (GN)}
    & \cellcolor[HTML]{CCCC58}0.3124 & \cellcolor[HTML]{CCCC58}0.4845 \\
    \addlinespace

    \multirow{3}{*}{\makecell[l]{\texttt{FFNO}\\\texttt{(10M)}}}
    & \texttt{MSE loss}
    & \cellcolor[HTML]{CCCC58}\underline{0.1661} & \cellcolor[HTML]{CCCC58}0.3790 \\
    & \texttt{Comp.loss (SA)}
    & \cellcolor[HTML]{BCBD22}\textbf{0.1612} & \cellcolor[HTML]{BCBD22}\textbf{0.3620} \\
    & \texttt{Comp.loss (GN)}
    & \cellcolor[HTML]{DBDB8D}0.1691 & \cellcolor[HTML]{DBDB8D}0.3794 \\

    \bottomrule
    \end{tabular}
\end{table}

\begin{table}[h!]
    \centering
    \scriptsize
    \setlength{\tabcolsep}{4pt}
    \caption{Physics-informed metrics reporting the nRMSE of integral quantities of interest on the 3D-SDBA (SSOOSS) test dataset. Lower values are better. Darker shade indicates lower value for each baseline within a group of training loss strategies. Bold values denote the lowest among all baselines for a given metric and the underlined values denote the second-lowest values.}
    \label{tab:3d_sdba_physics_informed}
    \begin{tabular}{@{}p{2.2cm}p{2.3cm}cccccc@{}}
    \toprule
    Model & Training loss & iqRMSE- & iqRMSE- & iqRMSE- & iqRMSE- & iqRMSE- & iqRMSE- \\
          &               & mass & momentum & momentum & com\_x & com\_y & enstrophy \\
          &               &  & \_x & \_y &  &  \\
    \midrule

    \multirow{3}{*}{\makecell[l]{\texttt{ConvNeXt}\\\texttt{(10M)}}}
    & \texttt{MSE loss}
    & \cellcolor[HTML]{DBDB8D}\texttt{>10} & \cellcolor[HTML]{DBDB8D}\texttt{>10} & \cellcolor[HTML]{DBDB8D}\texttt{>10} & \cellcolor[HTML]{DBDB8D}\texttt{>10} & \cellcolor[HTML]{DBDB8D}\texttt{>10} & \cellcolor[HTML]{CCCC58}0.9275 \\
    & \texttt{Comp.loss (SA)}
    & \cellcolor[HTML]{DBDB8D}\texttt{>10} & \cellcolor[HTML]{DBDB8D}\texttt{>10} & \cellcolor[HTML]{DBDB8D}\texttt{>10} & \cellcolor[HTML]{DBDB8D}\texttt{>10} & \cellcolor[HTML]{DBDB8D}\texttt{>10} & \cellcolor[HTML]{BCBD22}\underline{0.8792} \\
    & \texttt{Comp.loss (GN)}
    & \cellcolor[HTML]{DBDB8D}\texttt{>10} & \cellcolor[HTML]{DBDB8D}\texttt{>10} & \cellcolor[HTML]{DBDB8D}\texttt{>10} & \cellcolor[HTML]{DBDB8D}\texttt{>10} & \cellcolor[HTML]{DBDB8D}\texttt{>10} & \cellcolor[HTML]{DBDB8D}1.1215 \\
    \addlinespace

    \multirow{3}{*}{\makecell[l]{\texttt{FFNO}\\\texttt{(10M)}}}
    & \texttt{MSE loss}
    & \cellcolor[HTML]{DBDB8D}\texttt{>10} & \cellcolor[HTML]{DBDB8D}\texttt{>10} & \cellcolor[HTML]{DBDB8D}\texttt{>10} & \cellcolor[HTML]{DBDB8D}0.0985 & \cellcolor[HTML]{DBDB8D}0.0828 & \cellcolor[HTML]{BCBD22}\textbf{0.2710} \\
    & \texttt{Comp.loss (SA)}
    & \cellcolor[HTML]{DBDB8D}\texttt{>10} & \cellcolor[HTML]{DBDB8D}\texttt{>10} & \cellcolor[HTML]{DBDB8D}\texttt{>10} & \cellcolor[HTML]{CCCC58}\underline{0.0522} & \cellcolor[HTML]{CCCC58}\underline{0.0358} & \cellcolor[HTML]{CCCC58}\underline{0.3176} \\
\bottomrule
    \end{tabular}
\end{table}

\clearpage

\subsection{Generalization ablations}
\label{section:generalization ablations}

\begin{figure}[h!]
    \centering
    \includegraphics[width=0.95\textwidth]{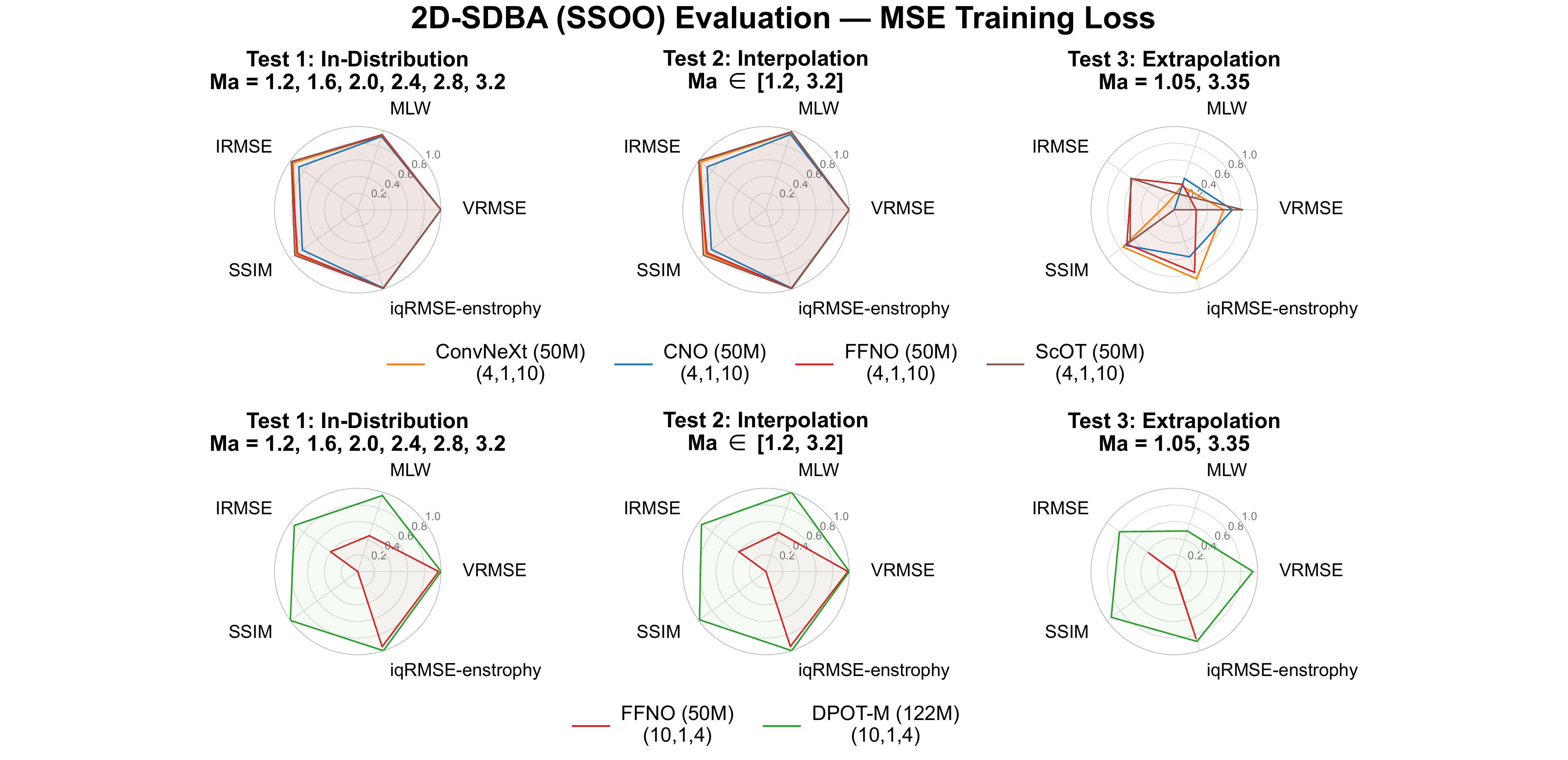}
    \caption{Radar plot showcasing performance of MSE-trained baselines across six different test datasets on five inference metrics: VRMSE, MLW, IRMSE, SSIM, and iqRMSE-enstrophy.}
    \label{fig:radar_chart_mse_training_ood_2d_sdba_ssoo}
\end{figure}

\begin{figure}[h!]
    \centering
    \includegraphics[width=0.95\textwidth]{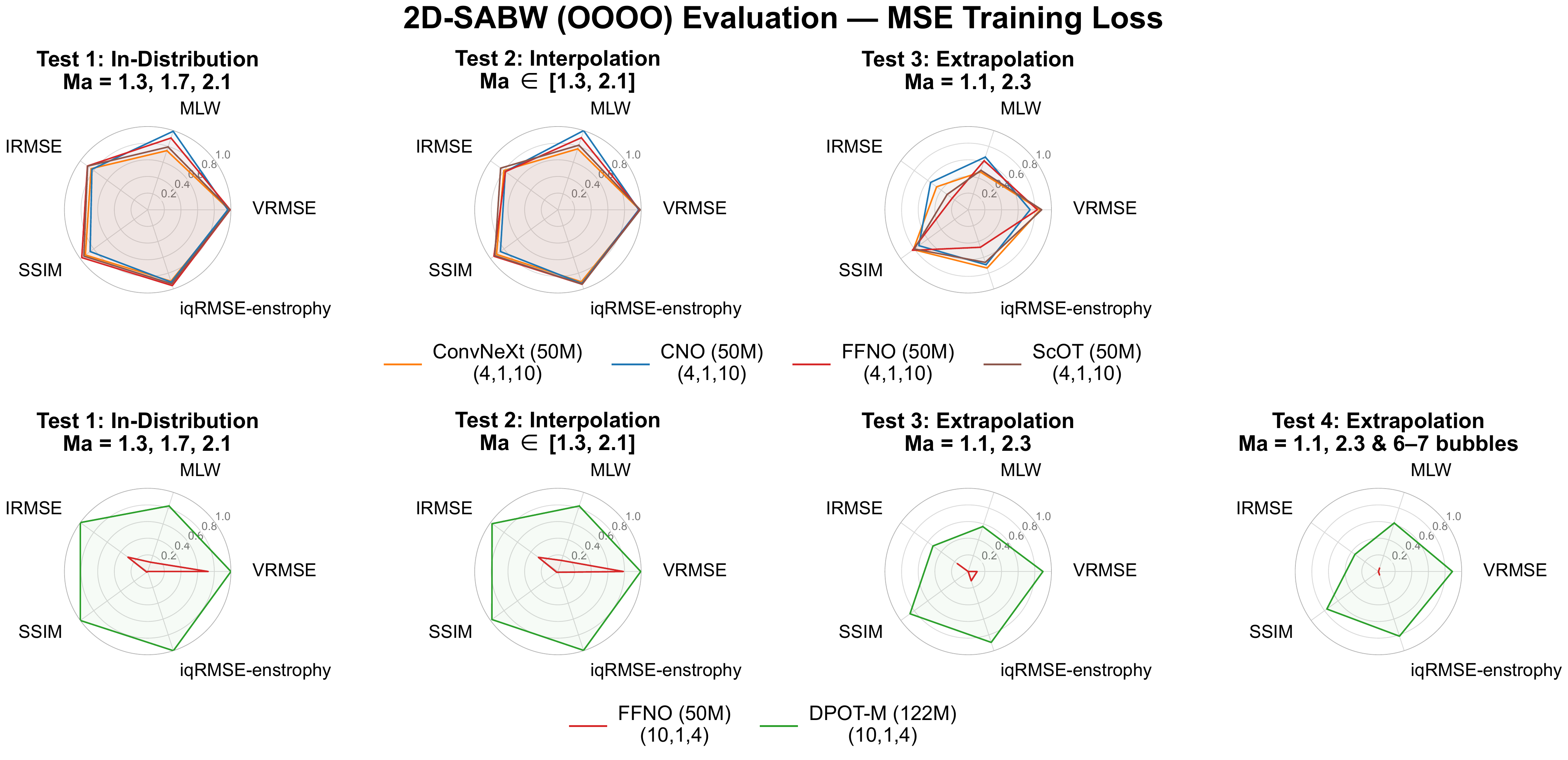}
    \caption{Radar plot showcasing performance of MSE-trained baselines across six different test datasets on five inference metrics: VRMSE, MLW, IRMSE, SSIM, and iqRMSE-enstrophy.}
    \label{fig:radar_chart_mse_training_ood_2d_sabw_oooo}
\end{figure}

We evaluate the generalization capabilities of the baseline models across multiple scenarios, focusing on Mach number interpolation between discrete training values and extrapolation beyond the training range. These experiments are conducted on the 2D Shock-induced Droplet Breakup with reflecting walls (2D-SDBA-SSOO) dataset, as shown in Figure \ref{fig:radar_chart_mse_training_ood_2d_sdba_ssoo}, and the 2D Shock-induced Bubble Collapse in Water with open boundaries (2D-SABW-OOOO) dataset, as illustrated in Figure \ref{fig:radar_chart_mse_training_ood_2d_sabw_oooo}.

The results demonstrate that the baseline models are highly robust to Mach-number interpolation, maintaining consistent performance relative to in-distribution test evaluations. However, their predictive accuracy deteriorates substantially when extrapolating beyond the training range. This highlights the challenges of generalizing to unseen flow regimes, in which the strongly nonlinear dynamics of shock-driven multiphase flows pose significant challenges for models. Similar results were obtained for models trained using the SoftAdapt and GradNorm strategies; therefore, these results are omitted for brevity.

\clearpage

\subsection{Rollout visualization for in-distribution test datasets}
\label{section:app_rollout_viz}
\begin{figure}[h!]
    \centering
    \includegraphics[angle=-90,width=0.55\textheight,keepaspectratio]{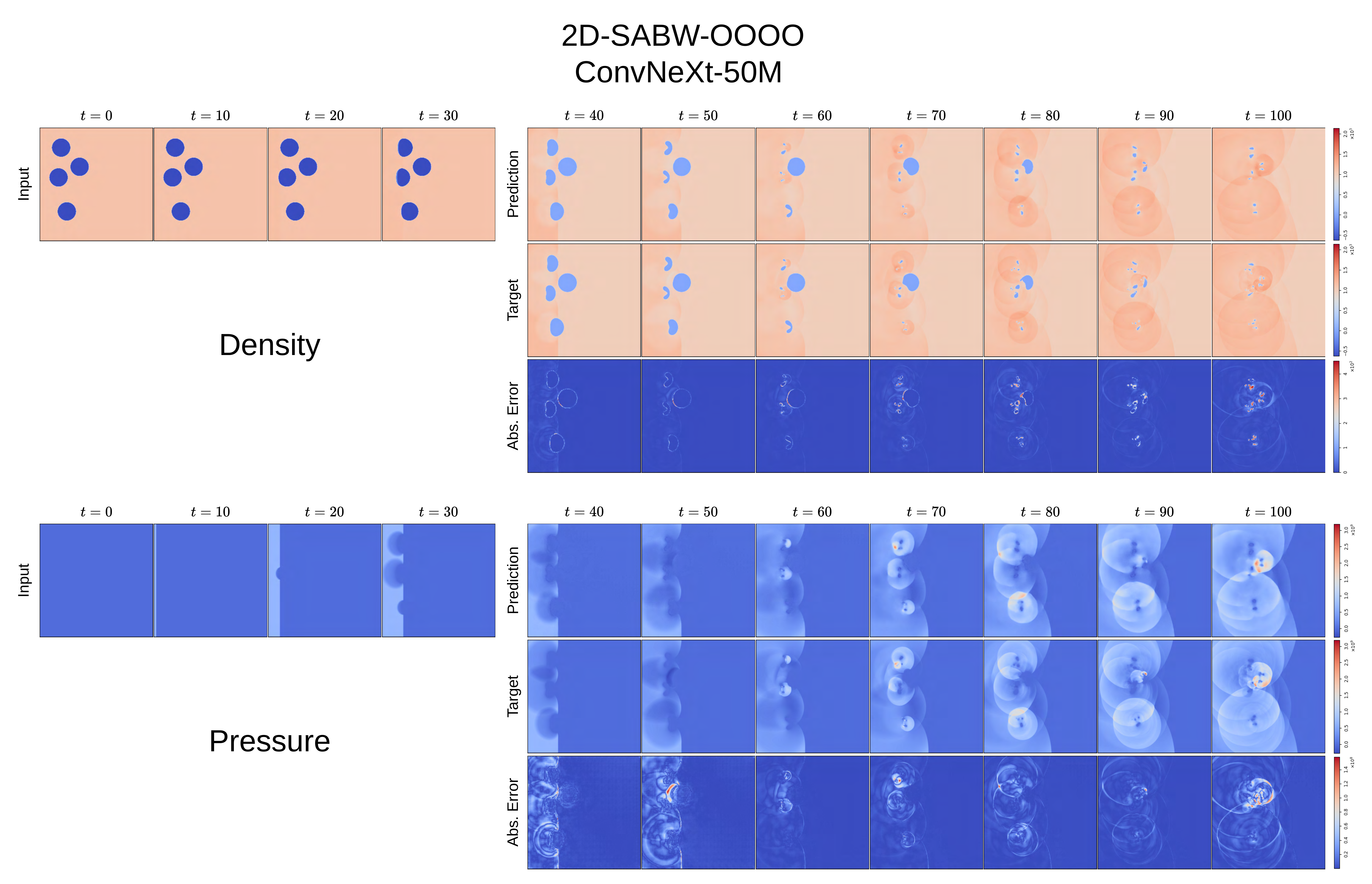}
    \caption{Rollout visualization of Density and Pressure fields for 2D Shock-induced Air Bubble Collapse in Water with Open Boundaries (2D-SABW-OOOO) using ConvNeXt-50M. }
    \label{fig:sb_abs_convnext_sa_1}
\end{figure}

\begin{figure}[h!]
    \centering
    \includegraphics[angle=-90,width=0.6\textheight,keepaspectratio]{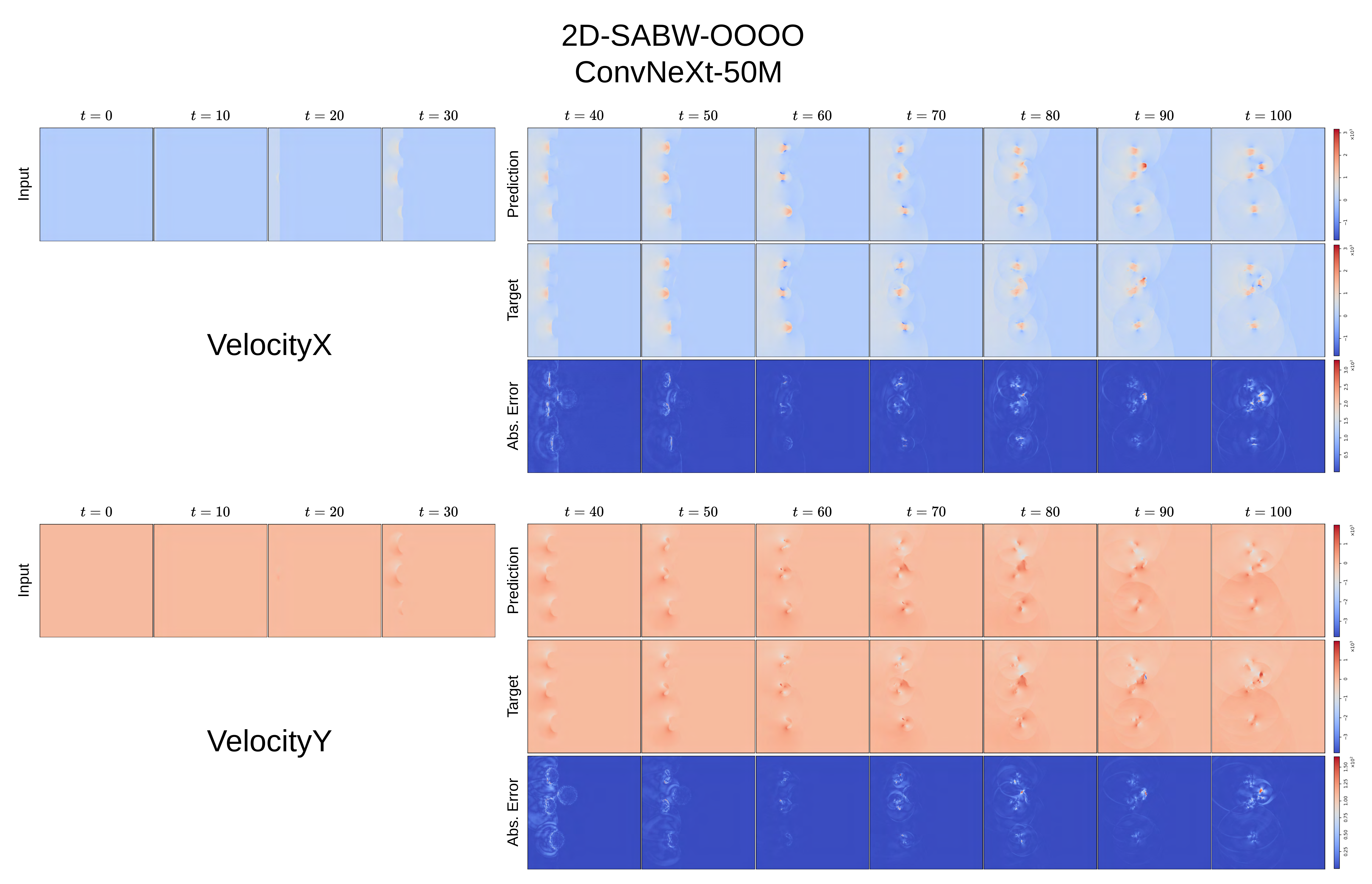}
    \caption{Rollout visualization of Velocity-X and Velocity-Y fields for 2D Shock-induced Air Bubble Collapse in Water with Open Boundaries (2D-SABW-OOOO) using ConvNeXt-50M. }
    \label{fig:sb_abs_convnext_sa_2}
\end{figure}

\begin{figure}[h!]
    \centering
    \includegraphics[angle=-90,width=0.55\textheight,keepaspectratio]{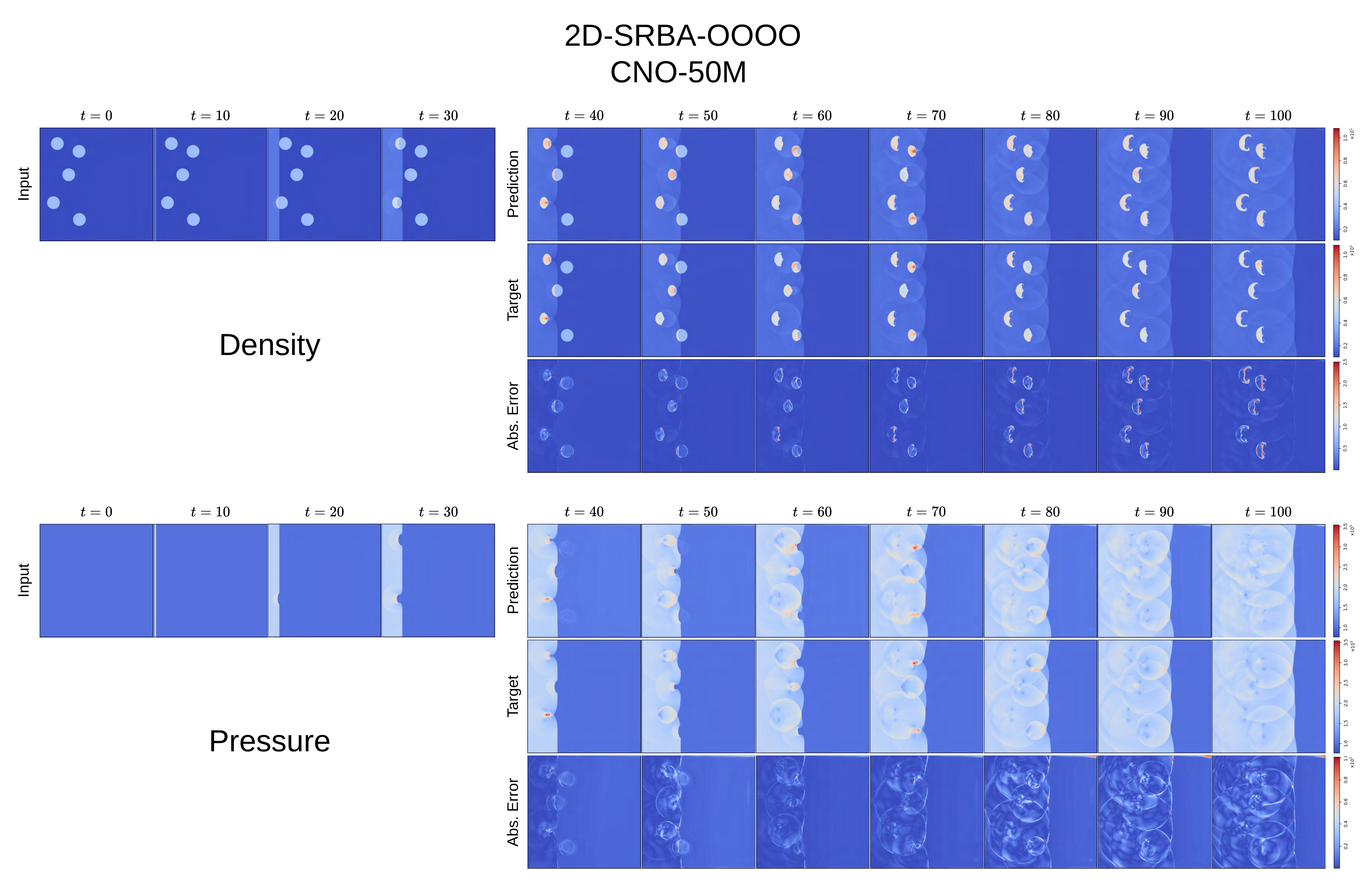}
    \caption{Rollout visualization of Density and Pressure fields for 2D Shock-induced R22 Bubble Collapse in Air with Open Boundaries (2D-SRBA-OOOO) using CNO-50M. }
    \label{fig:r22_abs_cno_sa_1}
\end{figure}

\begin{figure}[h!]
    \centering
    \includegraphics[angle=-90,width=0.6\textheight,keepaspectratio]{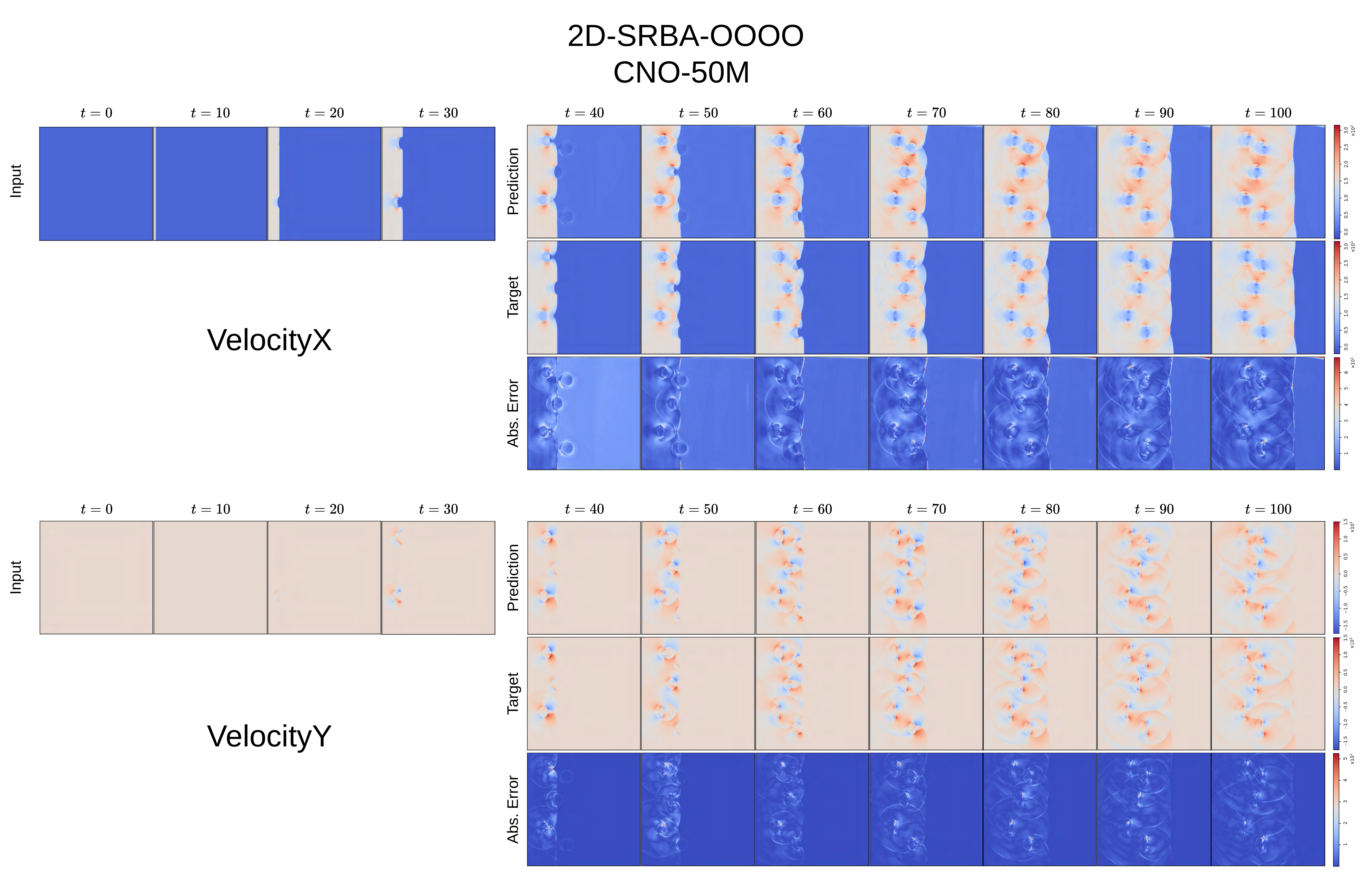}
    \caption{Rollout visualization of Velocity-X and Velocity-Y fields for 2D Shock-induced R22 Bubble Collapse in Air with Open Boundaries (2D-SRBA-OOOO) using CNO-50M. }
    \label{fig:r22_abs_cno_sa_2}
\end{figure}

\begin{figure}[h!]
    \centering
    \includegraphics[angle=-90,width=0.55\textheight,keepaspectratio]{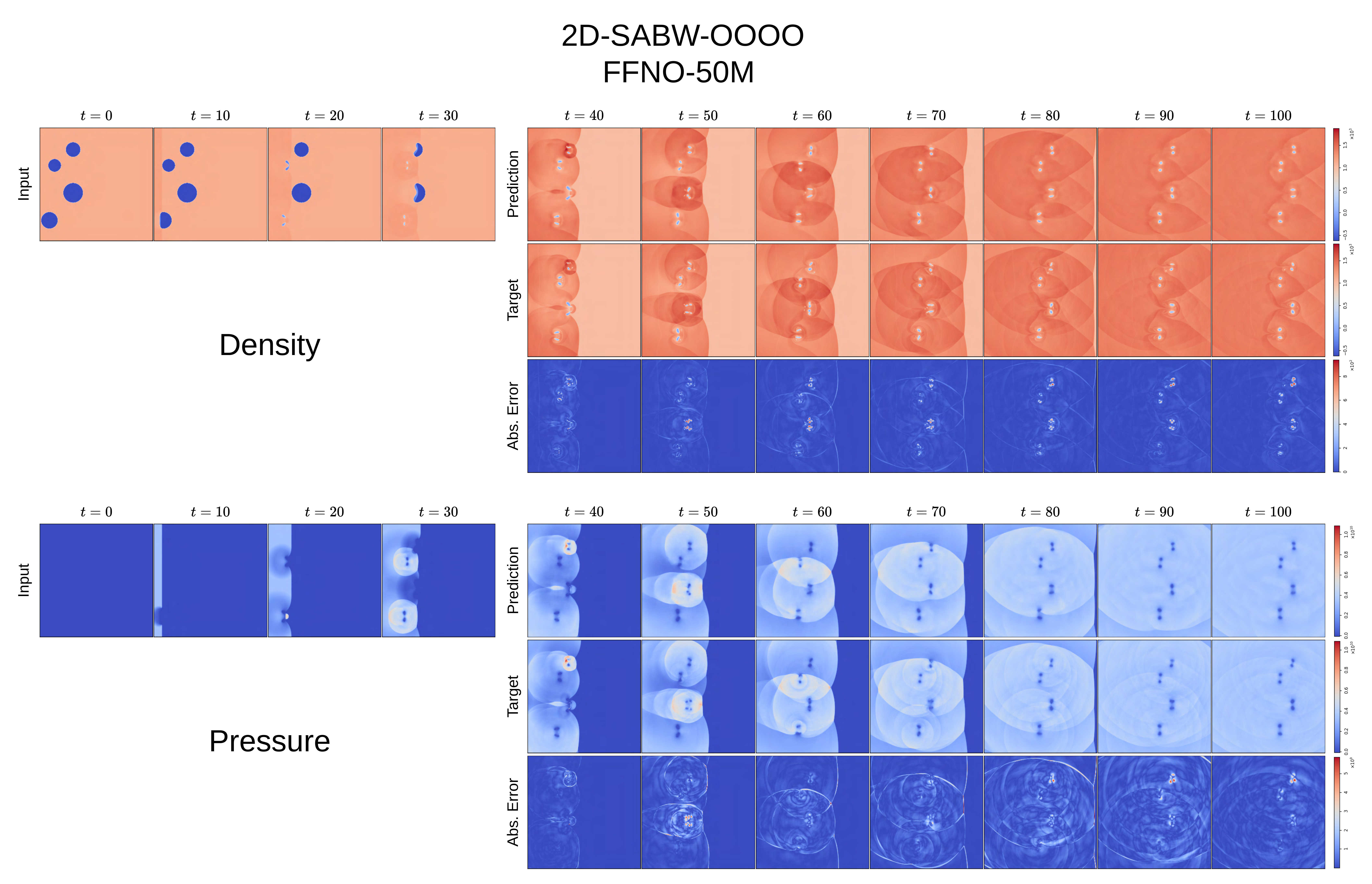}
    \caption{Rollout visualization of Density and Pressure fields for 2D Shock-induced Air Bubble Collapse in Water with Open Boundaries (2D-SABW-OOOO) using FFNO-50M. }
    \label{fig:sb_abs_ffno_sa_1}
\end{figure}

\begin{figure}[h!]
    \centering
    \includegraphics[angle=-90,width=0.6\textheight,keepaspectratio]{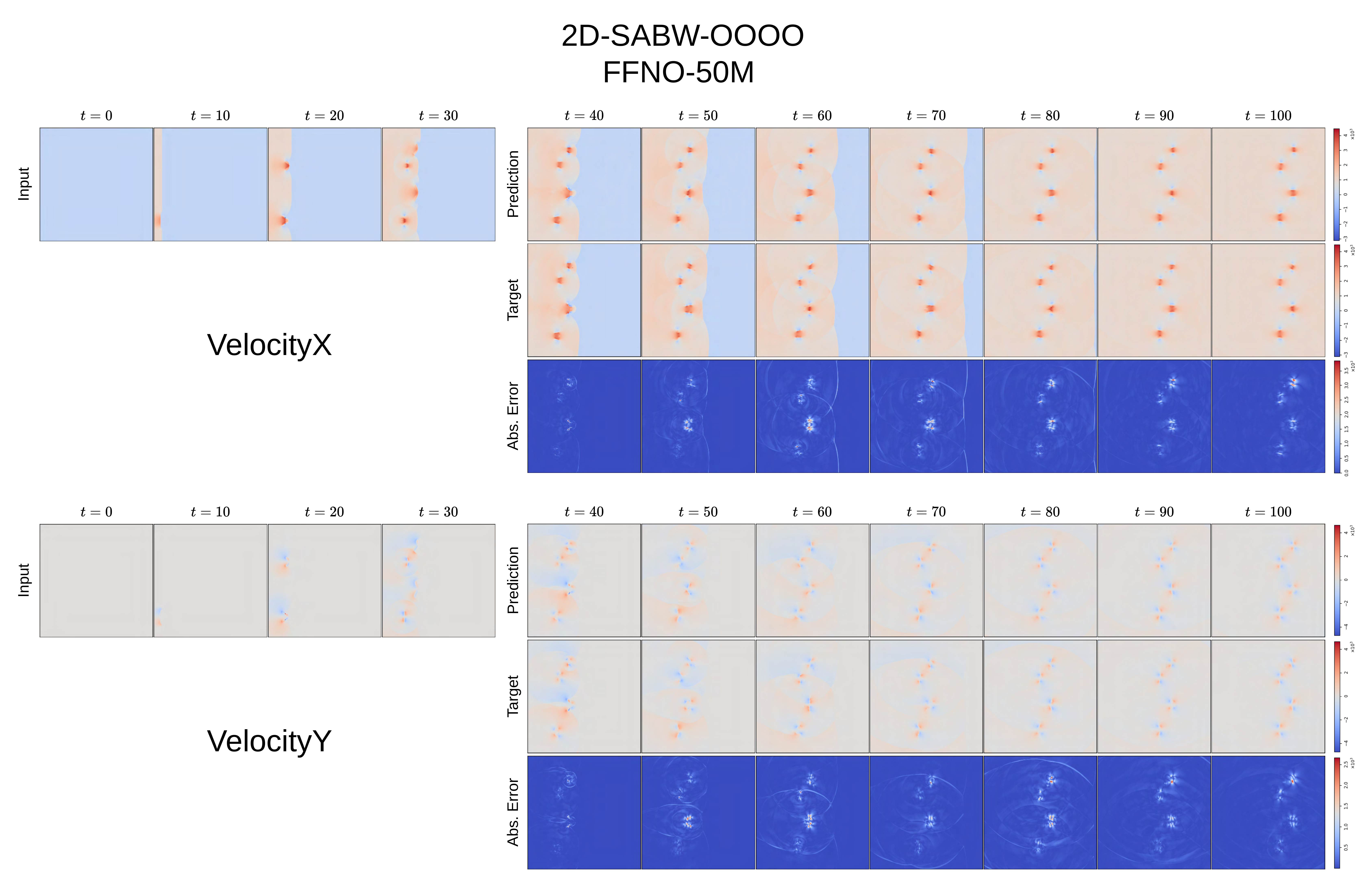}
    \caption{Rollout visualization of Velocity-X and Velocity-Y fields for 2D Shock-induced Air Bubble Collapse in Water with Open Boundaries (2D-SABW-OOOO) using FFNO-50M. }
    \label{fig:sb_abs_ffno_sa_2}
\end{figure}

\begin{figure}[h!]
    \centering
    \includegraphics[angle=-90,width=0.55\textheight,keepaspectratio]{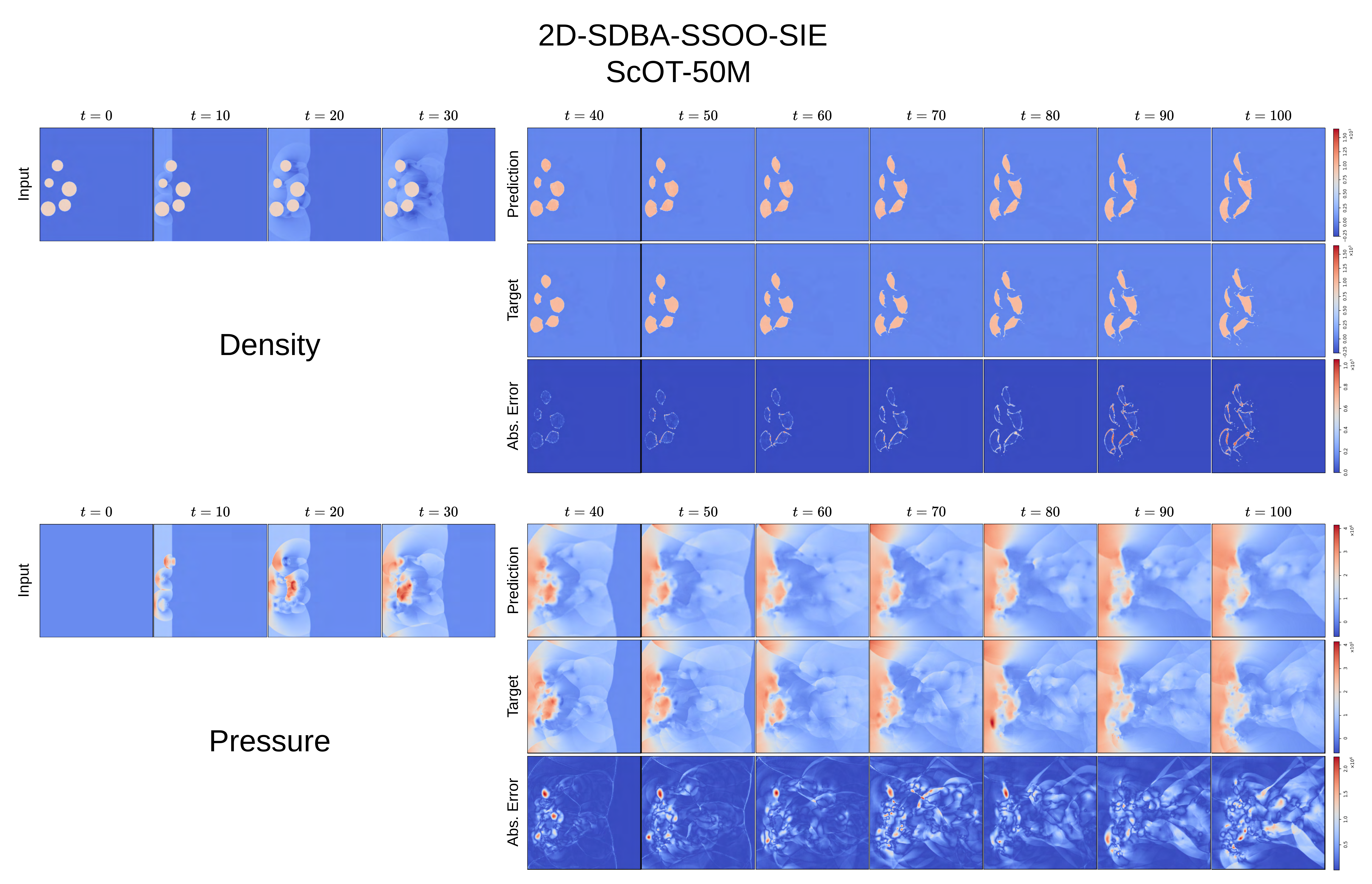}
    \caption{Rollout visualization of Density and Pressure fields for 2D Shock-induced Droplet Breakup in SIE regime with Symmetric Boundaries (2D-SDBA-SSOO) using ScOT-50M. }
    \label{fig:sd_sie_scot_sa_1}
\end{figure}

\begin{figure}[h!]
    \centering
    \includegraphics[angle=-90,width=0.6\textheight,keepaspectratio]{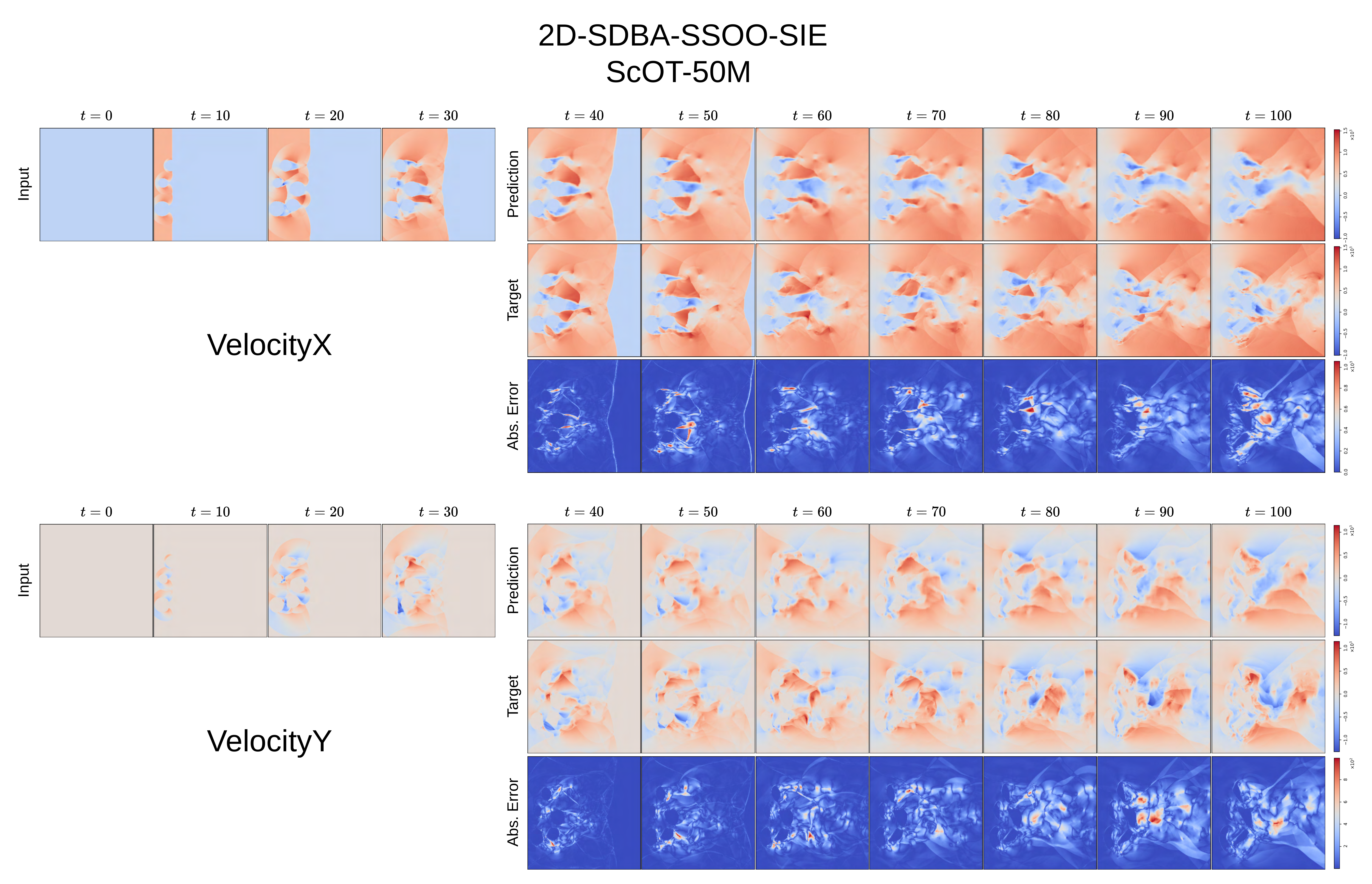}
    \caption{Rollout visualization of Velocity-X and Velocity-Y fields for 2D Shock-induced Droplet Breakup in SIE regime with Symmetric Boundaries (2D-SDBA-SSOO) using ScOT-50M. }
    \label{fig:sd_sie_scot_sa_2}
\end{figure}

\begin{figure}[h!]
    \centering
    \includegraphics[angle=-90,width=0.55\textheight,keepaspectratio]{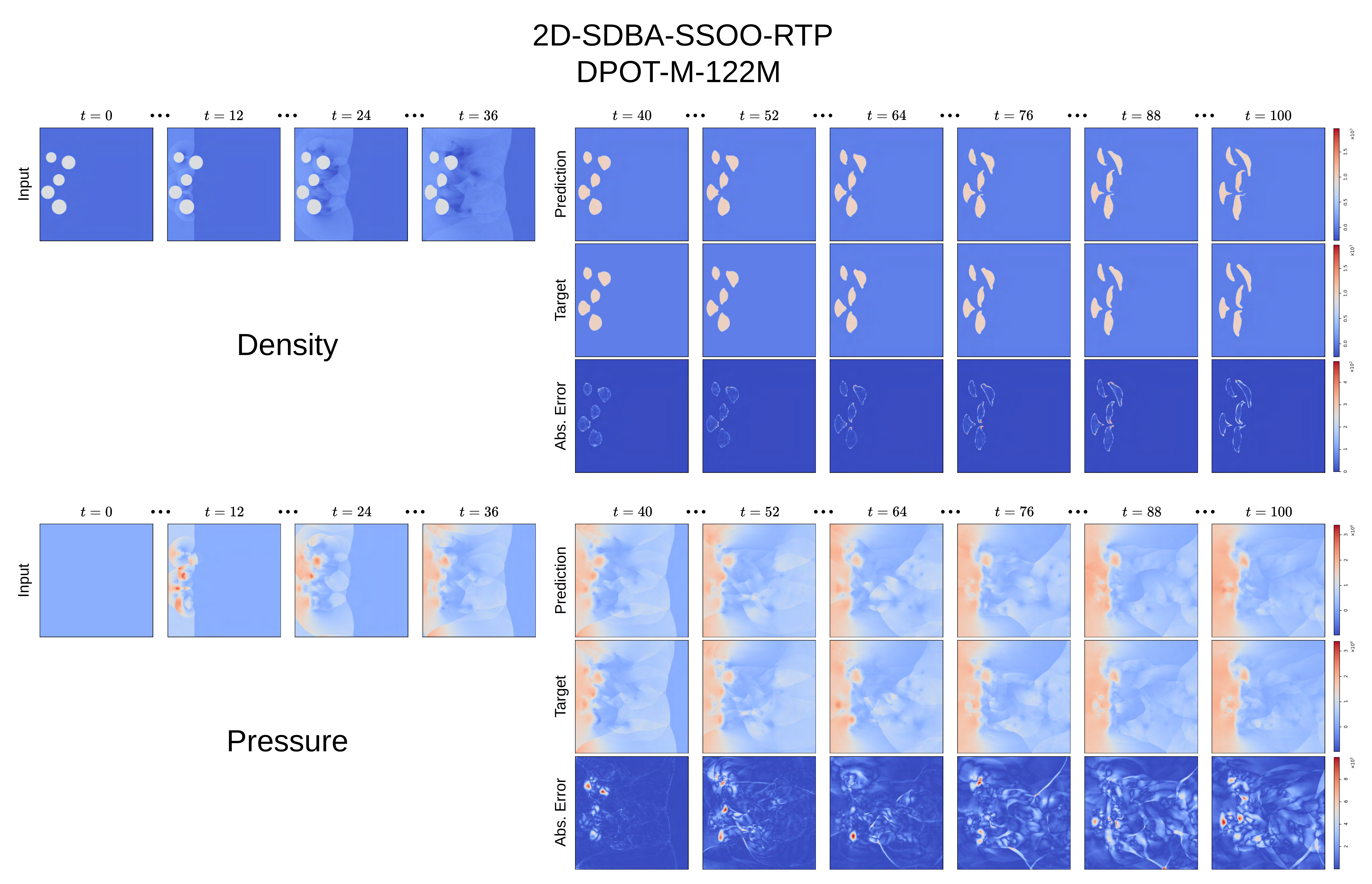}
    \caption{Rollout visualization of Density and Pressure fields for 2D Shock-induced Droplet Breakup in RTP regime with Symmetric Boundaries (2D-SDBA-SSOO) using DPOT-M-122M. }
    \label{fig:sd_rtp_dpot_sa_1}
\end{figure}

\begin{figure}[h!]
    \centering
    \includegraphics[angle=-90,width=0.6\textheight,keepaspectratio]{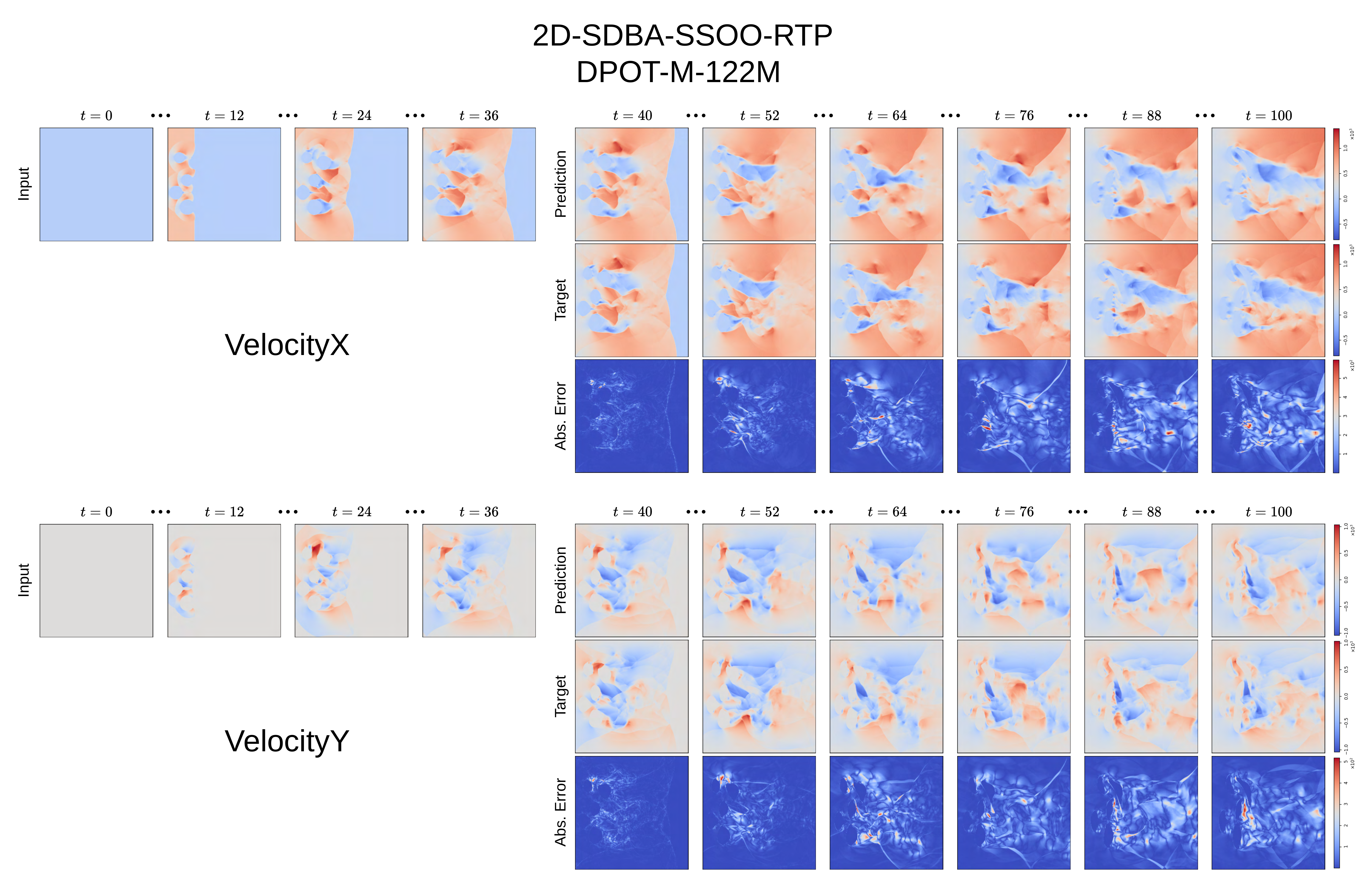}
    \caption{Rollout visualization of Velocity-X and Velocity-Y fields for 2D Shock-induced Droplet Breakup in RTP regime with Symmetric Boundaries (2D-SDBA-SSOO) using DPOT-M-122M. }
    \label{fig:sd_rtp_dpot_sa_2}
\end{figure}

\begin{figure}[h!]
    \centering
    \includegraphics[angle=-90,width=0.6\textheight,keepaspectratio]{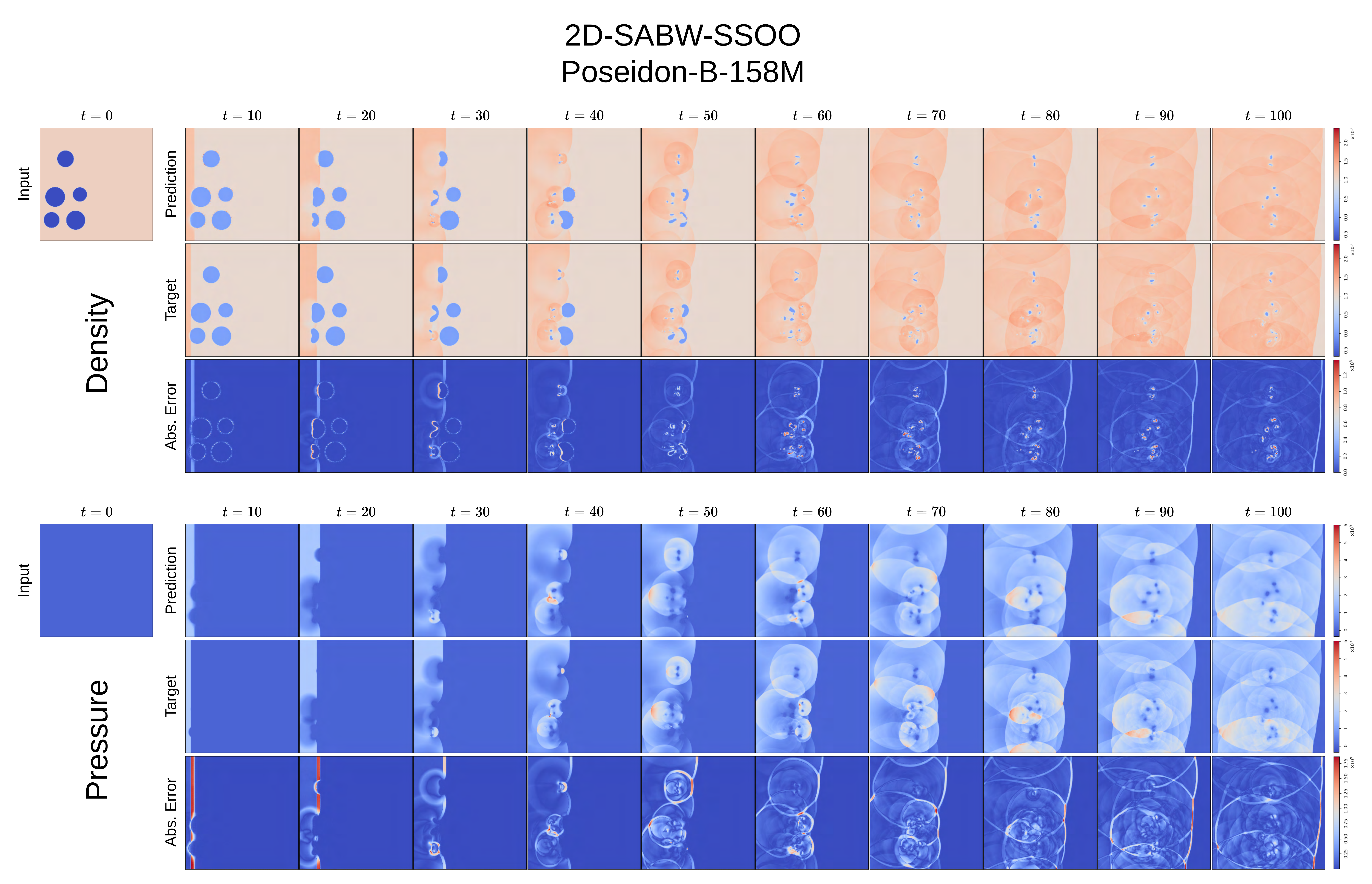}
    \caption{Rollout visualization of Density and Pressure fields for 2D Shock-induced Air Bubble Collapse in Water with Symmetric Boundaries (2D-SABW-SSOO) using Poseidon-B-158M. }
    \label{fig:sb_ref_poseidon_sa_1}
\end{figure}

\begin{figure}[h!]
    \centering
    \includegraphics[angle=-90,width=0.6\textheight,keepaspectratio]{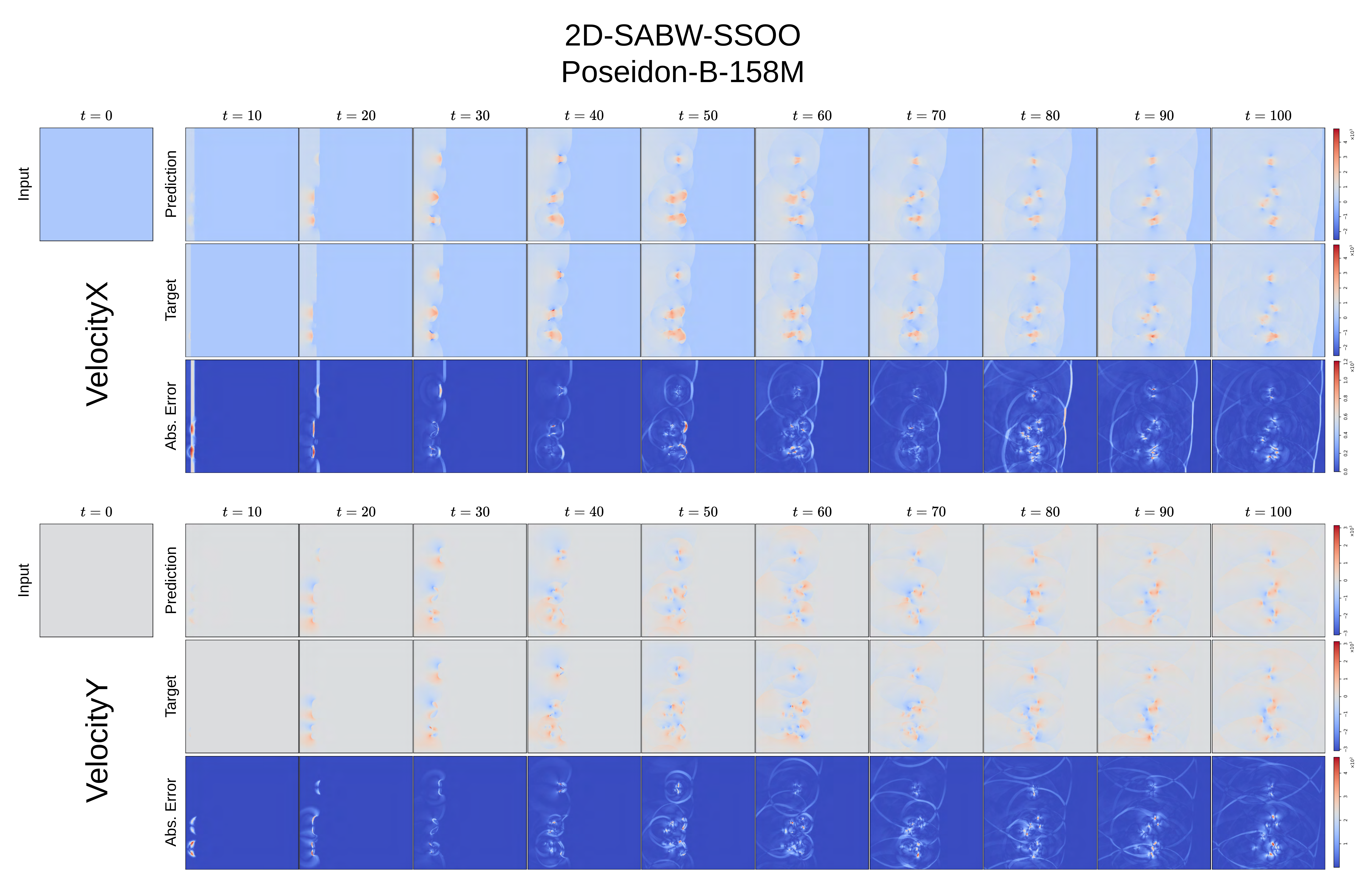}
    \caption{Rollout visualization of Velocity-X and Velocity-Y fields for 2D Shock-induced Air Bubble Collapse in Water with Symmetric Boundaries (2D-SABW-SSOO) using Poseidon-B-158M. }
    \label{fig:sb_ref_poseidon_sa_2}
\end{figure}


\newpage
\clearpage
\section*{NeurIPS Paper Checklist}

\begin{enumerate}

\item {\bf Claims}
    \item[] Question: Do the main claims made in the abstract and introduction accurately reflect the paper's contributions and scope?
    \item[] Answer: \answerYes{} 
    \item[] Justification: The claims made in the abstract are covered in sections \ref{section:main_datasets} and \ref{section:results_and_conclusion} of the main text. Further information are detailed in the appendix section \ref{section:app_extended_results}. 

\item {\bf Limitations}
    \item[] Question: Does the paper discuss the limitations of the work performed by the authors?
    \item[] Answer: \answerYes{} 
    \item[] Justification: We have provided the limitations clearly in the section \ref{section:conclustion} of the main text.

\item {\bf Theory assumptions and proofs}
    \item[] Question: For each theoretical result, does the paper provide the full set of assumptions and a complete (and correct) proof?
    \item[] Answer: \answerNA{} 
    \item[] Justification: The paper does not include theoretical results, but rather focuses on the empirical benchmarking of surrogate models for shock-induced multiphase flows. 

    \item {\bf Experimental result reproducibility}
    \item[] Question: Does the paper fully disclose all the information needed to reproduce the main experimental results of the paper to the extent that it affects the main claims and/or conclusions of the paper (regardless of whether the code and data are provided or not)?
    \item[] Answer: \answerYes{} 
    \item[] Justification: For generating the datasets, we provide all the details related to the physics and the solver in the metadata.json file. For reproducing the benchmarks we provide the model hyperparameters in section \ref{section:app_baseline_models}, training hyperparameters in section \ref{section:app_training_protocol} and hyperparameters related to metrics and loss-weighting strategies in section \ref{section:app_metrics}. Further the hardware specifications on which the training was carried out is specified in section \ref{section:app_computational_resources}. 

\item {\bf Open access to data and code}
    \item[] Question: Does the paper provide open access to the data and code, with sufficient instructions to faithfully reproduce the main experimental results, as described in supplemental material?
    \item[] Answer: \answerYes{} 
    \item[] Justification: The link to the open-sourced (anonymized) code and datasets is provided in the footnote of the first page of the main text. The README.md of the benchmarking repository contains all the necessary instructions to set up a virtual environment, train the baselines and generate metrics from the trained models during inference.

\item {\bf Experimental setting/details}
    \item[] Question: Does the paper specify all the training and test details (e.g., data splits, hyperparameters, how they were chosen, type of optimizer) necessary to understand the results?
    \item[] Answer: \answerYes{} 
    \item[] Justification: The data-splits are provided in Figure \ref{fig:dataset_overview}, the hyperparameters for the baselines in section \ref{section:app_baseline_models}, training hyperparameters in section \ref{section:app_training_protocol} and hyperparameters related to metrics and loss-weighting strategies in section \ref{section:app_metrics}

\item {\bf Experiment statistical significance}
    \item[] Question: Does the paper report error bars suitably and correctly defined or other appropriate information about the statistical significance of the experiments?
    \item[] Answer: \answerYes{} 
    \item[] Justification: We provide error bars in our metric evolution plots for the 2D and 3D baselines in section \ref{section:app_rollout_metrics}. 

\item {\bf Experiments compute resources}
    \item[] Question: For each experiment, does the paper provide sufficient information on the computer resources (type of compute workers, memory, time of execution) needed to reproduce the experiments?
    \item[] Answer:\answerYes{}, 
    \item[] Justification: The wall clock time required to generate the datasets are specified in Table \ref{tab:dataset_metadata}. The hardware specifications for generating the datasets and the models is specified in section \ref{section:app_computational_resources}. The peak memory requirements and the wall clock time for a single training epoch is presented in \ref{section:app_metric_performance_comparison}.
    
\item {\bf Code of ethics}
    \item[] Question: Does the research conducted in the paper conform, in every respect, with the NeurIPS Code of Ethics \url{https://neurips.cc/public/EthicsGuidelines}?
    \item[] Answer: \answerYes{} 
    \item[] Justification: We reviewed the NeurIPS Code of Ethics, and we confirm that our research conforms to it in every respect.

\item {\bf Broader impacts}
    \item[] Question: Does the paper discuss both potential positive societal impacts and negative societal impacts of the work performed?
    \item[] Answer: \answerNA{} 
    \item[] Justification:  The datasets and trained models are restricted to simplified benchmark configurations and are not directly deployable in real-world decision-making. As such, no immediate societal risks or negative impacts are anticipated, and the broader societal impact of this work is expected to remain primarily within the research community.

\item {\bf Safeguards}
    \item[] Question: Does the paper describe safeguards that have been put in place for responsible release of data or models that have a high risk for misuse (e.g., pre-trained language models, image generators, or scraped datasets)?
    \item[] Answer: \answerNA{} 
    \item[] Justification: The datasets and the trained surrogates provided in this paper are intended for scientific benchmarking and methodological evaluation of surrogate models rather than deployment in safety-critical or real-world operational systems

\item {\bf Licenses for existing assets}
    \item[] Question: Are the creators or original owners of assets (e.g., code, data, models), used in the paper, properly credited and are the license and terms of use explicitly mentioned and properly respected?
    \item[] Answer: \answerYes{} 
    \item[] Justification: The solver used to generate the datasets is open sourced and is cited in the paper and a link to the repository- \url{https://gitlab.lrz.de/nanoshock/ALPACA}. The license of the solver is GNU General Public License v3.0. 

\item {\bf New assets}
    \item[] Question: Are new assets introduced in the paper well documented and is the documentation provided alongside the assets?
    \item[] Answer: \answerYes{} 
    \item[] Justification: The datasets and benchmarking pipeline were developed by the authors. The datasets are publicly hosted on Hugging Face, while the benchmarking repository is maintained on GitHub. All required licensing information is clearly specified in the README.md files. The baseline models used for benchmarking are adapted from existing open-source implementations, and appropriate credits and citations are provided within the benchmarking repository.

\item {\bf Crowdsourcing and research with human subjects}
    \item[] Question: For crowdsourcing experiments and research with human subjects, does the paper include the full text of instructions given to participants and screenshots, if applicable, as well as details about compensation (if any)? 
    \item[] Answer: \answerNA{} 
    \item[] Justification: The paper does not involve crowdsourcing nor research with human subjects.

\item {\bf Institutional review board (IRB) approvals or equivalent for research with human subjects}
    \item[] Question: Does the paper describe potential risks incurred by study participants, whether such risks were disclosed to the subjects, and whether Institutional Review Board (IRB) approvals (or an equivalent approval/review based on the requirements of your country or institution) were obtained?
    \item[] Answer: \answerNA{} 
    \item[] Justification: The paper does not involve crowdsourcing nor research with human subjects.

\item {\bf Declaration of LLM usage}
    \item[] Question: Does the paper describe the usage of LLMs if it is an important, original, or non-standard component of the core methods in this research? Note that if the LLM is used only for writing, editing, or formatting purposes and does \emph{not} impact the core methodology, scientific rigor, or originality of the research, declaration is not required.
    \item[] Answer: \answerNA{} 
    \item[] Justification: LLMs were used only for writing, editing, and formatting purposes and did not impact the core methodology, scientific rigor, or originality of the research.

\end{enumerate}

\end{document}